\documentclass[usenatbib]{mn2e}
\usepackage{natbib}
\usepackage{url}
\usepackage{rotating}
\usepackage{subcaption}
\usepackage{amsmath}
\usepackage{amsfonts}
\usepackage{amssymb}
\usepackage{graphicx}
\usepackage{float}
\usepackage{aas_macros}
\usepackage{color}
\usepackage{ mathrsfs }

\def\gsim{\;\rlap{\lower 2.5pt
 \hbox{$\sim$}}\raise 1.5pt\hbox{$>$}\;}
\def\lsim{\;\rlap{\lower 2.5pt
   \hbox{$\sim$}}\raise 1.5pt\hbox{$<$}\;}

\usepackage{hyperref}
\hypersetup{
    pdfnewwindow=true,      % links in new window
   colorlinks=true,       % false: boxed links; true: colored links
    linkcolor=blue,          % color of internal links
    citecolor=blue,        % color of links to bibliography
    filecolor=blue,      % color of file links
    urlcolor=blue           % color of external links
}

\begin{document}

\title[Periodic Quasars in PTF]{A Population of Short-Period Variable Quasars from PTF as Supermassive Black Hole Binary Candidates}

\author[M. Charisi et al.]{M.~Charisi,$^{1}$\thanks{mc3561@columbia.edu} I.~Bartos,$^{2}$ Z.~Haiman,$^{1}$ A.~M.~Price-Whelan,$^{1}$  M.~J.~Graham,$^{3}$
 \newauthor E.~C.~Bellm,$^4$ R.~R.~Laher,$^5$ S.~M\'arka$^{2}$\\
$^1$Department of Astronomy, Columbia University, New York, NY 10027, USA \\
$^2$Department of Physics, Columbia University, New York, NY 10027, USA\\
$^3$Center for Data-Driven Discovery, California Institute of Technology, Pasadena, CA 91125, USA\\
$^4$Department of Astronomy, California Institute of Technology, Pasadena, CA 91125, USA\\
$^5$Spitzer Science Center, California Institute of Technology, Pasadena, CA 91125, USA}

\maketitle

\begin{abstract}
Supermassive black hole binaries (SMBHBs) at sub-parsec separations should be common in galactic nuclei, as a result of frequent galaxy mergers. 
Hydrodynamical simulations of circumbinary discs predict strong periodic modulation of the mass accretion rate on time-scales comparable to the orbital period of the binary. As a result, SMBHBs may be recognized by the periodic modulation of their brightness.
We conducted a statistical search for periodic variability in a sample of 35,383 spectroscopically confirmed quasars in the photometric database of the Palomar Transient Factory (PTF).
We analysed Lomb-Scargle periodograms and assessed the significance of our findings by modeling each individual quasar's variability as a damped random walk (DRW). 
We identified 50 quasars with significant periodicity beyond the DRW model, typically with short periods of a few hundred days.
We find 33 of these to remain significant after a re-analysis of their periodograms including additional optical data from the intermediate-PTF and the Catalina Real-Time Transient Survey (CRTS).
Assuming that the observed periods correspond to the redshifted orbital periods of SMBHBs, we conclude that our findings are consistent with a population of unequal-mass SMBHBs, with a typical mass ratio as low as $q\equiv M_2/M_1\approx0.01$.

\end{abstract}


\section{Introduction}
\label{section:Introduction}
Strong observational evidence suggests that every massive galaxy hosts a supermassive black hole in its nucleus \citep{2013ARA&A..51..511K}. The central black hole (BH) is an important component of the galaxy, since the BH mass is correlated with the global properties of the host galaxy, e.g., dispersion velocity, bulge luminosity, or bulge mass. Moreover, hierarchical models of structure formation predict that galaxies merge frequently \citep{2002MNRAS.336L..61H}, which naturally leads to the the formation of Supermassive black hole binaries (SMBHBs).

Following the merger, the BHs rapidly sink towards the centre of the common gravitational potential, under the effect of dynamical friction, and form a bound Keplerian binary. Subsequently, the binary orbit decays, as the BHs expel nearby stars in close three-body interactions, and/or as they interact with a gaseous circumbinary disc. At close separations, the binary is driven to coalescence by the emission of gravitational radiation \citep{1980Natur.287..307B}. Eventually, the emerging BH recoils \citep{2012AdAst2012E..14K}
to counterbalance the net linear momentum transferred by the gravitational waves (GWs). The strong gravitational radiation, emitted during the final stages of the inspiral, may be detectable by Pulsar Timing Arrays (North American Nanohertz Observatory for Gravitational waves; \citealt{2009arXiv0909.1058J}, Parkes Pulsar Timing Array; \citealt{2008AIPC..983..584M}, European Pulsar Timing Array; \citealt{2008AIPC..983..633J}) or future space-based interferometers, such as eLISA \citep{2013GWN.....6....4A}.

Whether three-body interactions can extract adequate energy to bring the binary to the GW regime is still unclear and it has been suggested that the binary may stall at parsec separations (\emph{final parsec problem}; see \citealt{2013ApJ...773..100K} or \citealt{2014SSRv..183..189C}, for a recent review).  
However, if cold gas exists in the central regions, the final parsec problem may be circumvented. This is motivated by simulations of gas-rich mergers, which indicate that copious amounts of gas are funneled into the nuclear regions \citep{1992ARA&A..30..705B}. The gas, following the formation of a Keplerian binary, settles in a rotationally supported disc \citep{2002MNRAS.333..481B}, and the binary hardens, as it dissipates energy and angular momentum to the gaseous disc. The exchange of angular momentum at this stage is slow and the binary is expected to spend a significant fraction of the fiducial $10^7-10^8$ yr lifetime at sub-parsec separations (e.g., \citealt{2009ApJ...700.1952H}; see also \citealt{2012MNRAS.427.2680K,2012MNRAS.427.2660K,2013ApJ...774..144R,2016arXiv160205206R} for the long-term evolution of a system of a SMBHB with a circumbinary disc).

%%%%%%%%%%%%%%%%%%%%%%%%%%%%%%%%%%%%%

Given the central role mergers play in galaxy formation and evolution and the uncertainties in binary evolution, identifying SMBHBs is of major importance. It would allow us to observationally constrain several key questions regarding the galaxy merger rate, as well as the processes and the environments involved in the path of SMBHBs to coalescence. It would also provide an estimate of the population of SMBHBs emitting GWs. Other important questions include the connection of SMBHBs with increased AGN activity \citep{1985Natur.315..386G} or triggering starburst activity \citep{1996ApJ...469..581T} and the growth of SMBHs through mergers. The significance of the above questions has prompted intensive efforts to detect SMBHBs (see \citealt{2006MmSAI..77..733K}, for a review).  

Several wide binaries (at kpc separation) have been spatially resolved in different bands, from X-rays to radio \citep{1985ApJ...294L..85O,2003ApJ...582L..15K,2008MNRAS.386..105B,2010ApJ...710.1578G,2011ApJ...735L..42K,2011Natur.477..431F,2011ApJ...737L..19C,2015ApJ...799...72F}, including a few in triple systems \citep{2011ApJ...736L...7L,2014Natur.511...57D}. At parsec separations, SMBHBs can be resolved only with radio interferometers (e.g., the Very Long Baseline Array; VLBA), if both BHs are radio loud. \citet{2006ApJ...646...49R}, for instance, identified a pair of active SMBHs separated by 7.3\,pc (the smallest separation that has been resolved) in the radio galaxy B3 0402+379. 
At sub-parsec separations, where binaries are expected to spend non-negligible time, resolving individual black holes is practically impossible, especially at cosmological distances. Therefore, observational efforts have focused on the effects of the binary on its environment as indirect probes of binaries.

If either of the BHs is associated with a radio jet, the orbital motion and/or the precession of the spin axis will be imprinted on the geometry of the emitted jet, resulting in radio jets with wiggles or knots \citep{1992A&A...254...96K}, or in helical jets with conical geometry \citep{1982ApJ...262..478G}, respectively. At the closest separations, when the orbital decay leads to an observable decrease in orbital period, the jets may exhibit a ``chirp" behavior \citep{KulkarniLoeb2016}
Helical or wiggled radio jets have been attributed to SMBHBs in several cases \citep{1993ApJ...409..130R,2000A&A...360...57R,2001A&A...374..784B,2005A&A...431..831L,2012MNRAS.421.1861V,2013MNRAS.428..280C,2015MNRAS.454.1290K}. Additionally, if gas is bound to one or both BHs, the spectral lines are expected to be noticeably Doppler shifted, reflecting the high orbital velocities in a close binary. Double peaked Balmer lines and displaced Broad Line Regions (BLR), relative to the galaxy's rest frame, have been identified in AGN and quasar spectra and were linked to SMBHBs (see \citealt{2012NewAR..56...74P}, for a review). However, the above candidates remain controversial, since alternative scenarios can also provide feasible explanations. For example, the morphology of a radio jet can be distorted due to Kelvin-Helmholtz instability, and spectral signatures can be produced from a single SMBH if the BLR has a complex geometry (e.g., \citealt{2010ApJ...709L..39C,2016ApJ...817...42L}).

%%%%%%%%%%%%%%%%%%%%%%%%%%%%%%%%%%%%%%%
Another proposed signature of SMBHBs is periodic modulation of the luminosity in AGNs or quasars (in optical, UV, or X-ray bands), induced by the orbital motion of the binary. From theoretical work on circumbinary discs, we expect that, if a SMBHB is embedded in a thin accretion disc, the torques exerted by the binary will expel the gas from the central region, leaving a central cavity almost devoid of gas \citep{1994ApJ...421..651A}. The orbital motion of the binary perturbs the inner edge of the cavity, pulling gaseous streams towards the BHs. The mass accretion rate, and possibly the brightness, is periodically modulated at the orbital period of the binary; a generic conclusion from several hydrodynamical simulations \citep{2007PASJ...59..427H,2008ApJ...672...83M,2012ApJ...755...51N,2012A&A...545A.127R,2013MNRAS.436.2997D,2014ApJ...783..134F,2014PhRvD..89f4060G}.

Periodic variability is a promising method to detect SMBHBs at very close separations.\footnote{In this regime, the separation of the binary is smaller than the size of the BLR and the gas is bound to both BHs. Therefore, the line profiles are very complex and discovery of SMBHBs through line diagnosis is challenging.} Multiple claims for periodic variability in blazars, AGNs and quasars have been reported in the literature, with periods ranging from a few days to a few decades \citep{2006MmSAI..77..733K}. A very compelling case is the BL Lac Object OJ287, which has been monitored for over a century. The optical light curve shows persistent twin outbursts with a period of 11.86\,yr. The outbursts are separated by $\sim$1\,yr, and the second outburst is also accompanied by enhanced radio emission \citep{2000A&AS..146..141P}. 

Nevertheless, the individual claims for detection of periodicity (and thus the binary nature of the individual sources) have been disputed. An alternative approach is to statistically identify a population of periodic AGNs and quasars. The period distribution of such a population can trace the evolution of SMBHBs, enabling us to study the physics of the orbital decay. More specifically, at shorter periods, the distribution is expected to follow a steep ($\propto t_{\rm orb}^{8/3}$) power law, indicating a rapidly decaying population dominated by the emission of GWs, whereas at longer periods, the distribution is less steep, signifying the slower evolution and slower acceleration, during which the binary exchanges angular momentum with the circumbinary disc \citep{2009ApJ...700.1952H}. The discovery of a population with the characteristic slope of GW decay would serve as the first indirect detection of low-frequency GWs.
At the same time, it would show that SMBHBs can produce bright electromagnetic emission even at the late stages of the merger \citep{2012ApJ...755...51N,2015MNRAS.447L..80F}.

Recently, \citet[][hereafter G15]{2015MNRAS.453.1562G}
reported the detection of a population of SMBHB candidates with optical periodicity, and suggested that the period distribution is consistent with a population of near-equal-mass SMBHBs in the GW-dominated regime. The search was conducted with data from the Catalina Real-Time Transient Survey (CRTS; \citealt{2009ApJ...696..870D,2011BASI...39..387M,2012IAUS..285..141D}), an all-sky, time domain survey, in unfiltered visible light, calibrated to Jonhson $V$-band (e.g., see \S~3 in \citealt{2013ApJ...763...32D}), with a limiting magnitude of 19-21.5.\footnote{CRTS combines data streams from three distinct Schmidt telescopes: (1) the 0.7\,m Catalina Sky Survey (CSS) telescope, (2) the 1.5\,m Mount Lemmon Survey (MLS) telescope (both located in Arizona) and (3) the 0.5\,m Siding Springs Survey (SSS) telescope, which was located in Australia and operated until July 2013. In 30\,s exposures, the telescopes achieve nominal $5 \sigma$ detection limits of 19.5, 21.5 and 19.0, respectively.}
A particularly compelling member of this sample, PG1302-102, is a bright (median V-band mag $\sim$15) quasar at redshift $z=0.2784$ with an observed period of $5.2\pm0.2$\,yr \citep{2015Natur.518...74G}.

\citet{2015ApJ...803L..16L} also identified a SMBHB candidate in the Panoramic Survey Telescope and Rapid Response System (Pan-STARRS; \citealt{2010SPIE.7733E..0EK}) Medium Deep Survey with an observed period of $542\pm15$\,d and an estimated separation of 7 Schwarzschild radii. PSO J3334.2028+01.4075 was not confirmed as a periodic quasar in G15, although the photometric precision of the two surveys is not comparable. We also note that, according to population models \citep{2009ApJ...700.1952H}, the discovery of such a close binary is extremely unlikely, for the small sample of quasars they analysed, due to the short ($<100$ year) lifetime of a massive binary at this separation.

\citet{2015arXiv151208730Z} reported the detection of a SMBHB candidate in the radio quiet quasar SDSS J0159+0105. The source was selected from the analysis of CRTS light curves in a small sample of $\sim$350 quasars in Stripe 82. This candidate shows two periodic components at $\sim$1500\,d and $\sim$740\,d, which were attributed to the redshifted orbital period of a putative SMBHB with separation of 0.013\,pc and half of the orbital period, respectively, as expected from hydrodynamical simulations. The quasar was not identified as a periodic source in the sample of G15; it is possible that the existence of the multiple periodic components decreased the significance of the primary component in the search developed by G15.

In the present paper, we perform a systematic search for periodically varying quasars in the photometric dataset from the Palomar Transient Factory (PTF), a large synoptic survey well suited for this search. PTF has a few advantages over the aforementioned surveys. With a $5\sigma$ limiting magnitude of $\sim$20.5, it allows the detection of fainter and hence more distant quasars compared to CRTS, while the higher cadence allows a search for periodicity at shorter time-scales.\footnote{G15 imposed a minimum requirement for the period at 400\,d.} Furthermore, PTF covered a much larger fraction of the sky ($\sim$3,000\,$\rm deg^2$) compared to the area from Pan-STARRS Medium Deep Survey (80\,$\rm deg^2$), thus offering the possibility to analyse a significantly larger sample. We identify periodic quasars via unusually high peaks in the Lomb-Scargle periodograms of their optical light curves. We then assess the statistical significance of periodic variability by simulating time series that exhibit stochastic damped random walk (DRW) variability. The DRW gives a good description of quasar variability in general \citep{2009ApJ...698..895K,2010ApJ...708..927K}, but our statistical analysis improves on previous work, by finding the best-fit DRW model to each individual quasar.

This paper is organized as follows: In \S~\ref{section:data}, we describe the PTF survey, the quasar sample, and the algorithm we use for the detection of periodic variability. In \S~\ref{section:results}, we present our findings. A discussion of our results follows in \S~\ref{section:Discussion}, and our findings and their implications are summarized in \S~\ref{section:Conclusions}. 

\section{AGN Sample and Methodology}
\label{section:data}

\subsection{Palomar Transient Factory}
\label{subsection:PTF}

The Palomar Transient Factory (PTF) was an optical time-domain survey designed to explore the transient and variable sky. The scientific phase of the survey lasted from 03/2009 to 12/2012. The observations were made at Palomar Observatory with the 48-inch Samuel Oschin Schmidt telescope, equipped with the CHF12K camera.\footnote{The camera was previously used at the Canada-Hawai-France Telescope and was modified for PTF \citep{2008SPIE.7014E..4YR}.} With 11 (out of initially 12) light-sensitive CCDs (2\,k$\times$4\,k pixels), the camera provided a wide field-of-view of 7.26\,$\rm deg^2$ and median seeing of 2$''$.   The scientific goals and the technical aspects of the PTF project are detailed in \citet{2009PASP..121.1334R} and \citet{2009PASP..121.1395L}.  Here we briefly recapitulate the main features of the survey relevant for the present paper.

The PTF survey was conducted mainly in Mould-$R$ and SDSS-$g$ bands. More specifically, initially, the majority of the images were taken using the $R$-filter, whereas, from 01/2011, the two filters were alternated between dark ($g$- band) and bright ($R$-band) nights. In 60\,s exposures, the camera achieved $5\sigma$ limiting magnitudes of $m_R\approx20.6$ and $m_{g}\approx21.3$.  PTF covered a total footprint of $\sim$8,000\,$\rm deg^2$, consisting of the entire Northern sky with declination $>-30^\circ$ with the exception of the Galactic plane. The observing time was mainly devoted to the \emph{transient search experiment}, which covered a large part of the sky ($\sim$2700\,$\rm deg^2$) with an average 5\,d cadence, and the \emph{dynamic cadence experiment}, which was activated at the detection of interesting transients for intensive follow-up. 

The data were stored and processed at the Infrared Processing and Analysis Centre (IPAC). For each image, the source positions were identified using \texttt{SExtractor}, a standard algorithm to generate source catalogues from images in large scale-surveys \citep{sextractor}. The flux of each source was calculated using aperture photometry, and the photometric measurements were calibrated by comparing the PTF magnitudes for a set of standard stars to the relevant SDSS values (for details, see \citealt{ofek12a, ofek12b, laher14}). When retrieving light curve data, we use a magnitude cut in both filters to remove saturated sources and sources well below the single-exposure, 5$\sigma$ detection limit. 
We additionally exclude data points with problematic photometry identified either by \texttt{SExtractor} or by the IPAC reduction software (e.g., see flags below). In detail, we select only the measurements that meet the following criteria:

\begin{enumerate}
\item The magnitude (in either filter) is in the range $12<\rm mag<22$.
\item The photometric error is less than 1\,$\rm mag$ ($\sigma<1$).
\item All \texttt{SExtractor} flags are off, except for {BIASED\_PHOTOM} or {SATURATED}.
\item All IPAC photometry flags are off, except for {HALO} or {GHOST}.
\end{enumerate}

\subsection{Light curves and sample selection}
\label{subsection:Sample}
We used the \emph{Half Million Quasars catalogue (HMQ)}\footnote{This is a subsample of the Million Quasar Catalogue v.4.4; \url{https://heasarc.gsfc.nasa.gov/W3Browse/all/milliquas.html} (HMQ; \citealt{2015PASA...32...10F})} as the input catalogue to select the sample of targets for our analysis. HMQ includes all the type-I quasars (QSOs), AGNs, and BL Lac objects published in the literature prior to January 15, 2015, including the most recent data release from SDSS (DR12; \url{http://www.sdss.org/dr12/}). 

We selected sources within a radius of 3$''$ from the input positions. For each source, we extracted two light curves ($R$-band and $g$-band), when available. From the 424,748 spectroscopically confirmed quasars in the catalogue, 278,740 were observed at least once in PTF. Among those, 99,630 were observed at least once in both filters, 172,829 (and 6,281) were observed in $R$-band, but not in $g$-band (and vice versa).

Although the selection of data points, described above, guarantees high-quality light curves, it is not surprising that some outliers are not flagged by the automated pipeline. We remove the remaining outliers with the following procedure: We apply a 3-point-median filter  to the light curve, and subsequently fit the filtered light curve with a $5^{th}$ order polynomial. We discard points that deviate by $\pm 3\sigma$ from the polynomial fit \citep{2011A&A...530A.122P}. 

As a result of the observing strategy, described in \S~\ref{subsection:PTF} (i.e. alternating between the regular 5\,d cadence survey and the intensive follow-up of transients), the extracted light curves show a very large diversity in terms of the total number of observations and the sampling rate. For instance, several fields were covered only a few times in the course of the 4\,yr survey ($\sim$25\% of the $R$-band light curves have between 1-10 data points), whereas fields in which a transient was detected were covered with high cadence, resulting in light curves with sampling rate as high as one observation every 2\,min and a maximum number of $\sim$6000 observations (see Figs.~2 and 3 in \citealt{pricewhelan14} for illustrations). 

Additionally, since the high-frequency (intra-night) variability is not relevant for our periodicity search,\footnote{We search for binaries with periods of several weeks. Given our sample size, the detection of any source with a shorter period (i.e. few days) would be extremely unlikely; see \S~\ref{subsection:PopulationOfSMBHBs} below.} we bin the observations taken within the same night.
More specifically, we replace all data points $y_{ij}$ taken during the $j^{th}$ night ($i=1,2,...N_j$, where $N_j$ is the total number of data points during the $j^{th}$ night), with their average $Y_j$, weighted by the inverse variance of the measurements:
\begin{equation}
\label{eq:BinY}
Y_j=\sum\limits_{i=1}^{N_j} w_{ij} y_{ij}.
\end{equation}
Here $j=1,2,...k$, with $k$ denoting the total number of nights,
during which at least one observation was taken.  We calculate the
weights $w_{ij}$ as follows:
 \begin{equation}
  w_{ij}=\frac{1}{W_j}\frac{1}{\sigma_{ij}^2}  \qquad\text{and}
  \qquad W_j\equiv\sum\limits_{i=1}^{N_j}\frac{1}{\sigma_{ij}^2},
 \end{equation}
where $\sigma_{ij}$ is the photometric error for the $i^{\rm th}$
datapoint during the $j^{\rm th}$ night.
We apply the same weights to
the times $t_{ij}$ of individual observations, whereas the photometric errors on the binned
fluxes are calculated via error propagation,
 \begin{equation}
 T_j=\sum\limits_{i=1}^{N_j} w_{ij} t_{ij} \qquad\text{and}\qquad \Sigma_j=\left(\sum\limits_{i=1}^{N_j} w_{ij}^2 \sigma_{ij}^2\right)^{1/2}.
 \end{equation}
The final light curve consists of $k$ data points $\{T_j,Y_j,\Sigma_j\}$.

\begin{figure}
\includegraphics[height=7cm,width=8.5cm]{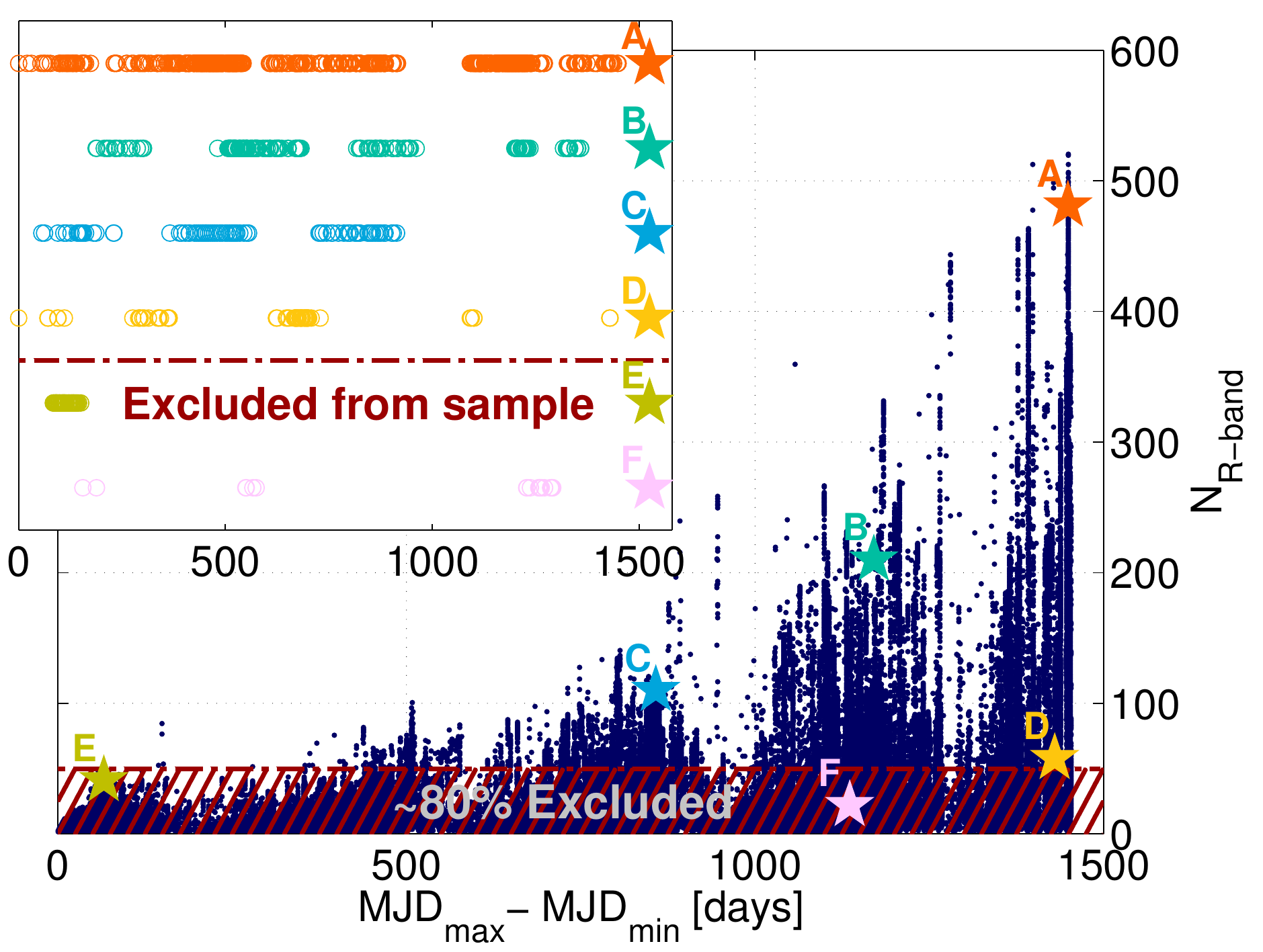}
\caption{Total number of data points (i.e. number of nights) versus the total length of the baseline of each binned R-band light curve. The hatched region shows the light curves that were excluded from the final sample, because the source was observed for fewer than 50 different nights. The time sampling of a few representative light curves is shown in the embedded panel.}
\label{Fig:NnightsVsBaseline}
\end{figure}

In Fig.~\ref{Fig:NnightsVsBaseline}, we illustrate the diversity of
the extracted light curves by showing the total number of data points
in the binned light curves (which practically represents the total
number of nights each quasar was observed) in $R$-band ($N_{R-band}$)
versus the baseline of the light curves (MJD$_{max}-$MJD$_{min}$,
where MJD is the Modified Julian Date of the observation). We also
illustrate the temporal sampling of a few representative light curves
in the embedded panel. We did not analyse light curves with fewer than
50 data points (i.e. quasars that were observed for fewer than a total
of 50 nights).  This resulted in a large cut, excluding $\sim$80\% of the initial
quasar sample, as shown in the hatched region of the main figure and below
the dashed-dotted line in the embedded panel. While the loss is significant,
we found that, for a reliable periodicity search, it is necessary to impose such a cut, since the number of independent frequencies in the periodogram is defined by the number of points in the light curve.

\begin{figure}
\includegraphics[height=7cm,width=8.5cm]{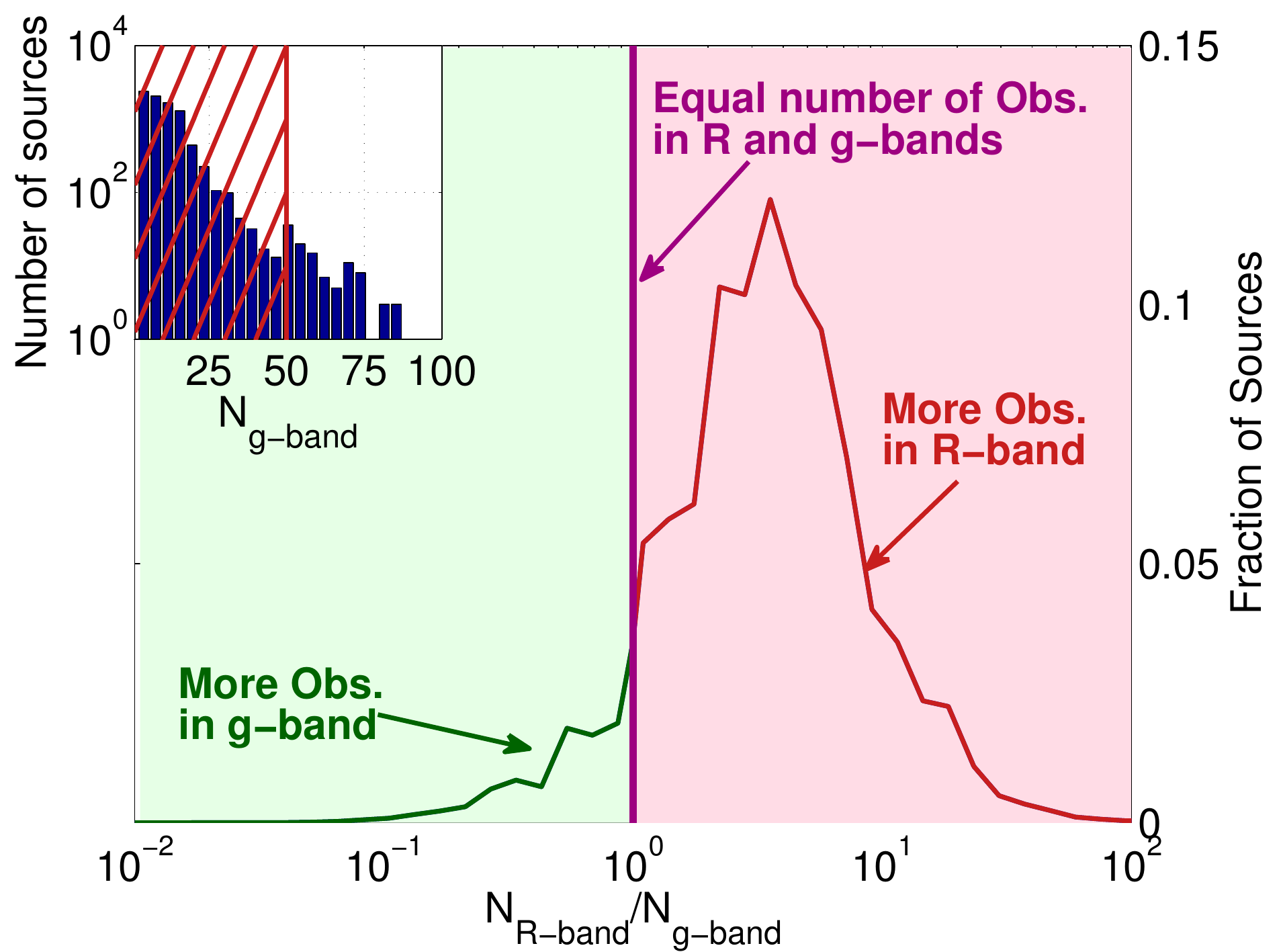}
\caption{Histogram of the total number of data points in the binned light curve in R-band versus g-band. For the small subsample of quasars that were observed more times in g-band, the histogram of the total number of data points (in the binned g-band light curves) is shown in the embedded panel, with the hatched region representing quasars that are not consistent with the minimum requirement of 50 distinct nights.}
\label{Fig:G_Versus_Rband}
\end{figure}

As mentioned above, most of the fields were more extensively covered in $R$-band, whereas the coverage in $g$-band is typically quite sparse.\footnote{This is also apparent from the number of quasars observed at least once in only one filter, stated above.} In Fig.~\ref{Fig:G_Versus_Rband}, we present the histogram of the fraction of the total number of data points in the binned light curve in $R$-band ($N_{R-band}$) over the relevant number for $g$-band ($N_{g-band}$), for the population of QSOs that were covered in both filters. We note that less than 5\% of the sources have more observations in $g$-band, compared to $R$-band. For this subsample, we plot the histogram of the number of data points in the binned $g$-band light curve, in the embedded panel; only a small number of sources (111 quasars) is in compliance with our minimum requirement of 50 data points (the quasars that are not consistent with our minimum requirement are shown in the hatched region of the embedded histogram). We conclude that it is more advantageous to focus only on the $R$-band light curves for the periodicity search. Therefore, our final sample consists of 35,383 QSOs, which have at least 50 data points in the binned $R$-band light curve.

\begin{figure}
\includegraphics[height=7cm,width=8.5cm]{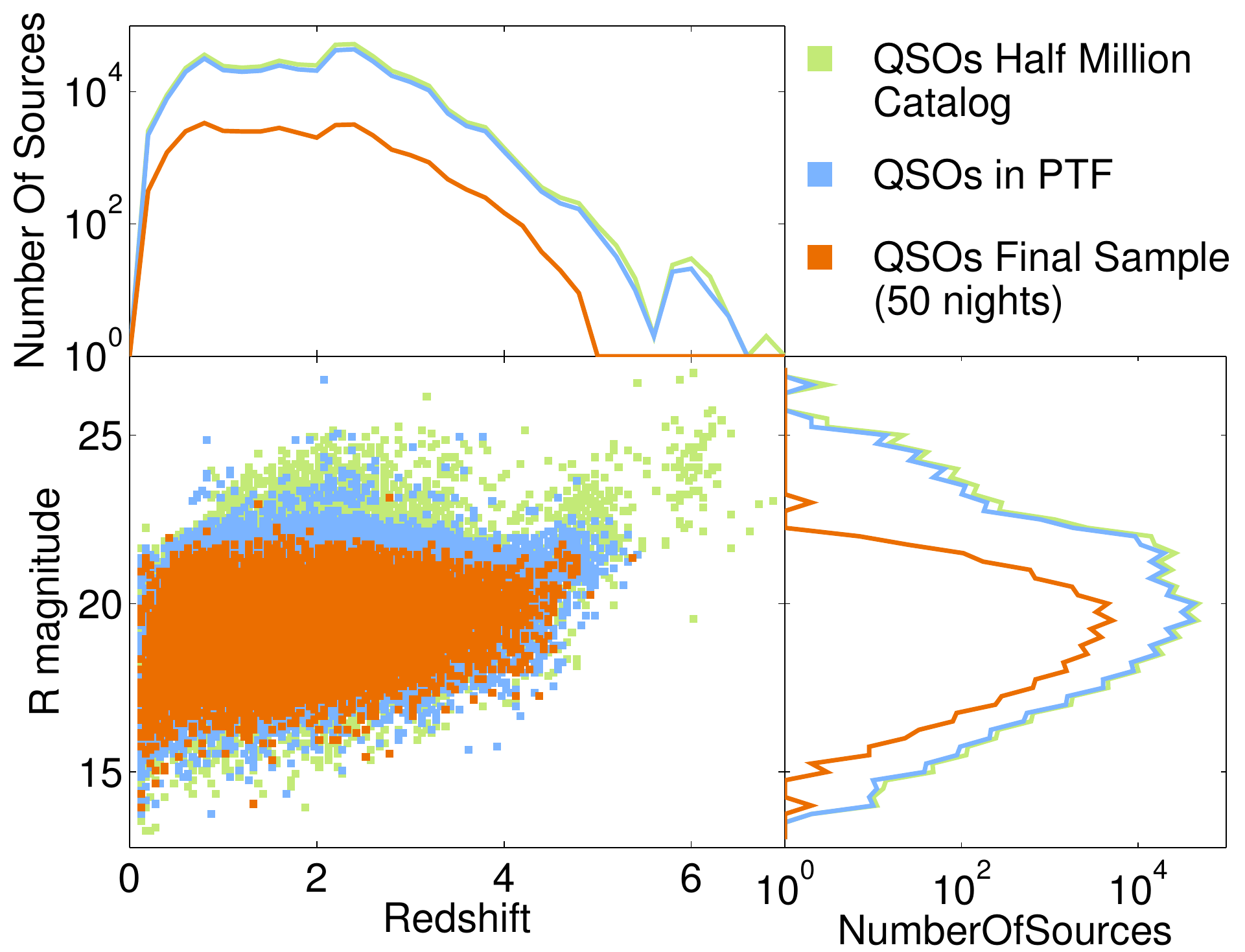}
\caption{Redshift and magnitude distribution for all the quasars in the HMQ (light green), the quasars that were observed at least once in R-band (light blue) and the quasars that were observed for at least 50 distinct nights and constitute our final sample (orange). The respective histograms are also shown with the same colour coding. The figure shows that the sample we analysed is representative of the entire population of quasars.}
\label{Fig:RedshiftMagnitude}
\end{figure}

We emphasize that there is no obvious selection effect, in terms of magnitude or redshift, for the sample we analysed, compared to the entire quasar sample. To illustrate this, in Fig.~\ref{Fig:RedshiftMagnitude} we show the redshift-magnitude distribution for all the quasars in the HMQ (light green points), the quasars that were observed at least once in $R$-band (light blue points) and the final sample of quasars with well-sampled light curves (orange). The side panels show the respective histograms of redshifts and magnitudes with the same colour coding. The most distant and faint quasars are outside of the detection capabilities of PTF and were not included in our sample. 

\begin{figure}
\includegraphics[height=7cm,width=8.5cm]{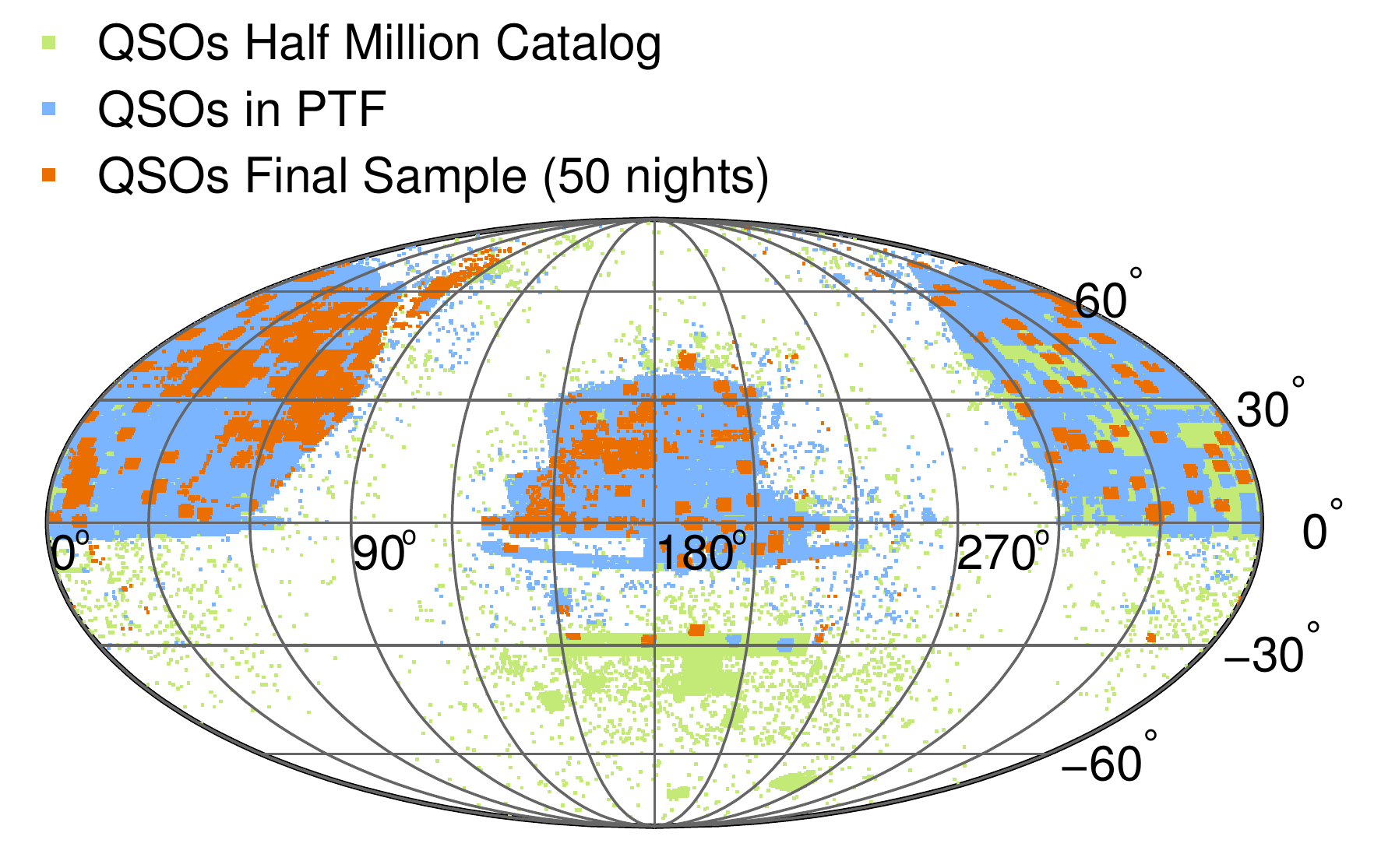}
\caption{Distribution of quasars on the sky in equatorial coordinates. The colour-coding is the same as in Fig.~\ref{Fig:RedshiftMagnitude}.}
\label{Fig:QuasarMap}
\end{figure}

Finally, in Fig.~\ref{Fig:QuasarMap} we show the distribution of quasars on the sky, with the same colour-coding as before. Most of the spectroscopically confirmed quasars were identified in the SDSS database;  the green points therefore roughly trace the SDSS footprint. Similarly, the blue points show the overlap of SDSS with the PTF footprint, and the orange points trace the PTF fields with significant coverage (i.e. our final sample of quasars with at least 50 nights of observations). 

\subsection{Statistical search methodology}
\label{subsection:algorithm}

We develop an automated algorithm to systematically analyse light curves and detect periodic variability, based on the generalized version of the Lomb-Scargle periodogram \citep{2009A&A...496..577Z}.\footnote{For the analysis, we use the astroML python package \citep{astroML,astroMLText}.} The Lomb-Scargle periodogram (\citealt{1976Ap&SS..39..447L}; \citealt{1982ApJ...263..835S}) is a standard method for detecting periodic signals in light curves with non-uniform temporal sampling. We calculate the periodogram for 1000 trial frequencies, uniformly distributed on a logarithmic frequency grid from $f_{min}=1/T_{data}$ to $f_{max}=1/T_{min}$, with $T_{data}=MJD_{max}-MJD_{min}$, the baseline of the light curve and $T_{min}=60\,d$.\footnote{The minimum period (or equivalently the maximum frequency) we probe is set to 60\,d. Given the sample size, the detection of a SMBHB with such a short period is unlikely, see \S~\ref{subsection:PopulationOfSMBHBs} below.}

A periodic signal is detected when a peak with significant power is identified in the periodogram. We use the power of the peaks as the statistic to test the null hypothesis of pure noise. More specifically, for every identified peak, we calculate the probability that a peak of similar power arises from the background by simulating light curves that mimic the quasar variability, and computing periodograms of repeated realizations of the simulated data.

Several authors \citep{2009ApJ...698..895K,2010ApJ...708..927K,2010ApJ...721.1014M} have suggested that the optical variability of quasars is successfully modeled by a damped random walk (DRW) process, described by an exponential covariance matrix
\begin{equation}
\label{eq:Covariance}
S_{ij}=\sigma^2\exp\left(-|t_i-t_j|/\tau\right),
\end{equation}
where $\sigma$ is the long-term variance of the variability, $\tau$ a characteristic time-scale and $t_i$, $t_j$ the different observing times. The power spectral density (PSD) of this model is given by 
\begin{equation}
\label{eq:PSD}
PSD(f)=4\sigma^2\tau/(1+4\pi^2f^2\tau^2).
\end{equation}
In this model, the power decreases with frequency for high frequencies, whereas for low frequencies ($f\ll 1/\tau$), the power spectrum becomes flat.\footnote{Recent work on quasar variability has shown that the variability may deviate from the DRW model, at high frequencies \citep{2014MNRAS.439..703G,2015MNRAS.451.4328K}. However, the deviations are expected to occur at frequencies outside our range of interest, and therefore are not significant for this work; see also \S~\ref{subsection:DerartureFromDRW} below.}

For each quasar, we identify the best-fit parameters ($\sigma$ and $\tau$) for the DRW model by directly fitting the light curve in time domain. For this purpose, we make use of the publicly available code Javelin v3.1 \citep{2013ApJ...765..106Z}. The algorithm employs a Gaussian likelihood associated with the covariance in eq. (\ref{eq:Covariance}) and samples the posterior distribution function of $\sigma$ and $\tau$ with a Markov Chain Monte Carlo (MCMC) sampler\footnote{The MCMC chain consists of 20,000 iterations in the $\sigma$-$\tau$ parameter space (10,000 iterations for the burn-in process and 10,000 iterations for the actual chain).} (see also \citealt{2013ApJ...765..106Z} and the documentation of the code\footnote{\url{http://www.andrew.cmu.edu/user//yingzu/codes.html}} for a detailed description). Moreover, we use log-normal priors for $\sigma$ and $\tau$ taking into account the scaling relations found by \citet{2010ApJ...721.1014M}.

In more detail, we use a prior distribution for each individual quasar according to its observed properties (e.g., magnitude, BH mass). The mean $\sigma$ and $\tau$ of these distributions are estimated from the fitting formulae \citep{2010ApJ...721.1014M}:
\begin{align}
\label{eq:avgTau}
\log\left(\tau_{RF}\right)= & 2.4+0.17\log \left(\frac{\lambda_{RF}}{4000\AA}\right)+0.03\left(M_i+23\right)\nonumber\\
						&+0.21\log\left(\frac{M_{\rm BH}}{10^9M_{\odot}}\right),\\
\label{eq:avgSigma}
\log\left(\sqrt{2}\sigma\right) =& -0.51-0.479\log \left(\frac{\lambda_{RF}}{4000\AA}\right)+0.131\left(M_i+23\right)\nonumber\\
&+0.18\log\left(\frac{M_{\rm BH}}{10^9M_{\odot}}\right)
\end{align}
where $\tau_{RF}$ is the characteristic time-scale $\tau$ in the rest frame of the quasar and $\lambda_{RF}$ is the effective wavelength of the $R$-band filter, $\lambda=6516.05\,\AA$, in which the observations were made, in the rest frame of the source (i.e. for a quasar at redshift z, $\tau_{RF}=\tau(1+z)^{-1}$ and $\lambda_{RF}=\lambda(1+z)^{-1}$). We calculate the absolute $i$-band magnitude $M_i$, $k$-corrected to $z=2$, from the mean apparent $R$-band magnitude in the HMQ, adopting the mean quasar spectral energy distribution (SED) from \citet{1994ApJS...95....1E} and an opacity model for the Ly$\alpha$ forest from \citet{1996MNRAS.283.1388M}. We adopt the virial black hole mass $M_{\rm BH}$ estimated from the width of broad lines \citep{2008ApJ...680..169S}. For quasars that do not have a mass estimate, we draw the mass from the expected Gaussian distribution given the absolute $i$-band magnitude (e.g., \citealt{2010ApJ...721.1014M} based on the results from \citealt{2008ApJ...680..169S}),
\begin{equation}
\label{eq:BHMass}
p\left(\log M_{\rm BH}|M_i\right)=\frac{1}{\sqrt{2\pi}s}\exp\left[-\frac{(\log M_{\rm BH}-m)^2}{2s^2}\right],
\end{equation}
with $m=2.0-0.27 M_i$ and $s=0.58+0.011 M_i$. The variance of the prior $\sigma$ and $\tau$ distributions, for each quasar, is determined by propagating the uncertainties of the fitting coefficients in eq. (\ref{eq:avgTau}) and (\ref{eq:avgSigma}),
\begin{align}
Var(\tau)&=\tau^2\log(10)^2 Var(\log\tau),\\
Var(\log\tau)&=0.2^2+0.02^2\log\left(\frac{\lambda_{RF}}{4000\AA}\right)^2\nonumber\\
+&0.04^2\left(M_i+23\right)^2+0.07^2\log\left(\frac{M_{\rm BH}}{10^9M_{\odot}}\right)^2,
\end{align}
and similarly for $\sigma$ (Table 1 in \citealt{2010ApJ...721.1014M}). We also note that the uncertainty in the BH mass measurement is included in the uncertainty of the fitting coefficients. 

Following the prescription from \citet{1995A&A...300..707T}, we generate evenly sampled light curves (temporal resolution $\Delta t$=1\,$\rm day$) that exhibit DRW variability (with the PSD in eq. \ref{eq:PSD}), fixing the values of $\sigma$ and $\tau$ at the median of the respective posterior distribution. Next, we downsample the time series to match the observing times of the quasar light curve under consideration.\footnote{Downsampling at the observation times ensures that aliasing peaks from the uneven sampling will not be falsely detected as periodic signals (see, e.g., \citealt{2015MNRAS.454L..21C} for a discussion of aliasing peaks in the periodogram of the quasar PG1302-102).} We add Gaussian deviates with zero mean and standard deviation equal to the photometric uncertainty of each point to incorporate the measurement errors, and shift the generated light curve by a constant to match the observed mean magnitude. Finally, we calculate the periodogram using the same frequency grid as for the actual observed time series.

Since the noise spectrum is frequency-dependent, it is more meaningful to assess the statistical significance of the identified peaks compared to the local background (i.e. within a relatively narrow frequency range). 
 Hence, we divide the frequency grid into 25 logarithmically spaced frequency bins, each containing 40 trial frequencies.\footnote{We choose relatively narrow frequency bins to ensure a fair comparison with the neighbouring frequencies. Within a narrow bin, the frequency dependence of the noise is not very pronounced and the noise locally resembles white noise.} For each frequency bin, we identify the peak with the maximum power and compare it to the distribution of maxima (within the same frequency bin), from the periodograms of the generated DRW light curves.

For each quasar, we simulate 250,000 DRW time series to account for the trial factors introduced by the number of frequency bins, and the sample size (\emph{look-elsewhere effect}).   
We define the P-value of a peak as the number of realizations with at least one peak with power exceeding the power of the peak under consideration divided by the total number of background realizations. A quasar is considered to show significant periodicity, when at least one peak (in any of the 25 frequency bins) is above our significance threshold (P-value$<$1/250,000). Finally, we fit a sine wave with frequency around the frequency of the significant peak to the observed light curve, and exclude candidates that are not observed for at least 1.5 cycles within our baseline.

\section{Results}
\label{section:results}
\subsection{Quasars with significant periodicity}
\label{subsection:PeriodicQuasars}

We analysed the periodograms of the 35,383 quasars which have been observed for at least 50 distinct nights. We identified 67 quasars with significant peaks, as defined in the previous section. Of these, 50 were consistent with our requirement for a minimum of 1.5 cycles within the data. If we increase the minimum requirement to at least 2 or 3 cycles within the baseline, the number of sources decreases to 42 and 25, respectively. 
Note that with the P-value threshold defined above, under the null-hypothesis of pure DRW noise, we expect to find, on average, $25\times35,383/250,000\sim 3.5$ such peaks by random chance.

For the identified population, it is crucial to assess the statistical significance for the ensemble of the sources, rather than the significance of individual findings.
For this purpose, we generated mock DRW light curves, with $\sigma$ and $\tau$ drawn from the distributions described in Section~\ref{subsection:algorithm}, downsampled at the observations times of quasars and repeated the entire automated analysis, from identifying the best--fit DRW parameters to calculating the P-value of the peaks. We identified 7 significant peaks in the DRW periodograms, which is a factor of two higher than the theoretically expected number of false positives (see \S~\ref{subsection:DRWFitting} for a possible explanation),
but with only one having at least 1.5 periods within our data.  It is clear that the DRW model alone cannot reproduce the set of significant periodicities found in the quasar sample and, thus, the identified sample of periodic quasars is statistically significant. We emphasize again that the statistical significance refers to the population of $\sim50$ candidates, and not necessarily to any of the individual quasars.

\begin{figure}
\includegraphics[width=8.5cm,height=7cm]{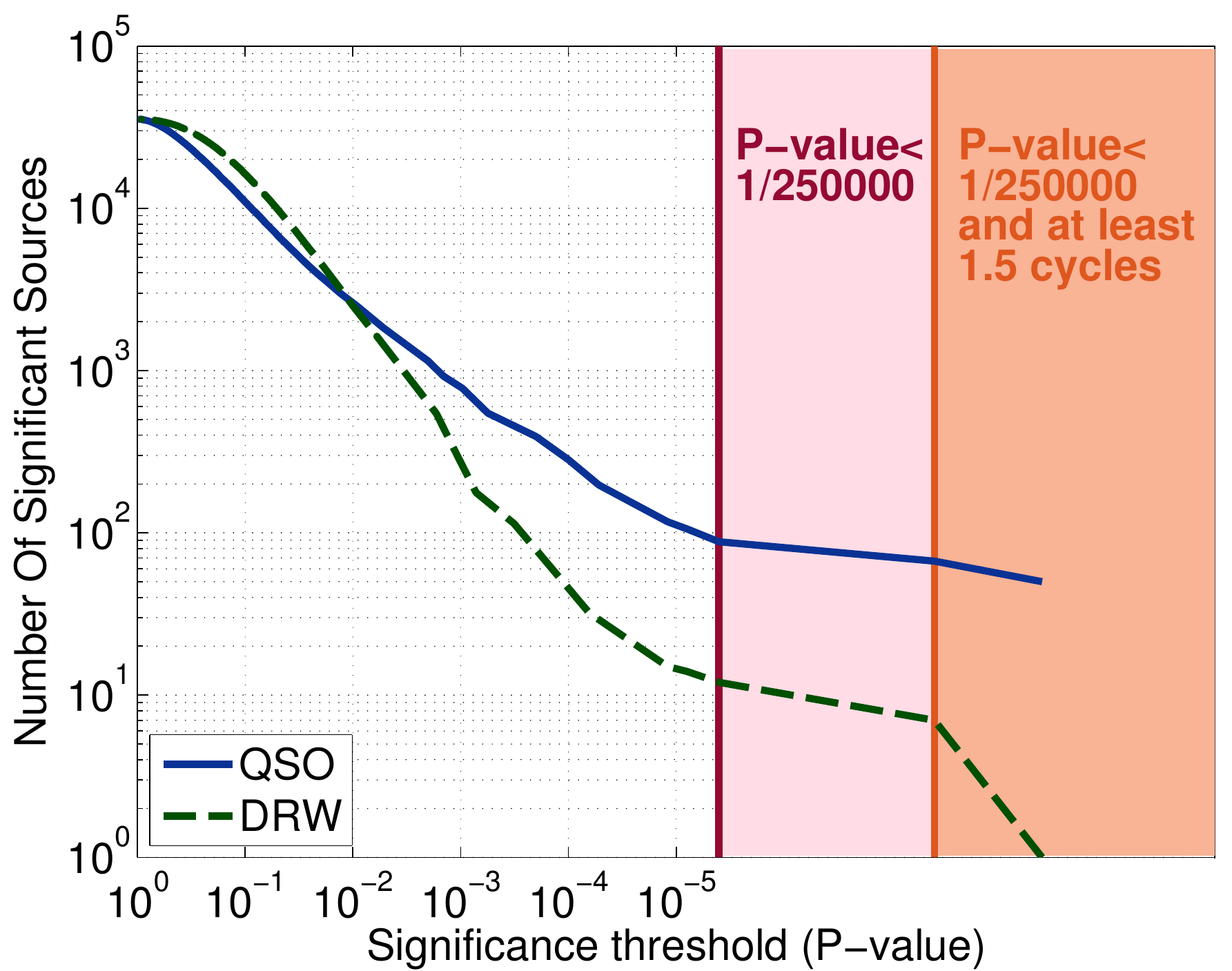}
\caption{Number of sources that would be identified as periodic versus the significance threshold (P-value). The solid line represents the observed quasar data and the dashed line represents the simulated DRW light curves downsampled at the observation times. The purple shaded region shows the number of candidates consistent with our final significance threshold (P-value$<$1/250,000). The shaded orange region shows the number of candidates that remain in the sample after we exclude sources that were not observed for at least 1.5 periods.}
\label{Fig:Pval}
\end{figure}

In Fig.~\ref{Fig:Pval}, we show the number of sources, both for the observed quasar data (solid curve) and for the mock DRW realizations (dashed curve), which would be identified by our procedure as periodic, as a function of the significance threshold.  The figure shows that at low thresholds ($P\lsim 10^{-2}$), the number of peaks in the quasar sample matches the expectations from pure DRW noise. 
However, for higher thresholds, we detect an increasingly larger excess of periodic sources in the real quasar sample, compared to the simulated DRW data. The shaded purple region on the right side of the figure highlights findings that are significant after 250,000 iterations (i.e. above the final threshold we considered), whereas the shaded orange region represents the final population (at the same significance level), after we excluded periodic sources with fewer than 1.5 observed cycles.

In Table~\ref{Table:CandidatesQuasarProperties}, we present the names, coordinates and the observed properties (redshift, average R-band apparent magnitude from HMQ) of the 50 quasars that were identified to show significant periodicity in PTF. The BH mass measurements from \citet{2008ApJ...680..169S} are shown. When the mass estimate was not available in the catalogue, we include the  value drawn from eq.~(\ref{eq:BHMass}), as well as the mean and standard deviation of the expected distribution, given the quasar magnitude.  In the Table, we also note whether the quasar has been associated with X-ray or radio emission, and the relevant catalogue in which the source was identified. 

For the 50 sources with significant periodicity, we extract additional photometric data from the intermediate-PTF (iPTF) and the CRTS, extending the available baseline by at least a factor of 2 (see \S~\ref{Subsection:ExtendedLC} for a discussion). In light curves from both iPTF and CRTS, we bin the observations taken within the same night, as described in \S~\ref{subsection:algorithm}. Additionally, since the data are obtained in different photometric systems (Mould-$R$ for PTF and iPTF and unfiltered $V$-band for CRTS), we calibrate the different datasets as follows: For each object, we first identify the maximum interval of temporal overlap between the two light curves $[MJD_0,  MJD_1]$.\footnote{For the calibration, we consider the data from PTF and iPTF as a single extended light curve.} In this interval, we interpolate each light curve using a non-parametric model (LOWESS regression) and calculate the offset between the light curves in 100 distinct points evenly spaced within the interval $[MJD_0,  MJD_1]$. Next, we shift the PTF light curve by a constant value defined as the median of the local offsets in the discrete points.

We analyze the periodograms of the extended light curves as before, using the same frequency grid (and frequency bins) as in PTF. We calculate the P-value of the previously identified period, which we present in Table~\ref{Table:BinaryProperties}. In the Appendix, we show the light curves of these candidates (red points for PTF observations, i.e. the light curves we analyzed initially and identified the periodicity, black points for iPTF observations and blue points for observations from CRTS), along with their best-fit sinusoids. In the right column of the figure, we include the light curves phase folded at the observed period. The P-values calculated from the analysis of the extended light curves are also included in the figures. (We note that three of the sources do not have additional data outside of PTF.)

\subsection{Periodic quasars as SMBHBs candidates}
\label{subsection:SMBHBCandidates}
Assuming that the observed periodicity is the redshifted orbital period of a SMBHB, we calculate several properties of the tentative binaries, which we include in Table~\ref{Table:BinaryProperties}. For instance, the separation is given from Kepler's law (assuming, for simplicity, a circular orbit):

\begin{equation}
R=\left(\frac{G M P_{orb}^2}{4\pi^2}\right)^{1/3},
\end{equation}
where $G$ is the gravitational constant, and $M$ and $P_{orb}$ are the total mass and the orbital period of the binary ($P_{orb}=P_{obs}(1+z)^{-1}$).
For reference, we also calculate the projected angular separation,
\begin{equation}
\theta\simeq R/D_{A},
\end{equation} 
where $D_{A}$ is the angular size distance for the standard cosmological parameters $H_0=67.8\,{\rm km~s^{-1}Mpc^{-1}}$ and $\Omega_{m}= 0.308$, and a spatially flat universe \citep{2015arXiv150201589P}.

The gravitational waves emitted by a close SMBHB affects the arrival time of radio-pulses, when the waves cross the line of sight between pulsars and the Earth, inducing a time residual $t_r$ in the arrival time. We estimate these time-residuals  (in ns) for the fiducial inferred SMBHB parameters of each quasar, following \citet{2015ASSP...40..147S}:
\begin{equation}
t_r\simeq30\left(\frac{\mathscr{M}}{10^9M_{\odot}}\right)^{5/3}\left(\frac{D_L}{100\,{\rm Mpc}}\right)^{-1}\left(\frac{f}{5\times10^{-8}\,{\rm Hz}}\right)^{-1/3}.
\end{equation}
Here $D_L$ is the luminosity distance to the quasar, $\mathscr{M}=(M_1 M_2)^{3/5}/(M_1+M_2)^{1/5}$ is the chirp mass of a binary with individual BHs of mass $M_1$, $M_2$ and $f=2/P_{orb}$. In Table~\ref{Table:BinaryProperties}, we include the maximum expected value for the time residuals induced by equal mass binaries ($q \equiv M_2/M_1=1$). We note that the estimated residuals are quite small compared to the sensitivity of current PTAs, for all of the tentative binaries. Even though individually undetectable, the binaries still contribute to the stochastic GW background, which was recently constrained by PTAs (e.g., \citealt{2015Sci...349.1522S}).

For each quasar, we also compute the inspiral time of the orbit due to losses to gravitational waves \citep{1964PhRv..136.1224P}
\begin{equation}
t_{GW}=\frac{5}{256}\frac{c^5}{G^3}\frac{R^4}{(M_1+M_2) (M_1 M_2)}.
\end{equation}
For reference, we show the inspiral time both for equal ($q=1$) and unequal mass ($q=0.01$) binaries.

Finally, for each SMBHB candidate, given its observed period, we calculate the \emph{residence time}, i.e. the time a binary is expected to spend at a given orbital period, or equivalently at a specific orbital separation. The residence time is determined by the rate of orbital decay $t_{\rm res}\equiv -R(dR/dt)^{-1}$. Following \citet{2009ApJ...700.1952H}, we assume that, at large separations, the orbital decay of the binary is dominated by the tidal-viscous exchange of angular momentum with a gaseous circumbinary disc, whereas at small separation the decay is dominated by the emission of gravitational waves. We adopt a standard geometrically thin, optically thick, radiatively efficient, steady-state accretion disc model for the circumbinary disc, coplanar with the orbit of the binary. For the disc parameters, we use the values of the fiducial model in \citet{2009ApJ...700.1952H} (e.g, viscosity parameter $\alpha=0.3$, and accretion rate at 10\% of the Eddington accretion rate $\dot m=\dot M/\dot M_{\rm Edd}=0.1$). In the above model, for fixed orbital period, the residence time depends only on the mass of the binary (which is either measured or estimated from the apparent magnitude of the quasar) and the (unknown) mass ratio $q$ of the binary.

\subsection{Expected SMBHB population}
\label{subsection:PopulationOfSMBHBs}

The residence time is a useful quantity to assess the feasibility of a tentative population of SMBHBs. If we attribute the bright phase of quasars to SMBHBs, we can derive the theoretically expected distribution of residence times for the analysed sample of quasars.
Since the bright phase of quasars ($t_{Q}\simeq$ few$\times10^7$\,yr, e.g., \citealt{2004cbhg.symp..169M})
 is comparable to the fiducial time-scale for the binary evolution from the outer edge of the circumbinary disc to coalescence \citep{2009ApJ...700.1952H}, quasars will harbor binaries with separations in this entire range, distributed according to their residence time (i.e. a larger fraction of sources at longer residence times, and a smaller fraction at shorter residence times). Therefore, we can express the expected fraction of SMBHBs at residence time $t_{\rm res}$, as a linear function of the residence time, $f(t_{\rm res})\sim t_{\rm res}/t_{Q}$. Accordingly, the total number of quasars $N$, identified as periodic in a sample with size $N_{sample}$, should scale linearly with the residence time. More specifically,
\begin{equation}
\label{eq:Ntres}
N(t_{\rm res})=f(t_{\rm res})\times N_{sample}=t_{\rm res}/t_{Q}\times N_{sample}
\end{equation}
For the sample in this paper ($N_{sample}=35,383$) and assuming $t_{Q}\sim3.5\times 10^7$\,yr, we expect about 1 quasar with $t_{\rm res}=10^3$\,yr.\footnote{It follows from the above that we should expect $\sim$0.1 SMBHBs with $t_{\rm res}=10^2$\,yr and $\sim$0.01 with $t_{\rm res}=10$\,yr. Therefore, it is unlikely that any identified periodicity corresponding to $t_{\rm res}\lsim 10^3$\,yr, is related to SMBHBs (see \S~\ref{subsection:PeriodicVariabilityAlternatives} below).}

The expectation above (eq. \ref{eq:Ntres}) is an upper limit, as it is derived for an idealized survey, without taking into account any observational limitations. An obvious such limitation is imposed by the finite baseline of the light curves. The cadence and length of the time series defines the range of periods we can identify, which in turn defines the range of residence times we can probe with the available data. This effect can be incorporated into our calculation by estimating the residence time intervals for each light curve. More specifically, if the baseline of the time series is $T_{data}$, we can search for signals with periods in the interval
\begin{equation}
\left[P_{obs}^{min},P_{obs}^{max}\right]=\left[T_{min}, \frac{2}{3} T_{data}\right]
\end{equation}
For a quasar at redshift $z$, this interval is translated into a range of orbital periods
\begin{equation}
\left[P_{orb}^{min},P_{orb}^{max}\right]= (1+z)^{-1} \times \left[T_{min}, \frac{2}{3} T_{data}\right]
\end{equation}
Given the models of binary evolution, discussed in \citet{2009ApJ...700.1952H}, this interval corresponds to a range of residence times $\left[t_{\rm res}^{min},t_{\rm res}^{max}\right]$.

\begin{figure}
\includegraphics[height=7cm,width=8.5cm]{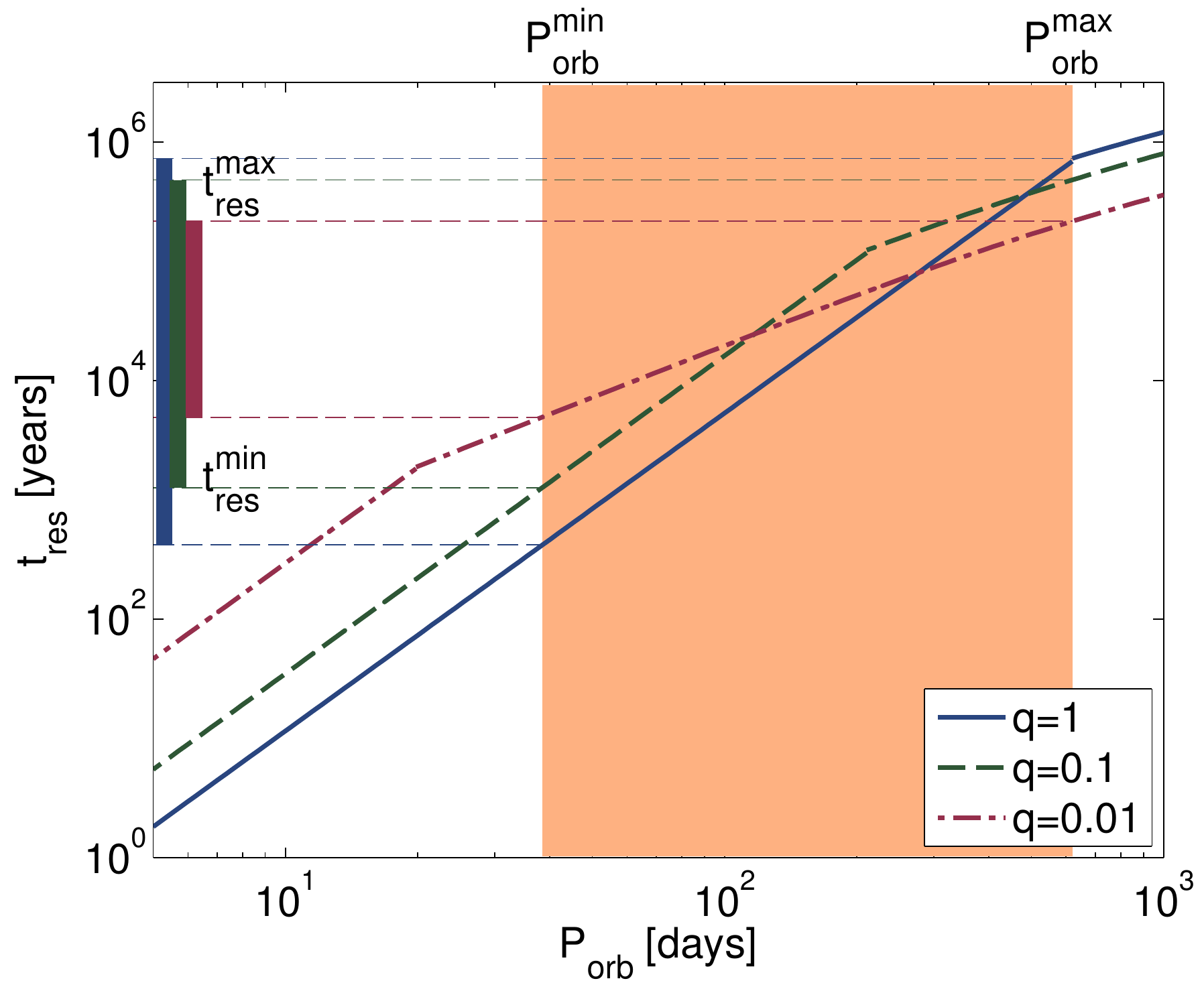}
\caption{Residence time of a SMBHB with a total mass of $M\approx10^8M_{\odot}$, as a function of the orbital period, based on the models in \citet{2009ApJ...700.1952H}, for mass ratio $q=1$ (blue solid line), $q=0.1$ (green dashed line) and $q=0.01$ (purple dashed-dotted line). The region highlighted in orange shows the accessible orbital periods, given the cadence and  finite baseline of the observed light curve. The corresponding residence time range is highlighted on the vertical axis (for different $q$ values, following the colour-coding of the lines).}
\label{Fig:ResidenceTimeRange}
\end{figure}

Fig.~\ref{Fig:ResidenceTimeRange} illustrates the above process for one of the quasars in our sample. The lines trace the evolution of a binary (i.e. the evolution of its residence time), with total BH mass $M\sim 10^8 M_{\odot}$, as the orbit decays from longer to shorter orbital periods, assuming three different mass ratios (blue solid line for $q=1$, green dashed line for $q=0.1$ and purple dashed-dotted line for $q=0.01$). The segments with the different slopes signify the distinct stages of the binary evolution. At long orbital periods, the binary evolution is slow and is dominated by angular momentum exchange with the circumbinary disc (shallower part of the evolutionary tracks), whereas at short orbital periods, the binary enters the GW-driven regime and the evolution is faster (steeper part). The transition between the two regimes occurs at different orbital periods, depending on the mass ratio. The orange shaded region highlights the parts of the binary evolution that are accessible for study, given the cadence and limited baseline of the observed data. The corresponding residence time window is highlighted on the vertical axis (colour-coded according to the mass ratio, as before).

We calculate the range of residence times $\left[t_{\rm res}^{min},t_{\rm res}^{max}\right]$ for each quasar in the sample, repeating the process described above. Next, for each value of the residence time $t_{\rm res}$, we calculate the fraction of the sources in the sample, for which the specific value of $t_{\rm res}$ is within the observable limits $\left[t_{\rm res}^{min},t_{\rm res}^{max}\right]$ and define this as the \emph{observable fraction} of the residence time $f_{o}(t_{\rm res})$.
Given the observable fraction, we can calculate the expected number of binaries $N_{e}$, accounting for observational limitations, by multiplying eq. (\ref{eq:Ntres}) with $f_{o}(t_{\rm res})$, 
\begin{equation}
\label{eq:Ne}
N_{e}(t_{\rm res})=N(t_{\rm res})\times f_{o}(t_{\rm res}).
\end{equation}

\begin{figure}
\includegraphics[height=7cm,width=8.5cm]{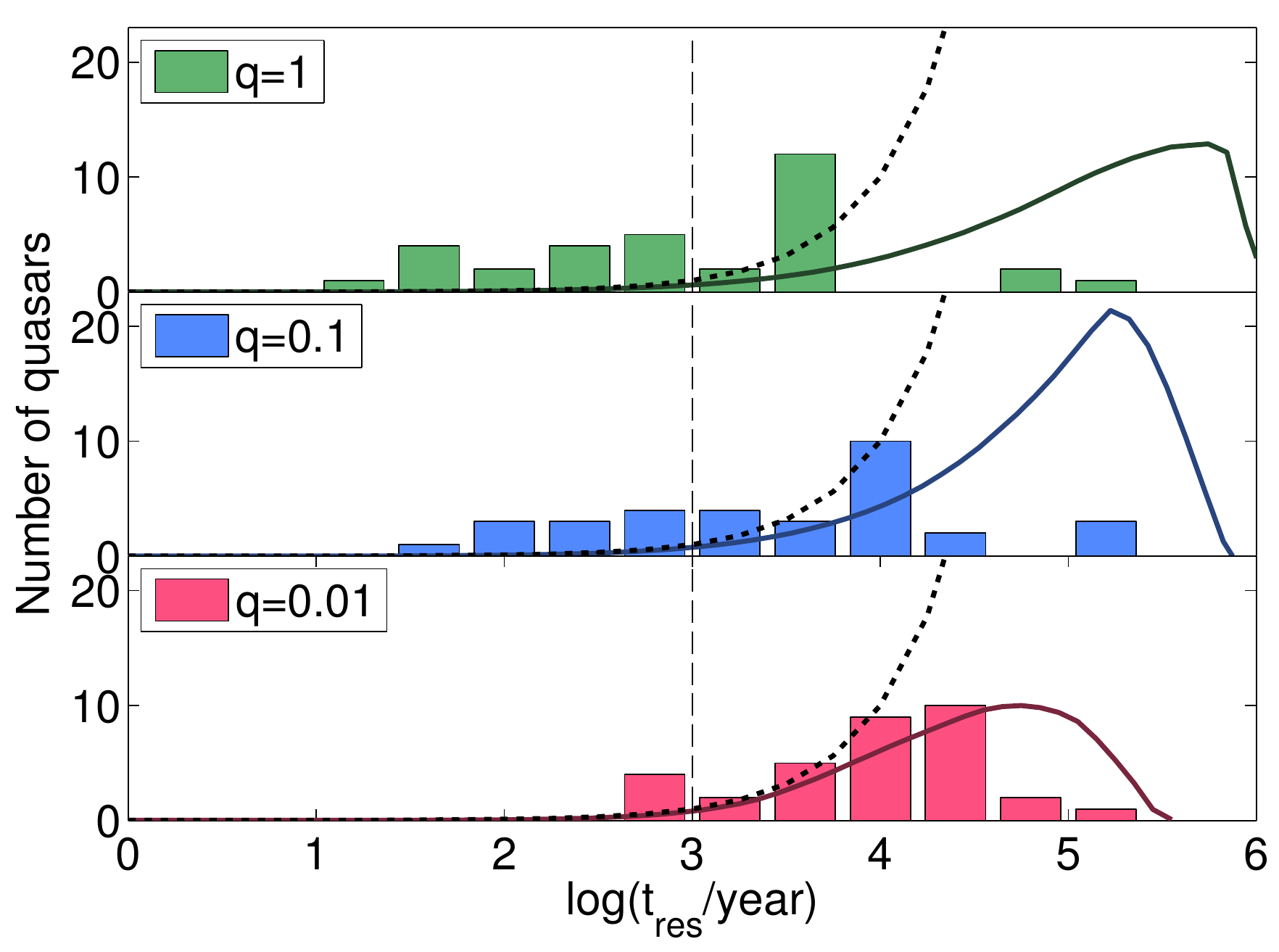}
\caption{Histograms of inferred residence time for the 33 periodic SMBHB candidates (out of the 50 candidates identified in PTF, 33 remain significant when we analyzed the extended light curves with data from iPTF and CRTS), assuming different mass ratios:  $q=1$ (top panel), $q=0.1$ (middle panel) and $q=0.01$ (bottom panel). The dotted curves show the distribution expected without considering observational effects, and the solid curves indicate these distributions, after taking into account the limitations imposed by the cadence and baseline of each quasar's observed light curve. The dashed vertical line corresponds to $t_{\rm res}=10^3$\,yr, below which it is unlikely to identify binaries in our sample.}
\label{Fig:TresidenceDistribution}
\end{figure}

In Fig.\ref{Fig:TresidenceDistribution}, we present the distributions of residence times for the sample of quasars with significant periodicity, calculated for the three different mass ratios  $q=1$ (top panel), $q=0.1$ (middle panel), and $q=0.01$ (bottom panel). From the distribution we exclude 17 sources, the peaks of which have high P-value in the periodograms of the extended light curves, e.g., P-value$>1\%$ (see \S~\ref{Subsection:ExtendedLC}). The corresponding expectations from eq.~(\ref{eq:Ne}) are shown by solid curves in each panel. For reference, the naively expected populations without taking into account the observable fractions (eq. \ref{eq:Ntres}) are also shown, by the dotted curves. The vertical dashed line represents $t_{\rm res}=10^3$\,yr, below which it is unlikely to identify SMBHBs in our sample. The figure shows that the observed periodic candidates match the theoretical expectation for the unequal-mass case ($q=0.01$) better.  Additionally, in this case, the number of unlikely findings (with $t_{\rm res}<10^3$\,yr) is smaller compared to $q=1$ or $q=0.1$. We also note that, for $q=0.01$, all of the candidates are in the gas-driven regime, whereas for the equal-mass case ($q=1$), the majority of the candidates 
%(39 out of 50)
 would be dominated by the emission of GWs. We emphasize that in reality, binaries will have a distribution of mass ratios, and not a fixed value; nevertheless, the distribution of the detected residence times favors a low typical mass ratio.

\section{Discussion}
\label{section:Discussion}

\subsection{DRW parameter biases}
\label{subsection:DRWFitting}

As discussed above, previous work has found that the optical variability of quasars, in general, is successfully described by a two-parameter stochastic noise model (DRW, eq. \ref{eq:PSD}). We assessed the statistical significance of the identified peaks in the periodograms by generating mock time series using this noise model. For each quasar, we identify the best-fit DRW parameters ($\sigma$ and $\tau$), using a Gaussian likelihood and sampling the posterior distribution with an MCMC sampler. Since the DRW model is an essential part of assigning a significance to observed periodicities, we tested the efficiency of recovering input DRW parameters in the PTF sample. For this purpose, we adopted the fitting algorithm from \citet{2013ApJ...765..106Z}, used to investigate light curves from the Optical Gravitational Lensing Experiment (OGLE). We generated DRW light curves with known input parameters $\sigma$ and $\tau$, downsampled at the observation times of the PTF light curves, and used these as inputs to the fitting algorithm. We found that the algorithm is very successful in recovering the input $\sigma$, whereas it typically underestimates $\tau$. 
This low-$\tau$ bias will generally tend to underestimate the noise, and over-estimate the significance of peaks. It is therefore a plausible explanation for why we identify slightly more (roughly twice as many) false positives than theoretically expected, when we analyse DRW light curves (e.g., see Fig.~\ref{Fig:Pval}).

\begin{figure}
\includegraphics[height=7cm,width=8.5cm]{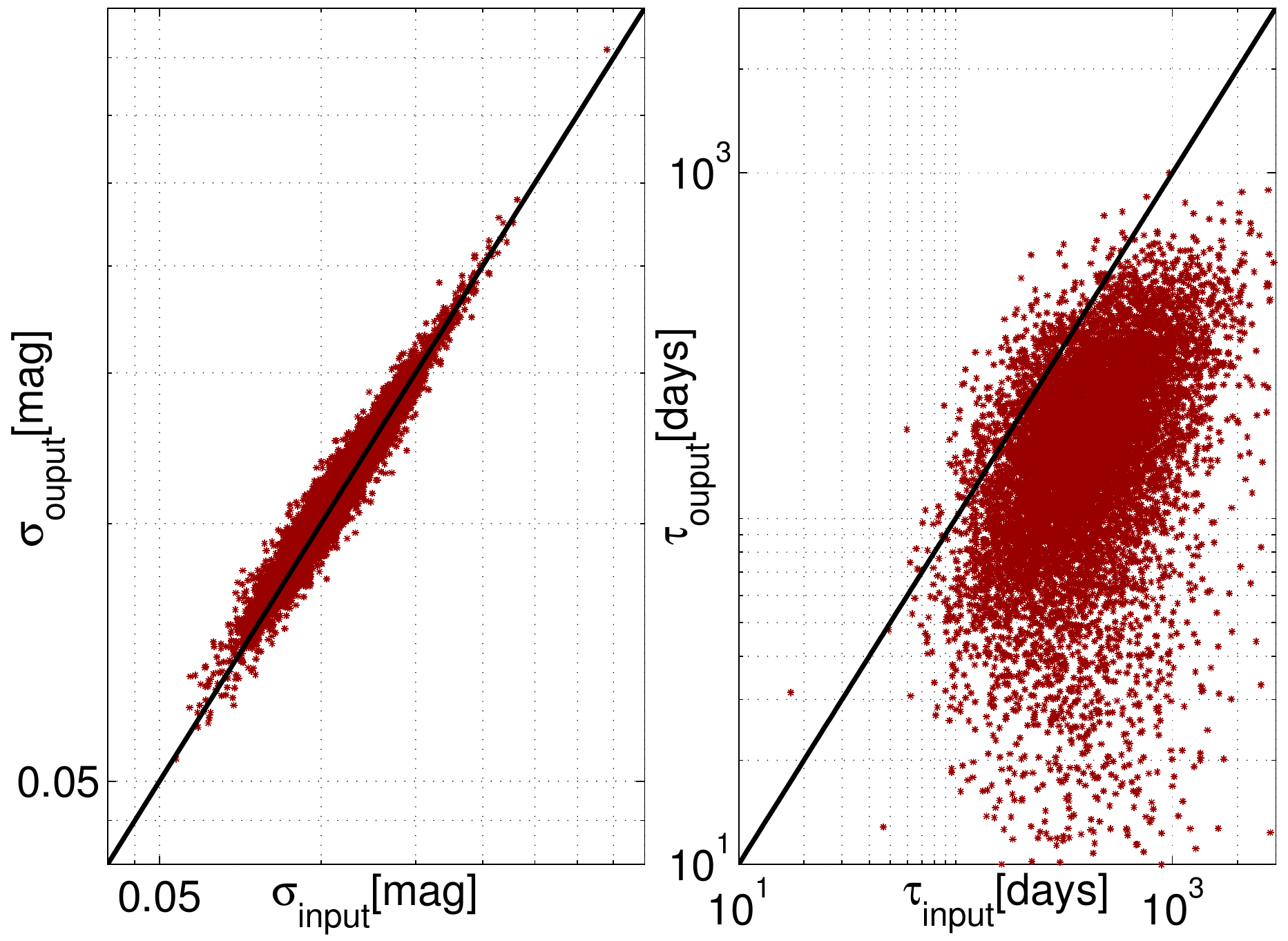}
\caption{Output parameters of the DRW fitting algorithm ($\sigma$ in the left panel and $\tau$ in the right panel) versus the relevant input values, for DRW realizations sampled at the observation times of 10,000 quasar light curves. The solid lines show the unbiased output values.}
\label{Fig:DRWParameters}
\end{figure}

In Fig.~\ref{Fig:DRWParameters}, we show the output $\sigma$ and $\tau$ versus the input values of these parameters, for 10,000 DRW realizations sampled at the observation times of quasars in our sample. The unbiased line is drawn in both cases for comparison. We see that for the PTF light curves, the estimates of $\tau$ are biased (a similar bias could of course arise when fitting any light curve with photometric errors and uneven sampling comparable to those in PTF). This bias is possibly the consequence of the following factors: (1) the light curves from PTF have relatively short baselines -- in the majority of cases, the input $\tau$ is a large fraction of the baseline, 
(2) the PTF observations are characterized by very uneven sampling, with periods of dense coverage and extensive gaps, (3) the photometric uncertainty can be comparable to the long-term variance $\sigma$, making it almost impossible to differentiate between DRW and white noise, which naturally results in underestimation of the parameter $\tau$, and (4) the DRW parameters are drawn from the prior distributions in \citet{2010ApJ...721.1014M}, which rely on estimates of the absolute $i$-band magnitude and BH mass of each quasar. 

The impact of the temporal sampling is also present in previous papers on quasar variability. For instance, \citet{2010ApJ...721.1014M} analysed light curves from SDSS Stripe82, which have long baselines and relatively sparse sampling and found that $\tau$ is overestimated for a fraction of their light curves (e.g., see their Fig.~11). They also identified typically longer time-scales $\tau$, compared to the sample in \citet{2010ApJ...708..927K}. The latter study analysed a sample of quasars from OGLE, with baselines comparable to Stripe82 and better sampling. \citet{2010ApJ...721.1014M} suggested that the lack of shorter $\tau$ in their sample could be explained either by the different sampling rate of the light curves in the two samples, or by a potential contamination of stars in the sample from \citet{2010ApJ...708..927K}. The above results, in combination with the underestimation of $\tau$ in the PTF light curves, shown in Fig.~\ref{Fig:DRWParameters}, provide a hint that the sampling may introduce a bias in the $\tau$ estimation, which may explain the discrepancy between the $\tau$ distributions from \citet{2010ApJ...708..927K} and \citet{2010ApJ...721.1014M}, although further investigation is required.

\begin{figure}
\includegraphics[height=7cm,width=8.5cm]{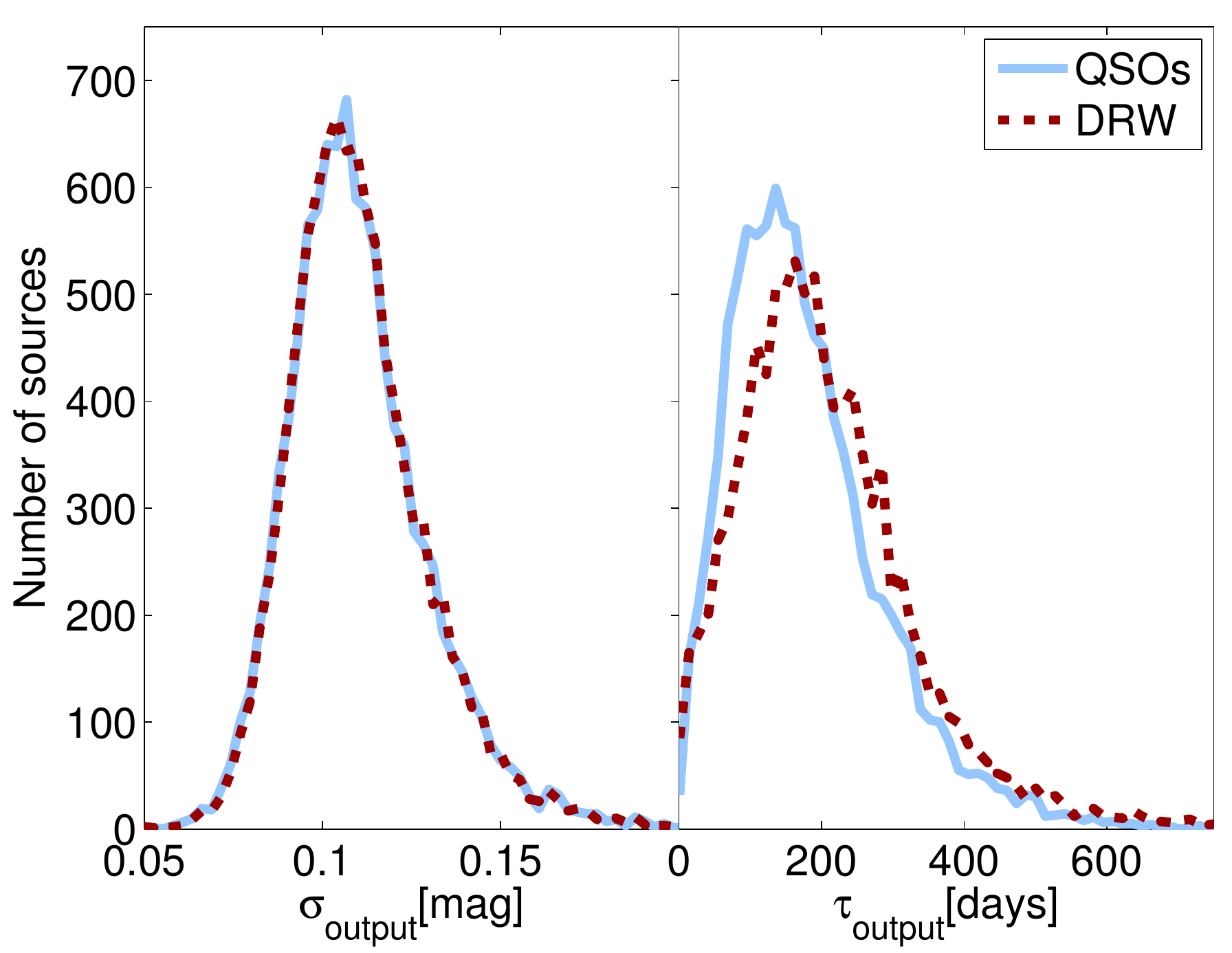}
\caption{Histograms of the best-fit parameters $\sigma$ (left) and $\tau$ (right) for 10,000 quasars (solid curves) and for mock DRW light curves sampled at the same observation times as the quasars (dotted curves). The parameters of the DRW realizations were drawn from the distributions in \citet{2010ApJ...721.1014M}.}
\label{Fig:OutputDistribution}
\end{figure}

We emphasize, however, that the biases in the inferred DRW parameters do not alter our conclusions about the identified periodicity.  We guard against this by analysing mock pure-noise realizations with the exact same algorithm that we apply to the quasar light curves: in other words, our null-hypothesis of pure noise will suffer from a similar bias.    To show this more explicitly, in Fig.~\ref{Fig:OutputDistribution}, we show the distributions of the best-fit $\sigma$ (left panel) and $\tau$ (right panel) for the real quasars (solid line) and for DRW realizations (dotted line).  These distributions are clearly similar. This also means that the distributions from which the input parameters for the DRW realizations were drawn \citep{2010ApJ...721.1014M} must be representative of the population, since the bias in the estimation is common both for the simulated and the real data. This also justifies our choice to use the distributions from \citet{2010ApJ...721.1014M} as priors in the estimation of the DRW parameters.

Most importantly, as shown in Fig.~\ref{Fig:Pval}, the mock data consisting of pure noise cannot produce the periodicities observed in the sample of quasars. Therefore, we conclude that the population of periodic quasars we identified is statistically significant compared to the DRW model.

\subsection{Departures from DRW quasar variability}
\label{subsection:DerartureFromDRW}
We statistically detected significant periodicity in a population of quasars, compared to the DRW variability. Even though we adopted the most widely accepted model for optical variability \citep{2009ApJ...698..895K,2010ApJ...708..927K,2010ApJ...721.1014M}, it is possible that a different process may provide a better description of the variability and then the identified periodicity may be less significant. For instance, \citet{2013A&A...554A.137A} performed a Bayesian comparison of $\sim$20 different stochastic and deterministic models in a sizable sample of 6304 quasars from Stripe82. Their results indicate that for a large fraction ($\sim$25\%) of these quasars, a combined model of DRW plus a sine wave is favored over pure DRW noise. Additionally, in a small number of quasars (29 out of 6304), they found decisive evidence against stochastic variability and a sinusoidal model is strongly preferred. 

Some recent studies have also suggested that the DRW model may be a simplistic description of quasar variability \citep{2011ApJ...743L..12M,2013ApJ...765..106Z,2014MNRAS.439..703G,2016A&A...585A.129S,}. These studies have reported deviations from the DRW model mainly on short time-scales, ranging from a few days up to $\sim$100\,d. Prompted by the above findings, \citet{2014ApJ...788...33K} introduced the Continuous-time Auto-Regressive Moving Average (CARMA) models to capture the variability features in quasar light curves. These models offer an extension to the DRW model, since they include higher-order derivatives in the differential equation that describes a stochastic process\footnote{The DRW model is also known as Continuous Auto-Regressive CAR(1) model, and considers only the first order terms of the stochastic differential equation.} and therefore allow greater flexibility overall (for details, see \citealt{2014ApJ...788...33K} and G15 for a discussion of CARMA models and periodic variability).

We note, however, that these reported deviations from DRW may not affect our results significantly, since they typically occur outside of the temporal window we analysed. We restricted our search to periods longer than 60\,d and identified only 7 candidates
 with periods shorter than 300\,d. Nevertheless, more generally, it is worth reiterating here that our statistical findings depend on the underlying variability model, and if quasars are proven to follow a more complex stochastic process, our results will need to be validated taking into account the new variability model and calculating the false alarm probability using the new variability model as the null hypothesis for pure noise. 

\subsection{Preference for low mass-ratio SMBHBs}

We have shown that the quasars with significant periodic variability are consistent with a population of SMBHBs with a low mass ratio.   Fig. \ref{Fig:TresidenceDistribution} indicates that the model with $q=0.01$ is preferred over models with $q=0.1$ or $q=1$.  G15 detected a sample of 111 periodic quasars in CRTS and suggested that their findings are consistent with a population of equal-mass ($q=1$) SMBHBs, the evolution of which is dominated by the emission of gravitational waves.   In order to see if there is any discrepancy between these two samples, we computed the distribution of residence times for their sample for $q=0.01$.  The important point is that if the quasars are indeed SMBHBs with this low mass ratio, then none of them are in the GW regime, which modifies the expected period distributions.   This exercise reveals that the G15 sample also prefers a low mass ratio, similarly to the one identified here.

\begin{figure}
\includegraphics[height=7cm,width=8.5cm]{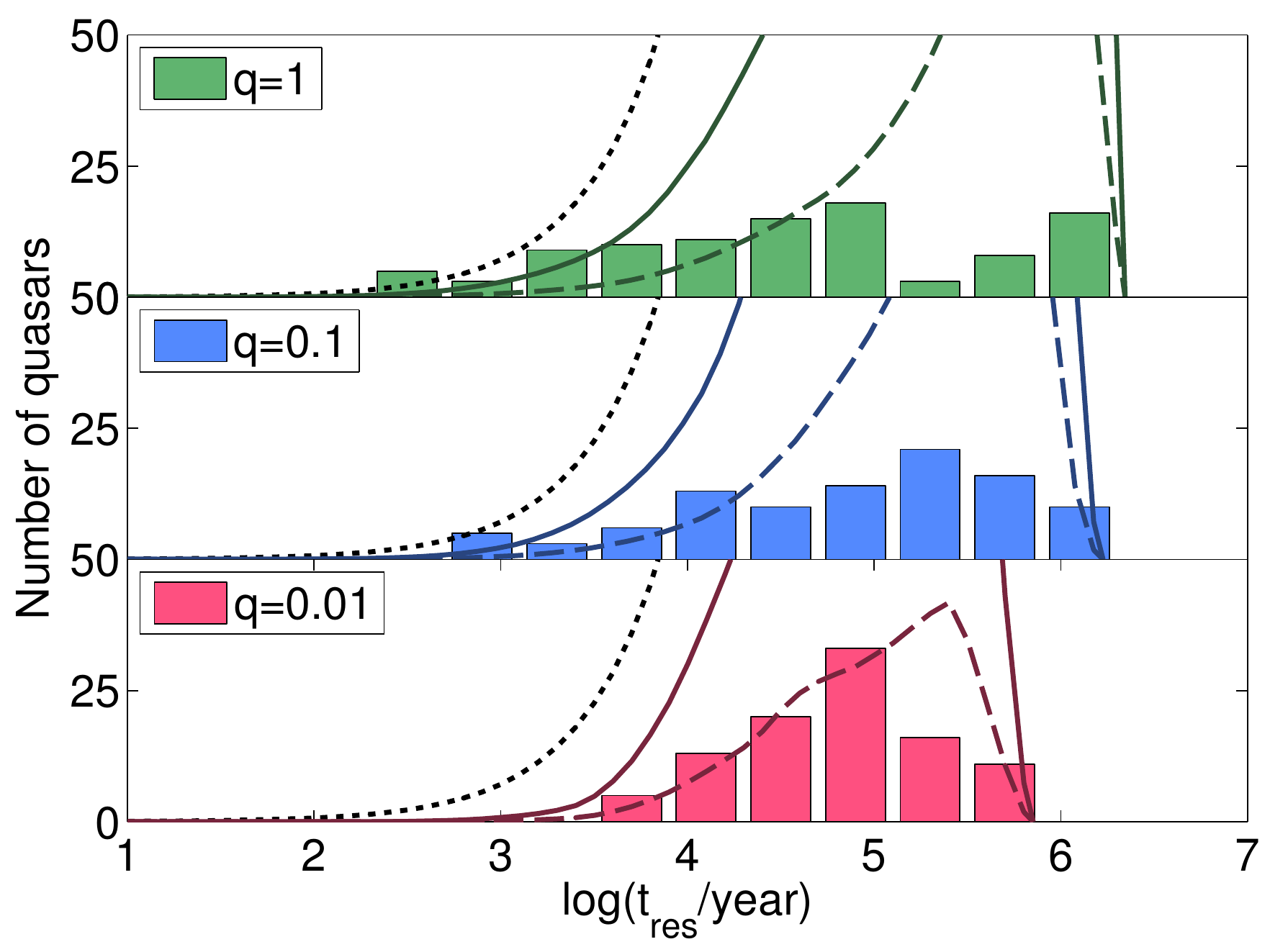}
\caption{Histograms of residence time inferred for 98 of the 111 SMBHB candidates identified by G15, assuming different mass ratios: $q=1, 0.1$ and $0.01$ (top, middle and bottom panel, respectively). 
The dotted curves show the distribution expected without accounting for observational limitations; the solid curves incorporate these observational constraints. The dashed curves are the same as the solid curves, scaled down by 25\%, which is the fraction of all quasars inferred to host SMBHB.}
\label{Fig:TresidenceDistributionGraham}
\end{figure}

More specifically, in Fig. \ref{Fig:TresidenceDistributionGraham}, we present the distribution of residence times for a subset (98 out of 111)\footnote{We considered only the quasars for which BH mass estimates were available, following G15, see their Fig.~9.} of the periodically varying quasars from G15 for mass ratios $q=1, 0.1$ and $0.01$ (top, middle and bottom panel, respectively). We also show, with dotted curves, the theoretically expected distribution from eq. (\ref{eq:Ntres}), for $N_{sample}$= 243,500. Given that G15 searched for periods between 400\,d and 6\,yr (at least 1.5 cycles within the 9-year baseline), and given the redshift, magnitude distributions of quasars analyzed in their sample, shown in Fig.~5 and Fig.~6 in G15, we can estimate the observable fraction $f_o(t_{\rm res})$ for this sample. The residence time distributions accounting for the finite baseline, expected from eq. (\ref{eq:Ne}), are shown by the solid curve in each panel. Moreover, as G15 pointed out, they identify $\sim$25\% of the theoretically expected quasars (which could be attributed to only a quarter of all quasars being activated by mergers). We scale the expected distribution by this factor to facilitate the comparison, shown by the dashed  curves. We see that the unequal-mass case ($q=0.01$) indeed fits the observed distributions better.

The figure above also reveals a discrepancy between the results from the two studies, in terms of the fraction of quasars that host a SMBHB. Our results indicate that all quasars may harbor a SMBHB, whereas the findings in G15 suggest that this fraction is $\sim$25\%. Nevertheless, the periodicity was identified in two distinct datasets using completely different search algorithms, making a direct comparison challenging. For instance, we note that there is a potential selection effect favoring the brighter quasars in the sample of G15. More specifically, if G15 had limited their analysis to the brighter end of the sample (e.g., quasars with mag$<$19), they would have identified 104 candidates in a sample of $\sim$78,000 quasars (see Fig.~5 and Table~2 in G15), resulting in a similar fraction of quasars hosting SMBHBs as in our sample. Therefore, it is possible that the decreased occurrence rate of periodic quasars in the sample of G15, and thus the discrepancy in the two samples, can be explained due to the limited photometric accuracy of CRTS at fainter magnitudes compared to PTF. We will address the question of the fraction of quasars with SMBHBs in a future study.

Mass ratios of SMBHBs have been discussed for a handful of individual candidates in the past, and have been inferred to be low. For instance, the variability of the well-studied SMBHB candidate OJ287 can be explained under the assumption of a very massive ($\sim10^{10} M_{\odot}$) primary BH with a  $\sim$100 times smaller secondary BH on a highly eccentric orbit, perpendicular to the accretion disc of the primary BH \citep{2012MNRAS.427...77V}. Furthermore, \citet{2015Natur.525..351D} proposed that the observed periodic variability of the recently identified SMBHB candidate PG1302-102 may be due to relativistic boosting of a steady accretion flow onto the rapidly moving secondary BH in a highly unequal-mass system, with $q\lsim 0.05$ favoured.
An unequal mass binary (with $q=0.026$) has also been proposed for the quasar Mrk 231 \citep{2015ApJ...809..117Y}. This candidate was identified from its peculiar spectrum, which the above authors interpreted as a signature of a circumbinary disc with a central cavity, cleared by the motion of the secondary BH (although we note that for such a low mass ratio, a cavity may not be present; \citealt{2015arXiv151205788D,2015MNRAS.447L..80F}). The recently detected SMBHB candidate SDSS J0159+0105 \citep{2015arXiv151208730Z}, which shows two periodic components in the optical variability, also requires the existence of an unequal-mass binary with $0.05<q<0.8$.
 
The above results raise the intriguing possibility that SMBHBs with low mass ratios may be more common than near-equal-mass binaries. 
This is not entirely surprising, in light of cosmological galaxy merger models, which predict that at moderate redshift (e.g., $z<10$), most BH pairs will have unequal masses \citep{2003ApJ...582..559V}. We note that unequal-mass binaries are fairly unexplored from a theoretical point of view. In particular, this low mass-ratio regime is typically ignored in hydrodynamical simulations of binaries with circumbinary gas discs, with only a handful of exceptions (\citealt{2013MNRAS.436.2997D, 2015MNRAS.447L..80F, 2015arXiv151205788D}; see also \citealt{ShiKrolik2015} for simulations of a $q=0.1$ binary).  

The above conclusion should be considered with caution, since it could be the result of a strong selection effect. Both the algorithm developed here and the one employed by G15 are optimized for detecting sinusoidal variations. Hence, they may be preferentially sensitive to binaries with significant Doppler boosting (e.g., see \citealt{2015Natur.525..351D}), which is more prominent for unequal-mass binaries. More nearly equal-mass binaries may produce more ``bursty" light curves, as a result of periodic fluctuations in the accretion rate onto the BHs \citep{2007PASJ...59..427H,2008ApJ...672...83M,2012A&A...545A.127R,2012ApJ...755...51N, 2013MNRAS.436.2997D, 2014ApJ...783..134F,2015MNRAS.447L..80F,ShiKrolik2015, 2015arXiv151205788D}. The latter are likely to remain undetected with the current searches for periodicity.

\subsection{Extended light curves}
\label{Subsection:ExtendedLC}
For the candidates identified in PTF, we extended the light curves adding points from iPTF and CRTS. The iPTF light curves are practically an extension of PTF, since the data are obtained with the same telescope and filter, following a similar observing strategy as in PTF. On the contrary, CRTS is a distinct survey; CRTS covers up to $\sim$2500\,$\rm deg{^2}$ per night, with 4 exposures per visit, separated by 10\,min, over 21 nights per lunation. The observations are obtained in unfiltered visual light and the depth of the survey is typically limited compared to PTF (as mentioned in \S~\ref{section:Introduction}, several telescopes are used for the survey, resulting in different limiting magnitudes for different parts of the sky). For simplicity, here we consider these additional data only for validation of the periodic candidates selected from PTF alone, rather than attempting to identify periodic candidates from a full combined dataset.\footnote{We intend to search for periodicity in the composite light curves from PTF, iPTF and CRTS in a future paper, since such a search will offer the combined benefits of the two surveys, e.g., the long CRTS baseline and the high temporal resolution from PTF and iPTF.}

From the extended light curves, we calculated the P-value of the period identified in the PTF light curves. We emphasize, however, that this P-value constitutes only a rough estimate. First, the photometric accuracy of the CRTS dataset is reduced compared to PTF. In several cases, the photometric errors, are comparable to the amplitude of the identified periodicity. Furthermore, despite our efforts to calibrate the PTF-iPTF photometry in order to match the CRTS, systematic effects (e.g., due to different photometric systems) are likely to be present. Third, our full analysis, in the case of the PTF data above, has demonstrated that the P-values can not be directly interpreted as true false alarm probabilities. For instance, when our analysis was applied to random DRW realizations, the number of false positives we identified was by a factor of two higher than the theoretically expected number given the trial factors (see Fig.~\ref{Fig:Pval} and the related discussion above).

Therefore, at this step, we only excluded sources with high P-values (P-value$\geq$1\%). This tentative significance threshold is justified, given that we only analyzed the selected 50 candidates and we restricted the search into one frequency bin (we calculate the significance of the previously identified period, not every possible peak in the periodogram). The number of trial factors is thus reduced compared to the initial PTF search. However, given the sparse PTF light curves from which the periodicity was selected and the fact that the extended data do not always provide a conclusive answer regarding the periodicity, it is crucial that the identified SMBHB candidates are further monitored with similar photometric precision in order to confirm that the periodicity persists for several cycles.

\begin{figure}
\includegraphics[height=7cm,width=8.5cm]{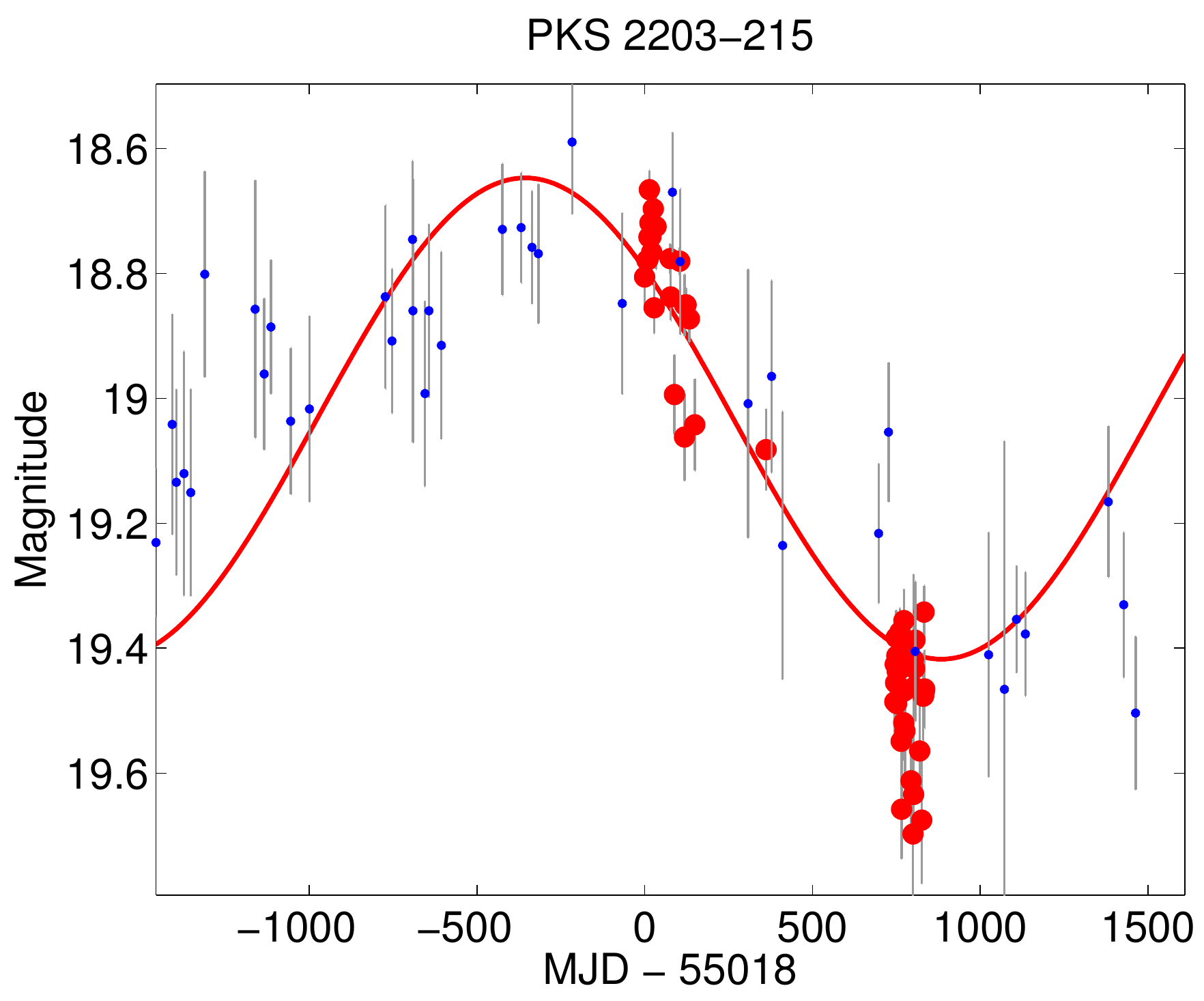}
\caption{The composite light curve of SMBHB candidate PKS 2203-103. Red points indicate PTF observations, blue points observations from CRTS and the red line shows the best fit sinusoid. A sinusoid with period 5 times longer than the one identified in PTF provides a better fit to the combined PTF and CRTS data.}
\label{Fig:PKS2203}
\end{figure}

We note that, in a few cases, even though the P-value from the extended light curve is below our threshold, the folded light curve looks inconsistent with the stated period. A possible explanation for this is that we detect higher harmonics of a true periodic signal with a longer period than the one identified in PTF. It is well known that the periodogram of a sine wave convolved with the periodogram of sampling function can introduce peaks at non-trivial frequencies \citep{1987AJ.....93..968R}. We stress that the aliasing peaks from the sampling pattern alone are taken into account in our analysis by generating mock light curves with the exact same time stamps as in the observed data. In Fig.~\ref{Fig:PKS2203}, we illustrate an example of this effect for quasar PKS 2203-215. The best fit sinusoid corresponding to the most significant periodogram peak within PTF has a period of 497\,d (see Appendix), whereas if we consider the extended light curve the period of the best fit sine wave is 2480\,d (5 times longer than the period in PTF). This low frequency peak is significant (P-value$<$1/250,000), if we examine the entire frequency range allowed by the composite light curve instead of limiting the search within the PTF baseline. We have also seen this effect in the periodogram of PG1302 \citep{2015MNRAS.454L..21C}. The periodogram shows a significant peak at ~300\,d, which coincides with a peak from a noiseless sinusoid with period of 1884\,d sampled at the observed times. However, if PG1302 was identified in a survey with a shorter baseline, comparable to PTF, the peak at 300\,d would be identified as the actual periodicity of the source. Although, falsely identifying a harmonic of the actual period can affect the interpretation of the population of quasars, from visual inspection of the phase folded light curves in our sample, we conclude that this effect is not dominant.

\subsection{Comparison with periodic quasars in CRTS}
\label{Subsection:CRTS}
G15 analysed a large sample of ($\sim 250,000$) quasars from CRTS. Although CRTS is an all-sky survey, most of the spectroscopically confirmed quasars are from SDSS and are spread mostly over the Northern Hemisphere (see Fig.~\ref{Fig:QuasarMap}). Therefore, there is a significant overlap between the quasars analysed in the two samples (75\% of the quasars in our sample were also included in the sample of G15). Despite this overlap, we do not identify any common periodic candidates. Here we clarify the reasons for this.

Among the 111 SMBHB candidates in the G15 sample, 101 were in the initial catalogue we used to extract sources from PTF. We examined a more recent version of the HMQ, and we did not include any low-luminosity AGN and blazars in our sample, since it is uncertain whether the DRW model can describe the variability of  these sources. As a result, of the 101 objects, only 77 were covered in PTF, and only 15 of these were consistent with our minimum requirement for 50 nights and were included in our final sample. 
The baseline of the PTF light curves is shorter than the periods identified by G15 for all of these 15 candidates. Therefore, it would be impossible to identify any of these objects as periodic in the sample we analysed.

Likewise, the fact that none of our candidates were identified in G15 is unsurprising. A large fraction of the sources we identified (28 out of 50)
 have periods below the 400\,d cutoff imposed by G15. Also the majority of the candidates are too faint for CRTS. There are only 7 out of 111 candidates with magnitude fainter than 19 mag in G15, even though the vast majority of the sources they analysed is below this magnitude. 
As mentioned in \S~\ref{Subsection:ExtendedLC}, the measurement uncertainty of the CRTS data points in several cases is comparable to the amplitudes of the sinusoids in the PTF sample. This is also obvious from the composite light curves in the Appendix.

\subsection{Periodic variability in quasars}
\label{subsection:PeriodicVariabilityAlternatives}

We have equated the observed optical periods with the (redshifted) orbital period of SMBHBs. According to several hydrodynamical simulations, the mass accretion rate onto the BHs is modulated at the orbital period of the binary \citep{2007PASJ...59..427H,2008ApJ...672...83M,2012ApJ...755...51N,2012A&A...545A.127R,2013MNRAS.436.2997D,2014ApJ...783..134F,2014PhRvD..89f4060G}.  In particular, our finding that low mass ratios are favoured support the identification of the optical period as the redshifted orbital period \citep{2013MNRAS.436.2997D,2014ApJ...783..134F,2015arXiv151205788D}.  A different scenario involves Doppler boosting of the emission arising in the mini-disc around the secondary BH, as it orbits with relativistic velocities. \citealt{2015Natur.525..351D} proposed this model to explain the optical and UV variability of PG1302.  In this case, the optical and orbital periods would again coincide.

Although the above is reassuring, it is worth noting that the optical periodicity does not necessarily reflect the orbital period. Hydrodynamical simulations of SMBHBs with higher mass ratios ($q>0.3$) predict the existence of several periodic components in the variability \citep{2012A&A...545A.127R, Shi+12, 2013MNRAS.436.2997D,2014ApJ...783..134F}. For instance, \citet{2015MNRAS.452.2540D} associated the observed period of PG1302-102 with the longer orbital period of a lump in the lopsided accretion disc and predicted that the orbital period may be 5-8 shorter than the observed, although the analysis of the periodogram did not reveal additional peaks \citep{2015MNRAS.454L..21C}. More generally, periodic variability of quasars does not necessarily require the presence of a binary: quasi-periodic modulations can arise around a single BH due to, e.g., Lense-Thiring precession, a warped accretion disc, or the precession of a jet (see G15 for an extended discussion).

\subsection{Selection Effects}
\label{subsection:selectionEffects}
\begin{figure}
\includegraphics[height=7cm,width=8.5cm]{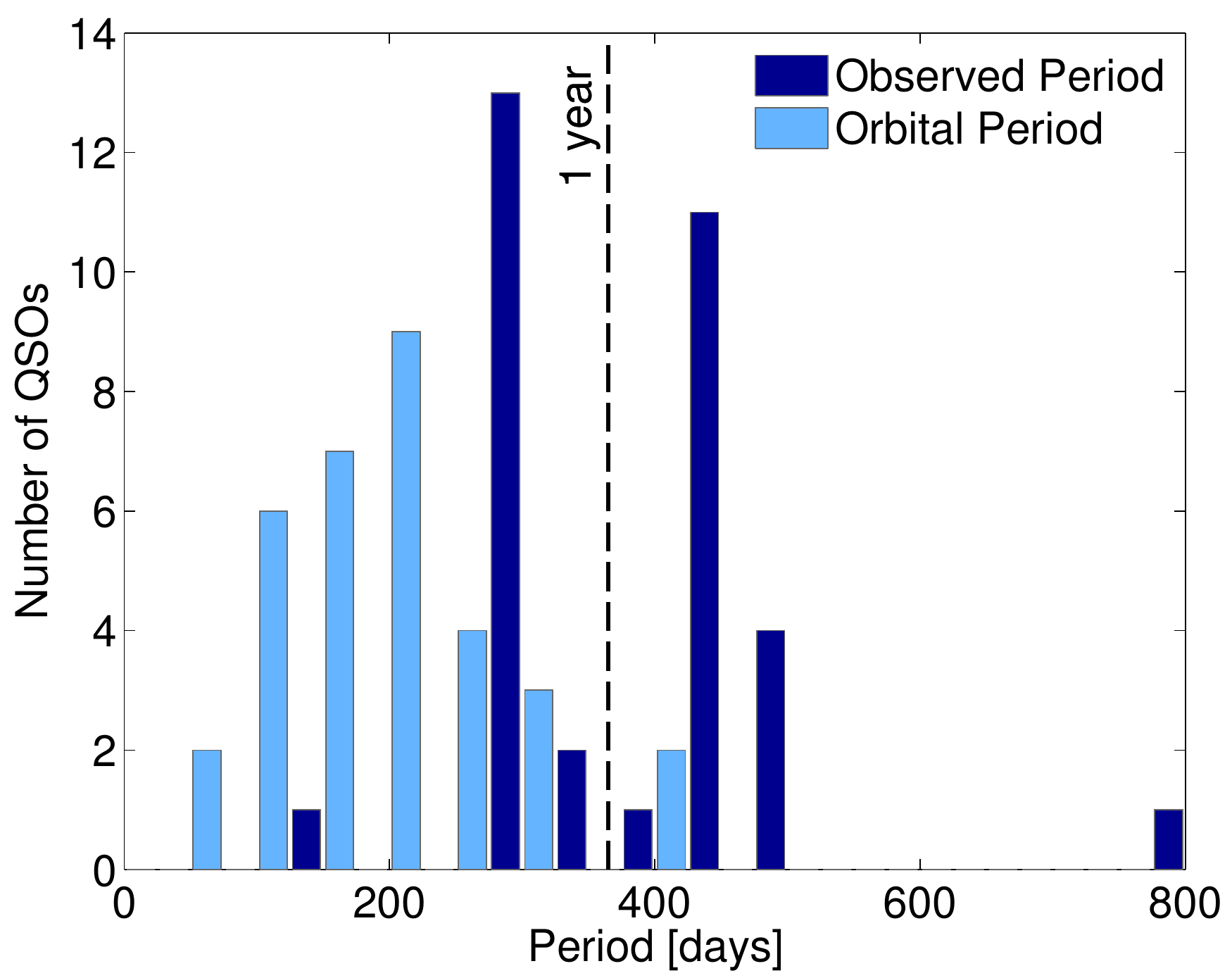}
\caption{Histogram of observed periods (in dark blue) and redshifted orbital periods (in light blue). The dashed line indicates a period of 1\,yr; a clear deficit of periods is observed at this timescale.}
\label{Fig:PeriodDistribution}
\end{figure}

We have analyzed a highly heterogeneous sample of light curves. Here, we explore the role of some potential selection effects and biases among the periodic sources we have identified, which are likely present in our sample. For instance, in Fig.~\ref{Fig:PeriodDistribution}, we show the histograms of the observed period, $P_{obs}$, in dark blue and the orbital period $P_{orb}= \left(1+z\right)^{-1}P_{obs}$, in light blue, for the periodic quasars in our sample. In the histogram of the observed period, we notice two prominent peaks at $\sim$300\,d and $\sim$450\,d, while a clear deficit of candidates with periods of $\sim$1 yr is also present. A possible explanation is that our algorithm is less sensitive at specific timescales (e.g., for periods of $\sim$1\,yr or for long periods $>500$\,d) and we miss some genuine periodic sources, leading to the observed scarcity of specific frequencies. Another explanation is that the peaks at $\sim$300\,d and $\sim$450\,d may reflect the identification of some aliasing peaks from real periodic signals with periods longer than the PTF baseline. An example of this is the periodogram of quasar PG1302-102; in our previous work, we have explicitly shown that at the particular time sampling of that source, a genuine 5.2\,yr sinusoid would introduce strong peaks at $\sim$300\,d and $\sim$500\,d in the periodogram (see Figure 1 in \citealt{2015MNRAS.454L..21C}). It is likely that similar aliasing results in the misidentification of some of our periods, as already shown for one source in Fig.~\ref{Fig:PKS2203}.

\begin{figure}
\includegraphics[height=7cm,width=8.5cm]{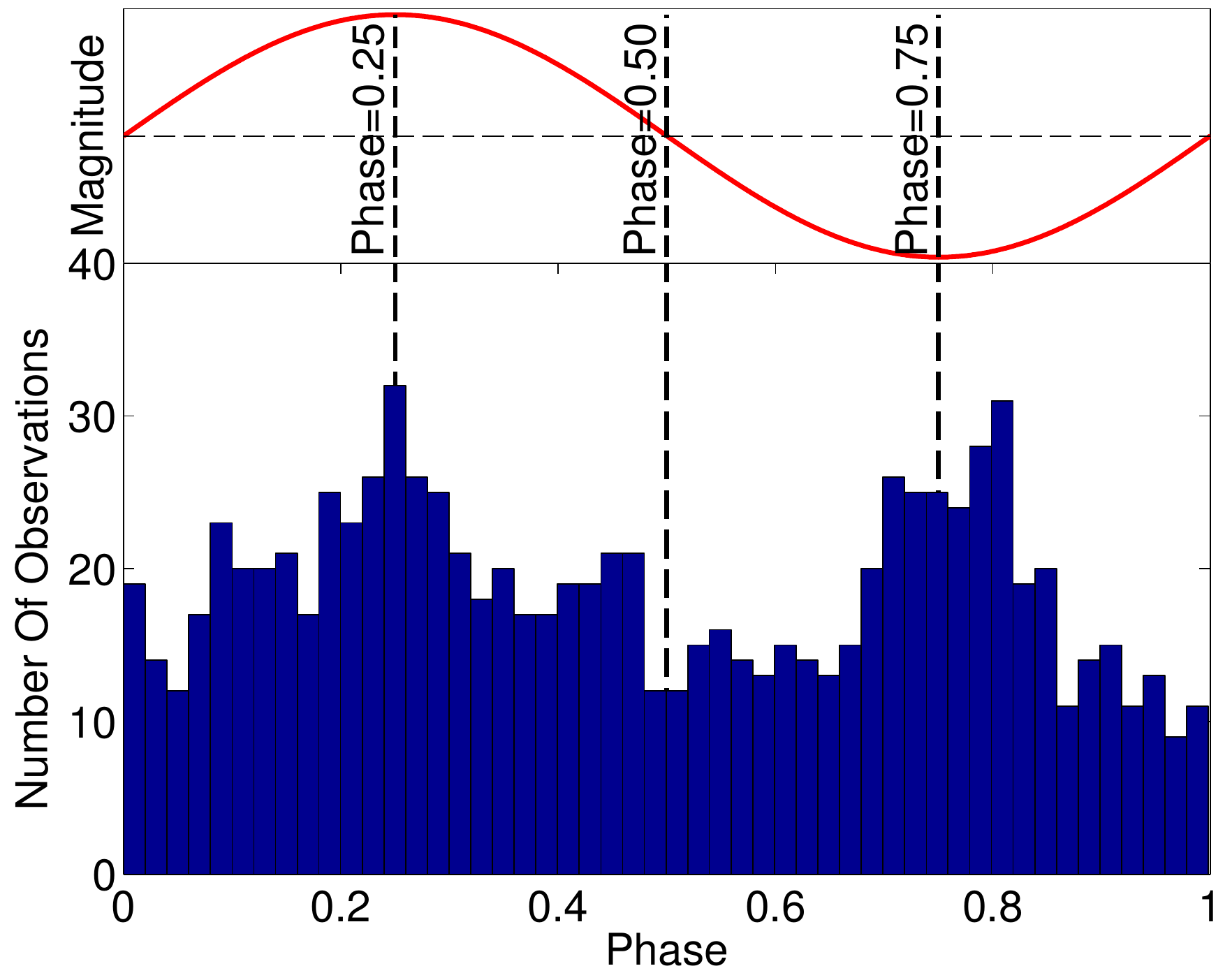}
\caption{Histogram of the phase for all the data points in the sample of quasars with significant periodicity. The histogram shows deviations from the expected uniform distribution. In the top panel, we illustrate a phase folded sinusoid.}
\label{Fig:PhaseDistribution}
\end{figure}

Another possible selection effect is illustrated in Fig.~\ref{Fig:PhaseDistribution}, in which we present the histogram of the phase of all the observations in the light curves of our periodic sample.
%, i.e. the distribution of the points shown in the right panels of Fig.~\ref{Fig:LightCurves}
 We notice that there is an excess of observations with phases of $\sim$0.25 and $\sim$0.75 (i.e. at the maxima and minima of the sinusoid, respectively). Although the phase distribution should be uniform in the ideal case of densely and uniformly sampled light curves, the observed distribution could be explained by the significant deviations from the ideal sampling of the PTF light curves (Fig~\ref{Fig:NnightsVsBaseline}). For instance, we consider how the two extreme cases of unfortunate sampling, which may be present in our data, can affect the phase distribution: (1) If a true sinusoid is sampled only around the mean (phases 0, 0.5 and 1), the Lomb-Scargle periodogram would fail to detect a peak, and the periodicity would be completely missed, leading to a scarcity of the relevant phases in the overall phase distribution, and (2) If only the minima and the maxima of the sinusoid were sampled, the power of the periodogram peak would be significant and the likelihood of detection is increased, leading to a potential excess of observations with phases $\sim$0.25 and $\sim$0.75. The combination of these effects would, therefore, translate to a deficit of phases near 0, 0.5, and 1, and an excess of phases near 0.25 and 0.75, matching the phase distribution we observe.

On the other hand, \citet{2016arXiv160602620V} generated mock DRW realizations to reassess the statistical significance of PG1302. They found that false positives, with periods of $\sim$1\,yr or less, were detected at an increased rate in cases when the simulated maxima and minima aligned with the quasi-periodic sampling. If a significant number of false detections was present in our sample due to the above effect, it could result in a phase distribution similar to Fig.~\ref{Fig:PhaseDistribution}. However, this is unlikely the case in our sample, because we have included the effects of the irregular sampling in the calculation of the false alarm probability. If the DRW model in combination with the irregular sampling could mimic the periodic variability, it would create peaks with significant power in the periodogram and thus the quasar would not be identified as periodic. Additionally, when we simulated DRW light curves and conducted the periodogram analysis over the entire sample (e.g., see Fig~\ref{Fig:Pval}), we did not detect false periodicity that shows clustering around the maxima and minima. In detail, among the 7 DRW realizations that were above our detection threshold, 6 had long periods and were not consistent with our requirement for a minimum of 1.5 cycles within the data. The only case that was consistent with all our requirements for detection and would be identified as a short period quasar did not show clustering of the observations around the minima and maxima of the best fit sinusoid. Finally, we note that the sampling of the PTF light curves is irregular and not quasi-periodic, like the sampling of PG1302.
 
In order to understand the underlying population of SMBHBs, it is important to understand and correct for the selection effects and biases discussed above. While this is beyond the scope of the present paper, these effects will be carefully explored in a separate follow-up study.

\section{Conclusions}
\label{section:Conclusions}
We developed a statistical search to identify periodicity in the optical variability of quasars. We analysed the data of 35,383 spectroscopically confirmed quasars from the Palomar Transient Factory, and assessed the statistical significance of our findings by simulating stochastic time series that mimic the quasar variability, which we modeled as a damped random walk process. Our main conclusions are the following:

\begin{itemize}
\item We detected a statistically significant population of 50 periodic quasars with at least 1.5 cycles within the PTF baseline. Of these periods, 33 remain significant even with the re-analysis of light curves including data from iPTF and CRTS. This identified population is significant as an ensemble of sources rather than as individual detections.
\item The periodic sources are characterized with typically short periods of a few hundred days and fainter magnitudes compared to the previous study in CRTS (G15). Our findings reflect the better photometry at fainter magnitudes and the higher temporal resolution of the PTF light curves compared to CRTS.
\item If the identified periodicities correspond to the redshifted orbital periods of SMBHBs, then the period-distribution of this population favors SMBHBs with a low mass ratio ($q\approx 0.01$).
\item We found a similar conclusion about the mass ratio for the population of periodic quasars from G15, which was identified in a separate sample with a different selection algorithm. Unequal masses ($q=0.01$) provide a better fit to their observed period-distributions, as well.
\end{itemize}

In this paper, we identified quasars with short-period optical variability as promising candidates for SMBHBs. However, given the relatively short PTF baseline and the limited photometric accuracy of the extended data from CRTS, it is essential to further monitor the candidates to confirm that the periodicity persists for several cycles. It is also important to search for additional signatures for the binary nature of the sources (e.g., wiggled radio jets, signs of relativistic boosting, etc). We also note that we have assessed the significance of the detected periodicity, compared to a null-hypothesis assuming that quasar variability is described by the DRW model. Even though this model has proven successful as a general description of quasar variability, the significance of our findings would need to be validated within any other variability model.

\section*{ACKNOWLEDGEMENTS}
We thank Jules Halpern and Daniel D'Orazio for useful discussions, Ying Zu for assisting with the Javelin code, and Marcel Agueros for providing the light curves and the stellar rotation periods from the Praesepe cluster for an initial test to our Lomb-Scargle algorithm. 

For our analysis, we made use of the High Performace Computing cluster Yeti at Columbia University. Additional resources supporting this work were provided by the NASA High-End Computing (HEC) Program through the NASA Advanced Supercomputing (NAS) Division at Ames Research Center. This work was supported in part by NASA grants NNX11AE05G and NNX15AB19G (to ZH). MJG acknowlegdes support from NSF grant AST-1518308. 

This paper is based on observations obtained with the Samuel Oschin Telescope as part of the Palomar Transient Factory project, a scientific collaboration between the California Institute of Technology, Columbia University, Las Cumbres Observatory, the Lawrence Berkeley National Laboratory, the National Energy Research Scientific Computing Center, the University of Oxford, and the Weizmann Institute of Science.   

The CSS survey is funded by the National Aeronautics and Space Administration under Grant No. NNG05GF22G issued through the Science Mission Directorate Near-Earth Objects Observations Program.  The CRTS survey is supported by the U.S.~National Science Foundation under grants AST-0909182 and AST-1313422.

This paper was approved for publication by the LIGO Scientific
Collaboration, with document number LIGO-P1600113.

\bibliography{PeriodicQSOsPTF}

\clearpage
\begin{table*}
\caption{Observational properties of quasars with significant periodicity.}
\centering
\begin{tabular}{|c| c| c| c| c| c| c| c| c|}
\hline
Name											&Ra						&Dec					&z		&R-mag	&$\log\left(M/M_{\odot}\right)$ &X-ray/radio catalogue\\
\hline\hline
\textbf{UM 269}$^{*}$ 									&10.8322440			&0.8542910			&0.308			&17.8		&(8.4079) 8.5337$\pm$0.3138$^{**}$	&3XMM(X),1RXS(X), FIRST(R) \\
\textbf{SDSS J005158.83-002054.1} 		&12.9951310		&-0.3483820		&1.047			&19.3		&8.783$^{***}$		&\\
\textbf{SDSS J005453.30-003258.3}		&13.7221040		&-0.0.5495320	&0.961			&19.8		&(7.9341)	8.6412$\pm$0.3094&\\
SDSS J023050.06+005843.1 		&37.7086065 		&0.9786625			&1.447			&17.9		&(10.02) 9.2635$\pm$0.2841&\\
\textbf{SDSS J024442.77-004223.2} 		&41.1782445		&-0.7064567		&0.628			&19.7		&7.926		&\\
\textbf{2QZ J095344.7+010354}     		&148.4366915		&1.0651954			&0.994			&19.3		&(8.3814) 8.7953$\pm$0.3032	&\\
SDSS J104648.62+513912.6 		&161.7026015		&51.6535127		&0.587			&18.7		&(8.853) 8.6646$\pm$0.3085&\\
SDSS J132815.49+361715.9 		&202.0645650		&36.2877730		&1.088			&17			&(9.6039) 9.4686$\pm$0.2757&1RXS(X), 2RXP(X), 1WGA(X)\\
\textbf{SDSS J133254.51+371735.5} 		&203.2271569		&37.2932211		&2.46			&19.3		&(8.7384) 9.3584$\pm$0.2802&\\
\textbf{SDSS J133840.66+315936.4} 		&204.6694431		&31.9934505		&2.944			&19.6		&(9.4063) 9.3868$\pm$0.2791&\\
\textbf{SDSS J134553.57+334336.0} 		&206.4732442		&33.7266720		&0.886			&19.3		&(9.0644)	 8.7308$\pm$0.3058& FIRST(R)\\
\textbf{SDSS J134556.16+343224.5} 		&206.4840056		&34.5401540		&0.874			&19.7		&(8.6547) 8.6152$\pm$0.3105&\\
SDSS J140929.76+535930.2 		&212.3740072		&53.9917232		&0.863			&19			&(8.5757)	 8.7972$\pm$0.3031&\\
\textbf{SDSS J141004.41+334945.5} 		&212.5184030		&33.8293160		&0.63			&18.2		&8.962		&\\
\textbf{SDSS J141244.09+421257.6} 		&213.1837413		&42.2160210		&0.805			&18.7		&(9.6913)	 8.8397$\pm$0.3013& CXOX(X), FIRST(R)\\
\textbf{SDSS J142339.44+471240.8} 		&215.9143459		&47.2113337		&1.24			&19.7		&(8.5389) 8.8182$\pm$0.3022&\\
TEX 1428+370				            	&217.6691030		&36.8177490		&0.566			&17.2		&(8.5293) 9.0492$\pm$0.2928& CXOX(X), FIRST(R), NVSS(R) \\
SDSS J143637.44+090155.5 		&219.1560102		&9.0320841			&0.568			&18.7		&(7.706) 8.6461$\pm$0.3092&\\
SDSS J145713.26+140334.1 		&224.3052882		&14.0594797		&2.926			&19.3		&(9.2106) 9.4641$\pm$0.2759&\\
SDSS J145859.07+153144.7 		&224.7461508		&15.5290922		&2.898			&17.2		&(9.9355)	 10.0252$\pm$0.2530&\\
\textbf{SDSS J150900.70+175114.3} 		&227.2529518		&117.8539974		&0.742			&19.6		&(8.6112)	 8.5516$\pm$0.3131&\\
FBQS J150911.2+215508    		&227.2968330		&21.919110			&0.438			&16.4		&(8.5381)	 9.1195$\pm$0.2899& FIRST(R), NVSS(R)\\
SDSS J150912.07+204004.6 		&227.3003130		&20.6679610		&0.339			&18.7		&(7.8945)	 8.3484$\pm$0.3214& 1RXS(X) \\
SDSS J151053.24+240943.3 		&227.7218528		&24.1620337		&0.807			&19.3		&(8.3276) 8.6791$\pm$0.3079&\\
SDSS J151243.67+195845.1 		&228.1819980		&19.9792200		&0.808			&18.1		&(8.8351)	 9.0038$\pm$0.2947&\\
\textbf{SDSS J151646.10+221724.7} 		&229.1921240		&22.2902190		&0.599			&18.9		&(8.6375) 8.6219$\pm$0.3102	&\\
\textbf{SDSS J152739.97+413234.6} 		&231.9165420		&41.5429560		&1.014			&17.8		&(9.232) 9.0332$\pm$0.2935		&1WGA(X) \\
SDSS J152903.11+223623.8 		&232.2629760		&22.6066240		&0.506			&18.5		&(8.4734) 8.6349$\pm$0.3097	&\\
\textbf{SDSS J153051.79+503440.1} 		&232.7158316		&50.5778224		&0.928			&19.4		&(8.8238)	 8.7296$\pm$0.3058&\\
\textbf{SDSS J153251.06+335852.2} 		&233.2127605		&33.9811873		&1.889			&18			&8.964		&\\
\textbf{PDS 898}						               	&234.2429720		&34.5304110		&0.886			&18			&(8.737) 8.8062$\pm$0.3027		&\\
SDSS J155308.65+501436.5 		&238.2860757		&550.2434900		&2.774			&19.4		&(9.2423) 9.4045$\pm$0.2783&\\
SDSS J160322.68+200535.2 		&240.8445362		&20.0931253		&2.337			&19			&(9.872) 9.4081$\pm$0.2782	&\\
\textbf{SDSS J160454.57+315733.5} 		&241.2273914		&31.9593278		&3.159			&20			&(9.2593) 9.3223$\pm$0.2817&\\
SDSS J162634.15+325032.6 		&246.6422999		&32.8423962		&0.858			&19.9		&(8.4876)	 8.5510$\pm$0.3131&\\
\textbf{SDSS J170942.58+342316.2} 		&257.4274310		&34.387850			&1.734			&19.4		&9.228		&CXOX \\
\textbf{SDSS J171122.67+342658.9} 		&257.8444609		&34.4497191		&2.132			&20.2		&(8.7648)	 9.0275$\pm$0.2937&\\
\textbf{SDSS J171617.49+341553.3} 		&259.0728780 	&34.2648190		&1.149			&18.5		&(9.1003) 9.0959$\pm$0.2909&\\
\textbf{SDSS J171909.93+344001.3} 		&259.7914157		&34.6670451		&2.155			&18.2		&(9.3335) 9.5742$\pm$0.2714&\\
\textbf{SDSS J212939.60+004845.5} 		&322.4150072		&0.8126513			&2.266			&19.8		&(9.2628)	 9.1732$\pm$0.2878&\\
\textbf{SDSS J214036.77+005210.1}		&325.1532346		&0.8694920			&0.92			&20.2		&(8.8142) 8.5087$\pm$0.3148	&\\
\textbf{SDSS J214225.29+001643.2} 		&325.6054161		&0.2786872			&1.26			&19.5		&(8.497) 8.7333$\pm$0.3057		&\\
\textbf{SDSS J214357.98+003349.6}		&325.9916181		&0.5638052			&2.338			&20.2		&9.355		&\\
\textbf{PKS 2203-215}				            	&331.6725000		&-21.3277778		&0.577			&18.8		&(8.907) 8.6280$\pm$0.3100	& XMMSL(X),1RXS(X), NVSS(R)  \\
SDSS J224008.39+263928.4 		&340.0349874		&26.6579024		&2.827			&19			&(9.2615) 9.5240$\pm$0.2735&\\
\textbf{SDSS J231733.66+001128.3} 		&349.3902730		&0.1912120			&0.841			&18.2		&8.898		&3XMM(X), 1RXS(X) \\
\textbf{SDSS J232135.73+173916.5} 		&350.3988982		&17.6546025		&0.842			&19.4		&(9.1391) 8.6755$\pm$0.3080	& 1SXPS(X) \\
\textbf{2QZ J235800.2-281429}		    	&359.5011112		&-28.2413889		&1.598			&19.1		&(8.9538) 9.1415$\pm$0.2891	&\\
\textbf{SDSS J235928.99+170426.9} 		&359.8708046		&17.0741612		&0.714			&18.6		&(8.995) 8.8004$\pm$0.3029		&\\
\textbf{SDSS J235958.72+003345.3} 		&359.9946880		&0.5625920			&1.694			&19.1		&9.076		&\\
\hline
\end{tabular}
\label{Table:CandidatesQuasarProperties}
\begin{flushleft}
$^{*}$We emphasize with bold, the candidates that remain significant after the re-analysis of the composite light curves.\\
$^{**}$ The BH mass in the parenthesis is the mass drawn from the distribution in eq.~(\ref{eq:BHMass}); The mean and the standard deviation of the distribution are also shown.\\
$^{***}$ BH mass without a parenthesis is taken from the catalog in \citet{2008ApJ...680..169S}, measured from the width of broad lines in quasar spectra.
\end{flushleft}

\end{table*}

\begin{table*}
\caption{SMBHB properties}
\label{Table:BinaryProperties}
\centering
\begin{tabular}{|c| c| c| c| c| c| c| c| c|c|}
\hline
Name		&P-value$^*$	&Observed 		&Variability				&Separation		&Angular  		&$\Delta$t	&	$t_{insp}$		&	$t_{insp}$ \\
				&					&Period	  		&Fraction$^{**}$				&						&Separation		&					&	($q=1$)		&	($q=0.01$) \\
				&					&[days]		  		&[$\%$]					& [mpc]			&[$\mu$as]		&[ns]			&	[years] 			&	[years] \\
\hline\hline
\textbf{UM 269}                  					&	5.4$\times 10^{-4}$		&490.5			&16.0	&3.13		&0.67		&0.044		&1.3$\times 10^4$		&	3.3$\times 10^5$\\
\textbf{SDSS J005158.83-002054.1} 	& 1.5$\times 10^{-3}$		&443.1		 	&9.6		&2.89		&0.35		&0.035		&7.4$\times 10^2$		&	1.9$\times 10^4$\\
\textbf{SDSS J005453.30-003258.3}	& 1.2$\times 10^{-3}$ 		&444.7			&14.4	&1.55		&0.19		&0.002		&2.2$\times 10^4$		&	5.6$\times 10^5$\\
SDSS J023050.06+005843.1 	& 1.5$\times 10^{-2}$		&445.2			&6.9		&6.65		&0.77		&2.570		&4.0$\times 10^0$		&	1.0$\times 10^2$\\
\textbf{SDSS J024442.77-004223.2} 	&	8.0$\times 10^{-6}$		&452.9			&10.5	&1.77		&0.25		&0.003		&3.9$\times 10^4$		&	9.8$\times 10^5$\\
\textbf{2QZ J095344.7+010354}     	&	4.0$\times 10^{-4}$		&308.0			&14.2	&1.70		&0.22		&0.007		&1.4$\times 10^3$		&	3.5$\times 10^4$\\
SDSS J104648.62+513912.6 	&	2.8$\times 10^{-1}$		&575.0			&9.6		&4.30		&0.63		&0.111		&2.2$\times 10^3$		&	5.5$\times 10^4$\\
SDSS J132815.49+361715.9 	&	3.5$\times 10^{-2}$		&301.1			&5.9		&4.14		&0.49		&0.685		&1.1$\times 10^1$		&	2.8$\times 10^2$\\
\textbf{SDSS J133254.51+371735.5} 	&	4.9$\times 10^{-3}$		&300.2			&17.0	&1.52		&0.18		&0.008		&7.7$\times 10^1$		&	1.9$\times 10^1$\\
\textbf{SDSS J133840.66+315936.4} 	&	5.9$\times 10^{-4}$		&301.3			&12.2	&2.33		&0.29		&0.077		&4.2$\times 10^0$		&	1.1$\times 10^2$\\
\textbf{SDSS J134553.57+334336.0} 	&	1.8$\times 10^{-3}$		&796.6			&19.0	&5.60		&0.70		&0.159		&1.5$\times 10^3$		&	3.8$\times 10^4$\\
\textbf{SDSS J134556.16+343224.5} 	&	2.0$\times 10^{-5}$		&400.1			&24.2	&2.60		&0.33		&0.027		&1.2$\times 10^3$		&	3.0$\times 10^4$\\
SDSS J140929.76+535930.2 	&	5.6$\times 10^{-1}$		&881.3			&4.5		&4.15		&0.53		&0.026		&1.3$\times 10^4$		&	3.3$\times 10^5$\\
\textbf{SDSS J141004.41+334945.5} 	&	3.6$\times 10^{-4}$		&509.4			&10.0	&4.24		&0.60		&0.147		&9.9$\times 10^2$		&	2.5$\times 10^4$\\
\textbf{SDSS J141244.09+421257.6} 	&	1.2$\times 10^{-5}$		&433.4			&26.6	&6.22		&0.80		&1.645		&3.0$\times 10^1$		&	7.6$\times 10^2$\\
\textbf{SDSS J142339.44+471240.8} 	&	$<$4.0$\times 10^{-6}$	&298.8			&13.6	&1.74		&0.20		&0.009		&5.2$\times 10^2$		&	1.3$\times 10^4$\\
TEX 1428+370             			&	7.3$\times 10^{-1}$		&288.3			&12.9	&2.14		&0.32		&0.027		&1.3$\times 10^3$		&	3.3$\times 10^4$\\
SDSS J143637.44+090155.5 	&	2.6$\times 10^{-2}$		&319.1			&11.5	&1.21		&0.18		&0.001		&3.9$\times 10^4$		&	9.8$\times 10^5$\\
SDSS J145713.26+140334.1 	&	2.0$\times 10^{-1}$		&321.1			&7.2		&2.10		&0.26		&0.037		&1.1$\times 10^1$		&	2.8$\times 10^2$\\
SDSS J145859.07+153144.7 	&	1.0$\times 10^{-2}$		&317.4			&6.0		&3.65		&0.46		&0.609		&6.5$\times 10^{-1}$	&	1.6$\times 10^{1}$\\
\textbf{SDSS J150900.70+175114.3} 	&	5.0$\times 10^{-3}$		&317.8			&14.3	&2.26		&0.30		&0.026		&9.1$\times 10^2$		&	2.3$\times 10^4$\\
FBQS J150911.2+215508    	&	6.5$\times 10^{-2}$		&314.4			&3.7		&2.41		&0.41		&0.040		&1.9$\times 10^3$		&	4.8$\times 10^4$\\
SDSS J150912.07+204004.6 	&	2.1$\times 10^{-2}$		&315.3			&9.6		&1.55		&0.31		&0.005		&2.8$\times 10^4$		&	7.1$\times 10^5$\\
SDSS J151053.24+240943.3 	&	3.2$\times 10^{-2}$		&712.2			&16.5	&3.04		&0.39		&0.010		&2.1$\times 10^4$		&	5.3$\times 10^5$\\
SDSS J151243.67+195845.1 	&	1.5$\times 10^{-1}$		&308.6			&5.3		&2.57		&0.33		&0.055		&3.2$\times 10^2$		&	8.1$\times 10^3$\\
\textbf{SDSS J151646.10+221724.7} 	&	3.9$\times 10^{-4}$		&309.7			&14.7	&2.40		&0.35		&0.038		&9.6$\times 10^2$		&	2.4$\times 10^4$\\
\textbf{SDSS J152739.97+413234.6} 	&	4.7$\times 10^{-3}$		&438.3			&7.2		&4.10		&0.50		&0.206		&1.3$\times 10^2$		&	3.3$\times 10^3$\\
SDSS J152903.11+223623.8 	&	1.2$\times 10^{-2}$		&310.4			&14.4	&2.21		&0.35		&0.026		&2.1$\times 10^3$		&	5.3$\times 10^4$\\
\textbf{SDSS J153051.79+503440.1} 	&	$<$4.0$\times 10^{-6}$	&429.2			&28.9	&3.04		&0.38		&0.048		&6.8$\times 10^2$		&	1.7$\times 10^4$\\
\textbf{SDSS J153251.06+335852.2} 	&	3.1$\times 10^{-3}$		&436.2			&7.8		&2.61		&0.30		&0.030		&1.4$\times 10^2$		&	3.5$\times 10^3$\\
\textbf{PDS 898}                  					&	3.0$\times 10^{-4}$		&436.2			&12.1	&2.92		&0.37		&0.037		&1.1$\times 10^3$		&	2.8$\times 10^4$\\
SDSS J155308.65+501436.5 	&	1.2$\times 10^{-2}$		&438.2			&13.1	&2.72		&0.34		&0.050		&2.4$\times 10^1$		&	6.1$\times 10^2$\\
SDSS J160322.68+200535.2 	&	8.0$\times 10^{-1}$		&237.5			&15.0	&3.18		&0.38		&0.592		&5.8$\times 10^{-1}$	&	1.5$\times 10^{1}$\\
\textbf{SDSS J160454.57+315733.5} 	&	4.4$\times 10^{-5}$		&307.3			&16.5	&2.03		&0.26		&0.040		&6.7$\times 10^0$		&	1.7$\times 10^2$\\
SDSS J162634.15+325032.6 	&	4.6$\times 10^{-2}$		&297.5			&13.5	&1.89		&0.24		&0.013		&1.0$\times 10^3$		&	2.5$\times 10^4$\\
\textbf{SDSS J170942.58+342316.2} 	&	3.6$\times 10^{-4}$		&455.2			&7.2		&3.42		&0.39		&0.096		&6.7$\times 10^1$		&	1.7$\times 10^3$\\
\textbf{SDSS J171122.67+342658.9} 	&	3.4$\times 10^{-4}$		&285.5			&13.6	&1.60		&0.19		&0.010		&7.9$\times 10^1$		&	2.0$\times 10^3$\\
\textbf{SDSS J171617.49+341553.3} 	&	4.4$\times 10^{-3}$		&130.7			&5.5		&1.58		&0.19		&0.070		&7.4$\times 10^0$		&	1.9$\times 10^2$\\
\textbf{SDSS J171909.93+344001.3} 	&	7.4$\times 10^{-4}$		&292.6			&5.6		&2.50		&0.29		&0.090		&9.3$\times 10^0$		&	2.3$\times 10^2$\\
\textbf{SDSS J212939.60+004845.5} 	&	----									&313.0			&9.8		&2.43		&0.29		&0.066		&1.3$\times 10^1$		&	3.3$\times 10^2$\\
\textbf{SDSS J214036.77+005210.1} 	&	7.4$\times 10^{-3}$		&315.8			&7.2		&2.47		&0.32		&0.042		&3.1$\times 10^2$		&	7.8$\times 10^3$\\
\textbf{SDSS J214225.29+001643.2} 	&	8.3$\times 10^{-3}$		&316.7			&13.3	&1.74		&0.20		&0.008		&7.0$\times 10^2$		&	1.8$\times 10^4$\\
\textbf{SDSS J214357.98+003349.6} 	&	7.1$\times 10^{-3}$		&456.0			&10.1	&3.30		&0.39		&0.101		&2.4$\times 10^1$		&	6.1$\times 10^2$\\
\textbf{PKS 2203-215}             			&	1.6$\times 10^{-3}$		&497.0			&17.7	&4.08		&0.60		&0.133		&1.3$\times 10^3$		&	3.3$\times 10^4$\\
SDSS J224008.39+263928.4 	&	5.4$\times 10^{-1}$		&314.1			&3.8		&2.19		&0.27		&0.047		&8.9$\times 10^0$		&	2.2$\times 10^2$\\
\textbf{SDSS J231733.66+001128.3} 	&	5.7$\times 10^{-3}$		&467.3			&10.8	&3.51		&0.45		&0.076		&7.3$\times 10^2$		&	1.8$\times 10^4$\\
\textbf{SDSS J232135.73+173916.5} 	&	2.2$\times 10^{-4}$		&337.4			&22.1	&3.40		&0.43		&0.171		&1.2$\times 10^2$		&	3.0$\times 10^3$\\
\textbf{2QZ J235800.2-281429}     		&	----									&306.1			&13.5	&2.20		&0.25		&0.033		&7.6$\times 10^1$		&	1.9$\times 10^3$\\
\textbf{SDSS J235928.99+170426.9} 	&	8.8$\times 10^{-5}$		&330.3			&15.4	&3.15		&0.43		&0.122		&2.4$\times 10^2$		&	6.1$\times 10^3$\\
\textbf{SDSS J235958.72+003345.3} 	&	----									&486.3			&10.3	&3.21		&0.37		&0.056		&1.5$\times 10^2$		&	3.8$\times 10^3$\\
\hline
\end{tabular}
\begin{flushleft}
$^*$ The P-values shown here are calculated from the composite light curves (with data from PTF, iPTF and CRTS). We note that all the quasars shown in the tables have P-value$<$1/250,000, when only the PTF data are taken into account.\\
$^{**}$We calculated the variability fraction as $\frac{F_{max}-F_{min}}{2 F_{mean}}\times100\%=\frac{10^{-\frac{m-A/2}{2.5}}-10^{-\frac{m+A/2}{2.5}}}{2\times10^{-\frac{m}{2.5}}}\times 100\%$, where m is the mean magnitude of the quasar and A the amplitude of the best fit sinusoid.
%=\frac{10^{-\frac{m_{mean}-Amp/2}{2.5}}-10^{-\frac{m_{mean}+Amp/2}{2.5}}}{2\times10^{-\frac{m_{mean}}{2.5}}$
\end{flushleft}
\label{table:results}
\end{table*}

\clearpage
%\section*{APPENDIX}
\begin{figure*}
%\section*{APPENDIX}
\begin{subfigure}{.45\textwidth}
\centering
\includegraphics[width=8cm,height=4.5cm]{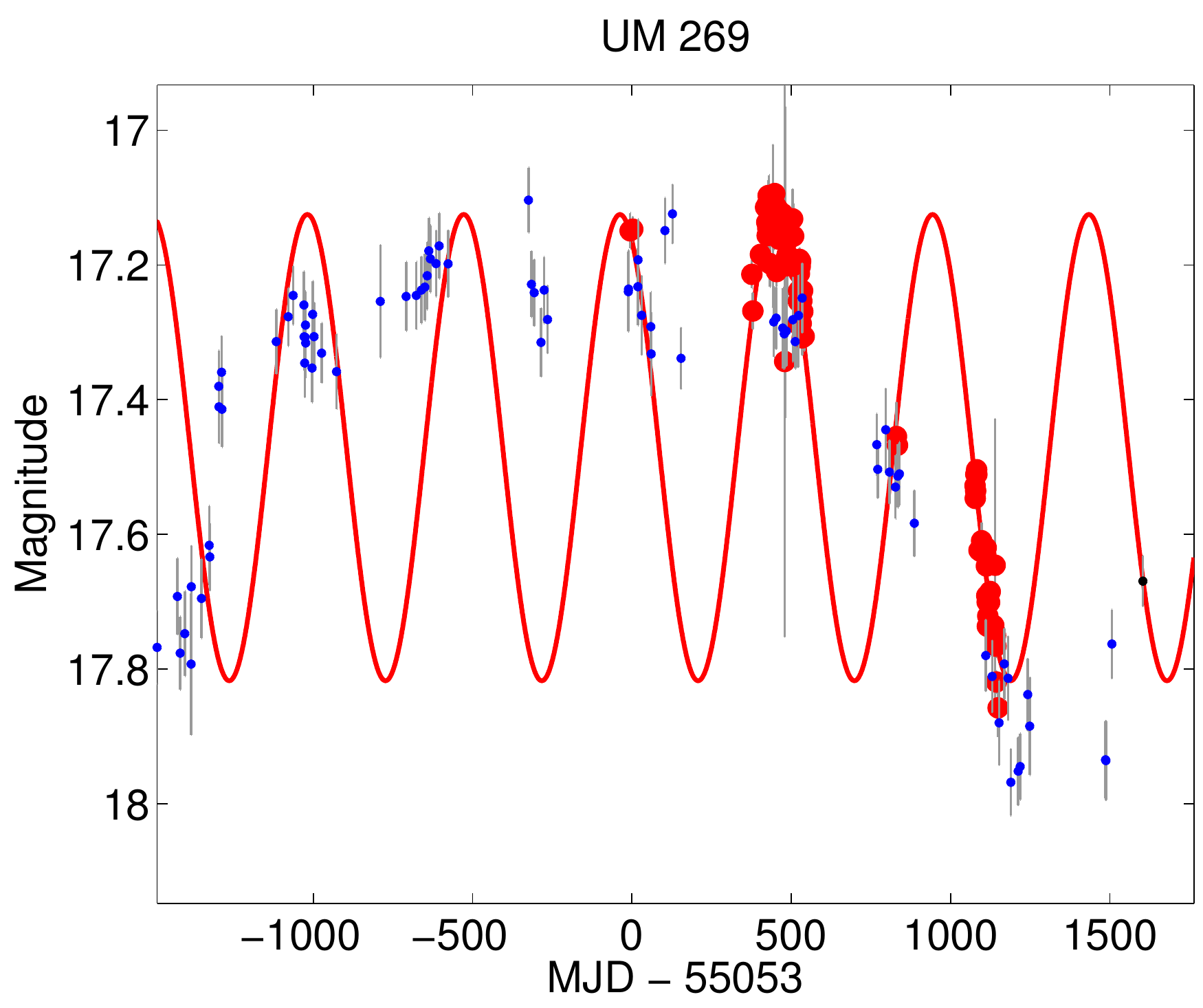}
\end{subfigure} \hspace{0.2cm}
\begin{subfigure}{.45\textwidth}
\centering
\includegraphics[width=8cm,height=4.5cm]{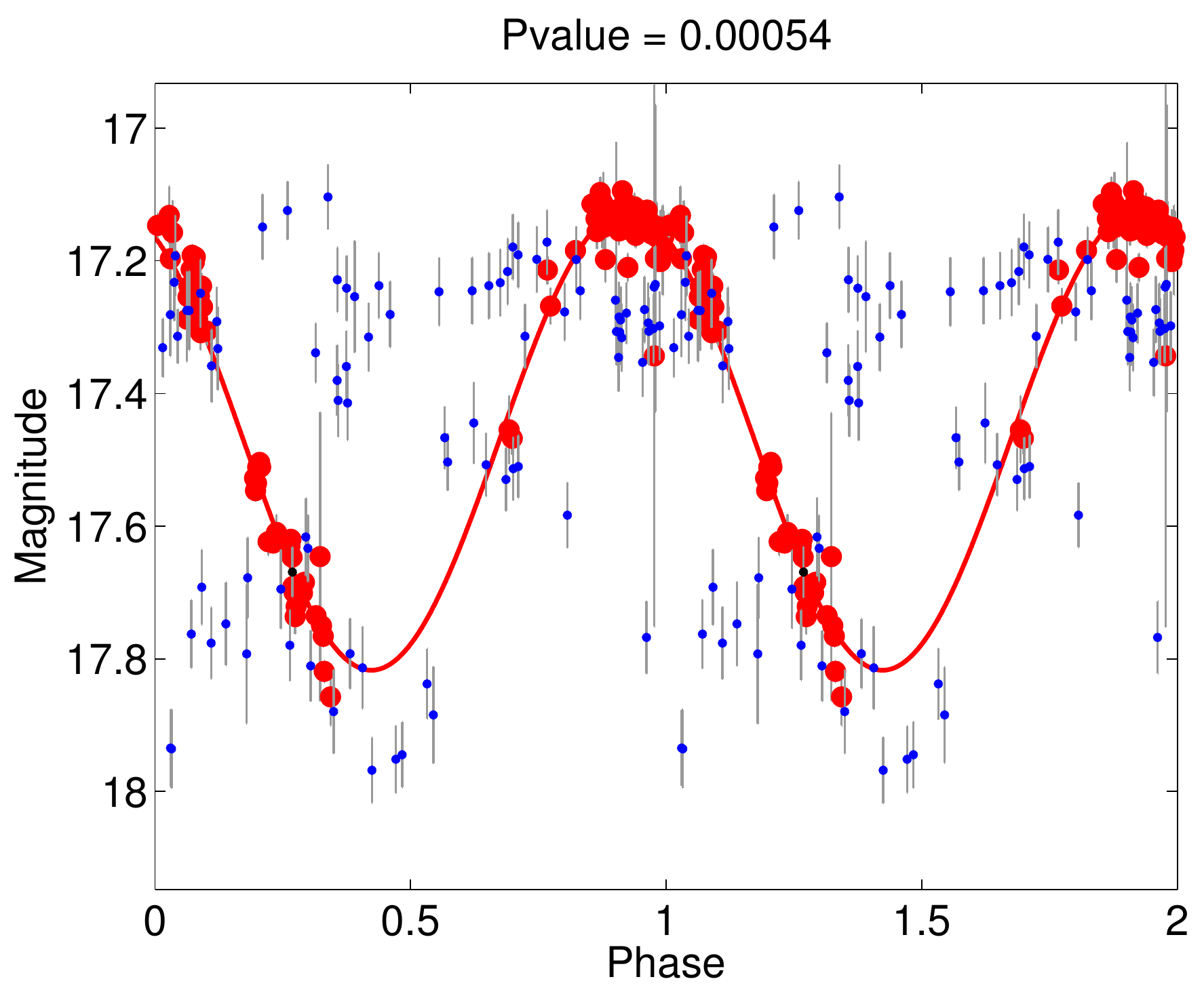}
\end{subfigure} \hspace{0.2cm}
\begin{subfigure}{.45\textwidth}
\centering
\includegraphics[width=8cm,height=4.5cm]{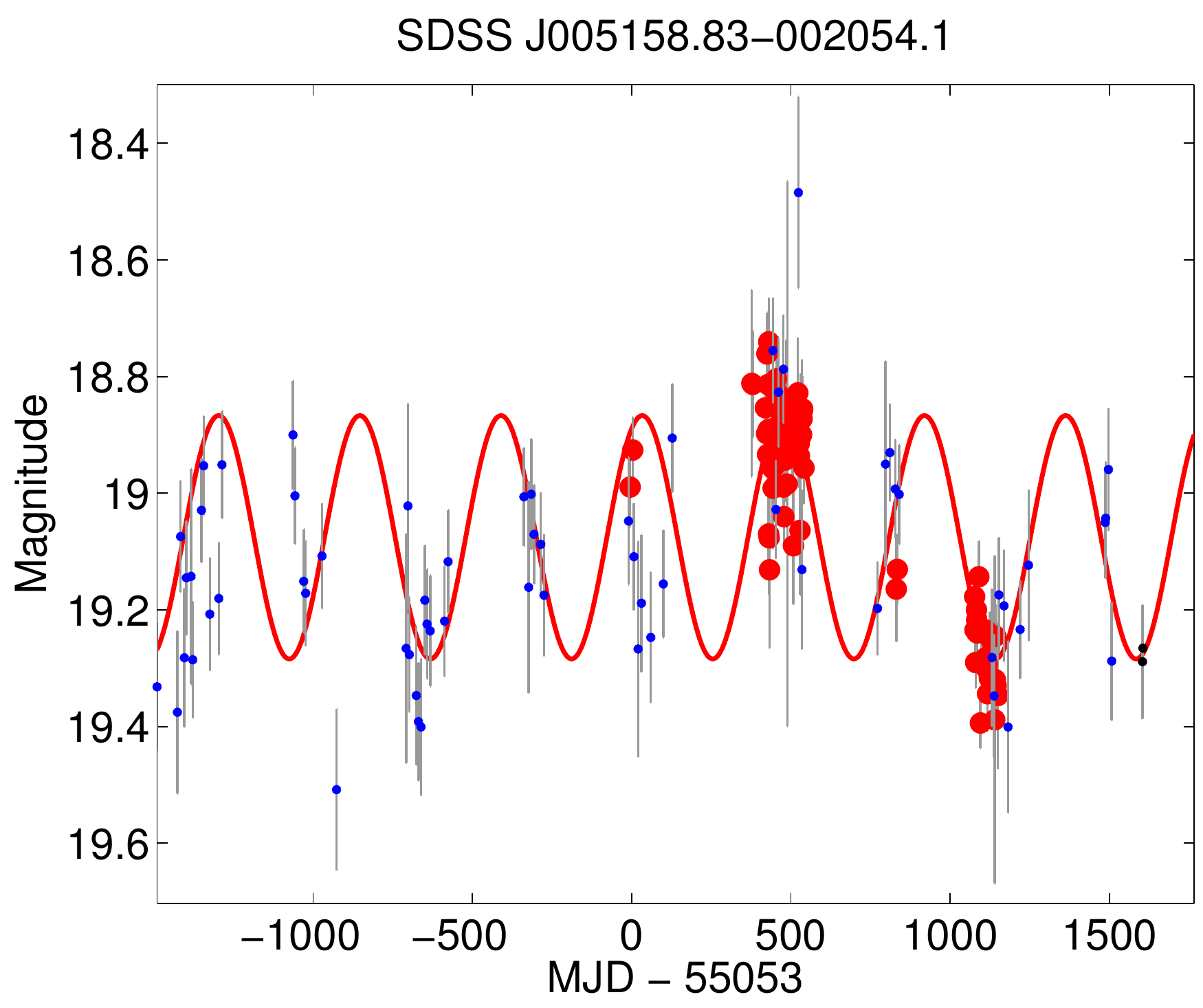}
\end{subfigure} \hspace{0.2cm}
\begin{subfigure}{.45\textwidth}
\centering
\includegraphics[width=8cm,height=4.5cm]{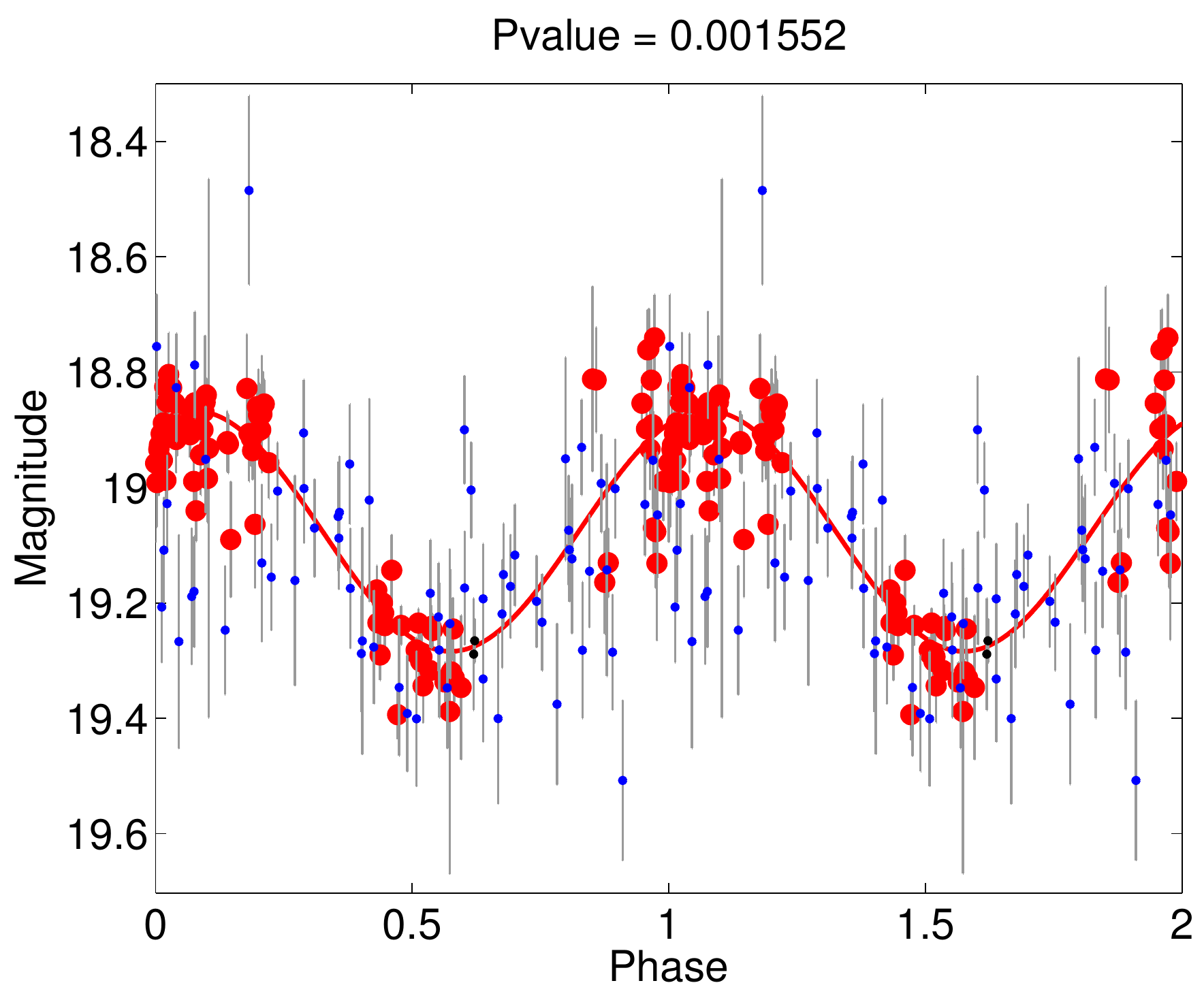}
\end{subfigure} \hspace{0.2cm}
\begin{subfigure}{.45\textwidth}
\centering
\includegraphics[width=8cm,height=4.5cm]{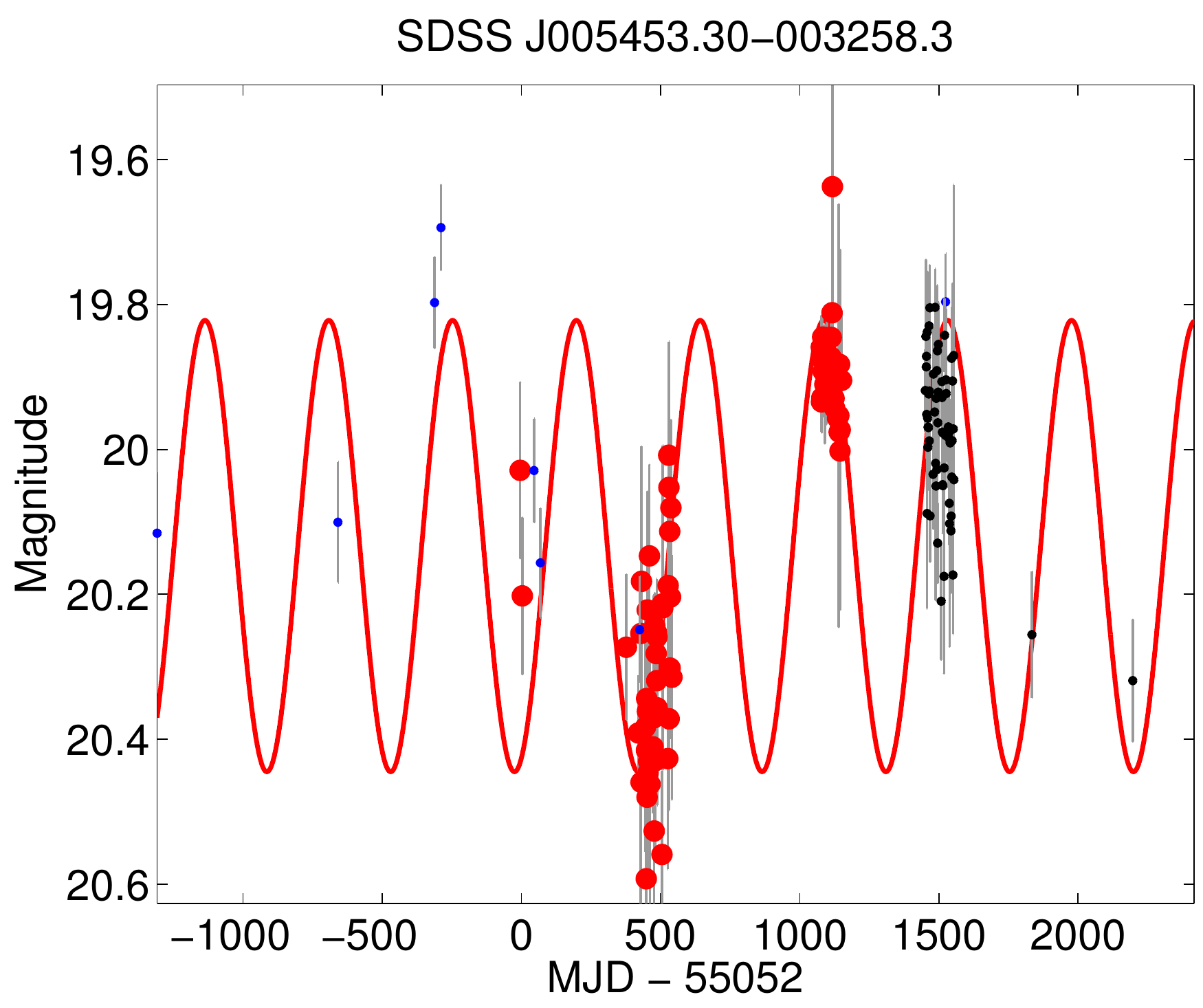}
\end{subfigure} \hspace{0.2cm}
\begin{subfigure}{.45\textwidth}
\centering
\includegraphics[width=8cm,height=4.5cm]{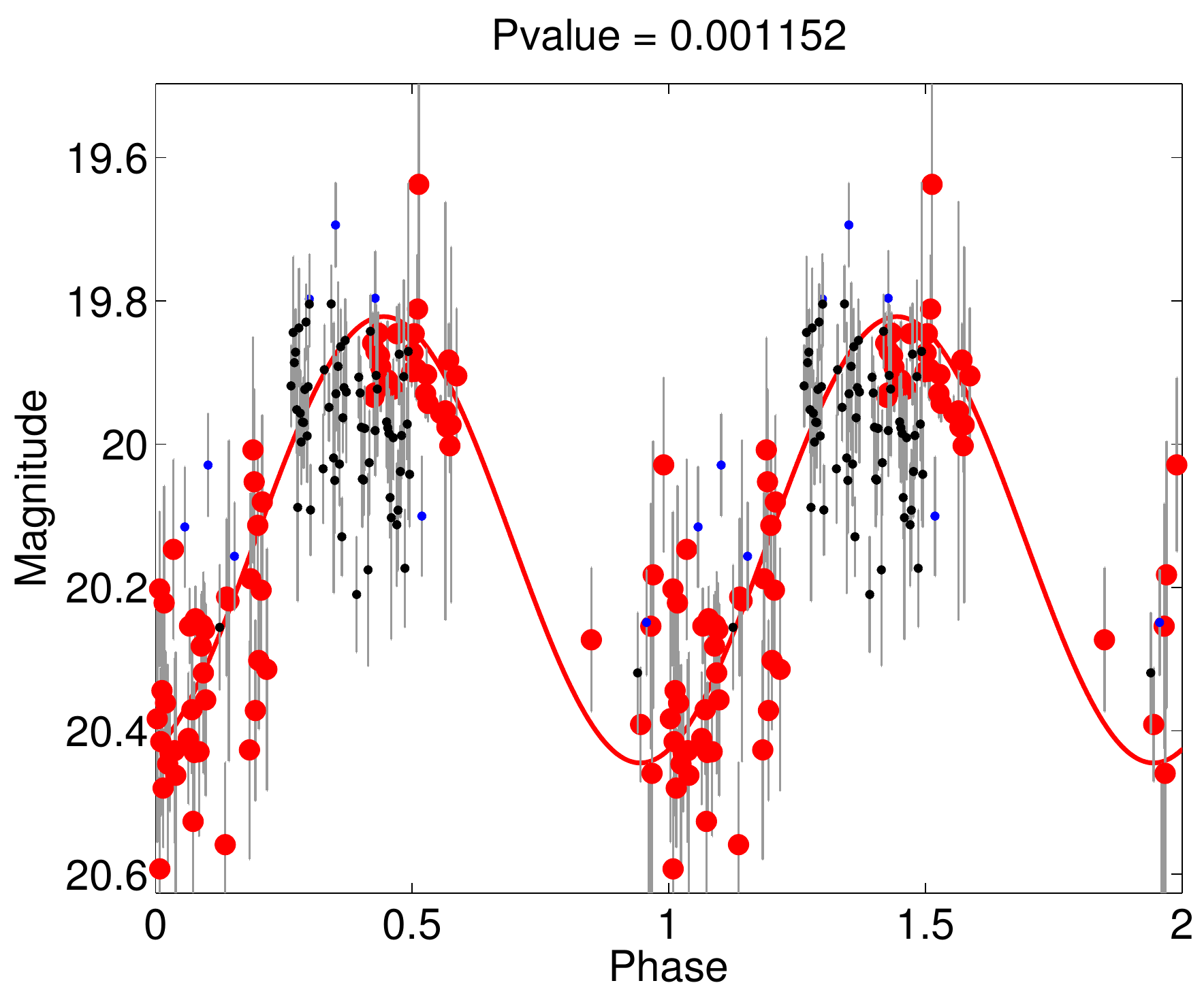}
\end{subfigure} \hspace{0.2cm}
\begin{subfigure}{.45\textwidth}
\centering
\includegraphics[width=8cm,height=4.5cm]{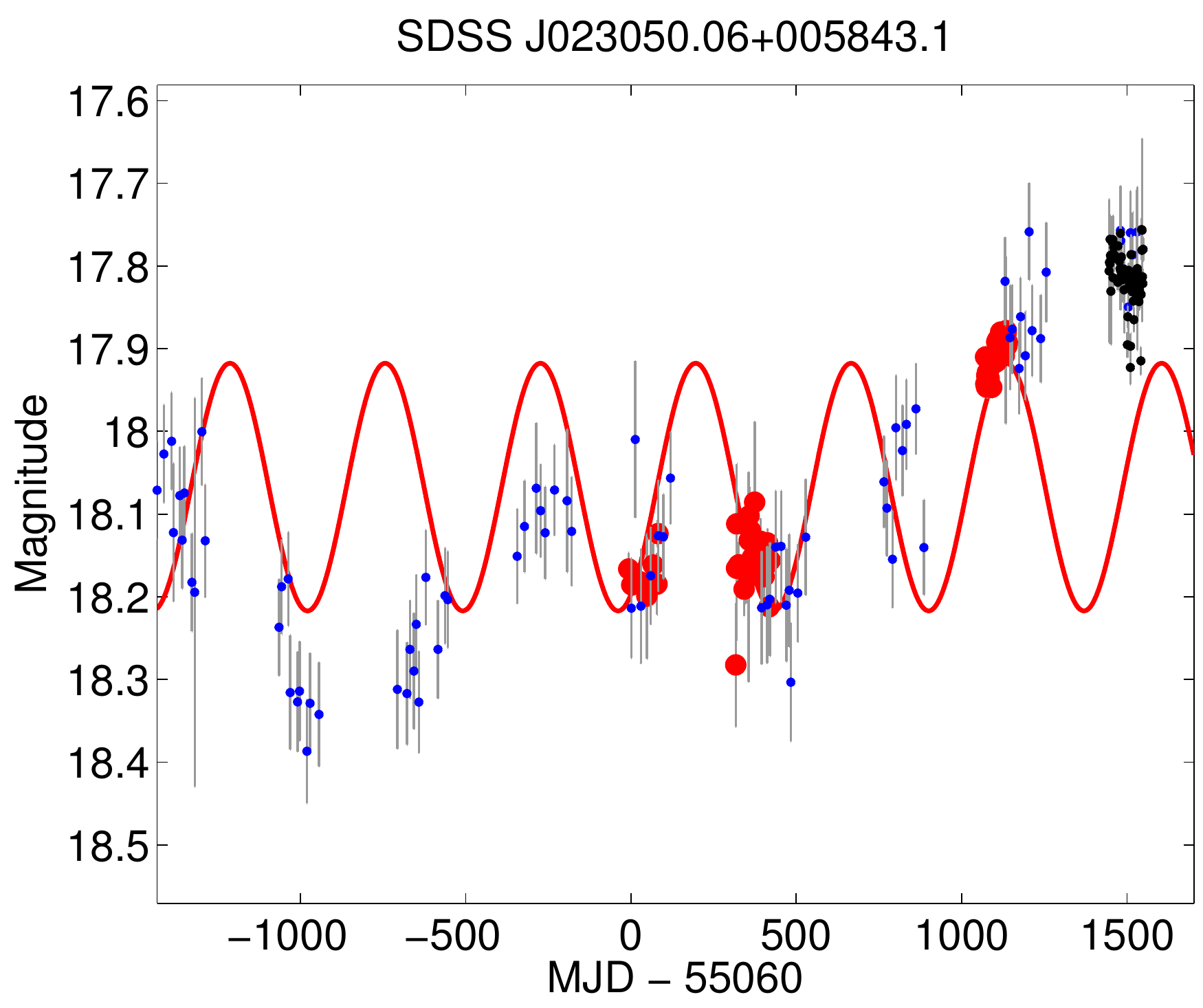}
\end{subfigure} \hspace{0.2cm}
\begin{subfigure}{.45\textwidth}
\centering
\includegraphics[width=8cm,height=4.5cm]{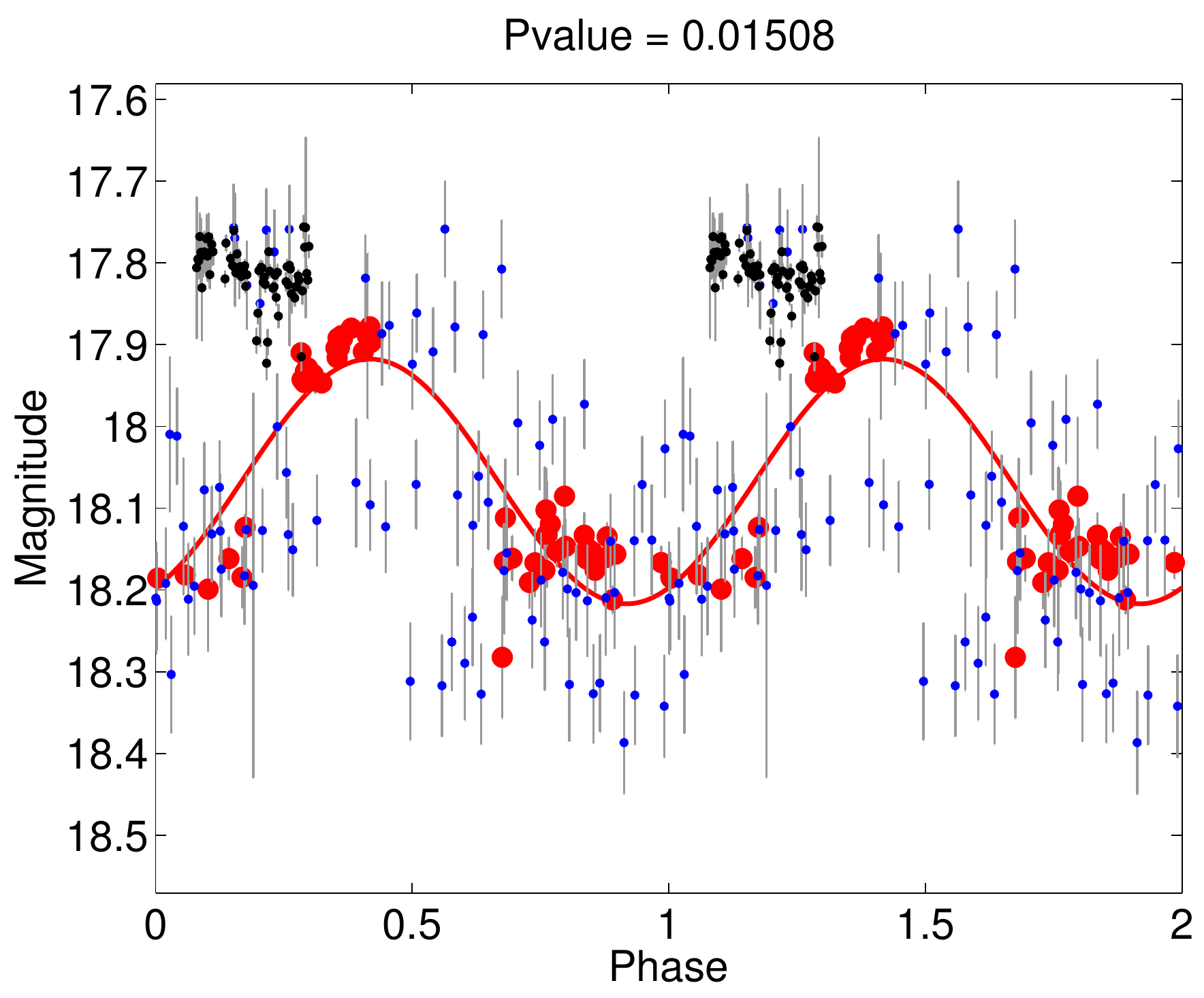}
\end{subfigure} \hspace{0.2cm}
\begin{subfigure}{.45\textwidth}
\centering
\includegraphics[width=8cm,height=4.5cm]{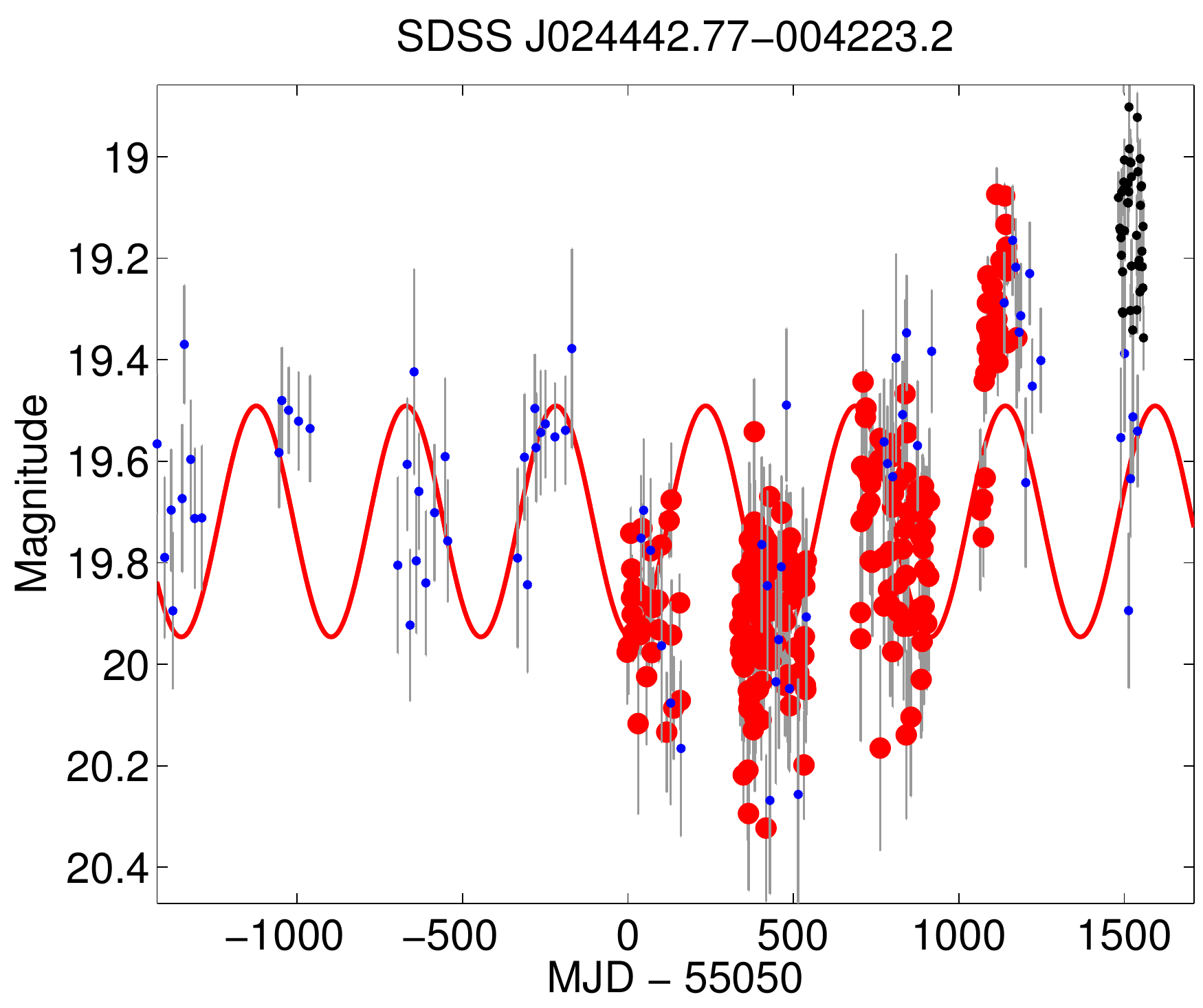}
\end{subfigure} \hspace{0.2cm}
\begin{subfigure}{.45\textwidth}
\centering
\includegraphics[width=8cm,height=4.5cm]{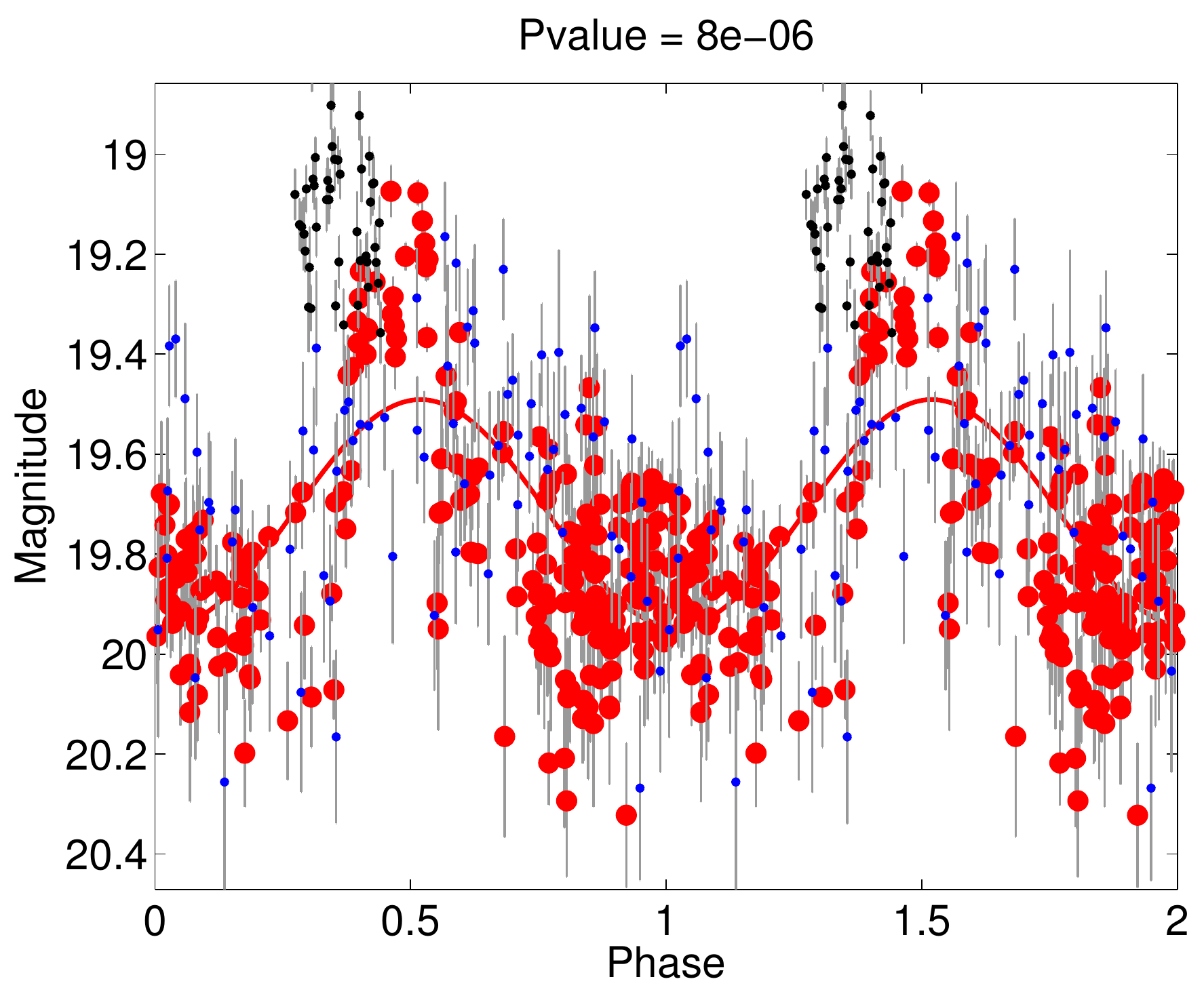}
\end{subfigure} \hspace{0.2cm}
\phantomcaption
\end{figure*}
\begin{figure*}
\ContinuedFloat
\begin{subfigure}{.45\textwidth}
\centering
\includegraphics[width=8cm,height=4.5cm]{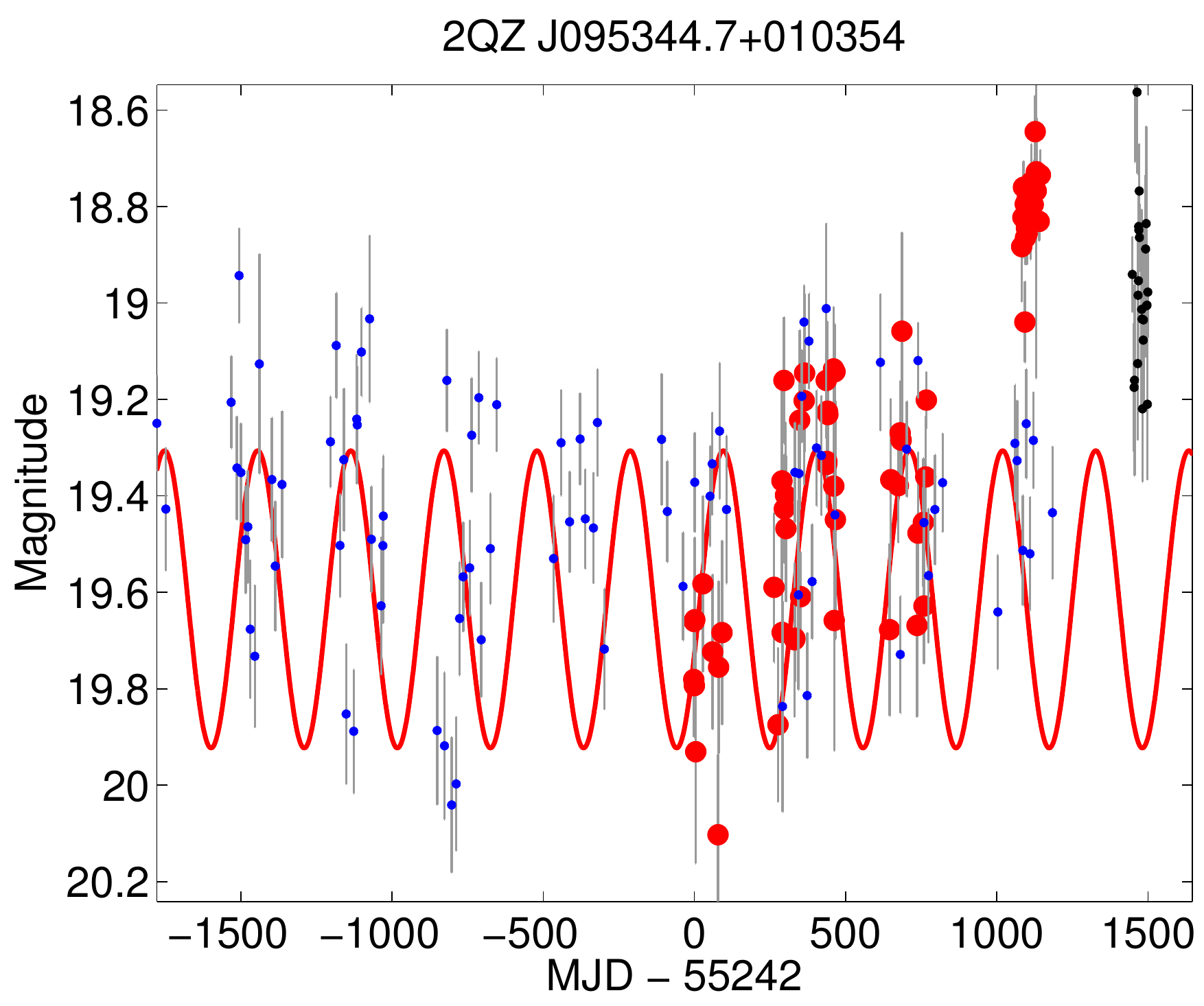}
\end{subfigure} \hspace{0.2cm}
\begin{subfigure}{.45\textwidth}
\centering
\includegraphics[width=8cm,height=4.5cm]{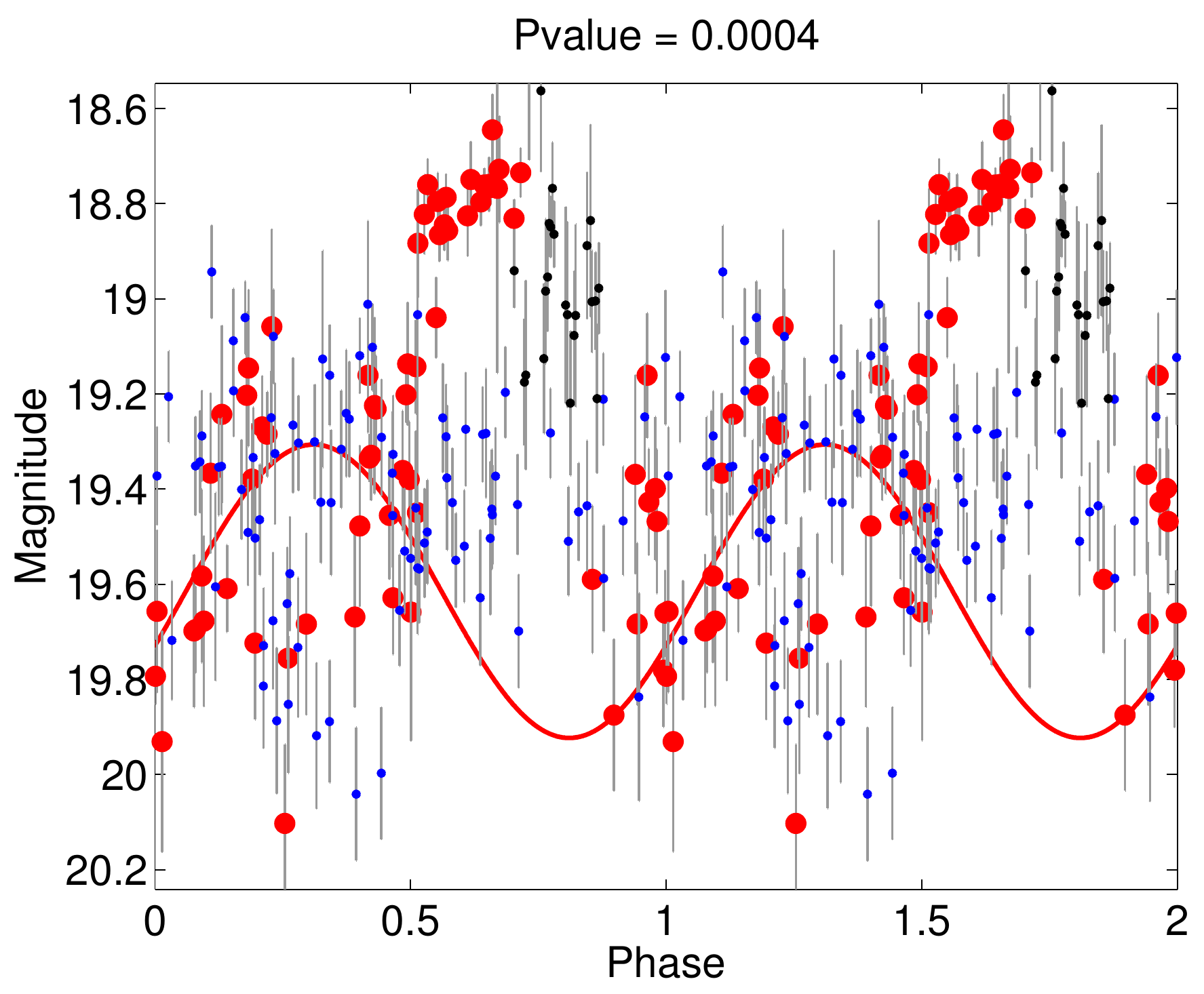}
\end{subfigure} \hspace{0.2cm}
\begin{subfigure}{.45\textwidth}
\centering
\includegraphics[width=8cm,height=4.5cm]{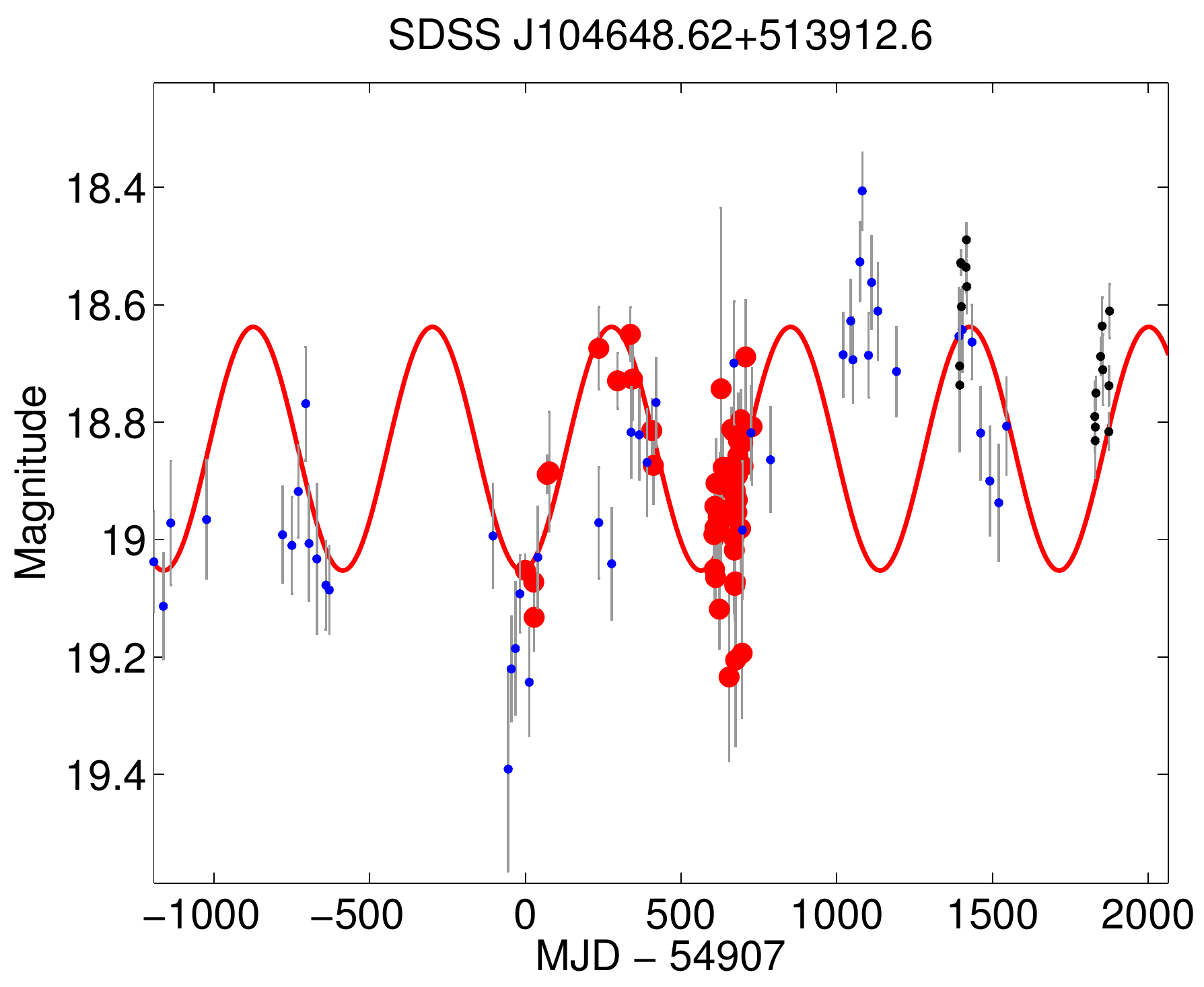}
\end{subfigure} \hspace{0.2cm}
\begin{subfigure}{.45\textwidth}
\centering
\includegraphics[width=8cm,height=4.5cm]{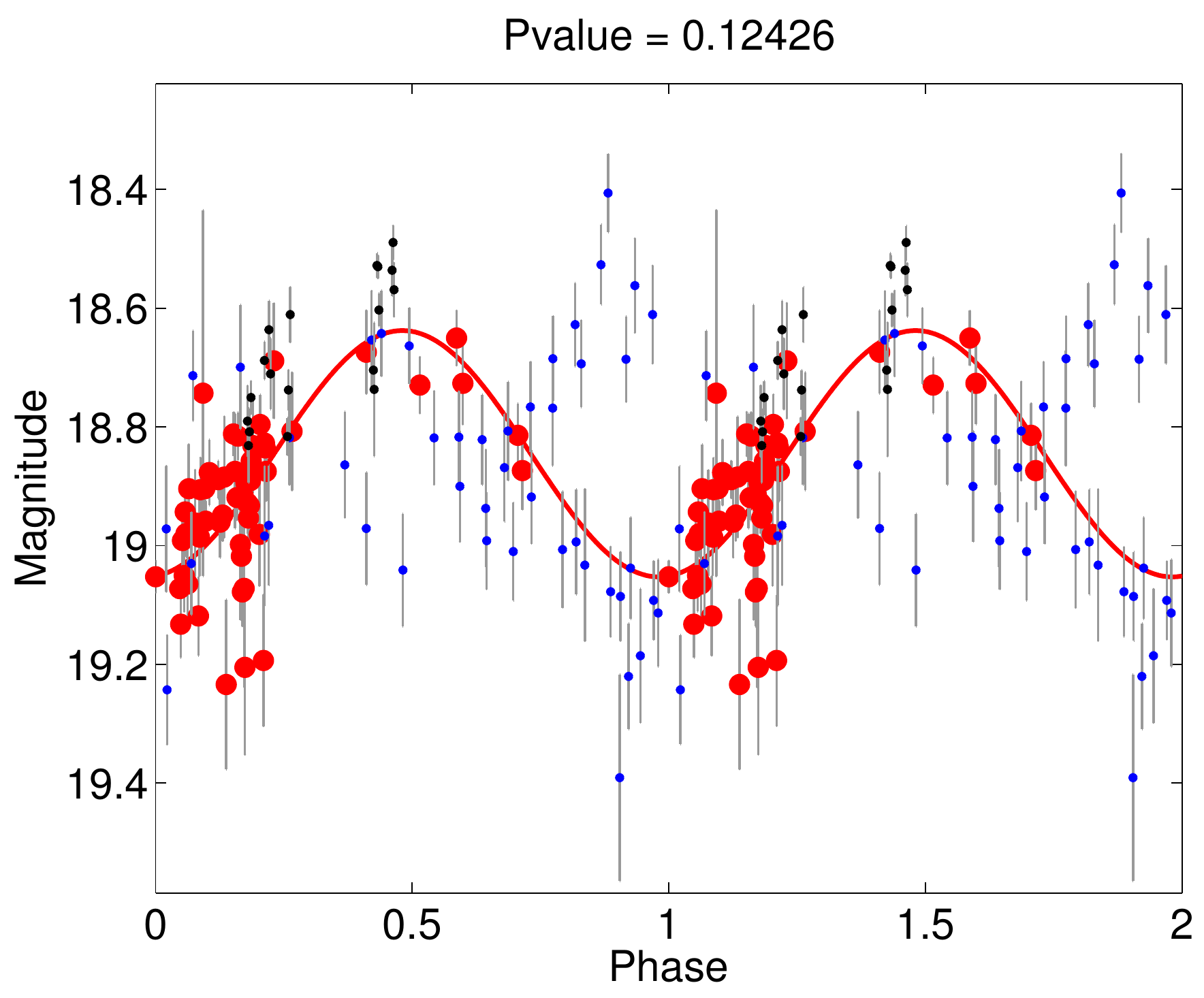}
\end{subfigure} \hspace{0.2cm}
\begin{subfigure}{.45\textwidth}
\centering
\includegraphics[width=8cm,height=4.5cm]{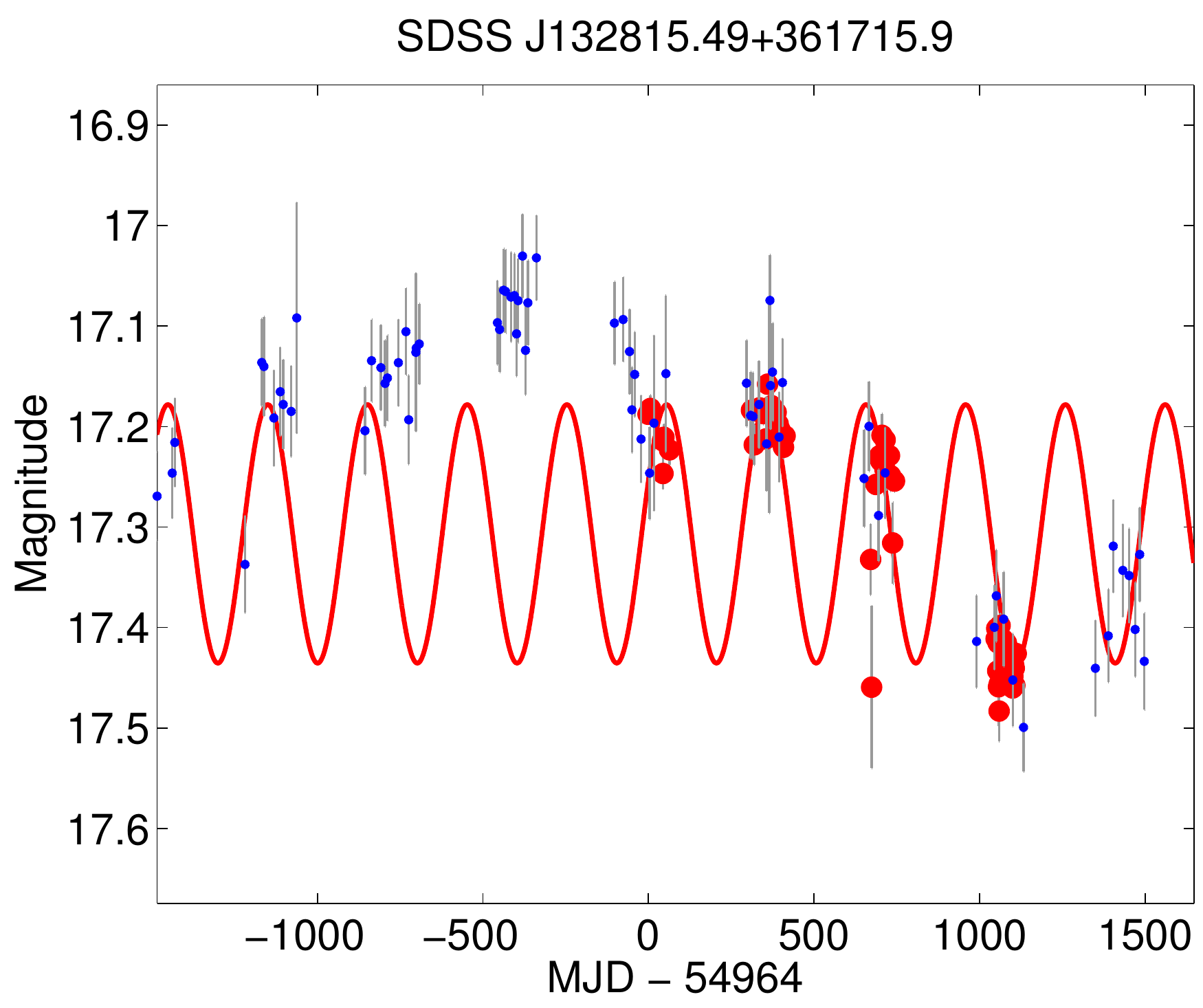}
\end{subfigure} \hspace{0.2cm}
\begin{subfigure}{.45\textwidth}
\centering
\includegraphics[width=8cm,height=4.5cm]{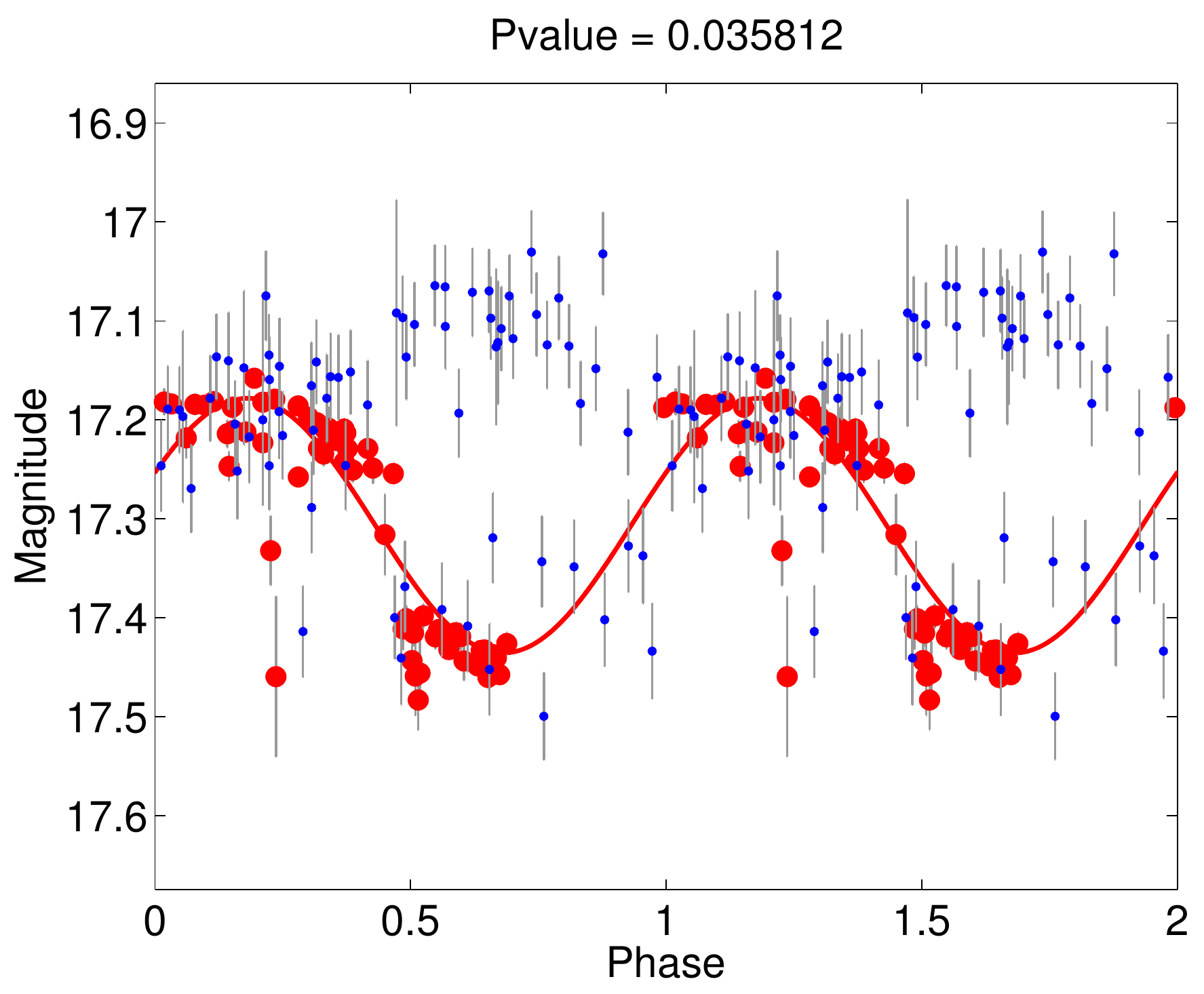}
\end{subfigure} \hspace{0.2cm}
\begin{subfigure}{.45\textwidth}
\centering
\includegraphics[width=8cm,height=4.5cm]{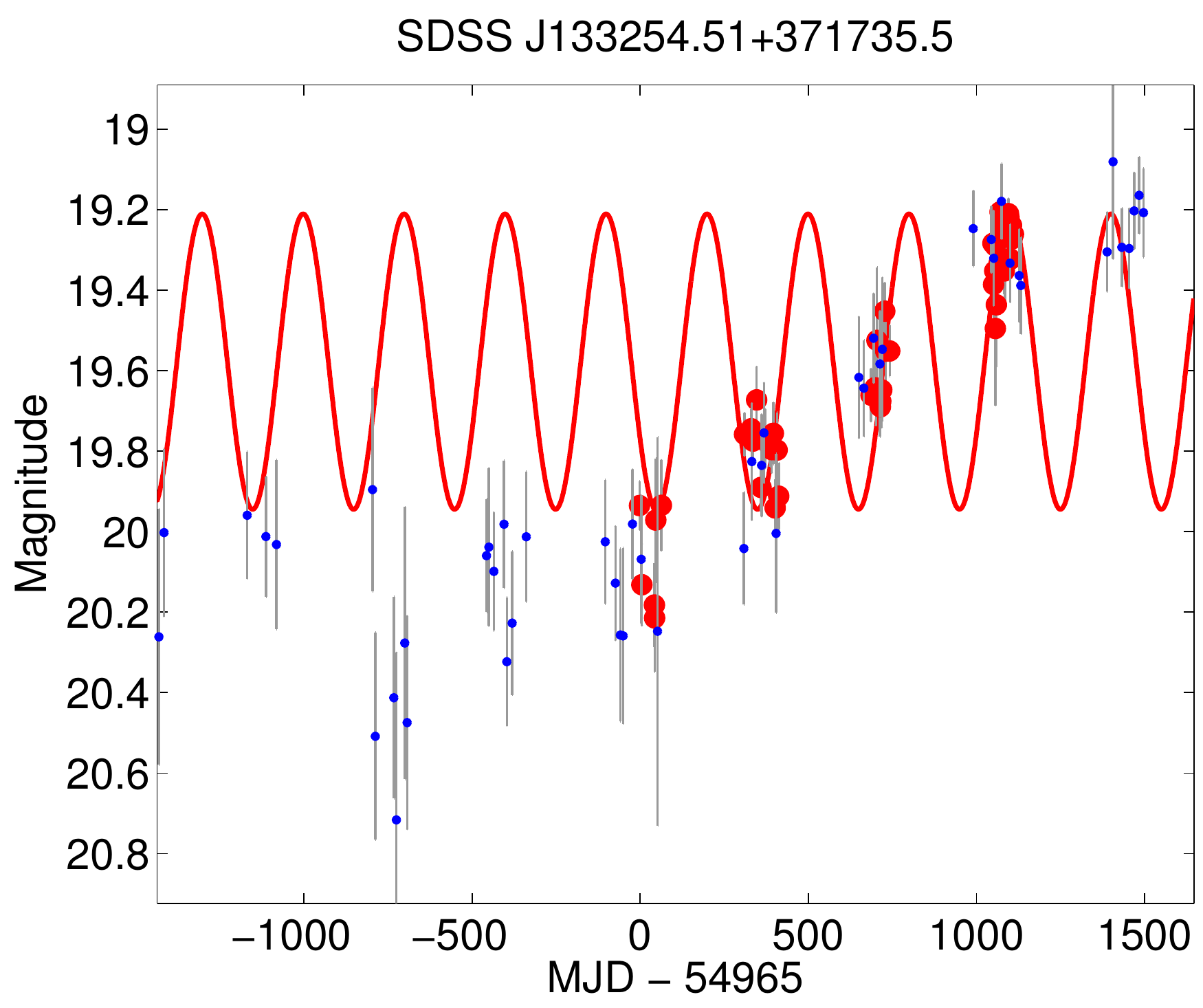}
\end{subfigure} \hspace{0.2cm}
\begin{subfigure}{.45\textwidth}
\centering
\includegraphics[width=8cm,height=4.5cm]{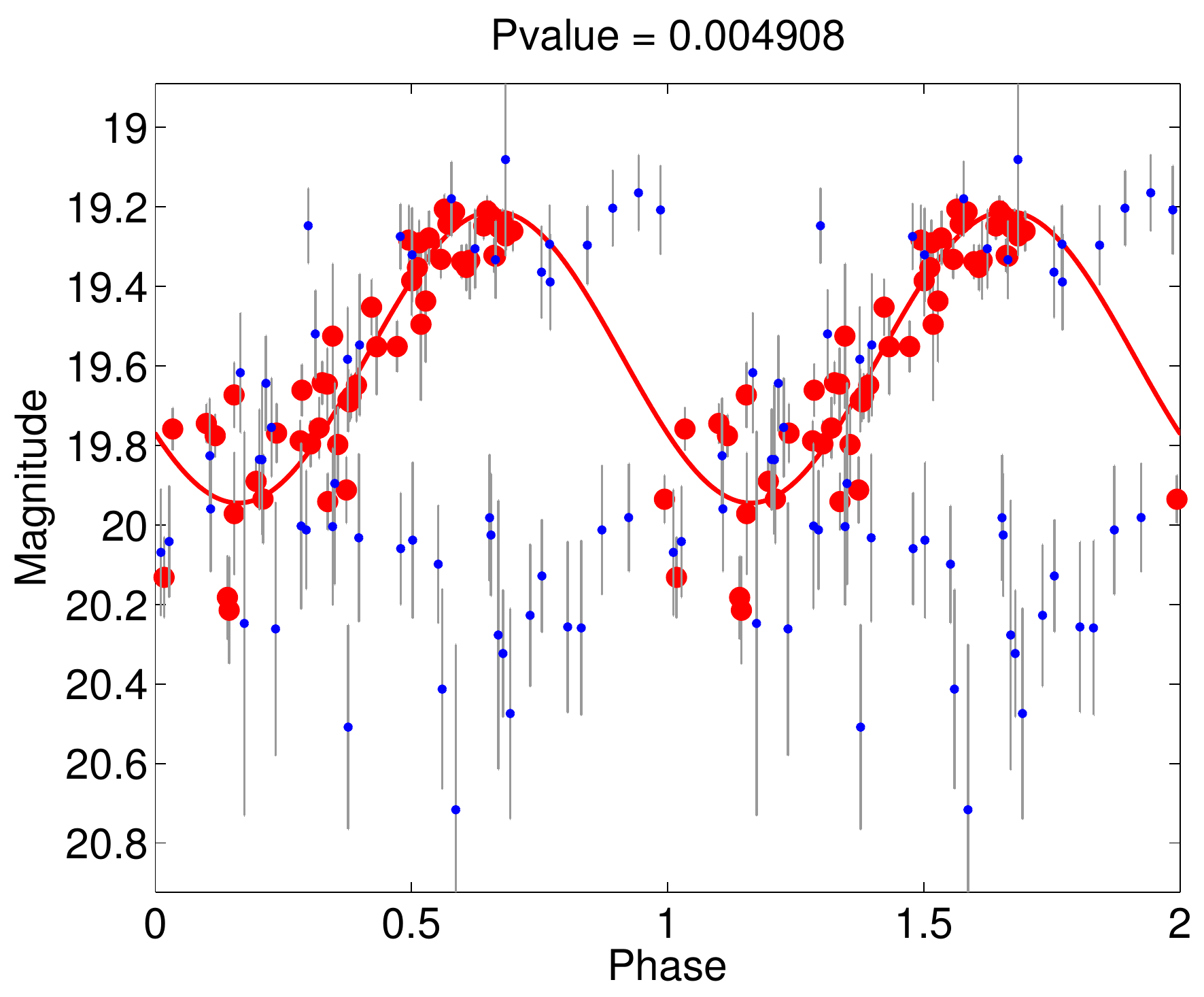}
\end{subfigure} \hspace{0.2cm}
\begin{subfigure}{.45\textwidth}
\centering
\includegraphics[width=8cm,height=4.5cm]{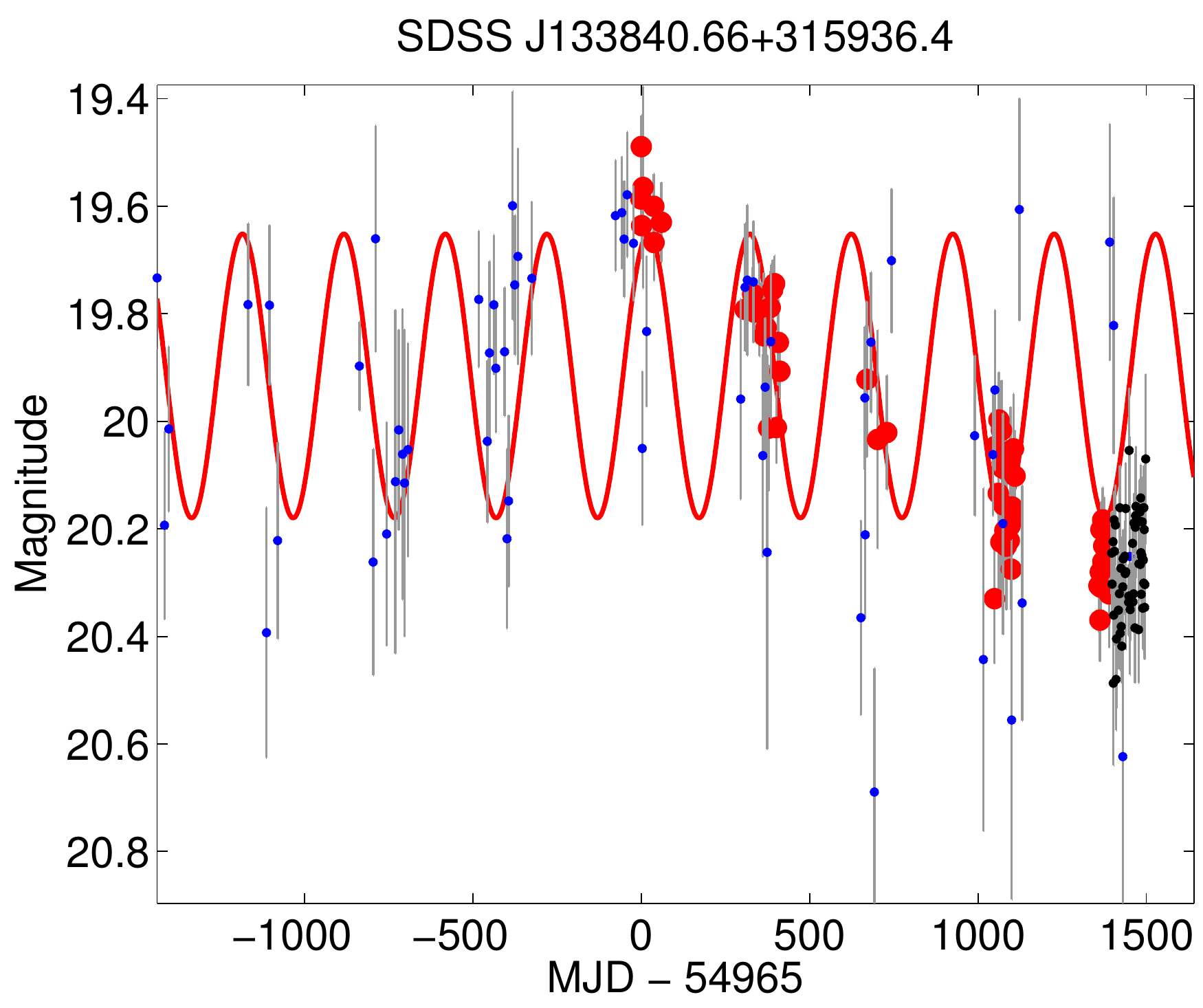}
\end{subfigure} \hspace{0.2cm}
\begin{subfigure}{.45\textwidth}
\centering
\includegraphics[width=8cm,height=4.5cm]{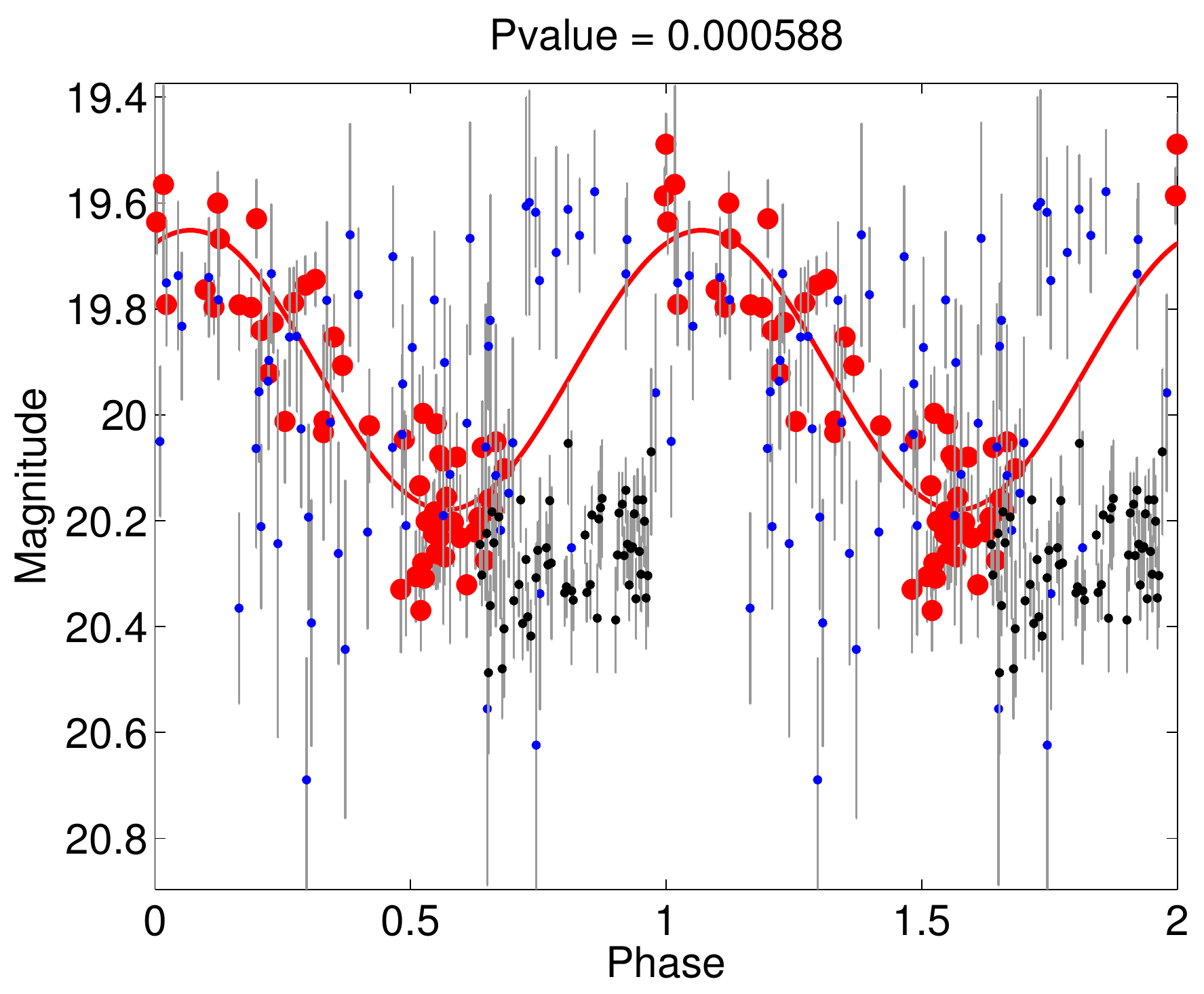}
\end{subfigure} \hspace{0.2cm}
\phantomcaption
\end{figure*}
\begin{figure*}
\ContinuedFloat
\begin{subfigure}{.45\textwidth}
\centering
\includegraphics[width=8cm,height=4.5cm]{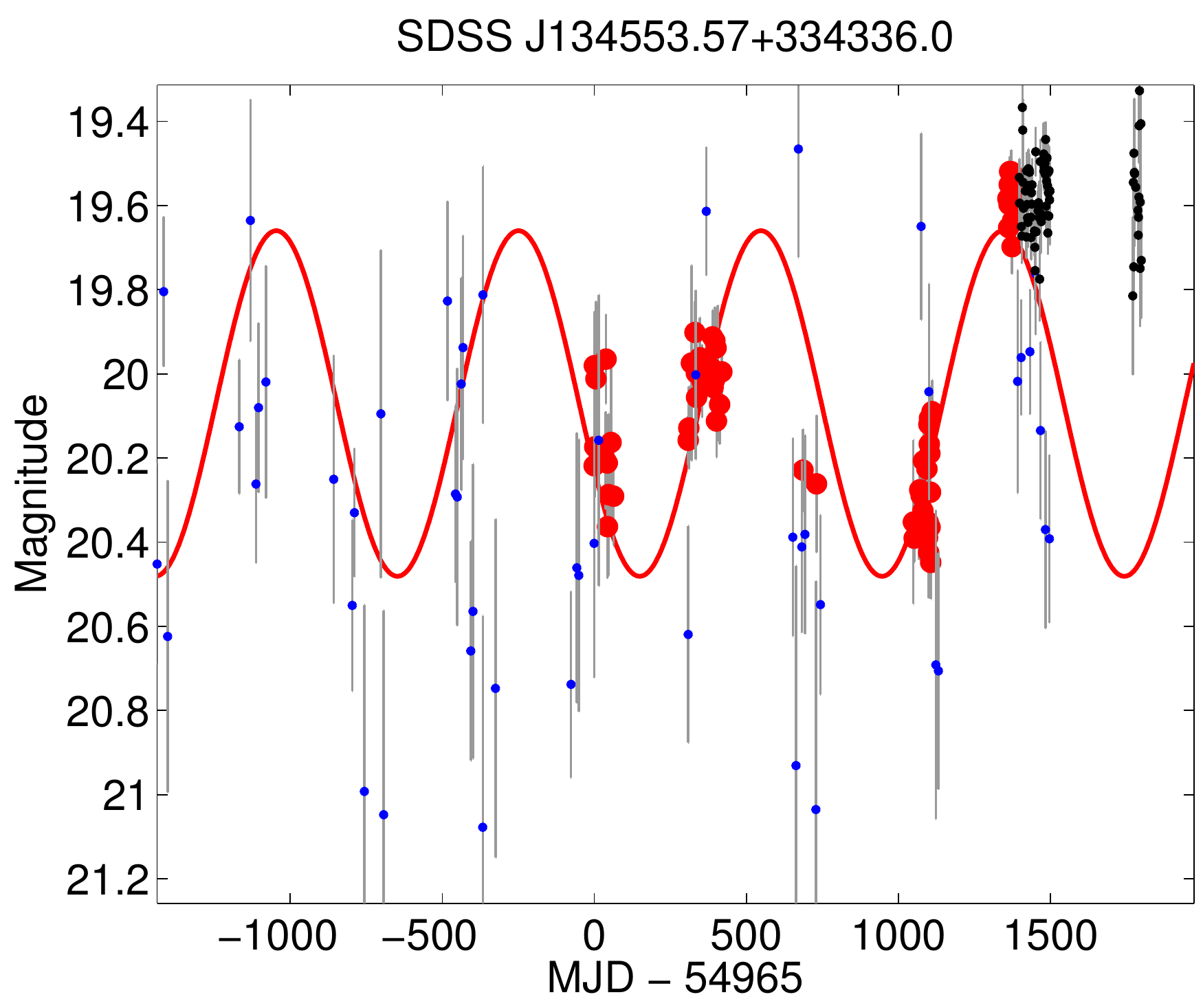}
\end{subfigure} \hspace{0.2cm}
\begin{subfigure}{.45\textwidth}
\centering
\includegraphics[width=8cm,height=4.5cm]{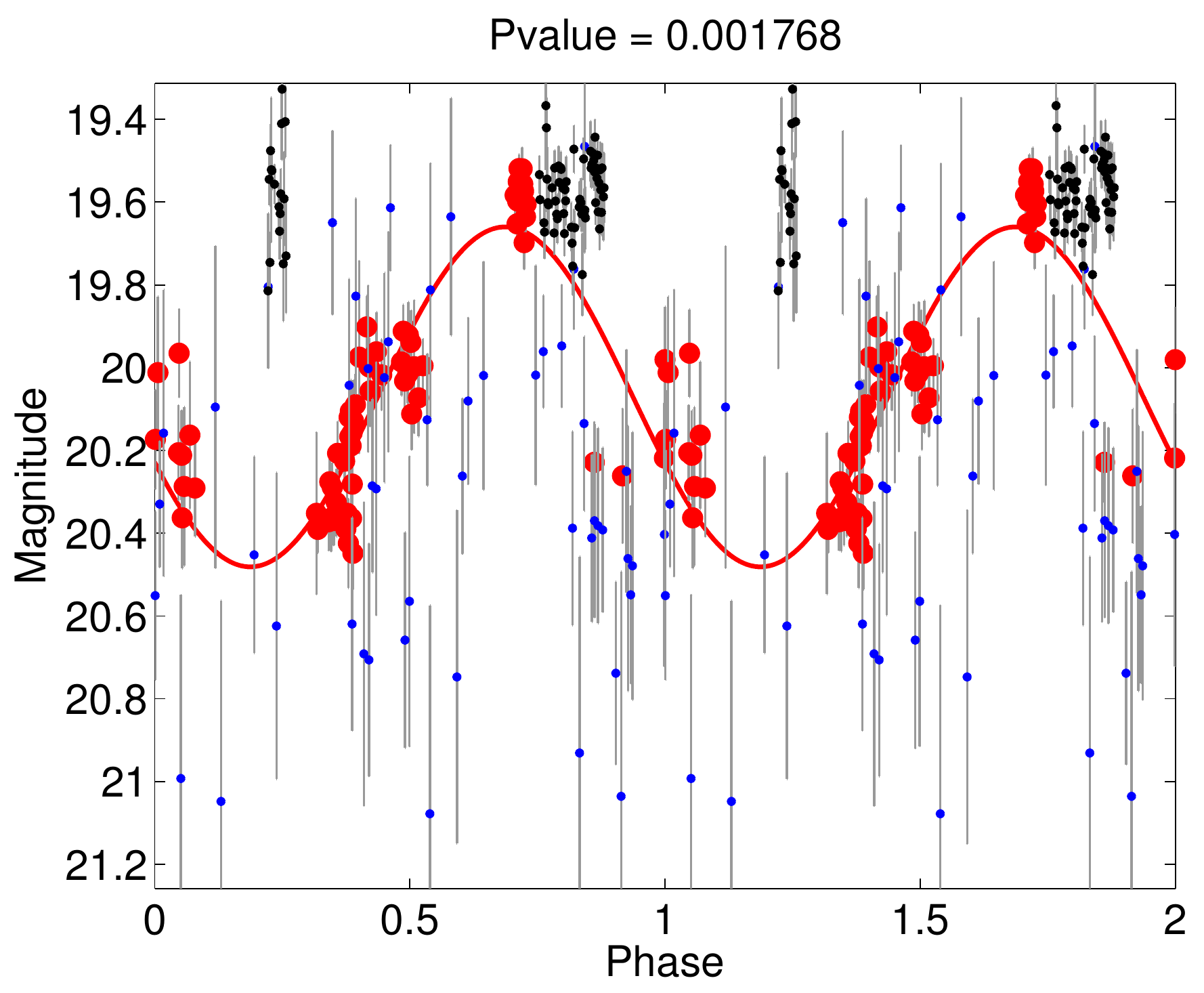}
\end{subfigure} \hspace{0.2cm}
\begin{subfigure}{.45\textwidth}
\centering
\includegraphics[width=8cm,height=4.5cm]{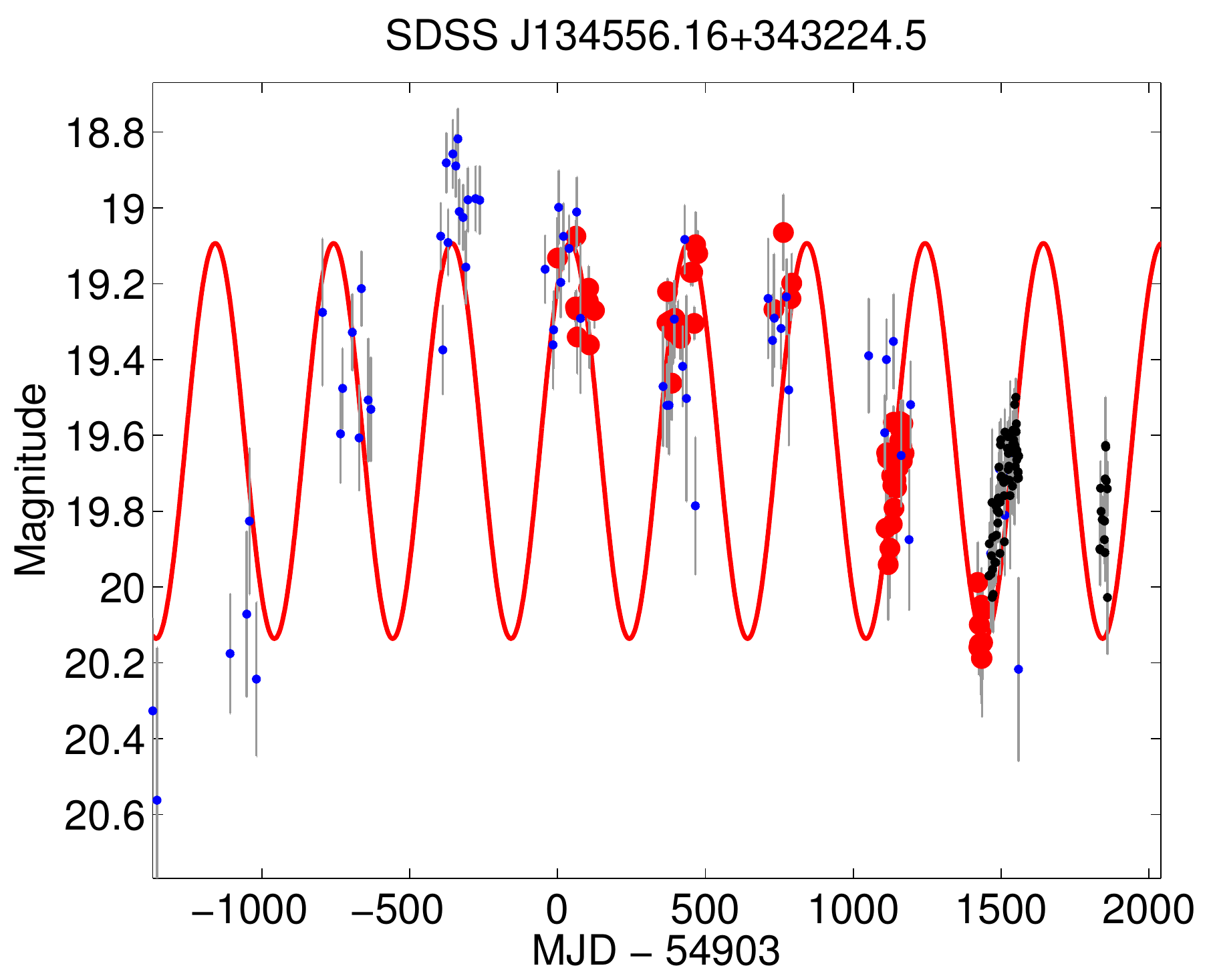}
\end{subfigure} \hspace{0.2cm}
\begin{subfigure}{.45\textwidth}
\centering
\includegraphics[width=8cm,height=4.5cm]{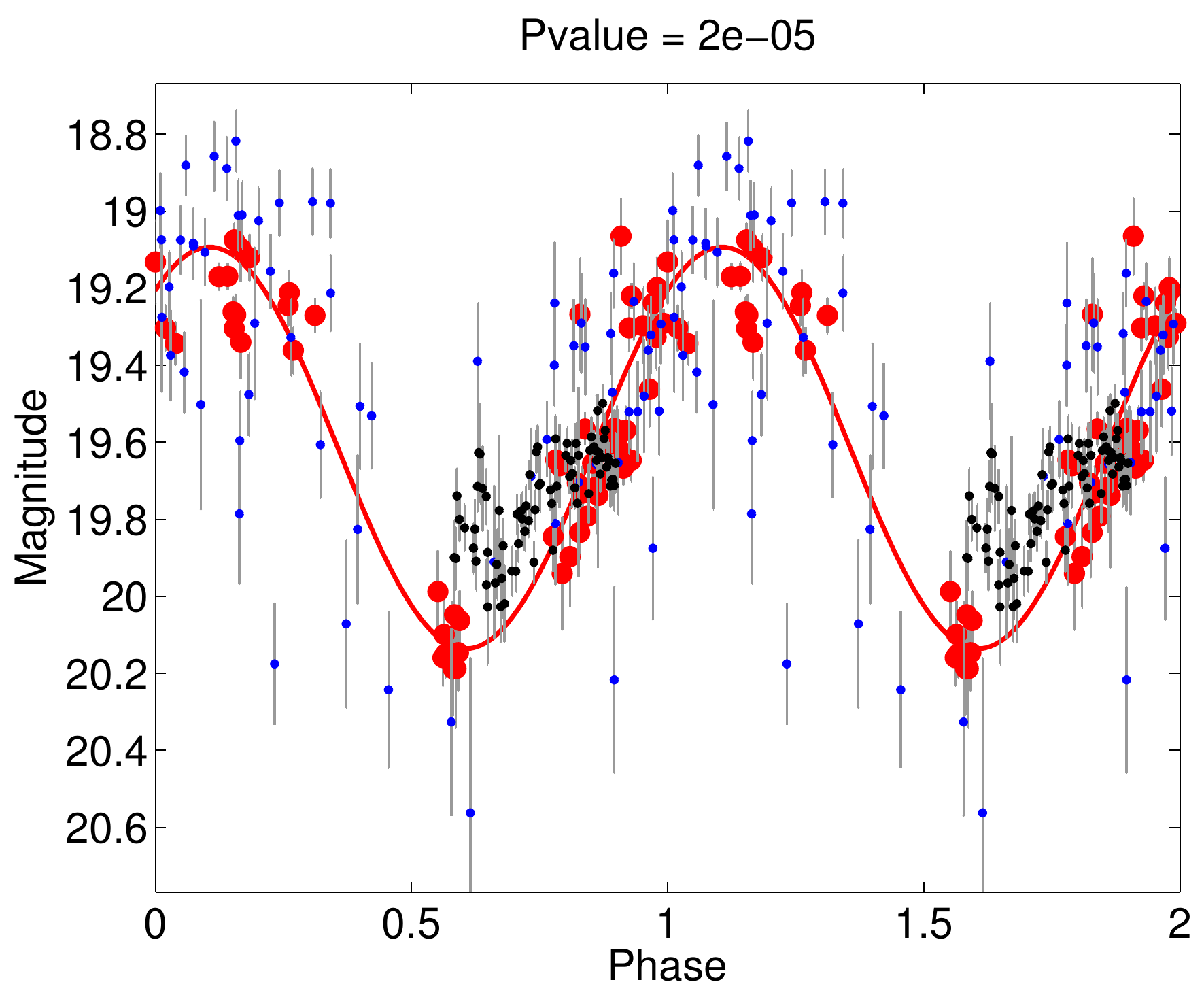}
\end{subfigure} \hspace{0.2cm}
\begin{subfigure}{.45\textwidth}
\centering
\includegraphics[width=8cm,height=4.5cm]{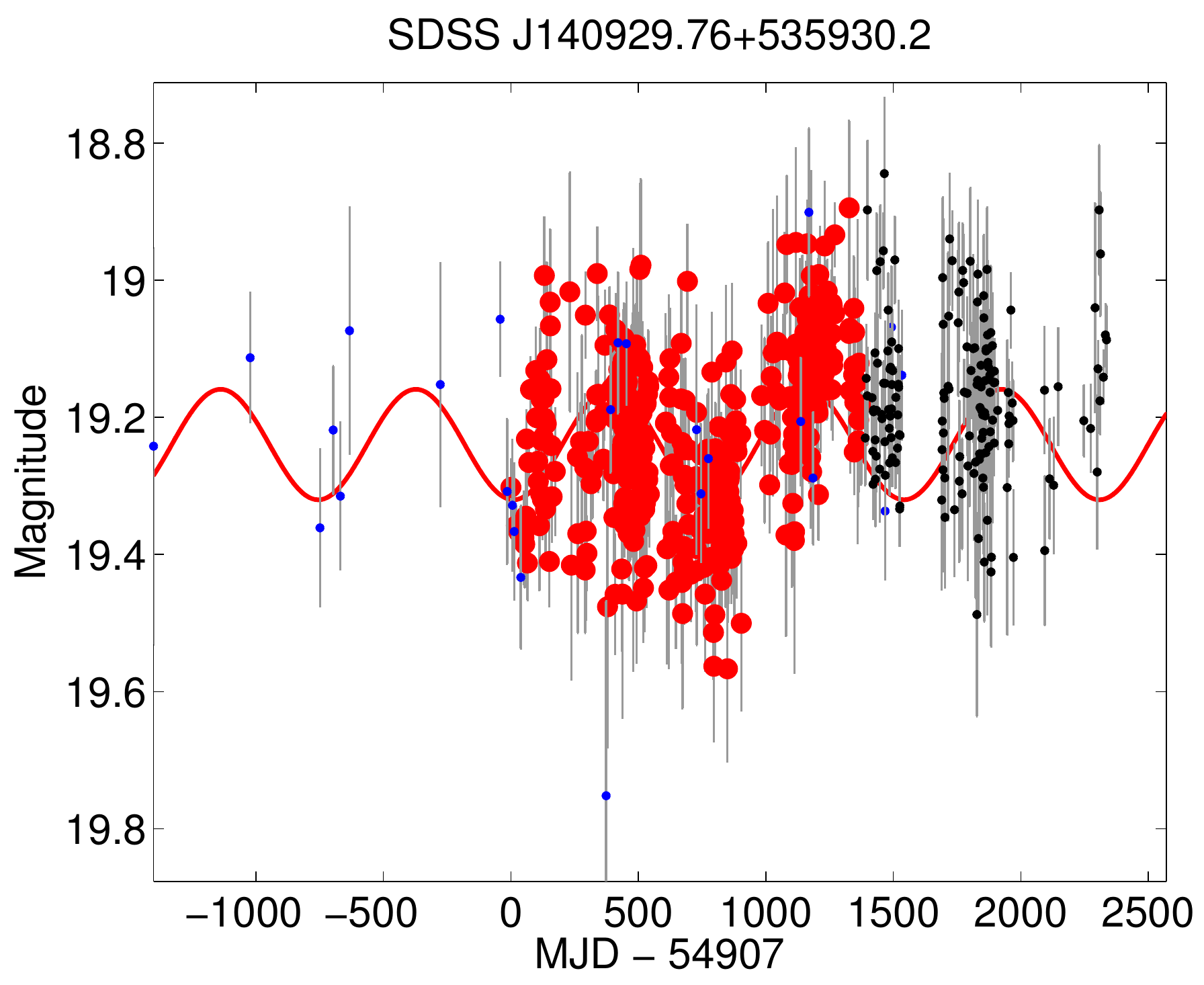}
\end{subfigure} \hspace{0.2cm}
\begin{subfigure}{.45\textwidth}
\centering
\includegraphics[width=8cm,height=4.5cm]{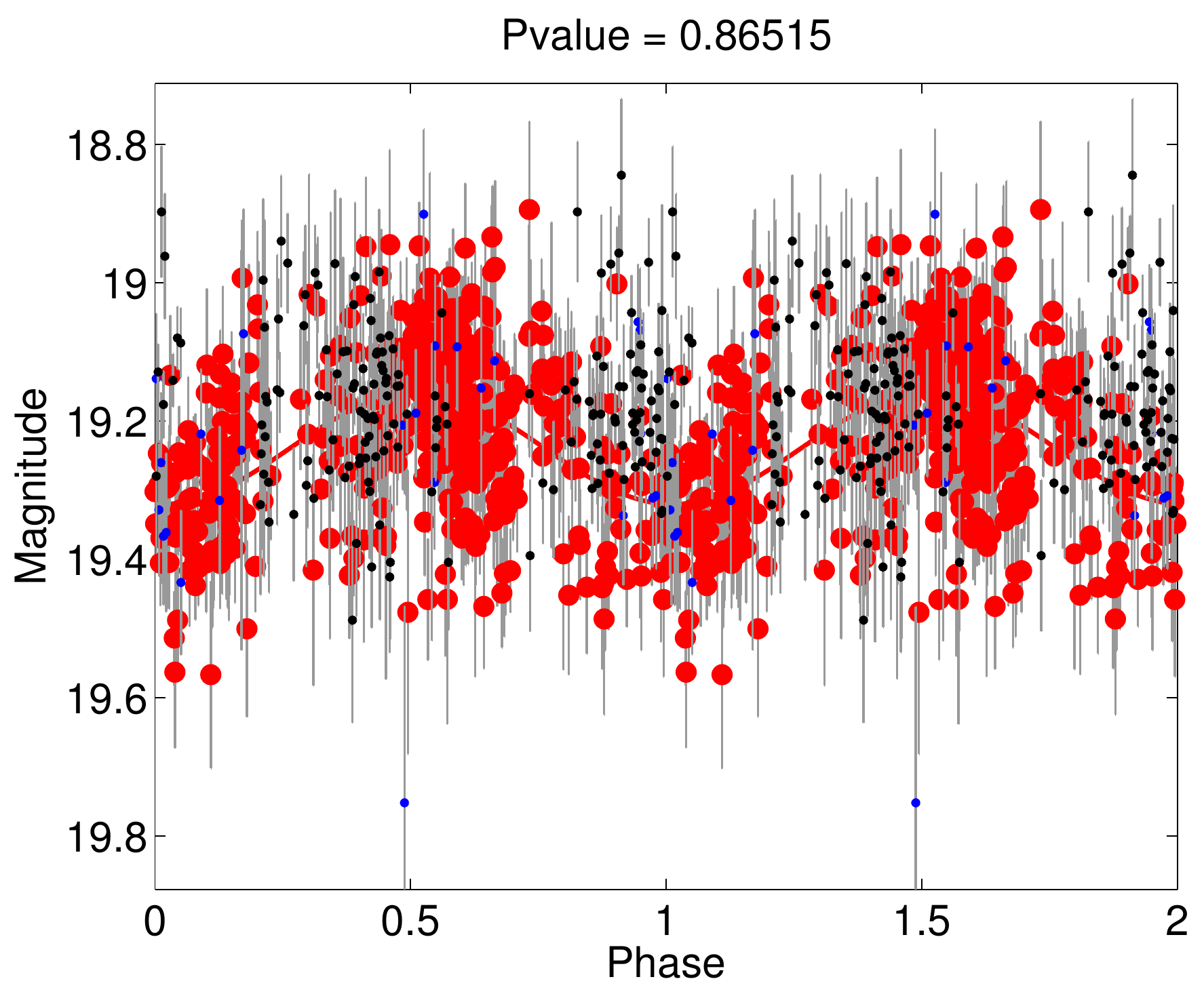}
\end{subfigure} \hspace{0.2cm}
\begin{subfigure}{.45\textwidth}
\centering
\includegraphics[width=8cm,height=4.5cm]{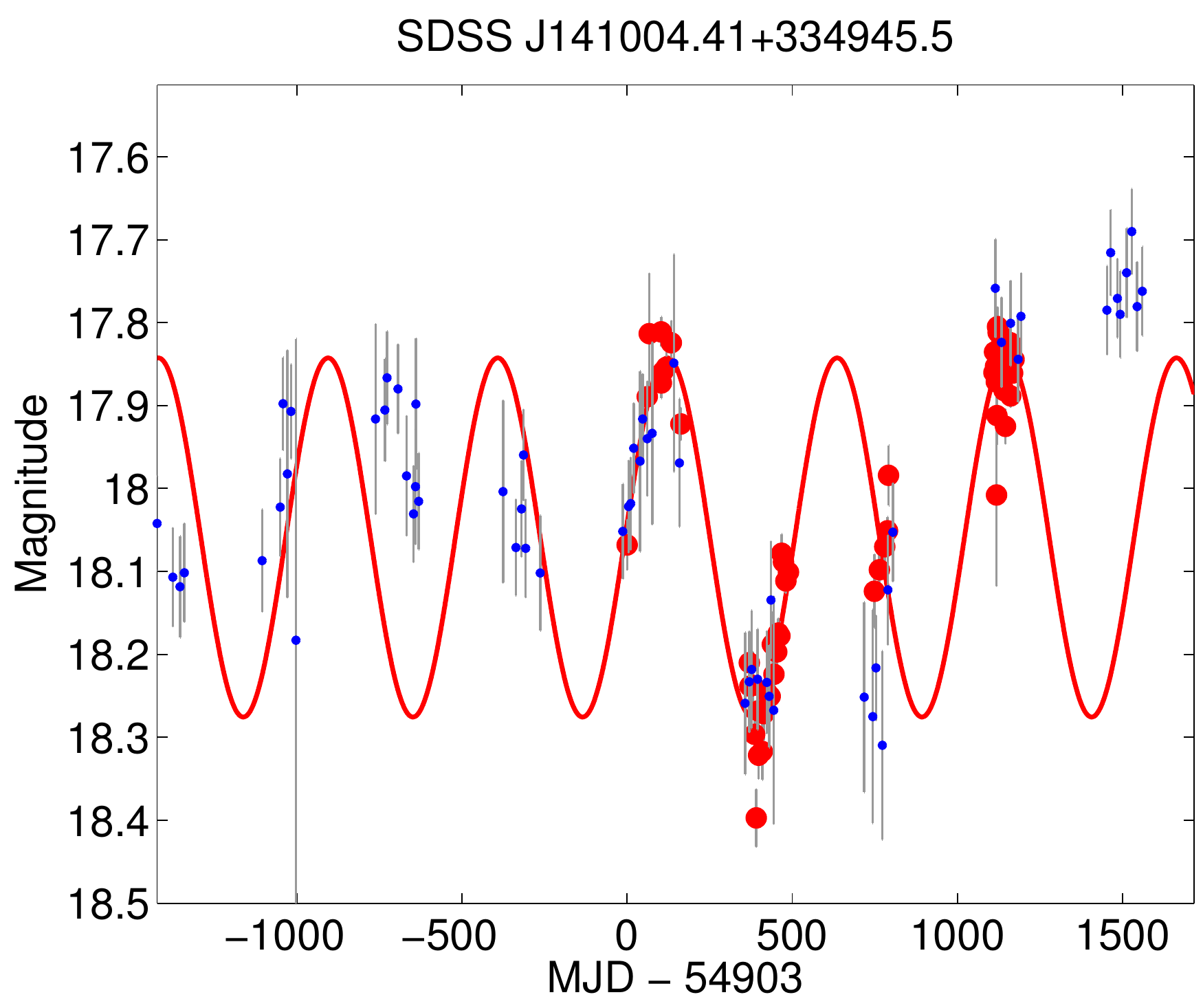}
\end{subfigure} \hspace{0.2cm}
\begin{subfigure}{.45\textwidth}
\centering
\includegraphics[width=8cm,height=4.5cm]{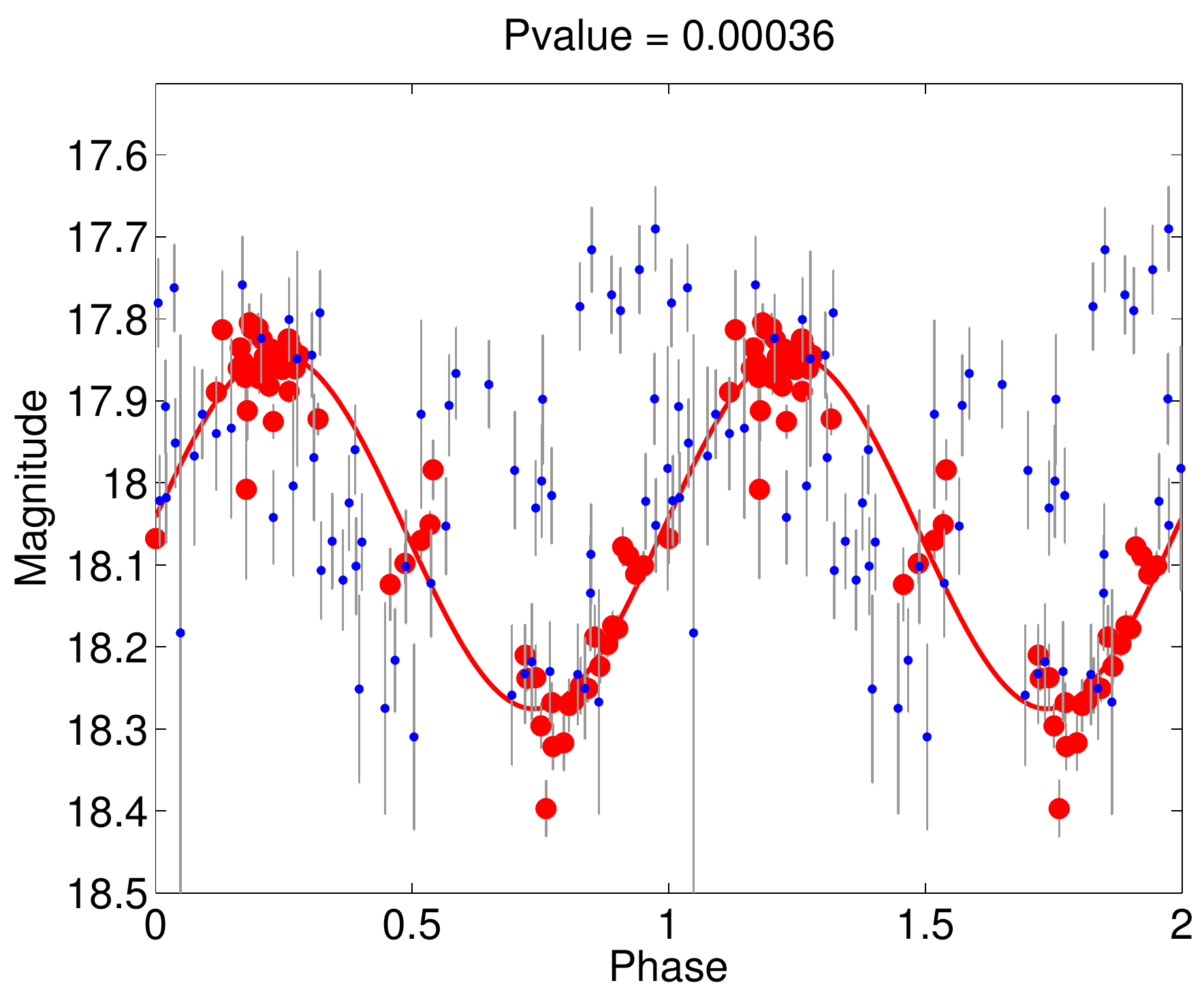}
\end{subfigure} \hspace{0.2cm}
\begin{subfigure}{.45\textwidth}
\centering
\includegraphics[width=8cm,height=4.5cm]{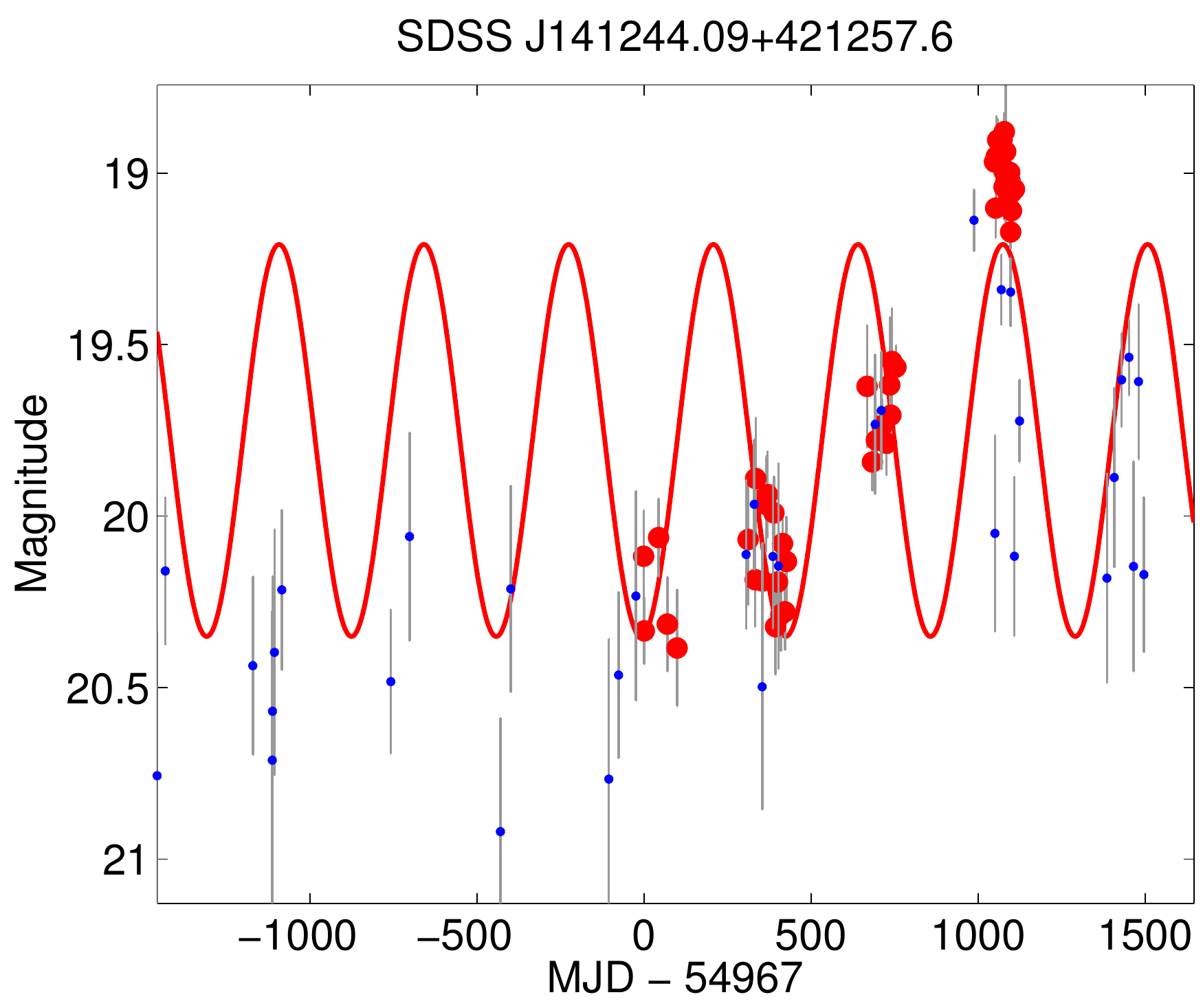}
\end{subfigure} \hspace{0.2cm}
\begin{subfigure}{.45\textwidth}
\centering
\includegraphics[width=8cm,height=4.5cm]{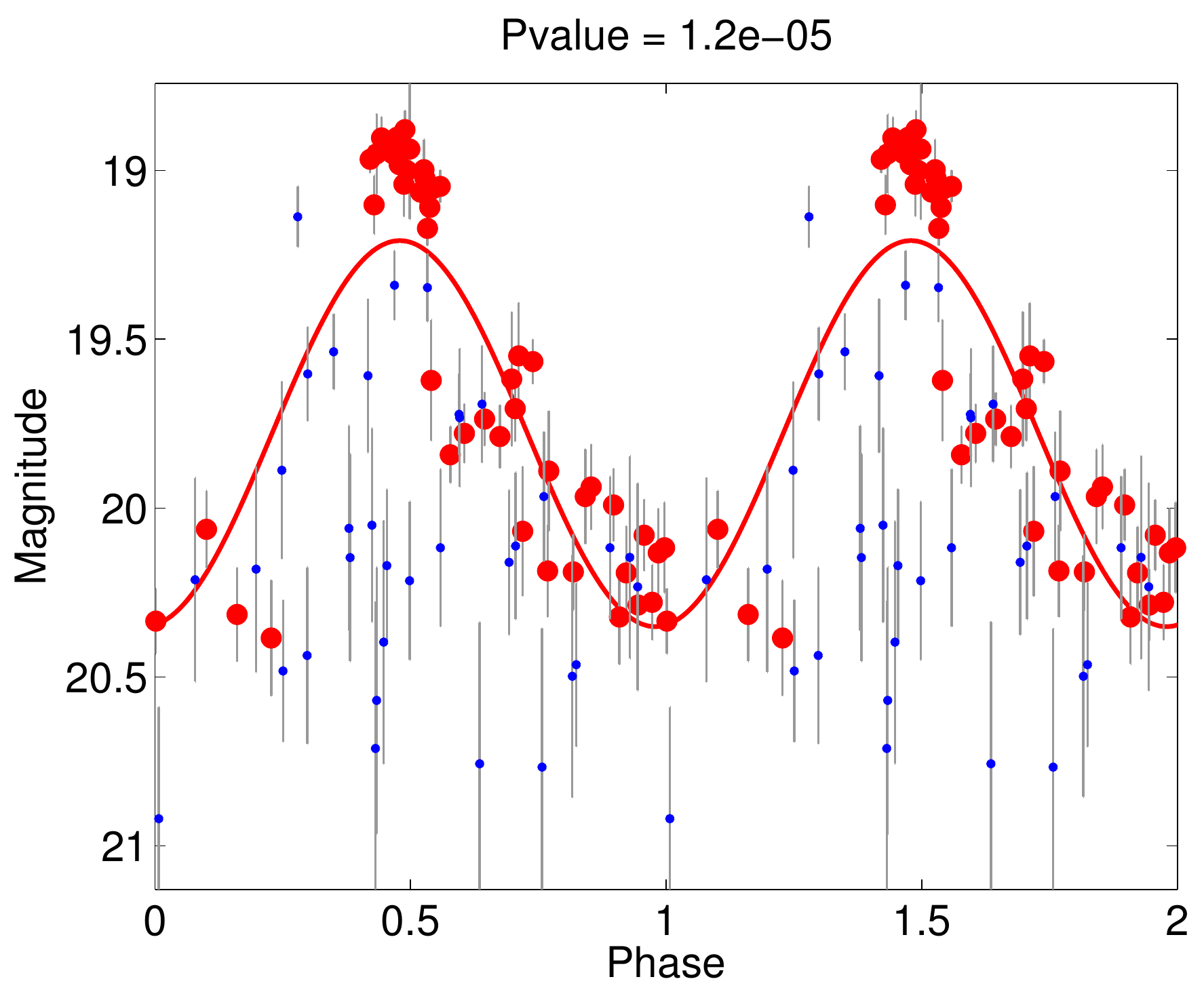}
\end{subfigure} \hspace{0.2cm}
\phantomcaption
\end{figure*}
\begin{figure*}
\ContinuedFloat
\begin{subfigure}{.45\textwidth}
\centering
\includegraphics[width=8cm,height=4.5cm]{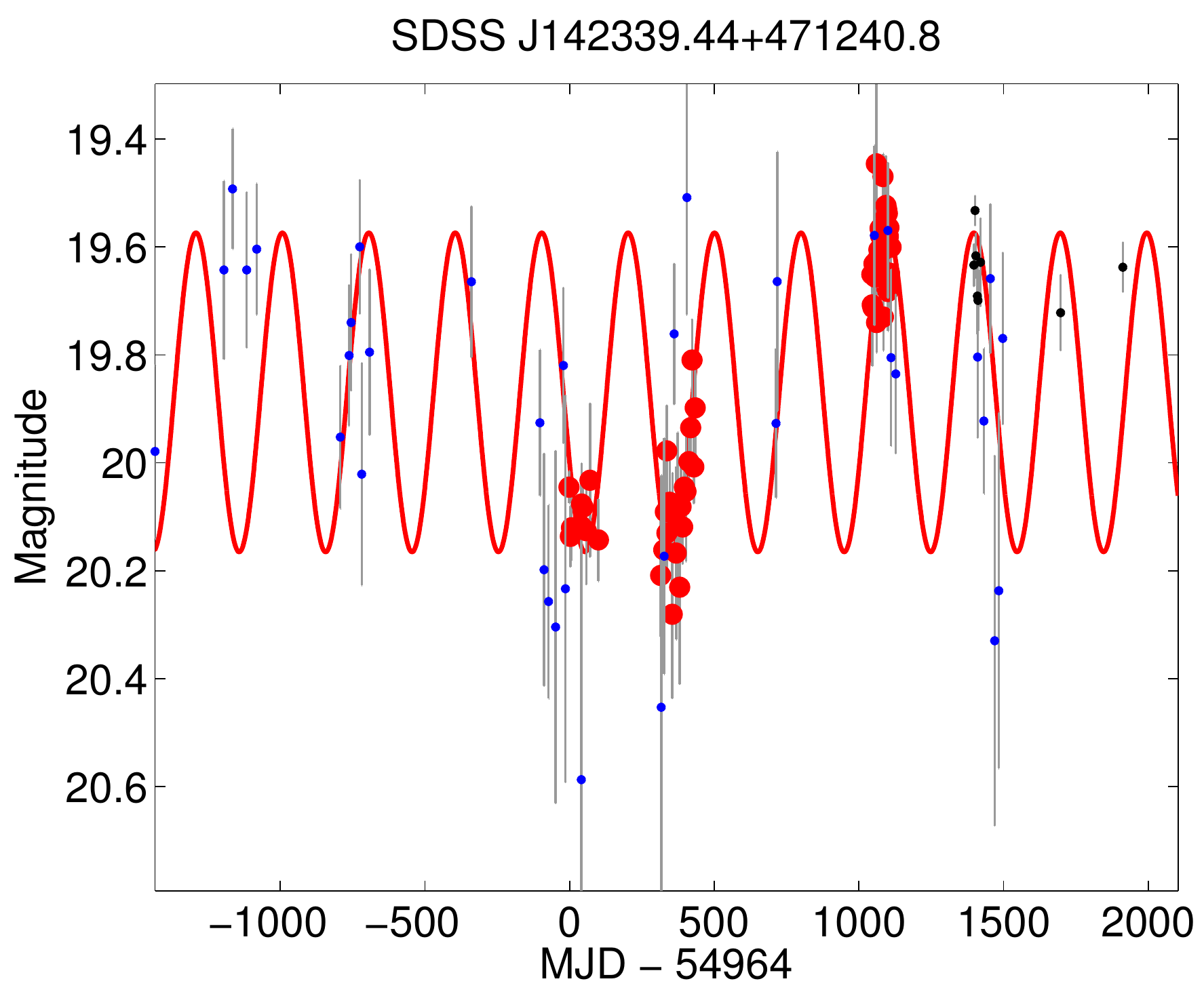}
\end{subfigure} \hspace{0.2cm}
\begin{subfigure}{.45\textwidth}
\centering
\includegraphics[width=8cm,height=4.5cm]{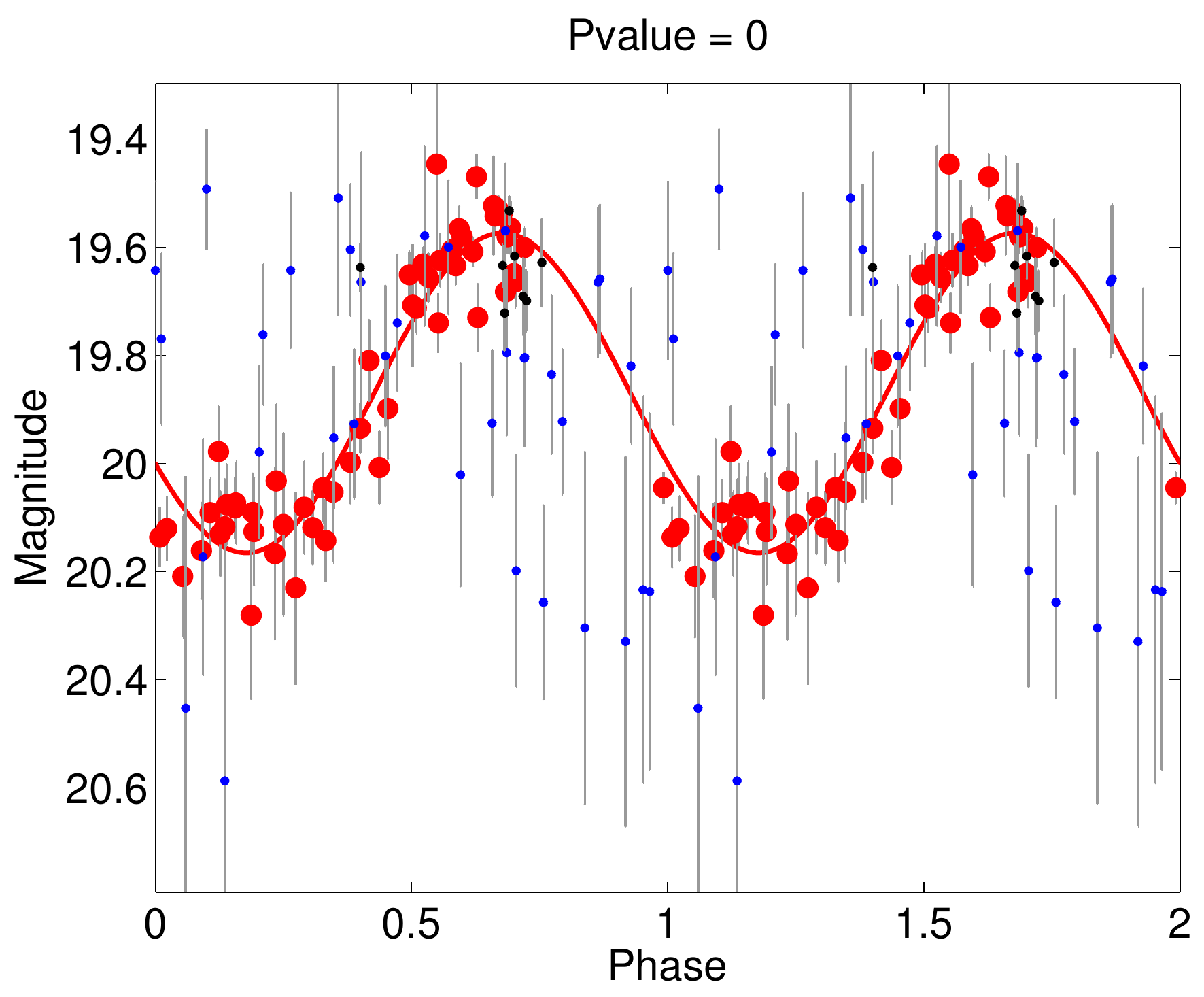}
\end{subfigure} \hspace{0.2cm}
\begin{subfigure}{.45\textwidth}
\centering
\includegraphics[width=8cm,height=4.5cm]{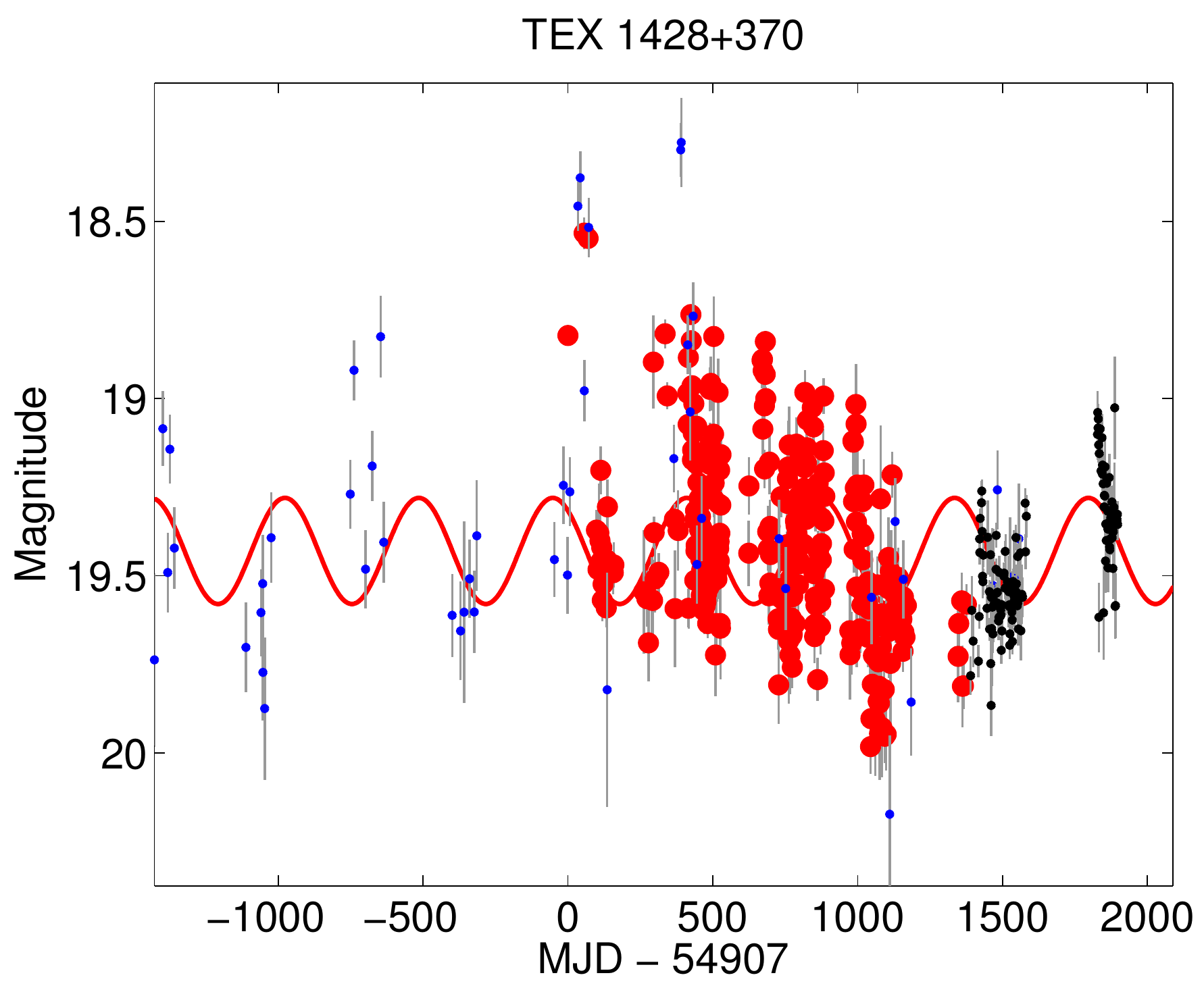}
\end{subfigure} \hspace{0.2cm}
\begin{subfigure}{.45\textwidth}
\centering
\includegraphics[width=8cm,height=4.5cm]{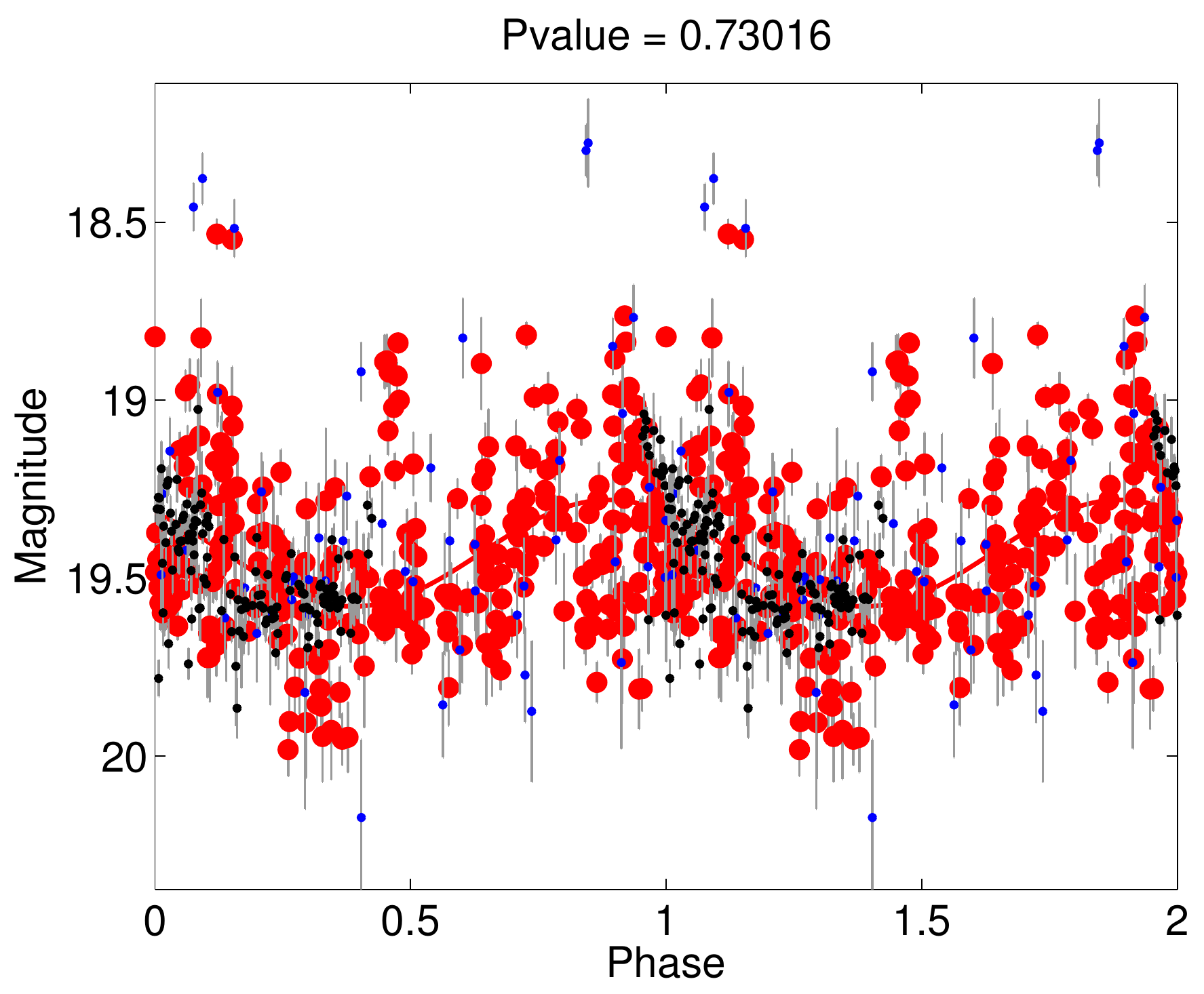}
\end{subfigure} \hspace{0.2cm}
\begin{subfigure}{.45\textwidth}
\centering
\includegraphics[width=8cm,height=4.5cm]{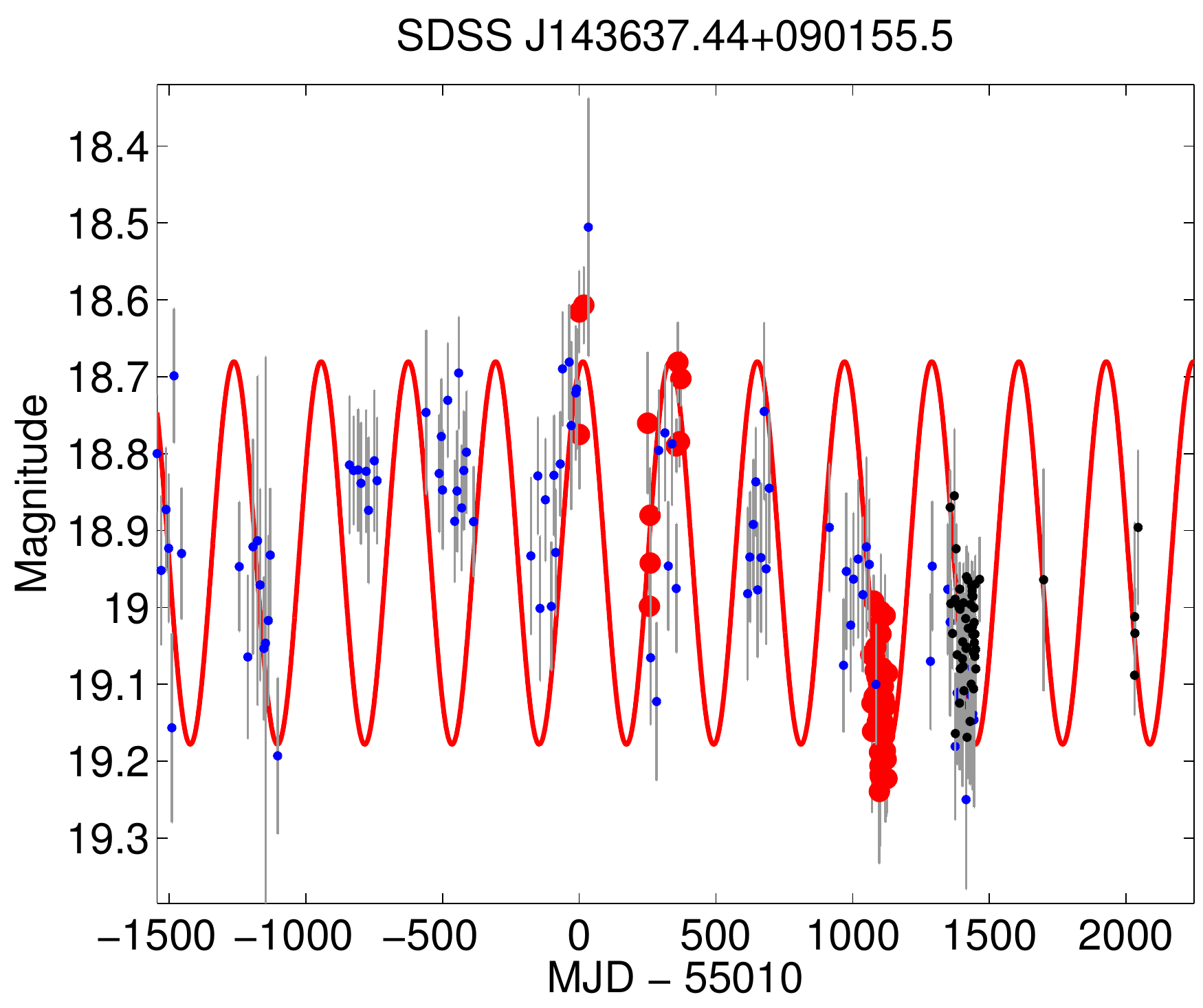}
\end{subfigure} \hspace{0.2cm}
\begin{subfigure}{.45\textwidth}
\centering
\includegraphics[width=8cm,height=4.5cm]{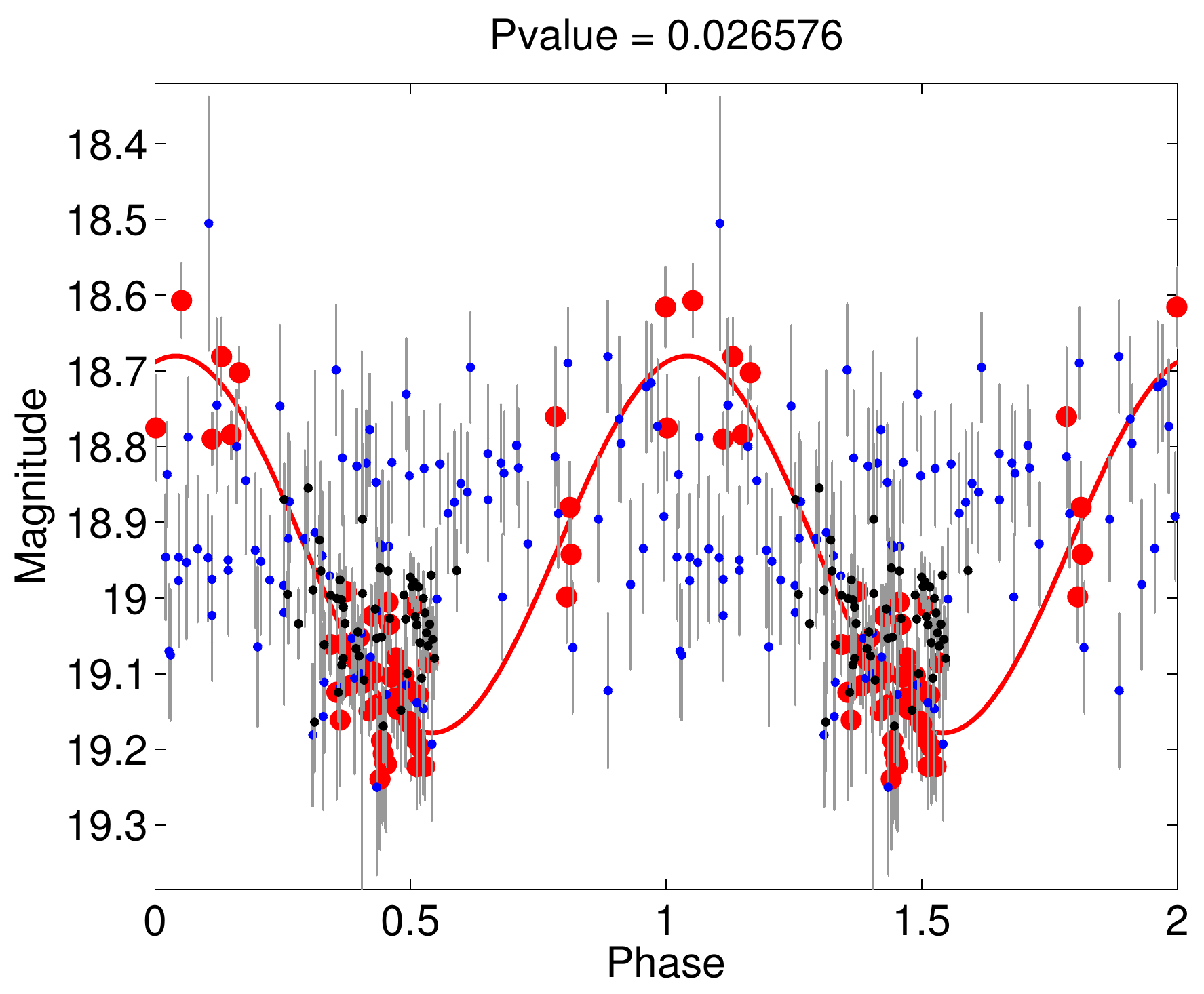}
\end{subfigure} \hspace{0.2cm}
\begin{subfigure}{.45\textwidth}
\centering
\includegraphics[width=8cm,height=4.5cm]{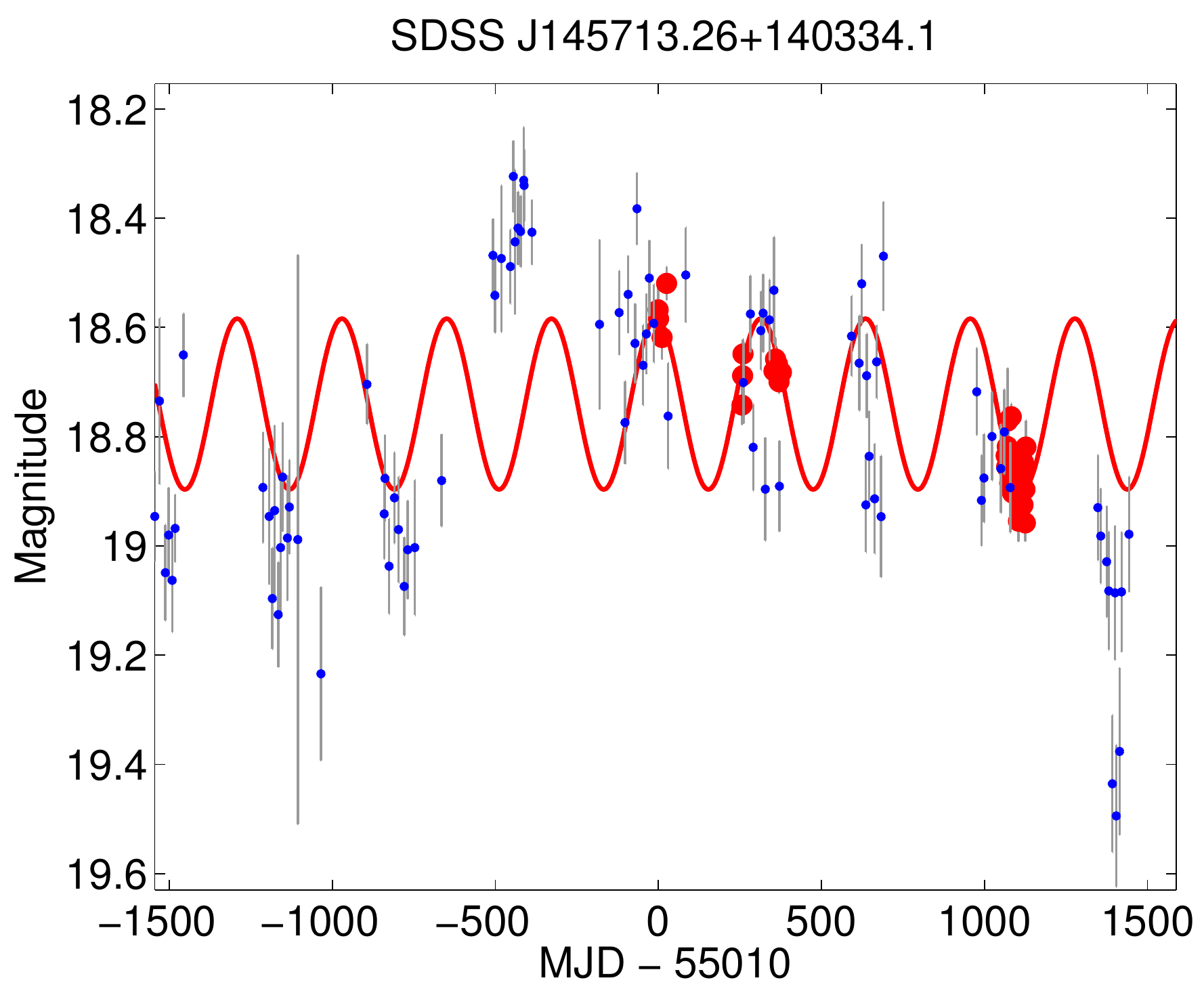}
\end{subfigure} \hspace{0.2cm}
\begin{subfigure}{.45\textwidth}
\centering
\includegraphics[width=8cm,height=4.5cm]{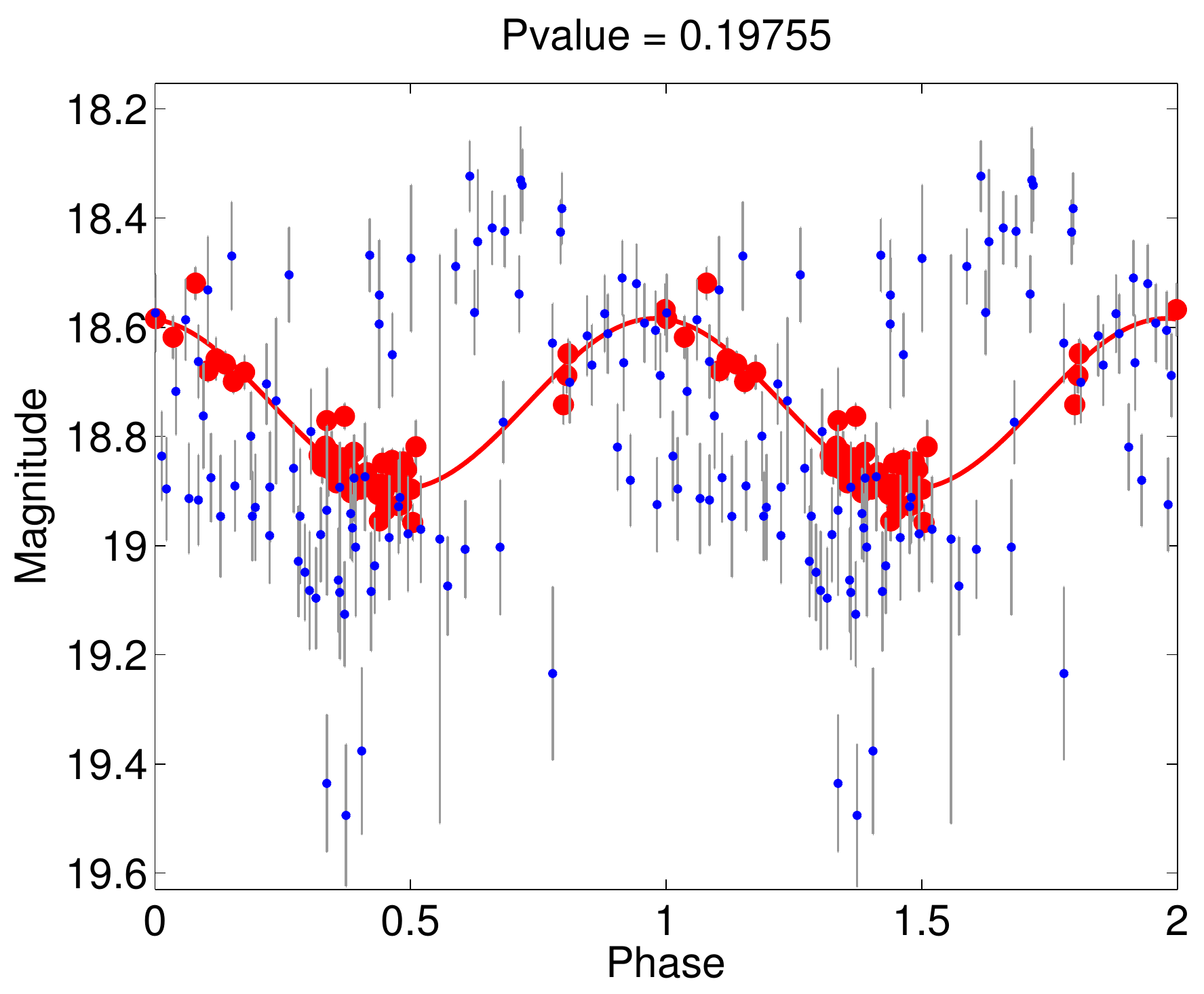}
\end{subfigure} \hspace{0.2cm}
\begin{subfigure}{.45\textwidth}
\centering
\includegraphics[width=8cm,height=4.5cm]{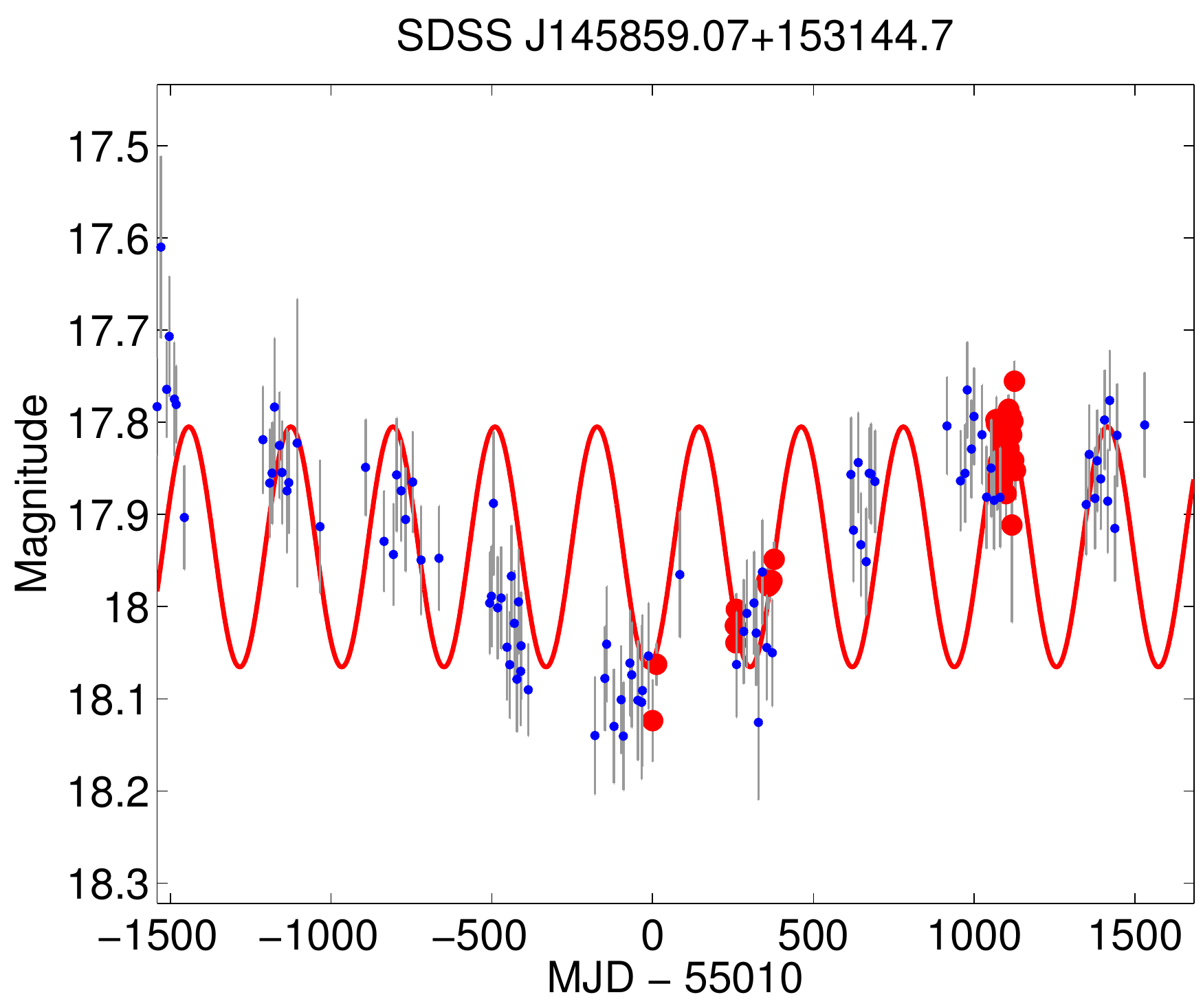}
\end{subfigure} \hspace{0.2cm}
\begin{subfigure}{.45\textwidth}
\centering
\includegraphics[width=8cm,height=4.5cm]{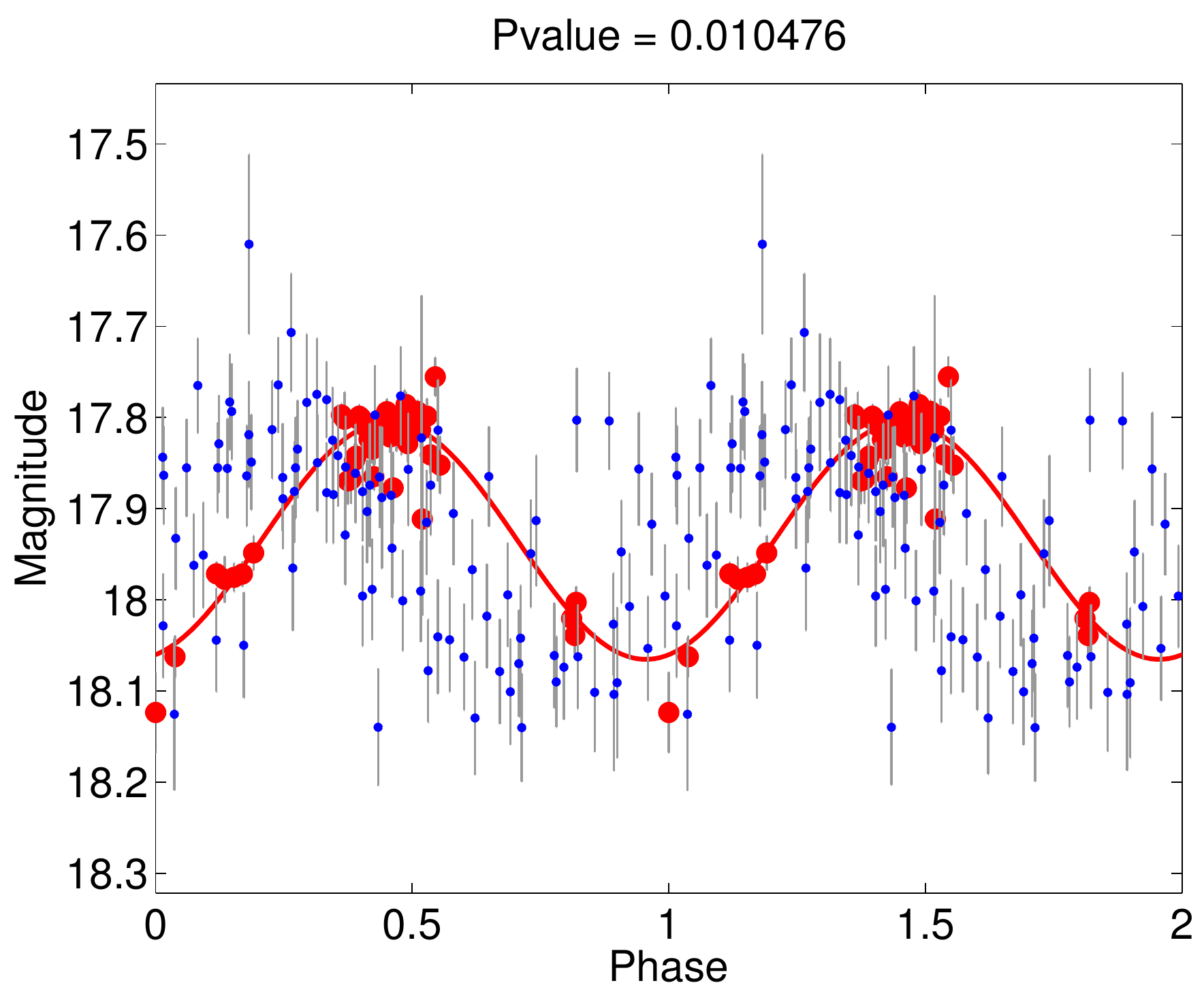}
\end{subfigure} \hspace{0.2cm}
\phantomcaption
\end{figure*}
\begin{figure*}
\ContinuedFloat
\begin{subfigure}{.45\textwidth}
\centering
\includegraphics[width=8cm,height=4.5cm]{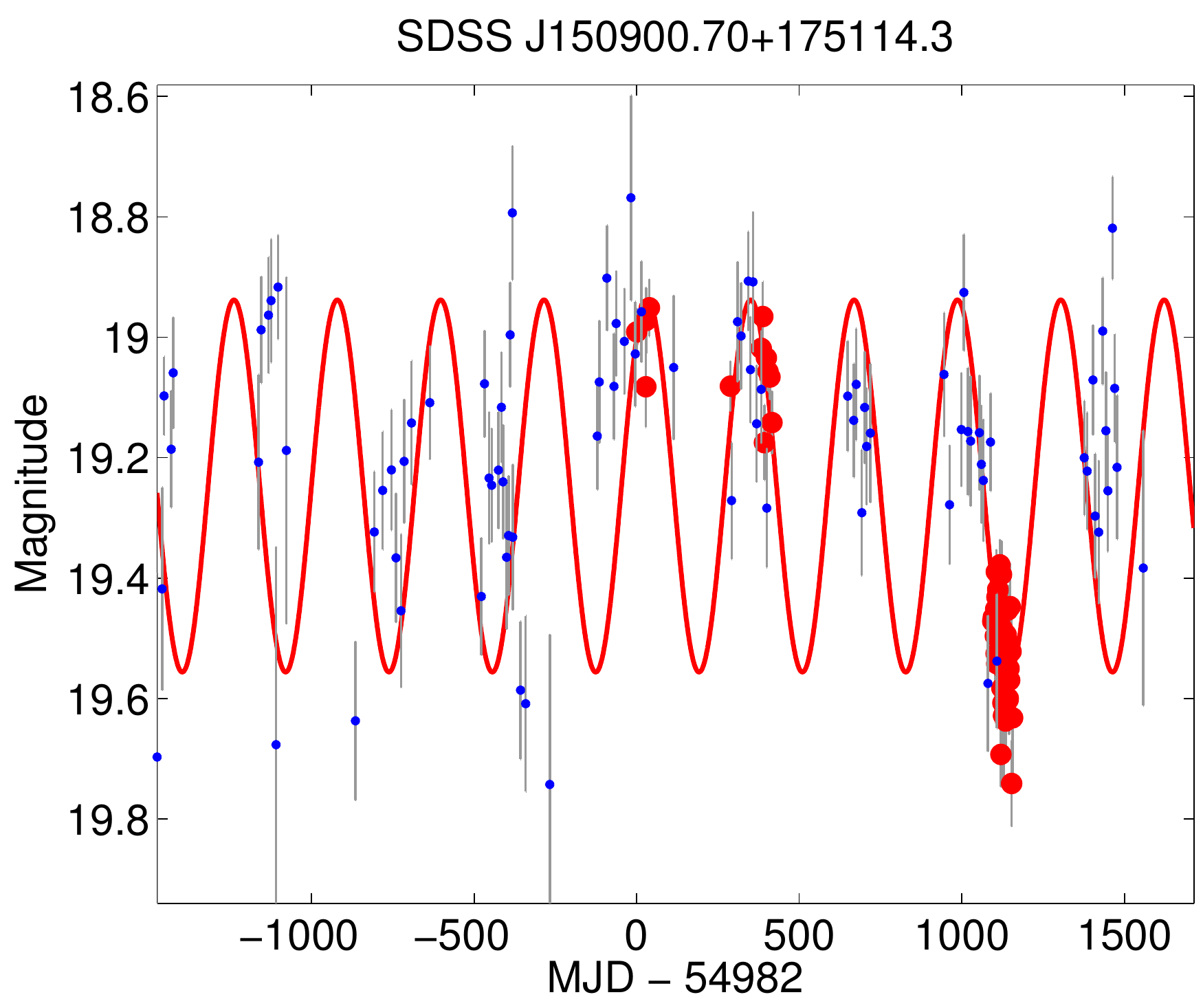}
\end{subfigure} \hspace{0.2cm}
\begin{subfigure}{.45\textwidth}
\centering
\includegraphics[width=8cm,height=4.5cm]{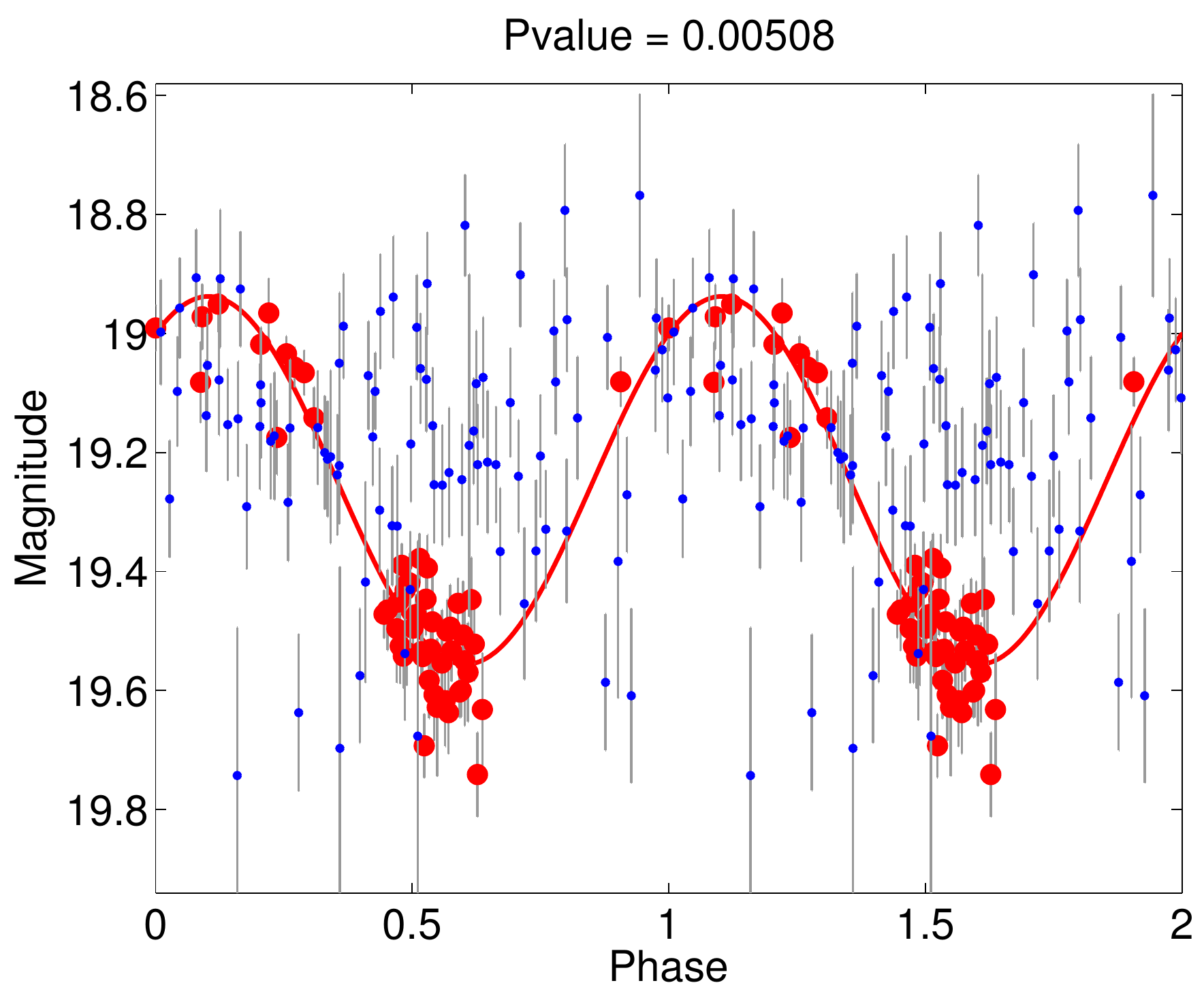}
\end{subfigure} \hspace{0.2cm}
\begin{subfigure}{.45\textwidth}
\centering
\includegraphics[width=8cm,height=4.5cm]{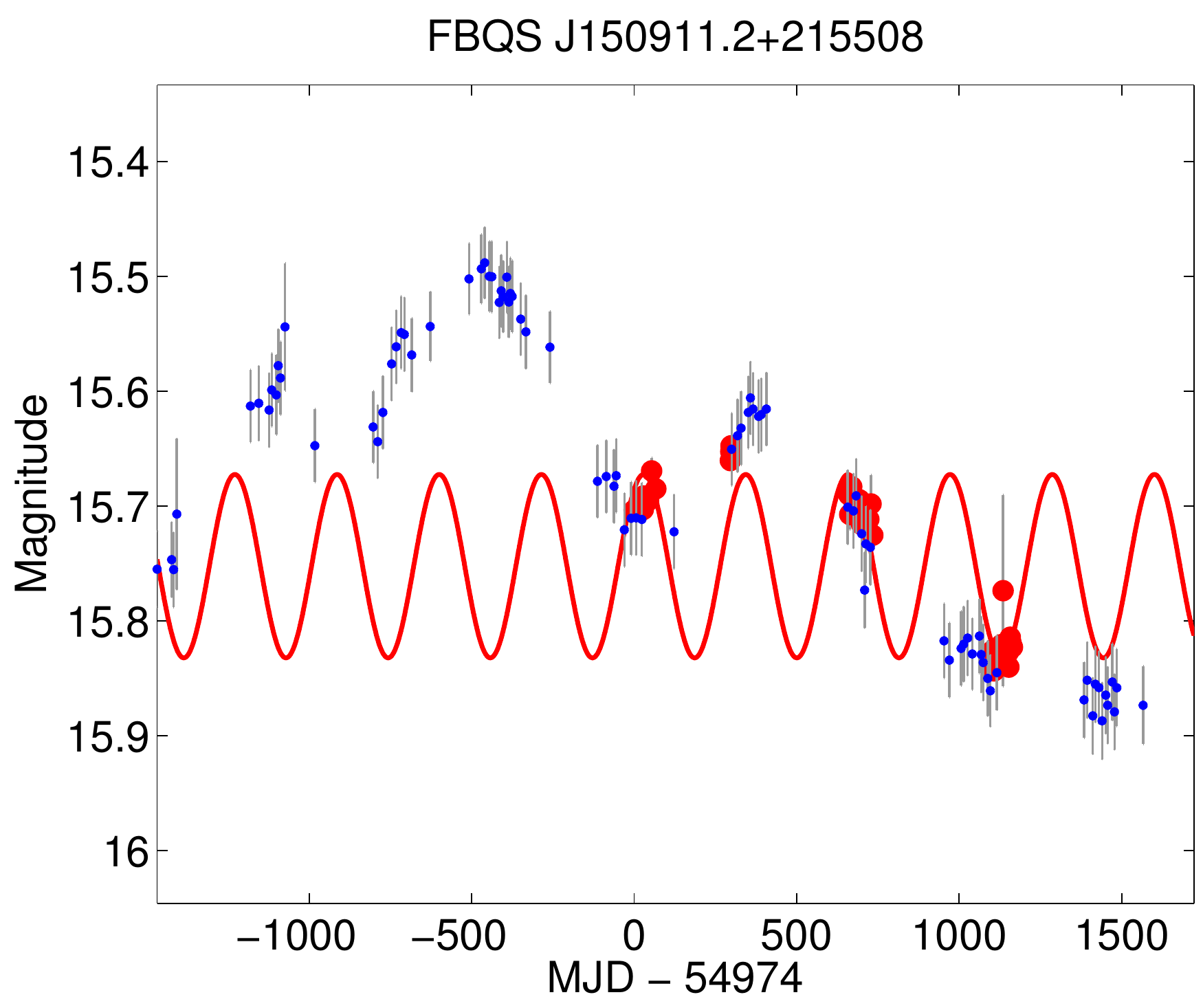}
\end{subfigure} \hspace{0.2cm}
\begin{subfigure}{.45\textwidth}
\centering
\includegraphics[width=8cm,height=4.5cm]{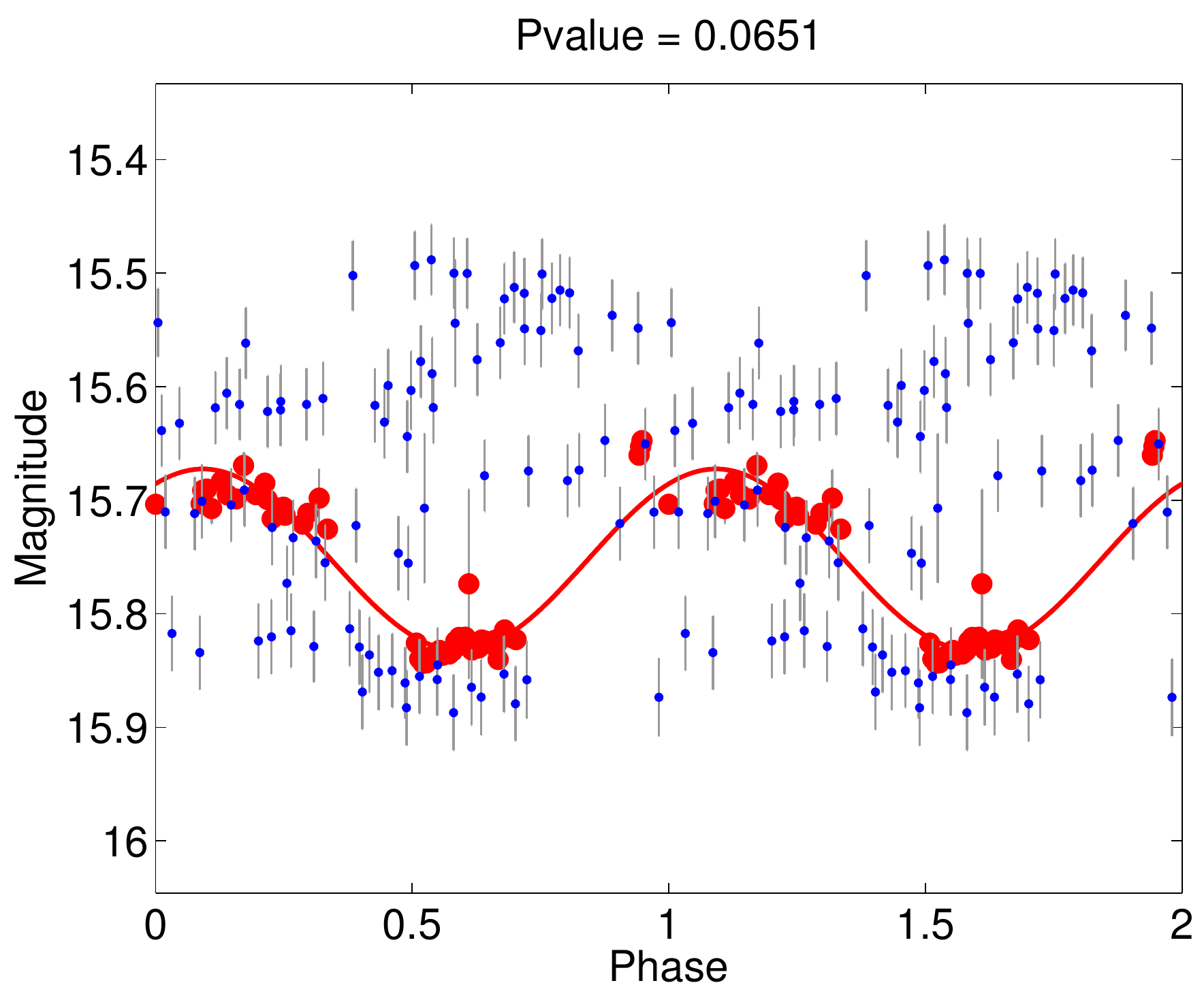}
\end{subfigure} \hspace{0.2cm}
\begin{subfigure}{.45\textwidth}
\centering
\includegraphics[width=8cm,height=4.5cm]{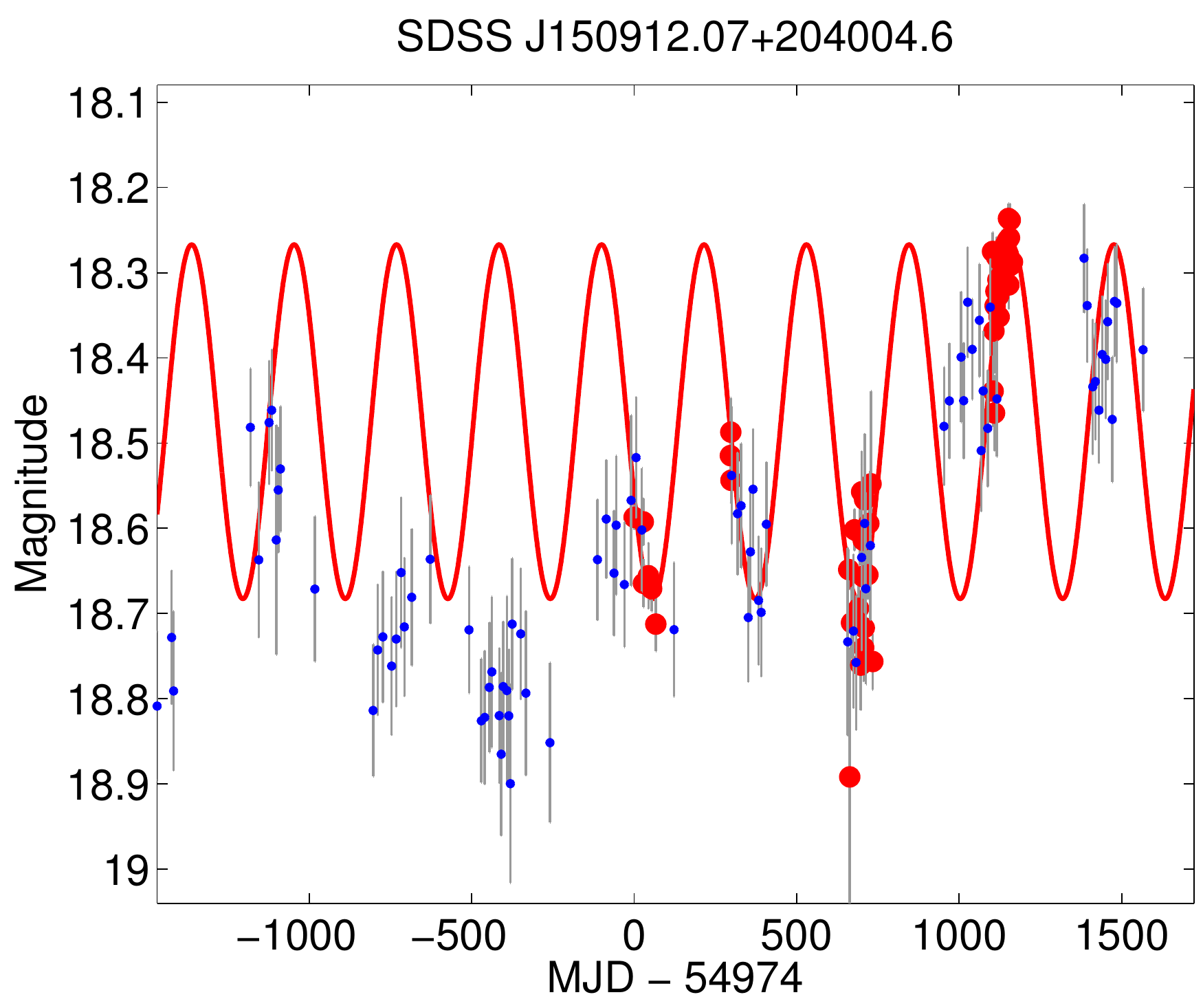}
\end{subfigure} \hspace{0.2cm}
\begin{subfigure}{.45\textwidth}
\centering
\includegraphics[width=8cm,height=4.5cm]{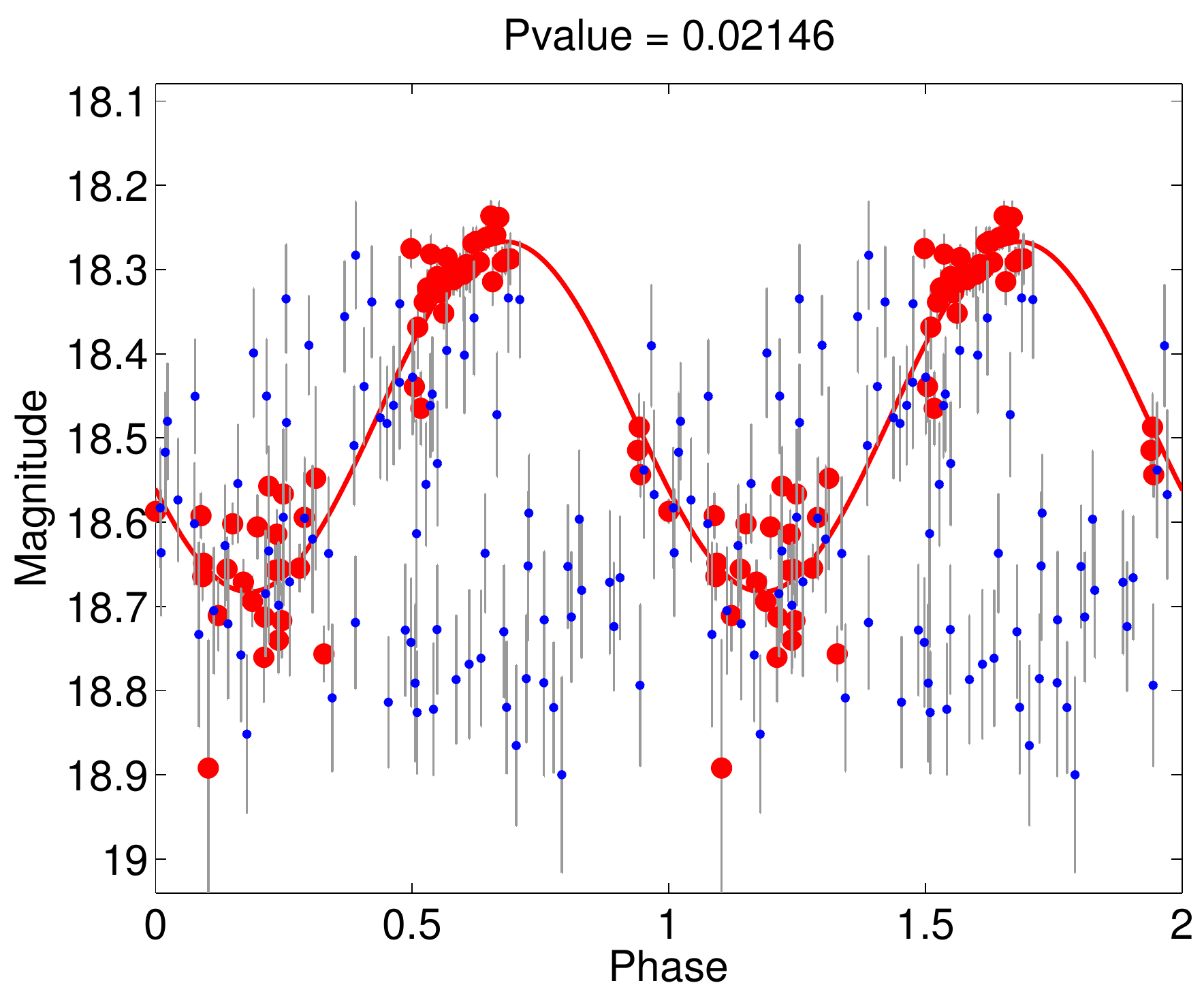}
\end{subfigure} \hspace{0.2cm}
\begin{subfigure}{.45\textwidth}
\centering
\includegraphics[width=8cm,height=4.5cm]{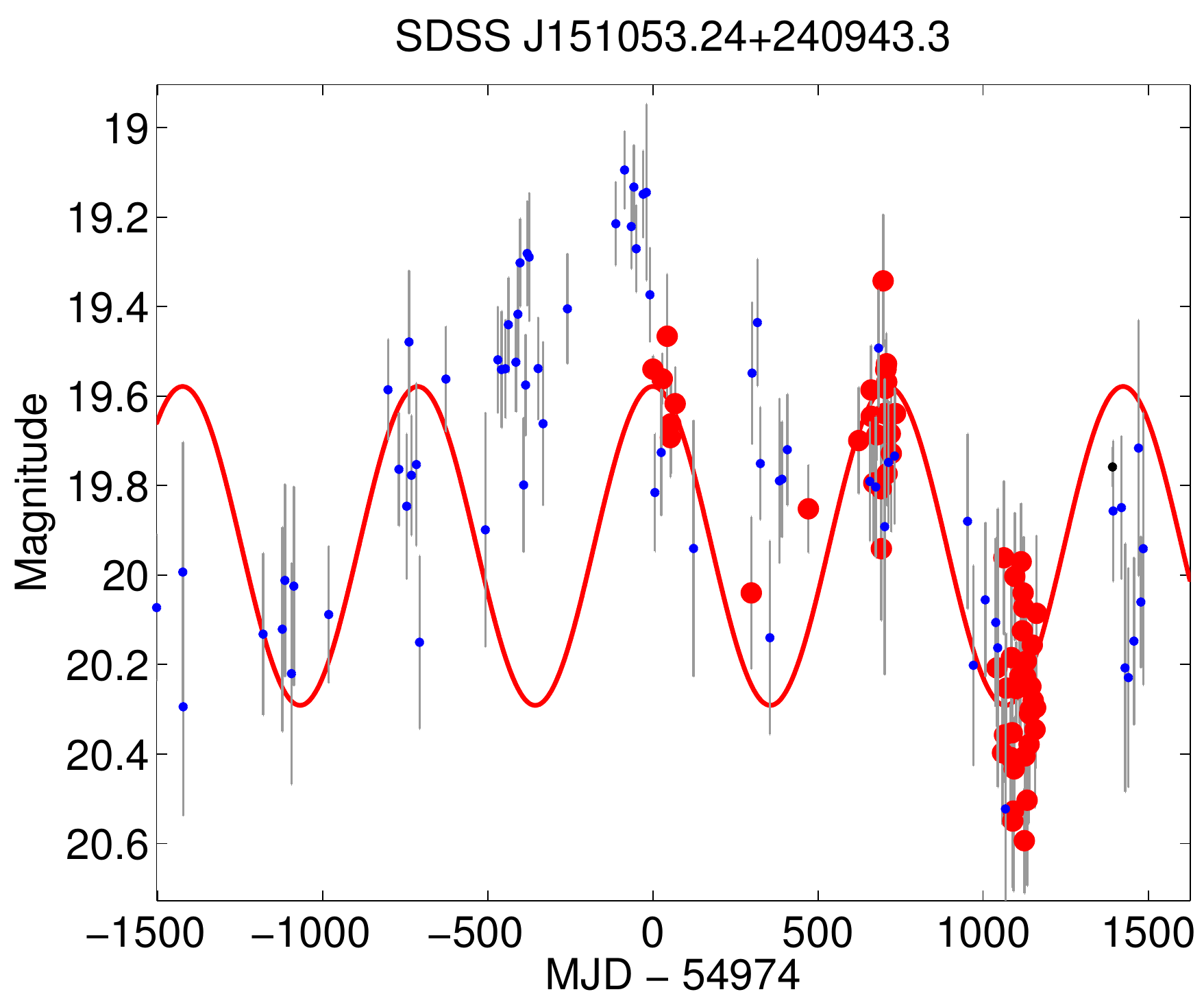}
\end{subfigure} \hspace{0.2cm}
\begin{subfigure}{.45\textwidth}
\centering
\includegraphics[width=8cm,height=4.5cm]{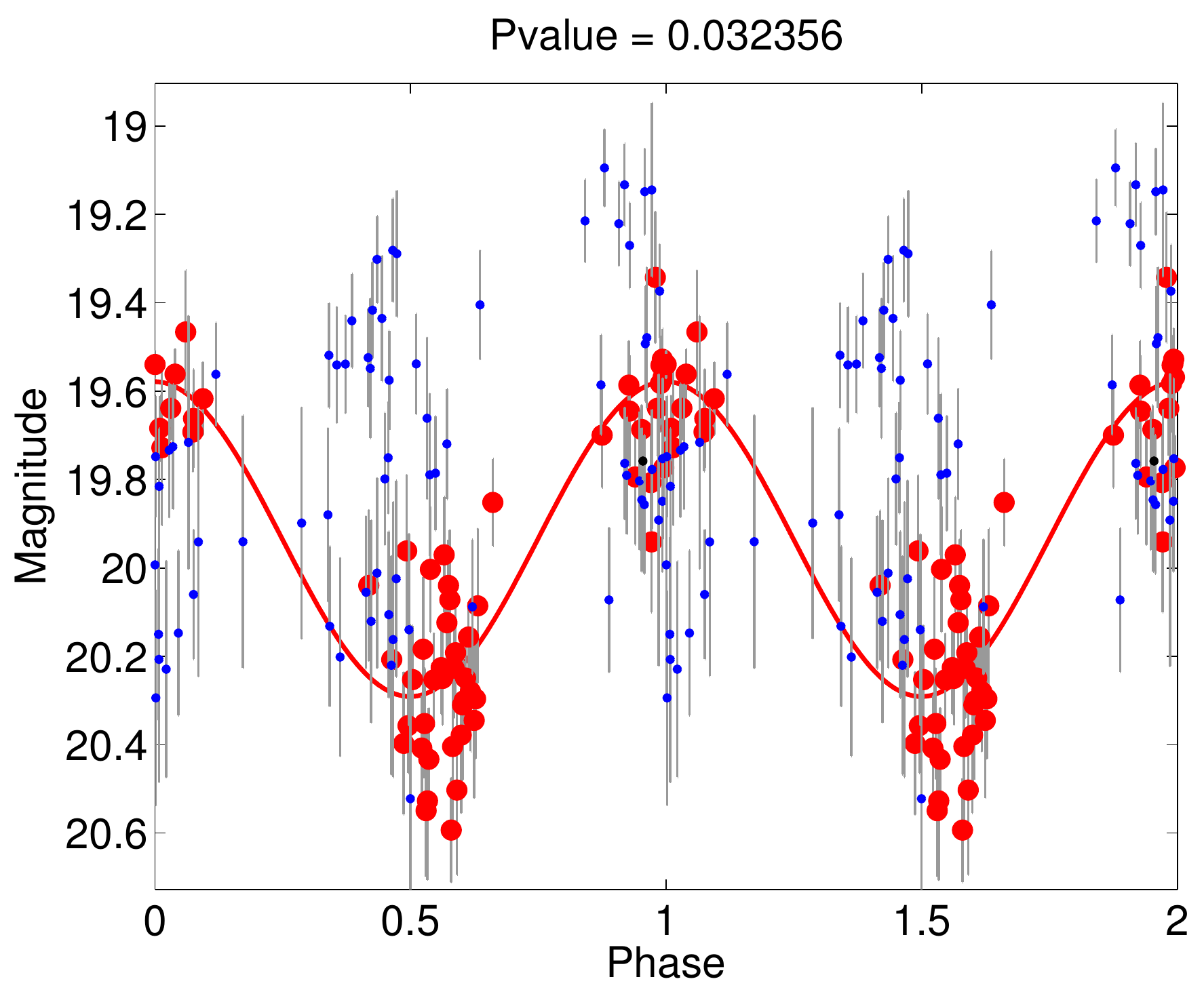}
\end{subfigure} \hspace{0.2cm}
\begin{subfigure}{.45\textwidth}
\centering
\includegraphics[width=8cm,height=4.5cm]{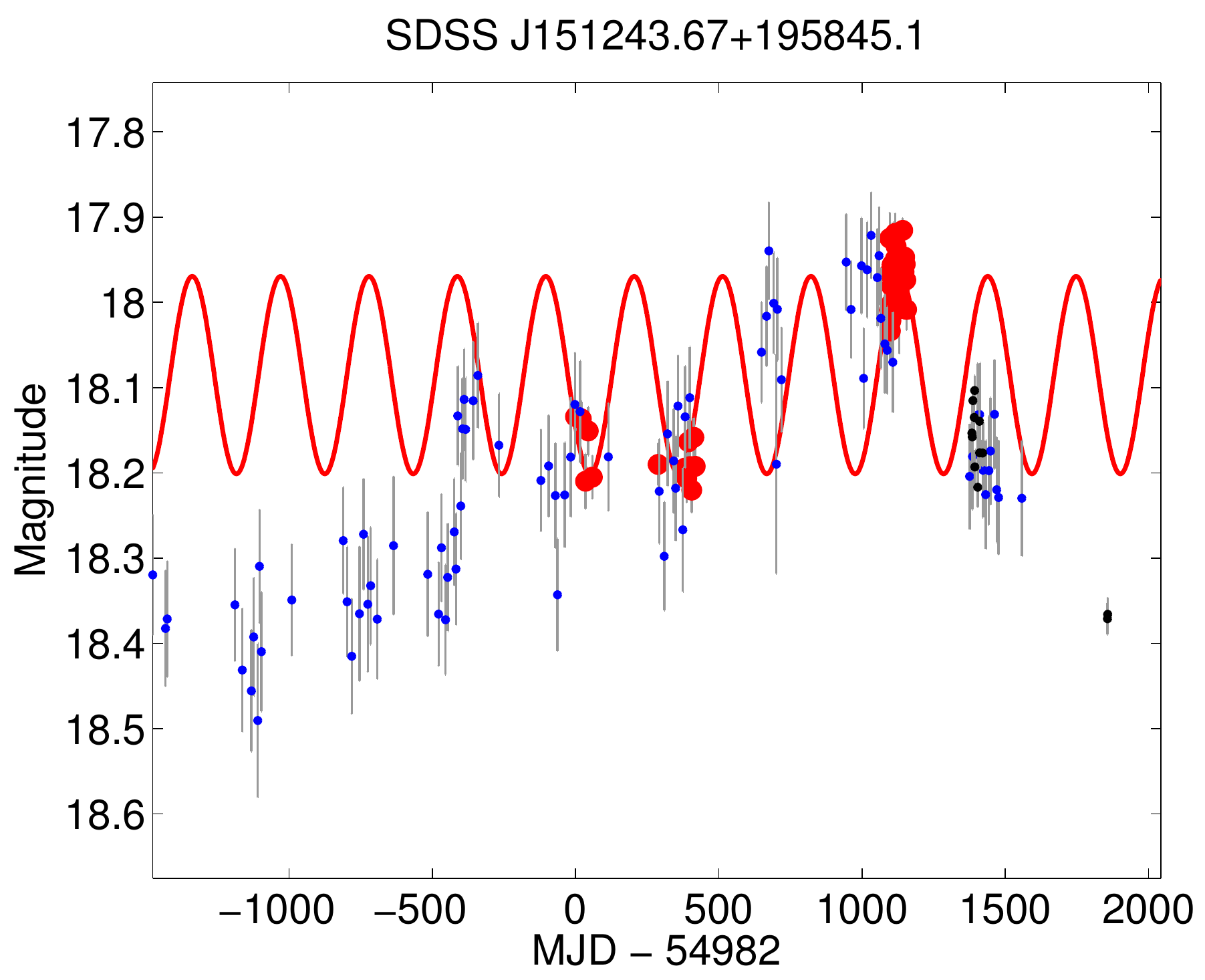}
\end{subfigure} \hspace{0.2cm}
\begin{subfigure}{.45\textwidth}
\centering
\includegraphics[width=8cm,height=4.5cm]{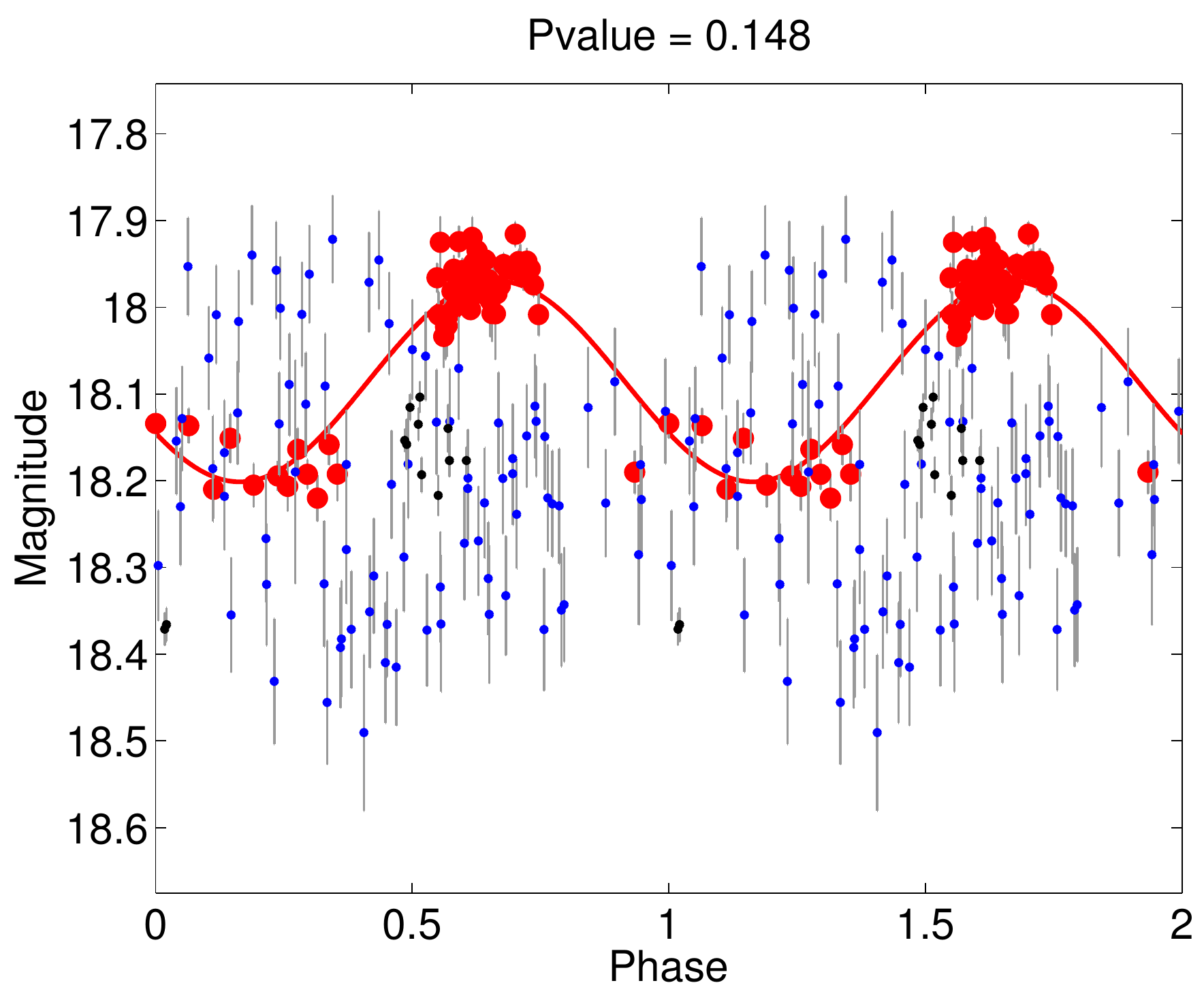}
\end{subfigure} \hspace{0.2cm}
\phantomcaption
\end{figure*}
\begin{figure*}
\ContinuedFloat
\begin{subfigure}{.45\textwidth}
\centering
\includegraphics[width=8cm,height=4.5cm]{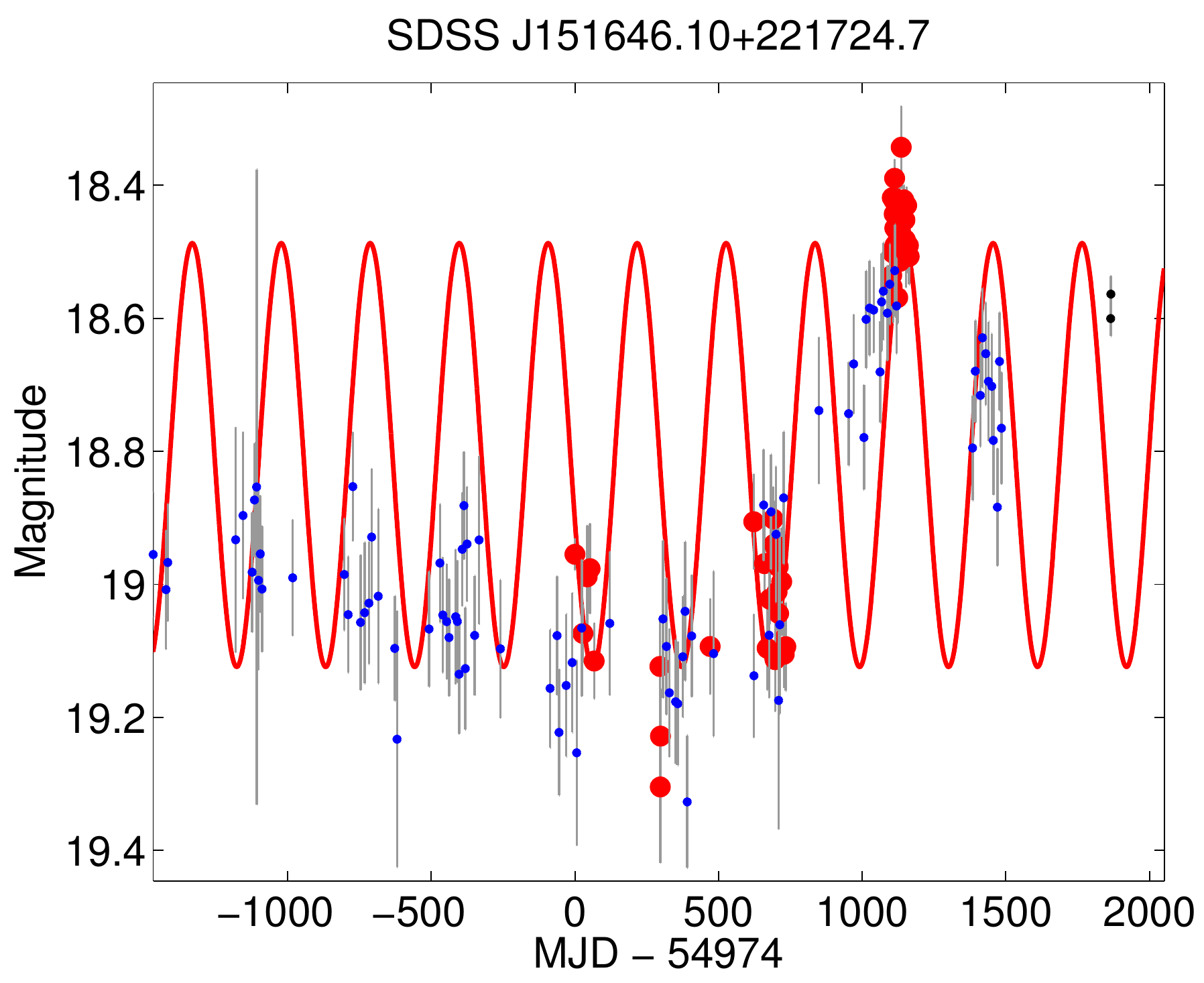}
\end{subfigure} \hspace{0.2cm}
\begin{subfigure}{.45\textwidth}
\centering
\includegraphics[width=8cm,height=4.5cm]{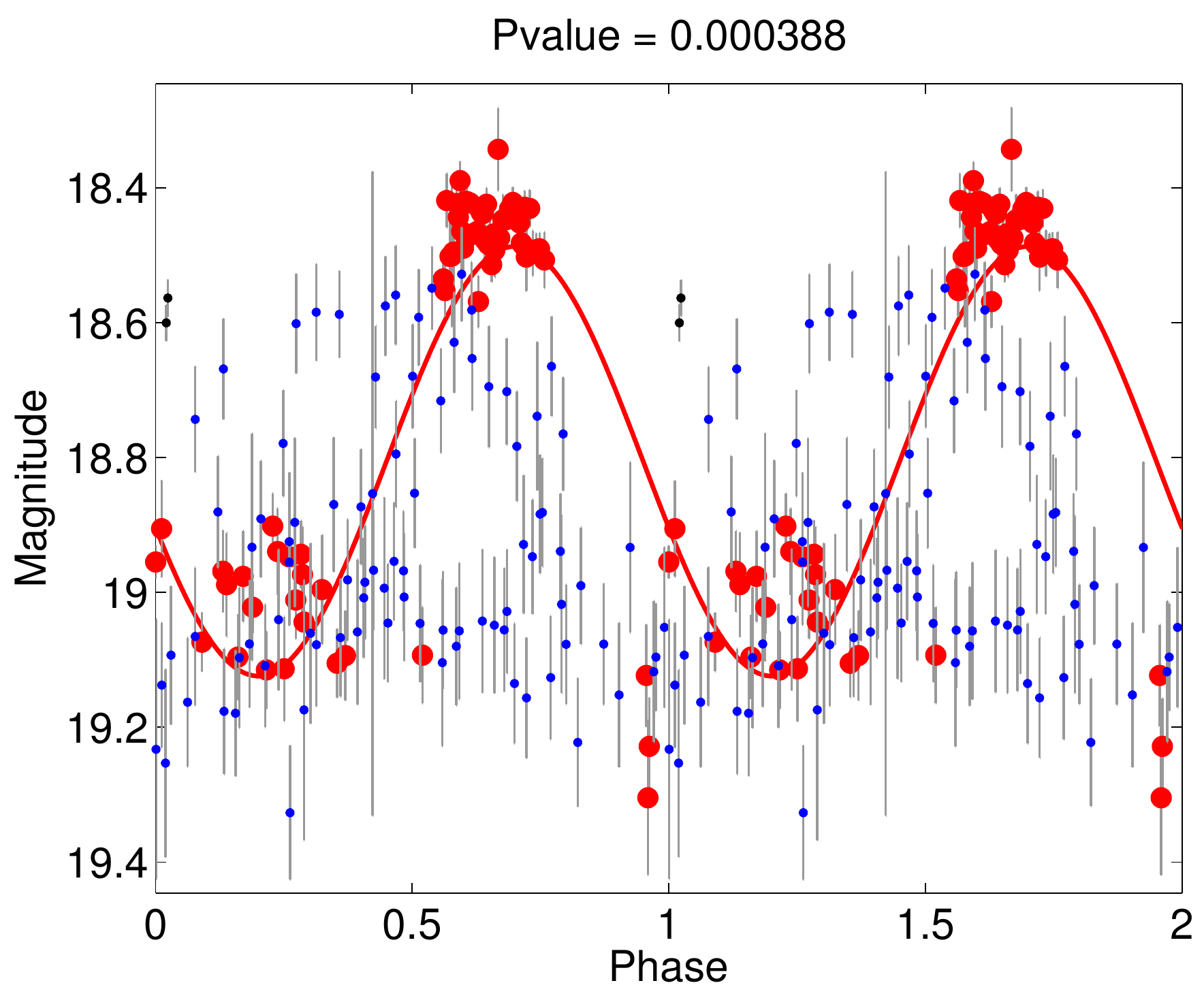}
\end{subfigure} \hspace{0.2cm}
\begin{subfigure}{.45\textwidth}
\centering
\includegraphics[width=8cm,height=4.5cm]{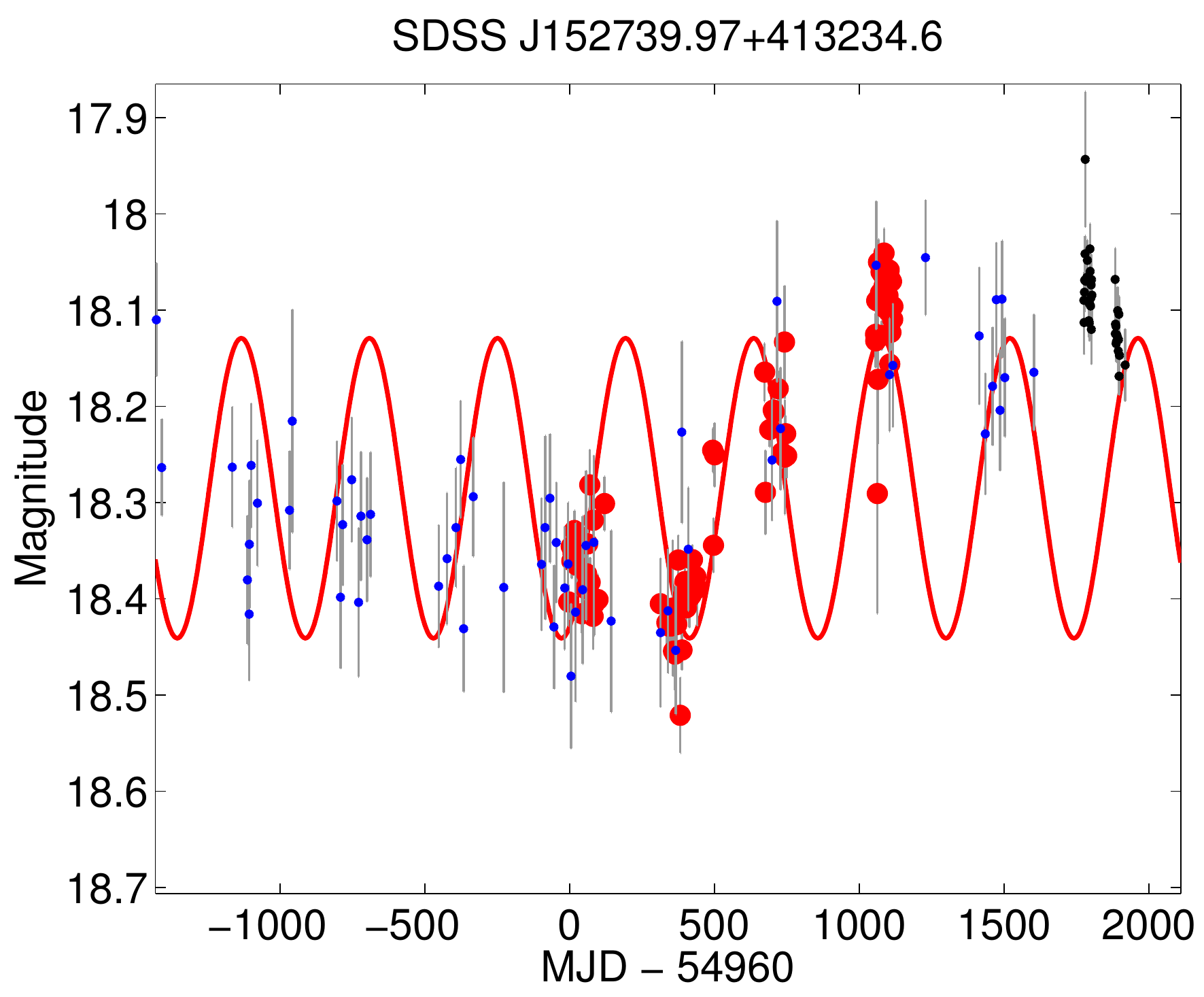}
\end{subfigure} \hspace{0.2cm}
\begin{subfigure}{.45\textwidth}
\centering
\includegraphics[width=8cm,height=4.5cm]{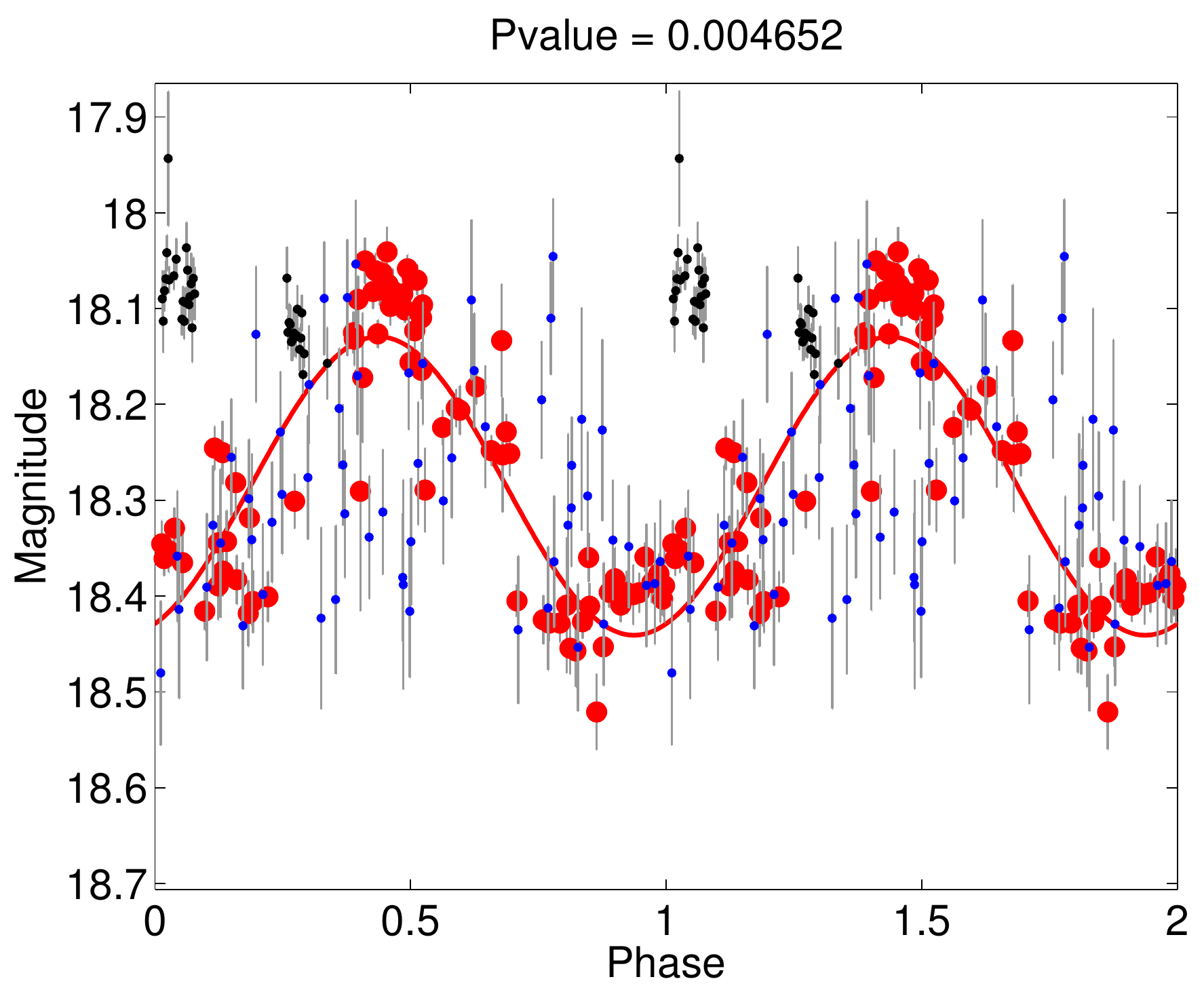}
\end{subfigure} \hspace{0.2cm}
\begin{subfigure}{.45\textwidth}
\centering
\includegraphics[width=8cm,height=4.5cm]{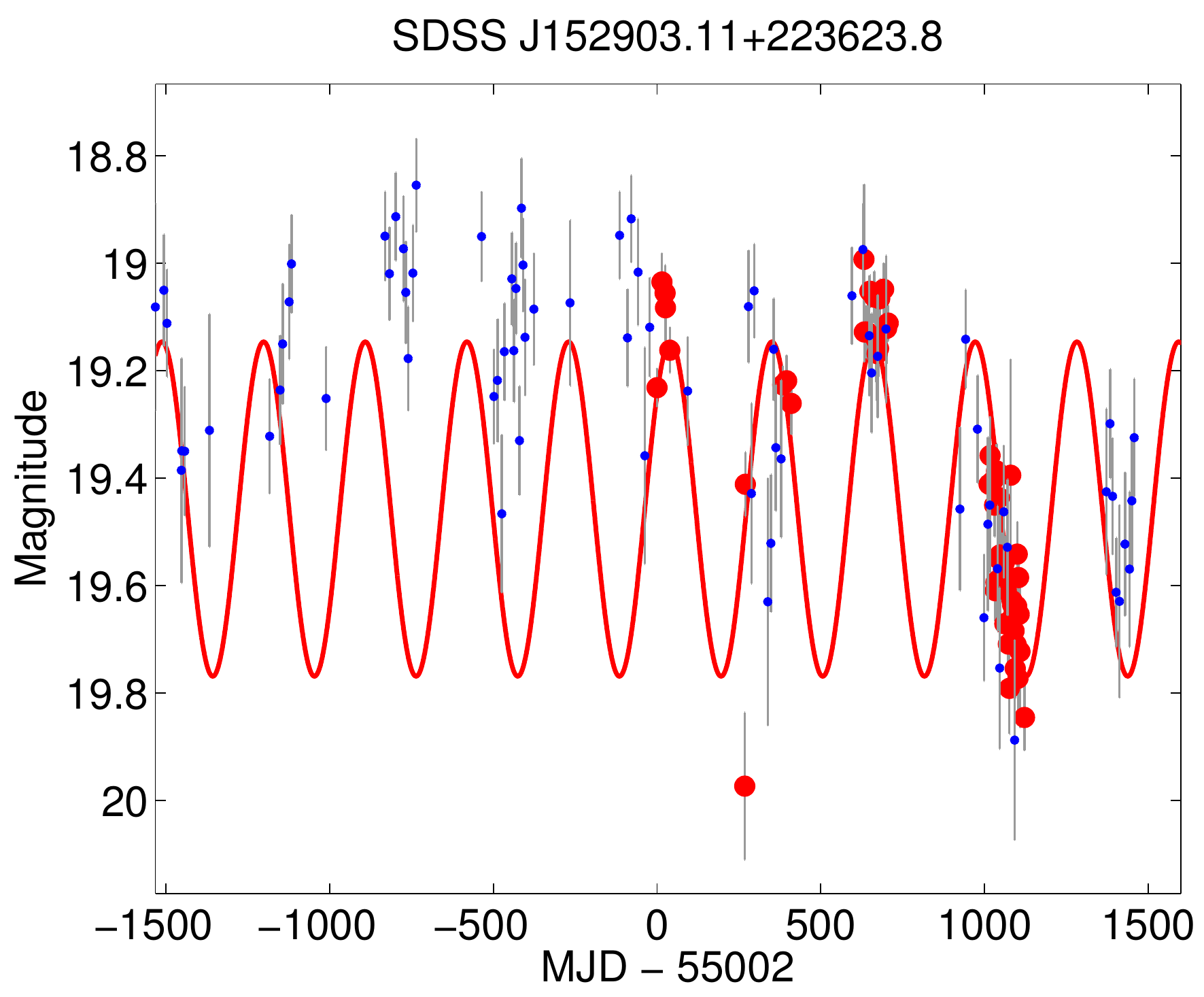}
\end{subfigure} \hspace{0.2cm}
\begin{subfigure}{.45\textwidth}
\centering
\includegraphics[width=8cm,height=4.5cm]{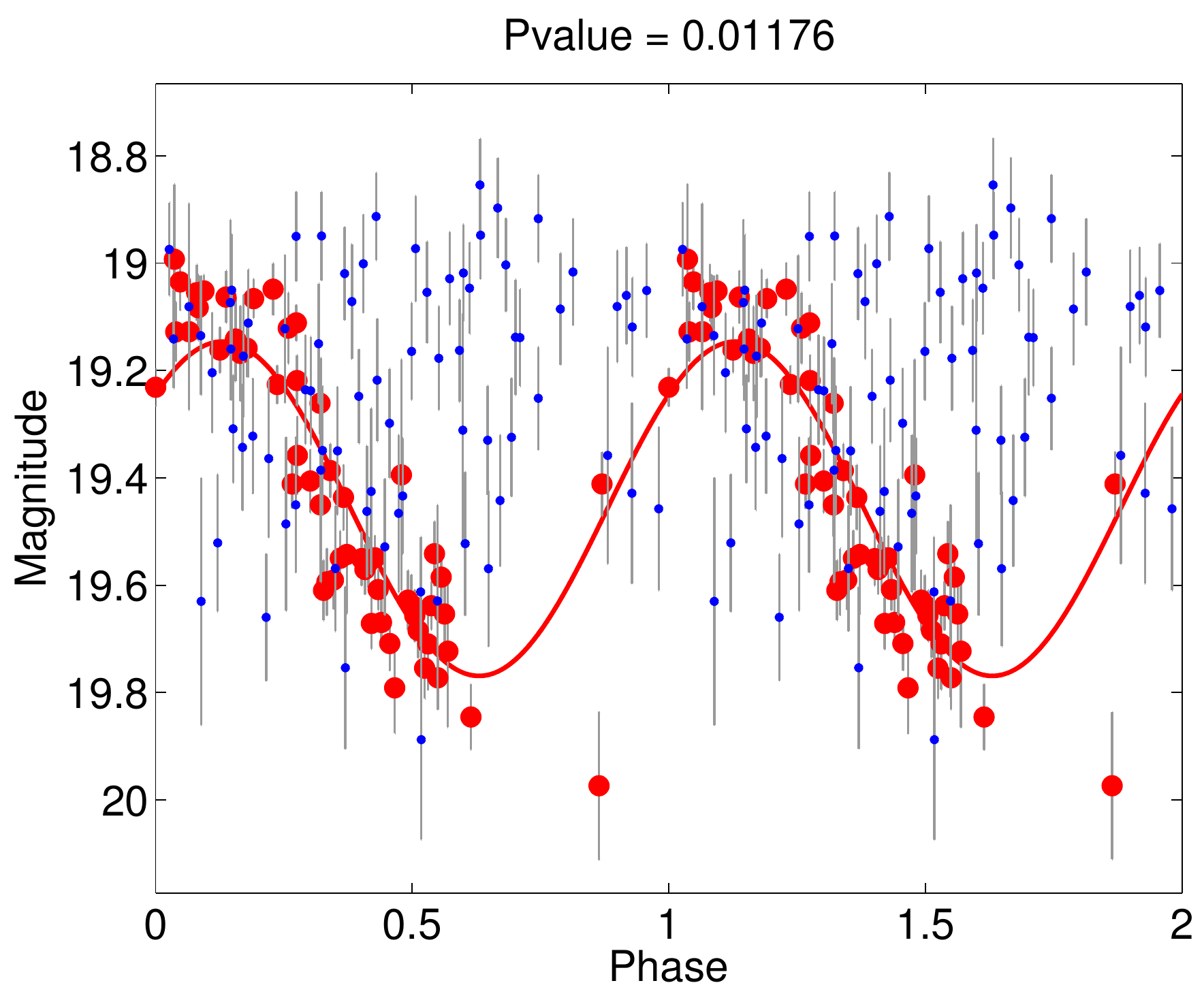}
\end{subfigure} \hspace{0.2cm}
\begin{subfigure}{.45\textwidth}
\centering
\includegraphics[width=8cm,height=4.5cm]{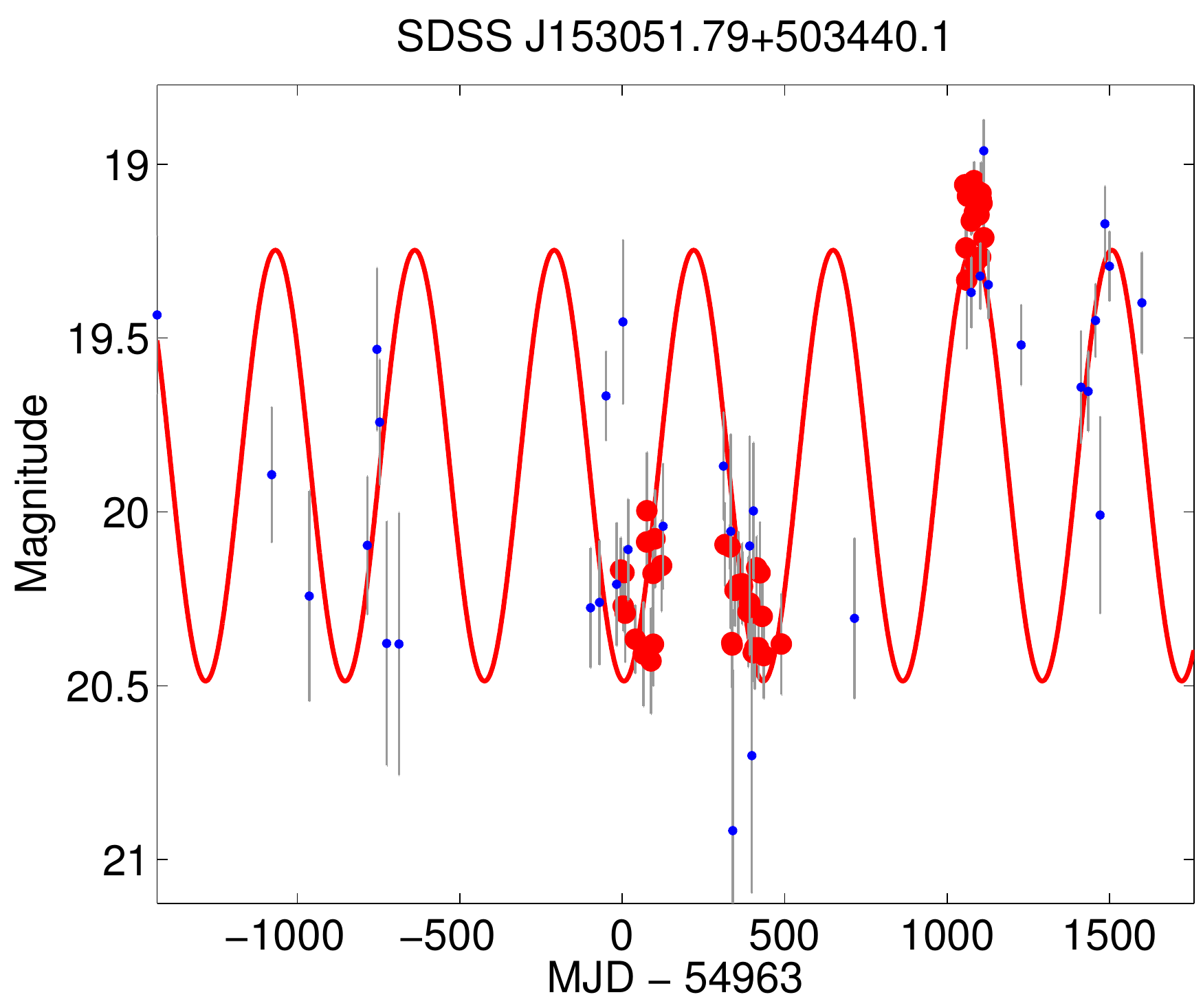}
\end{subfigure} \hspace{0.2cm}
\begin{subfigure}{.45\textwidth}
\centering
\includegraphics[width=8cm,height=4.5cm]{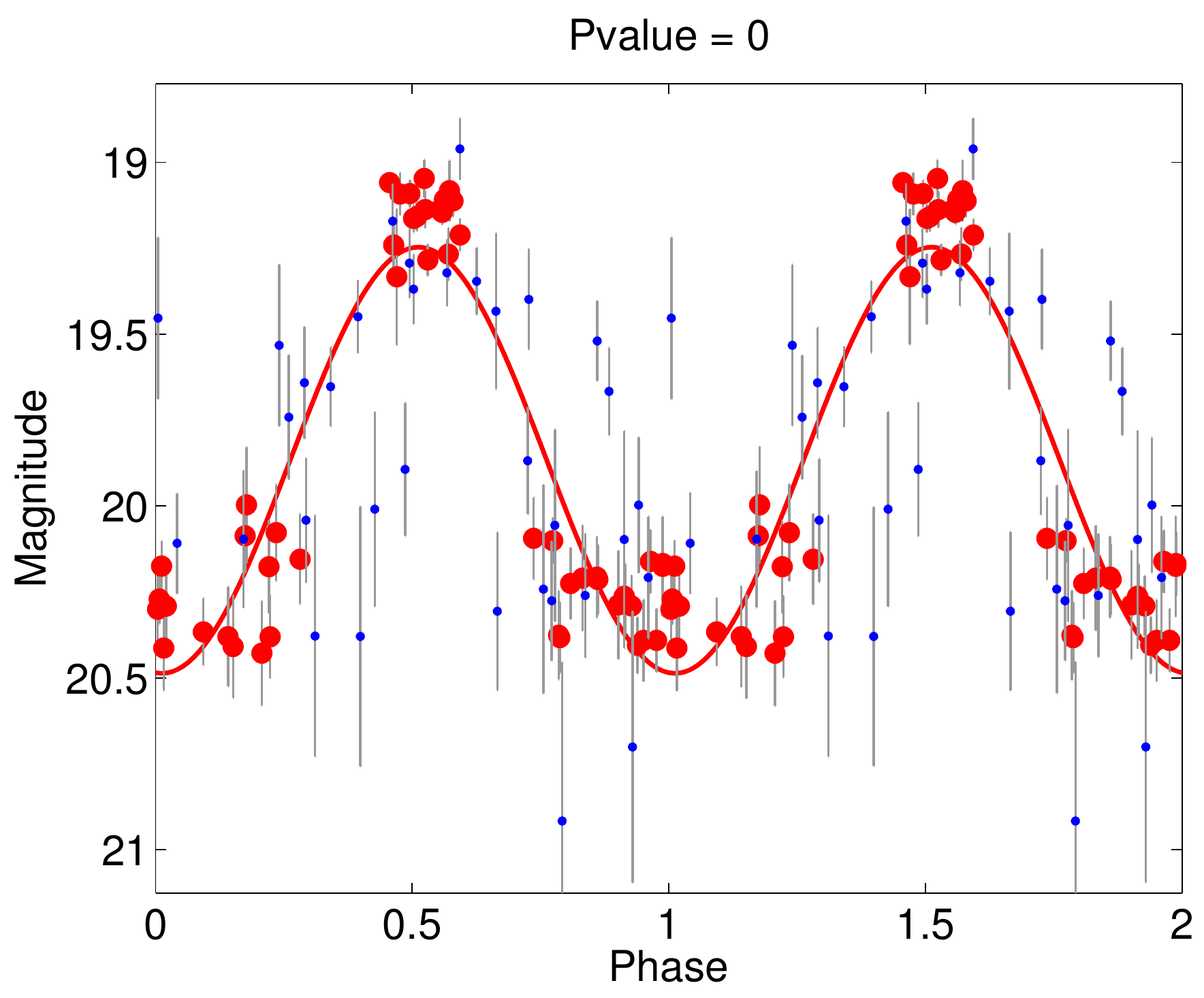}
\end{subfigure} \hspace{0.2cm}
\begin{subfigure}{.45\textwidth}
\centering
\includegraphics[width=8cm,height=4.5cm]{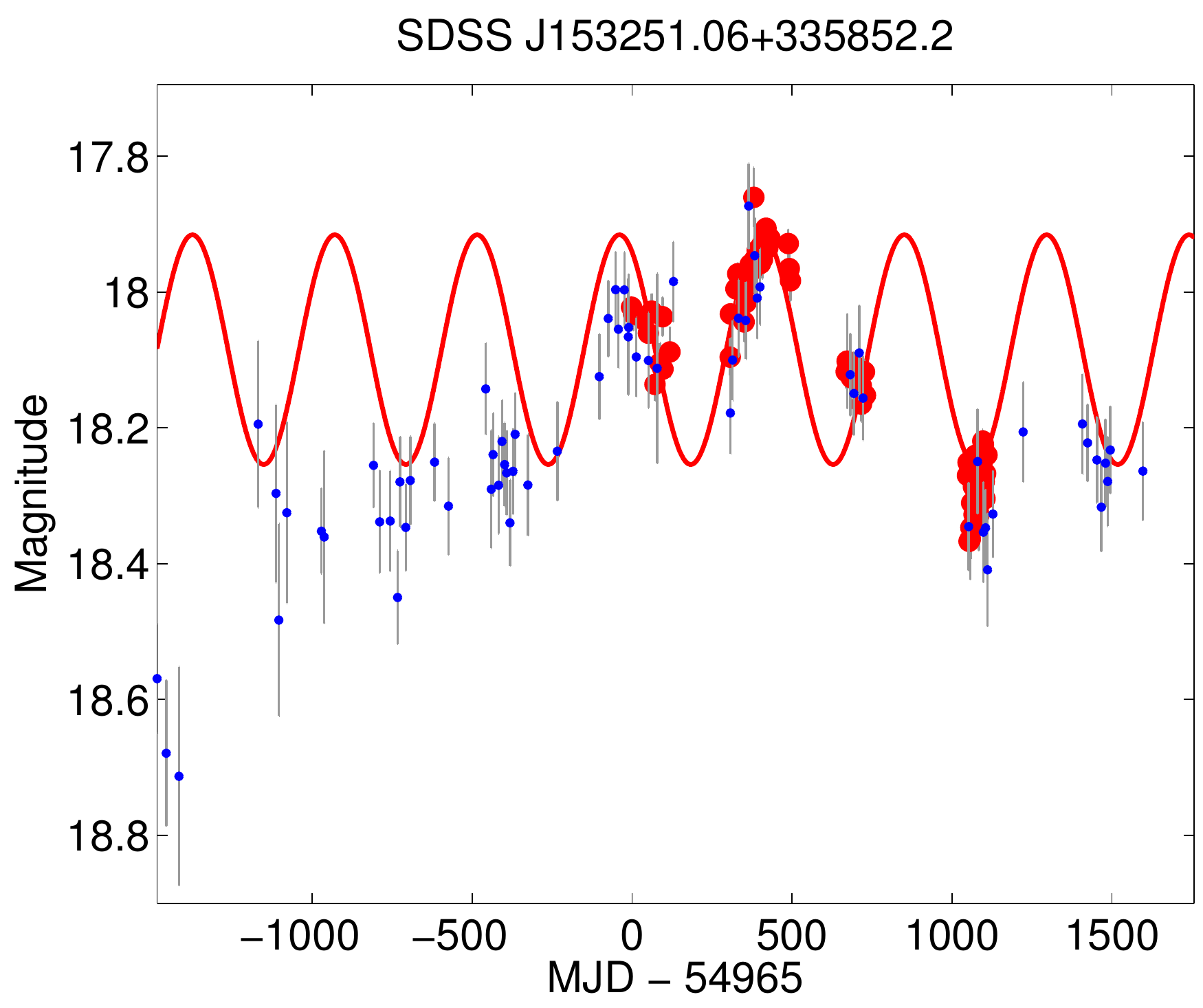}
\end{subfigure} \hspace{0.2cm}
\begin{subfigure}{.45\textwidth}
\centering
\includegraphics[width=8cm,height=4.5cm]{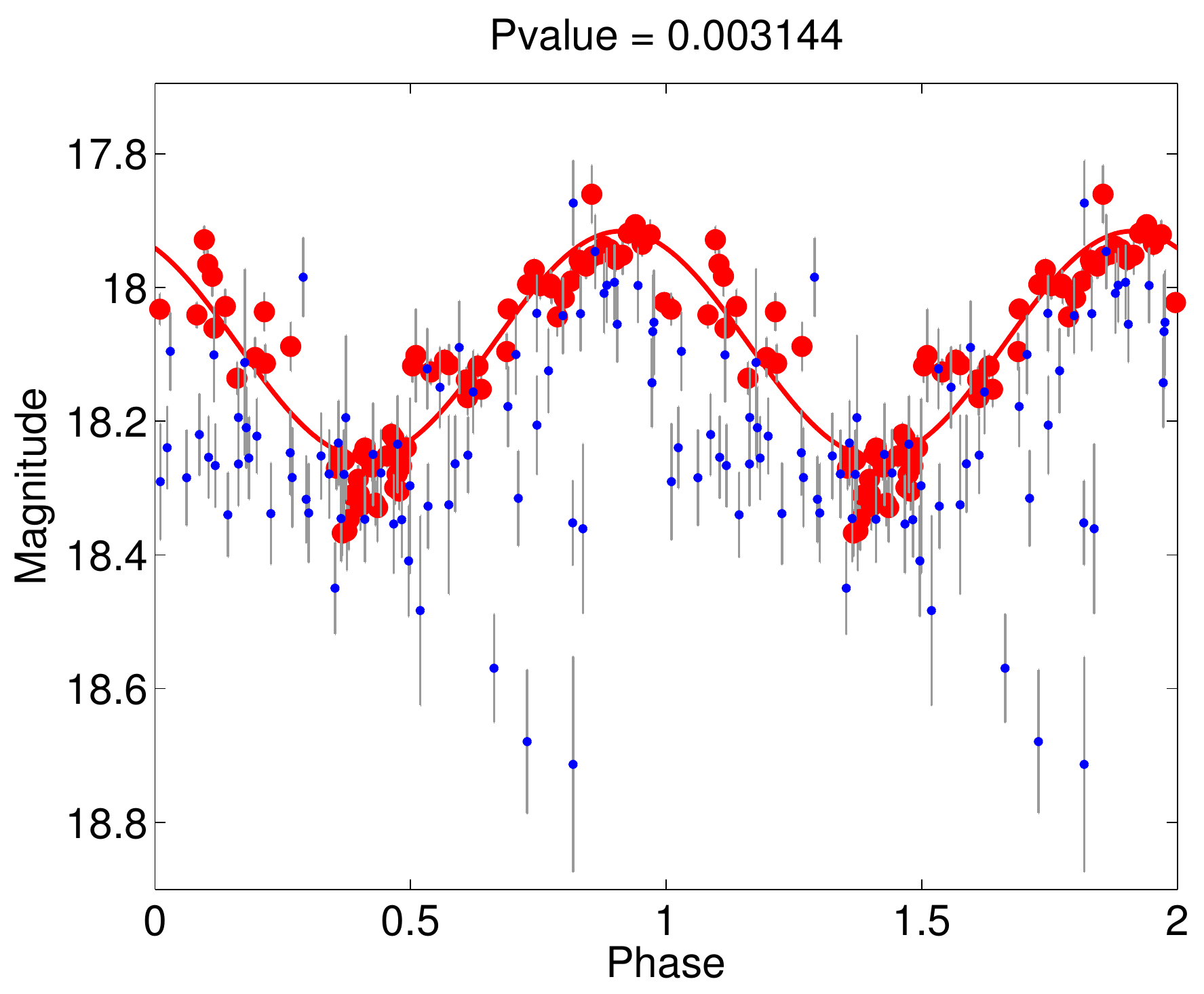}
\end{subfigure} \hspace{0.2cm}
\phantomcaption
\end{figure*}
\begin{figure*}
\ContinuedFloat
\begin{subfigure}{.45\textwidth}
\centering
\includegraphics[width=8cm,height=4.5cm]{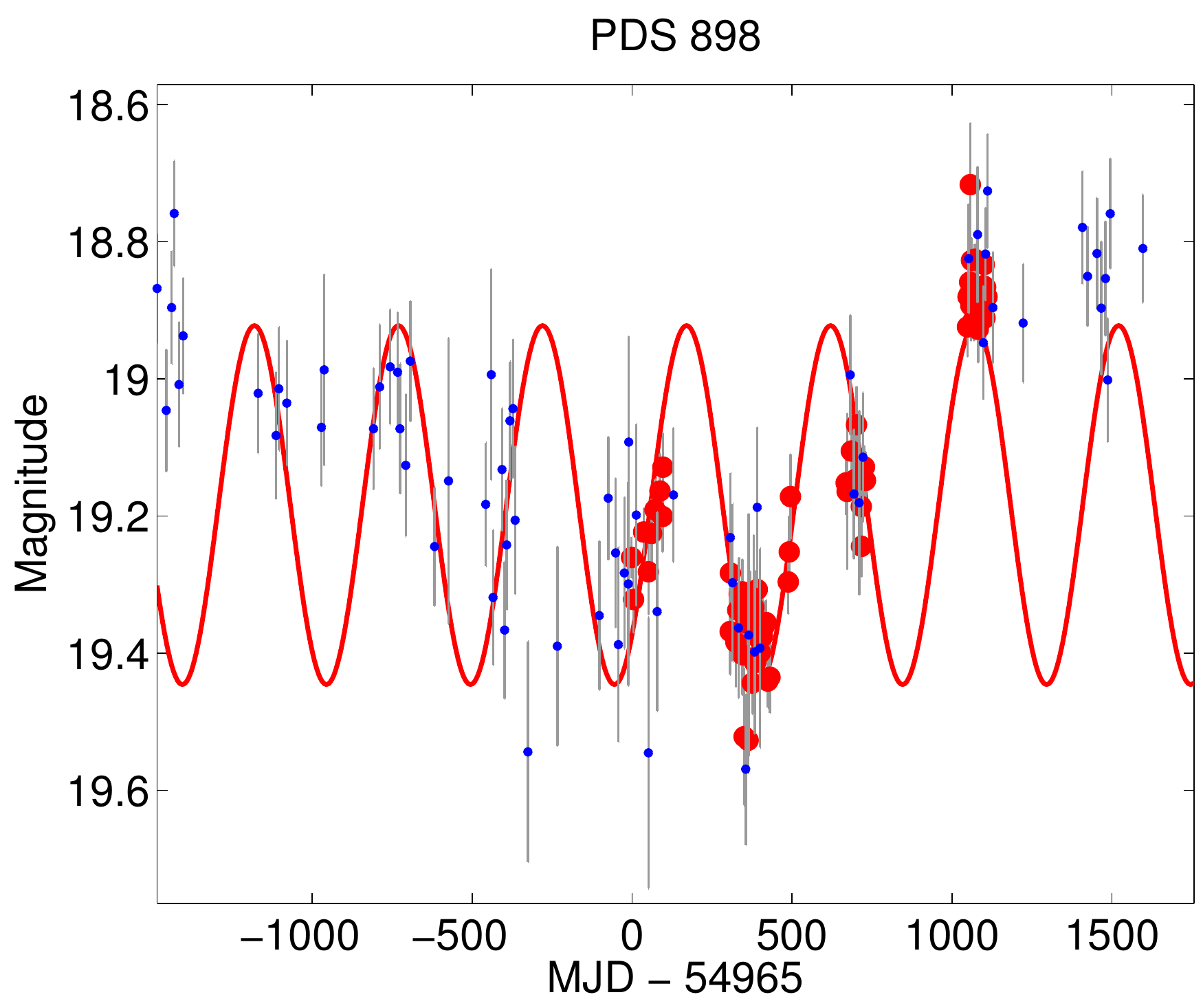}
\end{subfigure} \hspace{0.2cm}
\begin{subfigure}{.45\textwidth}
\centering
\includegraphics[width=8cm,height=4.5cm]{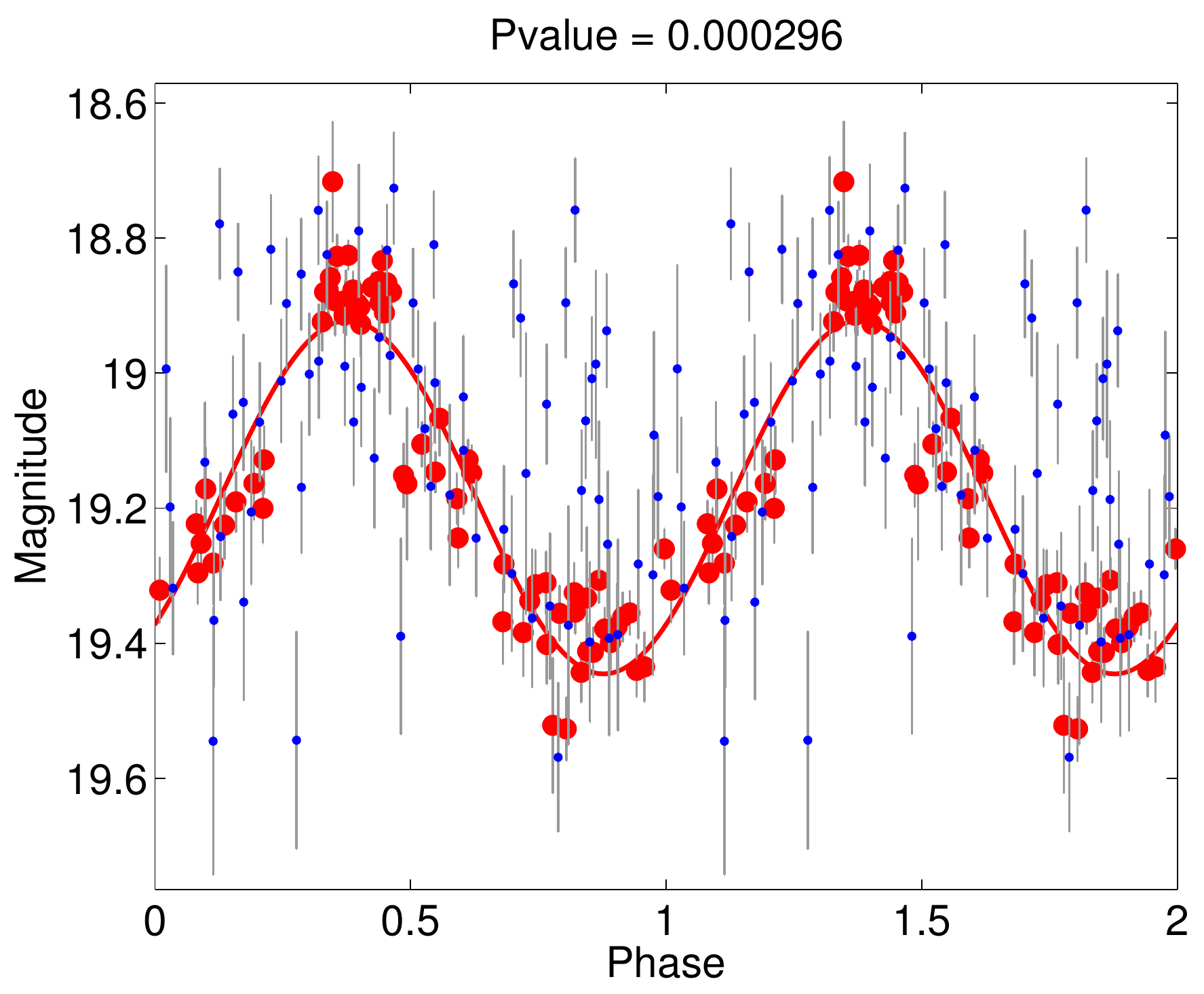}
\end{subfigure} \hspace{0.2cm}
\begin{subfigure}{.45\textwidth}
\centering
\includegraphics[width=8cm,height=4.5cm]{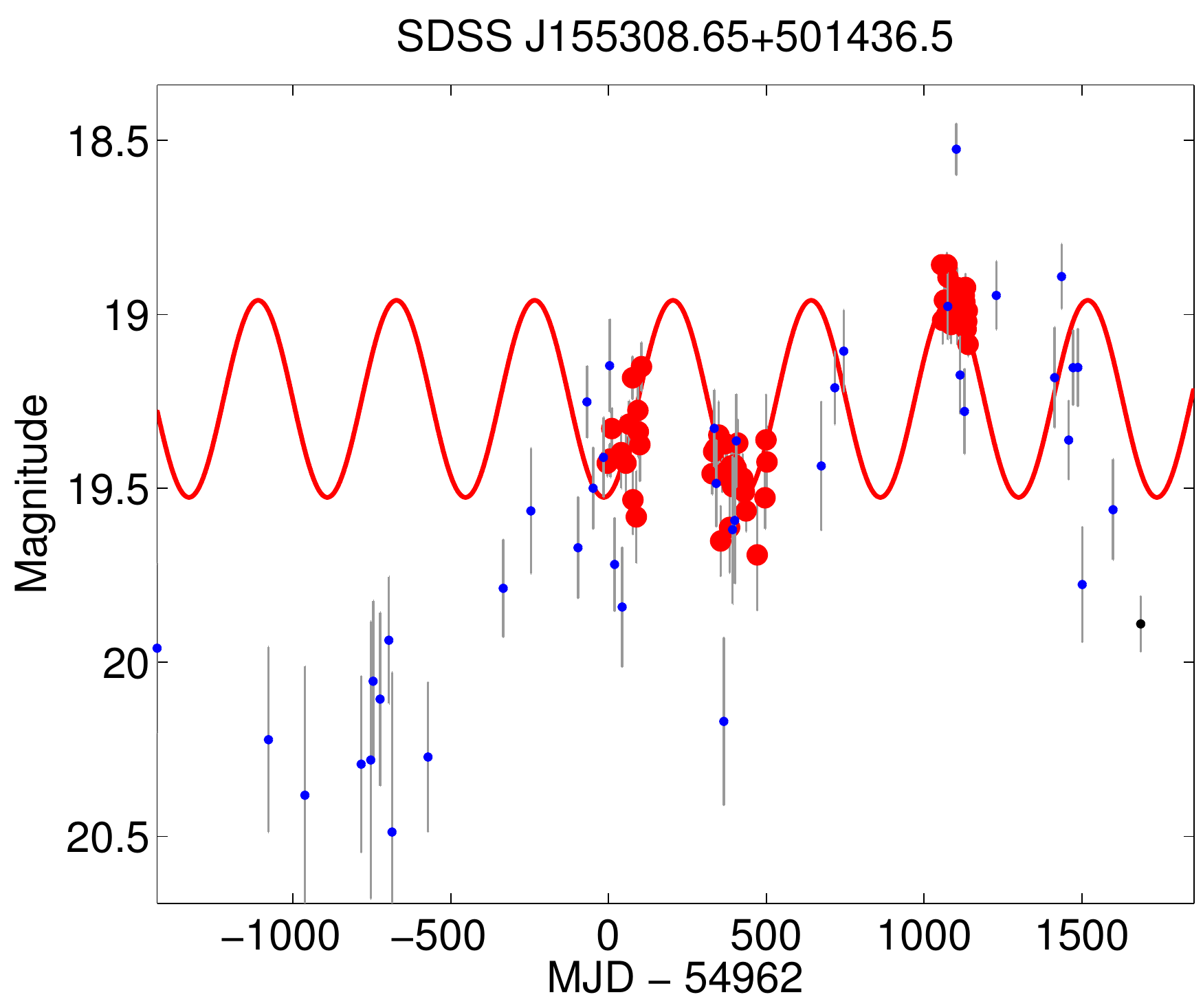}
\end{subfigure} \hspace{0.2cm}
\begin{subfigure}{.45\textwidth}
\centering
\includegraphics[width=8cm,height=4.5cm]{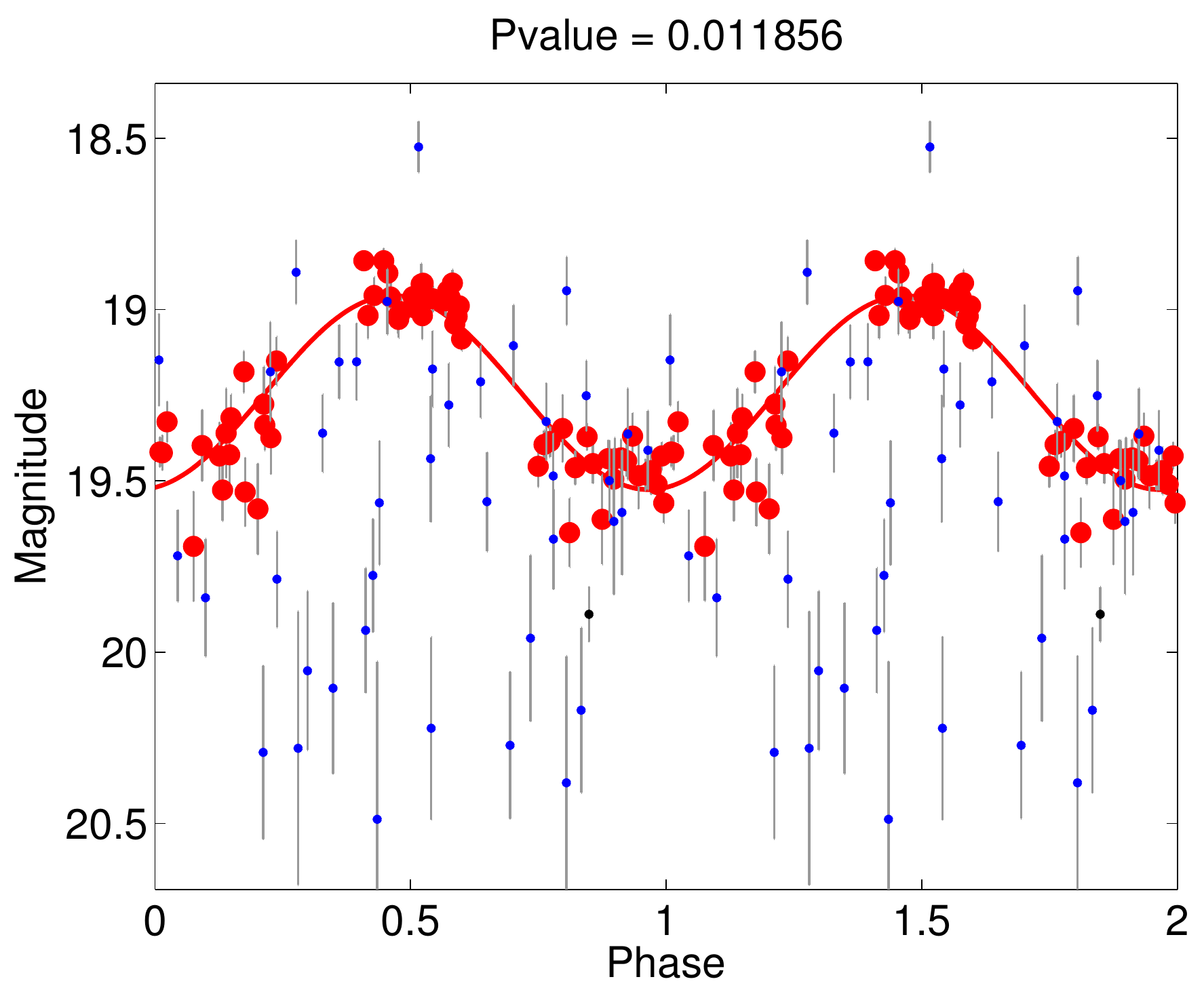}
\end{subfigure} \hspace{0.2cm}
\begin{subfigure}{.45\textwidth}
\centering
\includegraphics[width=8cm,height=4.5cm]{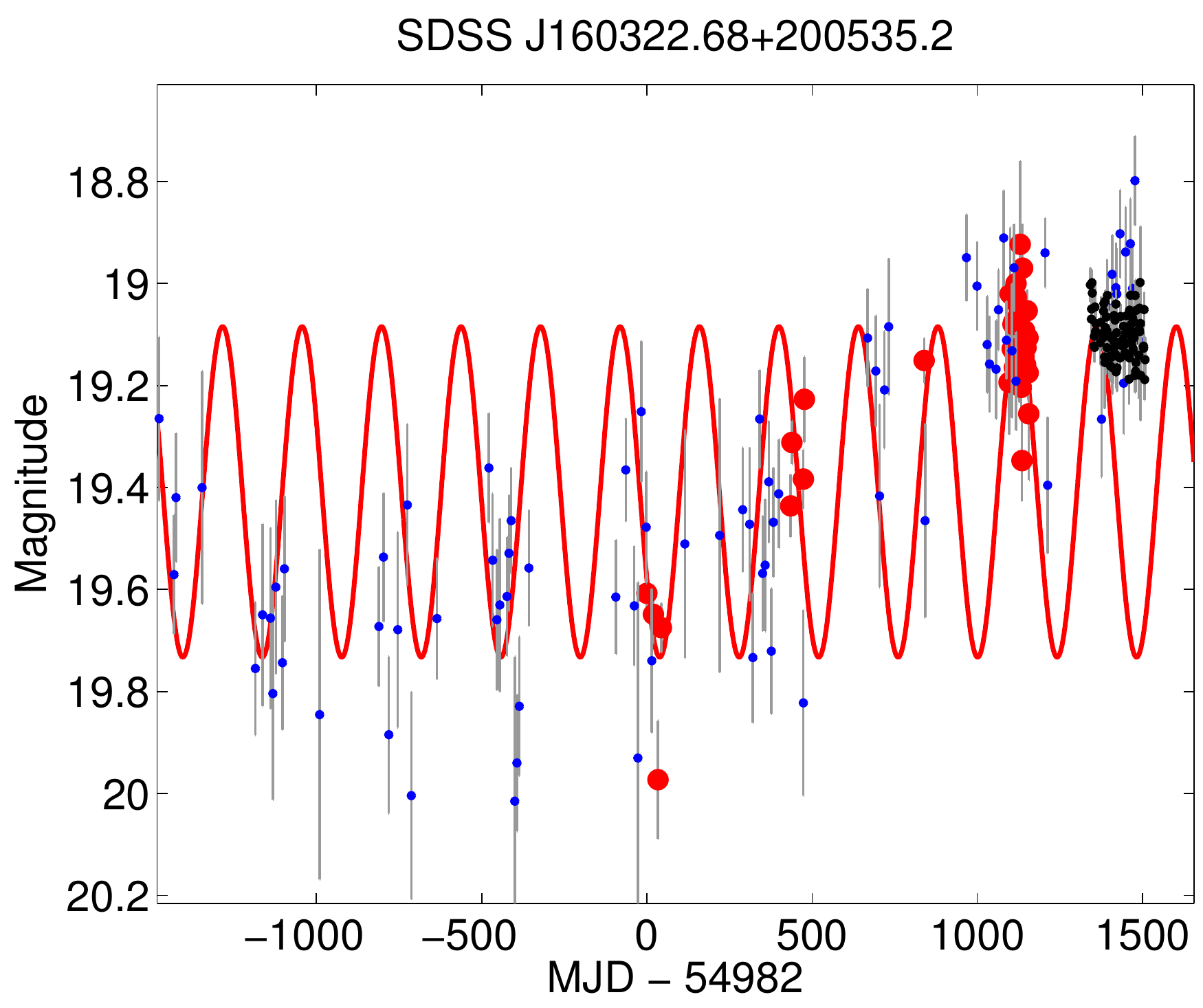}
\end{subfigure} \hspace{0.2cm}
\begin{subfigure}{.45\textwidth}
\centering
\includegraphics[width=8cm,height=4.5cm]{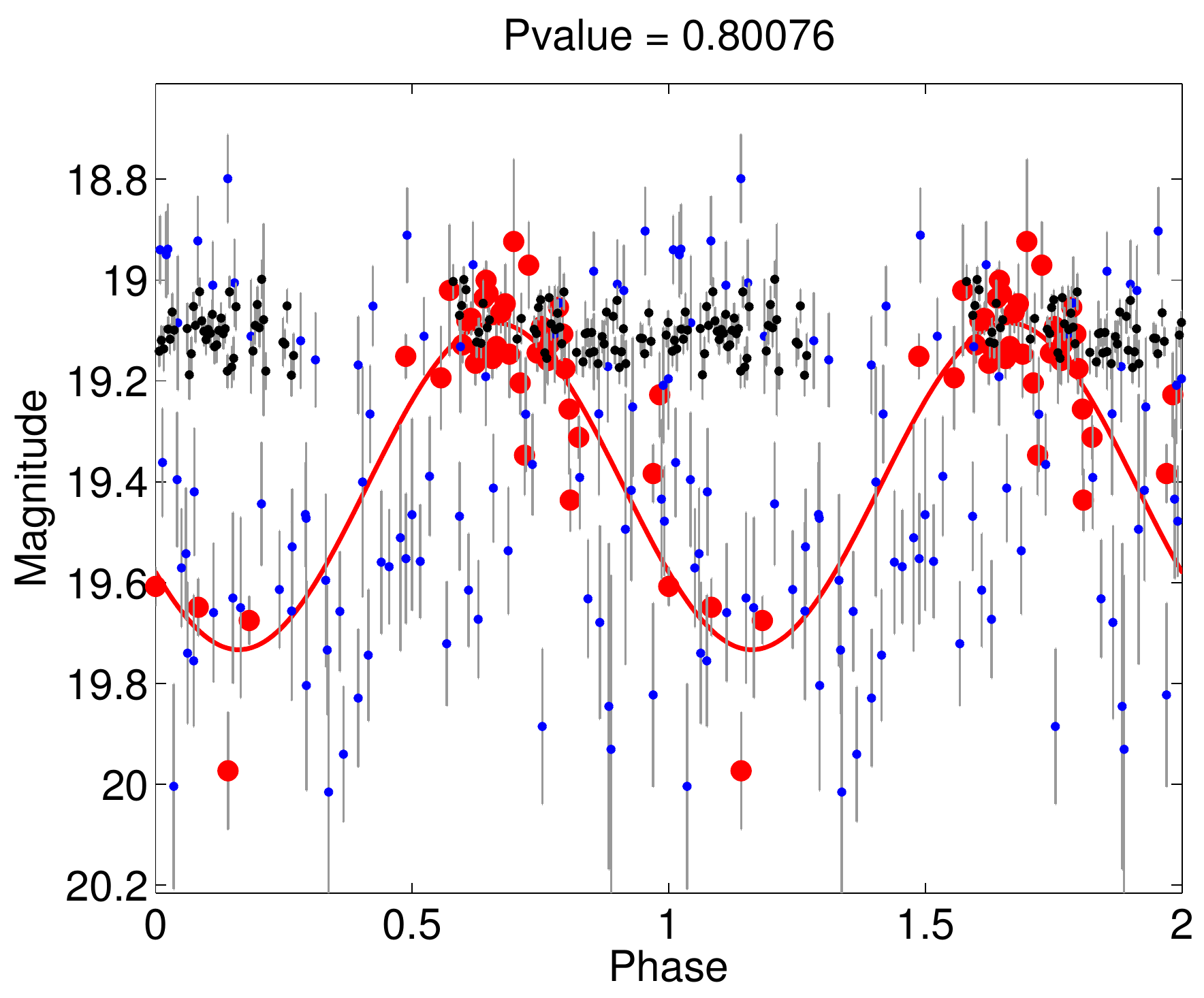}
\end{subfigure} \hspace{0.2cm}
\begin{subfigure}{.45\textwidth}
\centering
\includegraphics[width=8cm,height=4.5cm]{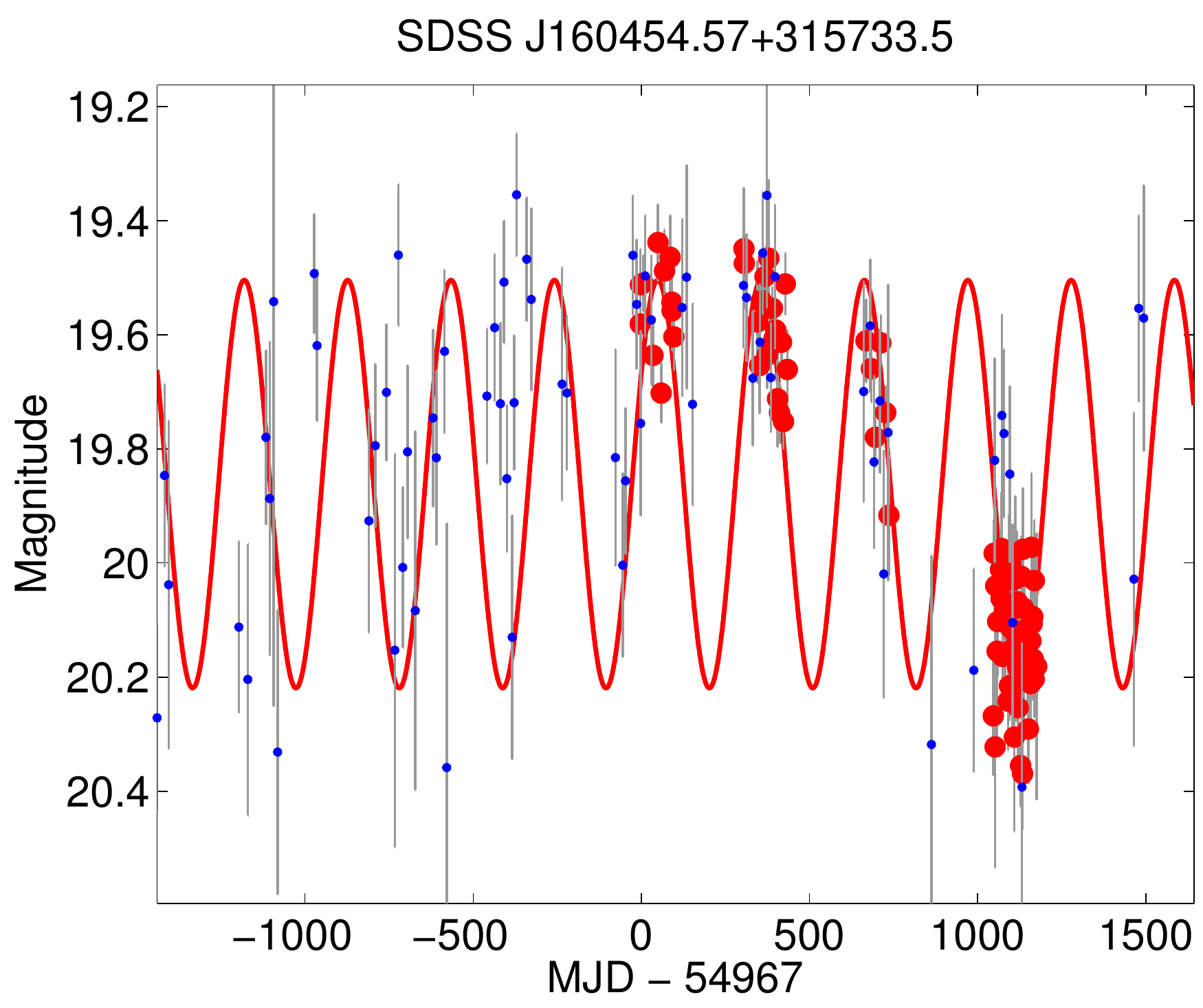}
\end{subfigure} \hspace{0.2cm}
\begin{subfigure}{.45\textwidth}
\centering
\includegraphics[width=8cm,height=4.5cm]{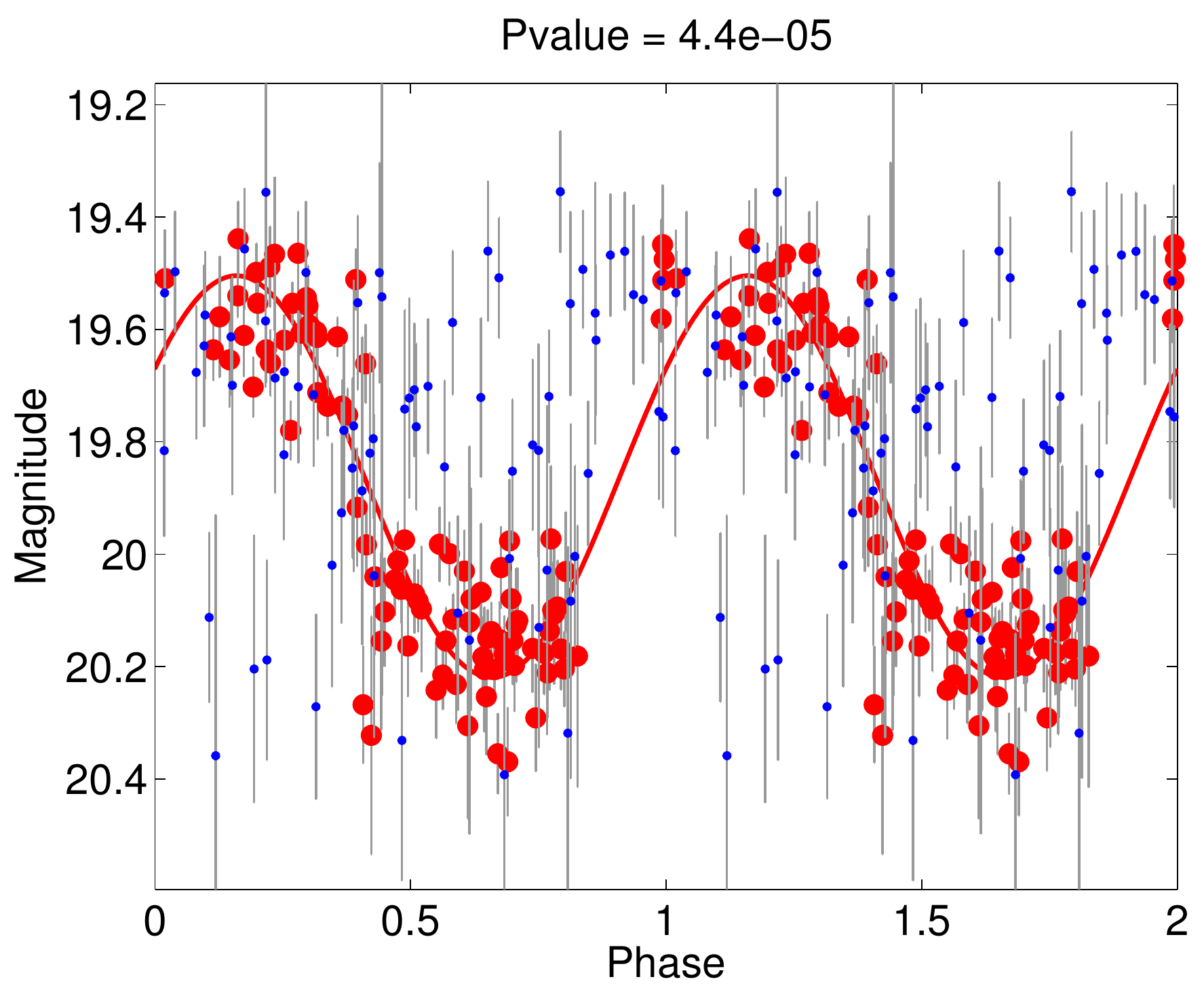}
\end{subfigure} \hspace{0.2cm}
\begin{subfigure}{.45\textwidth}
\centering
\includegraphics[width=8cm,height=4.5cm]{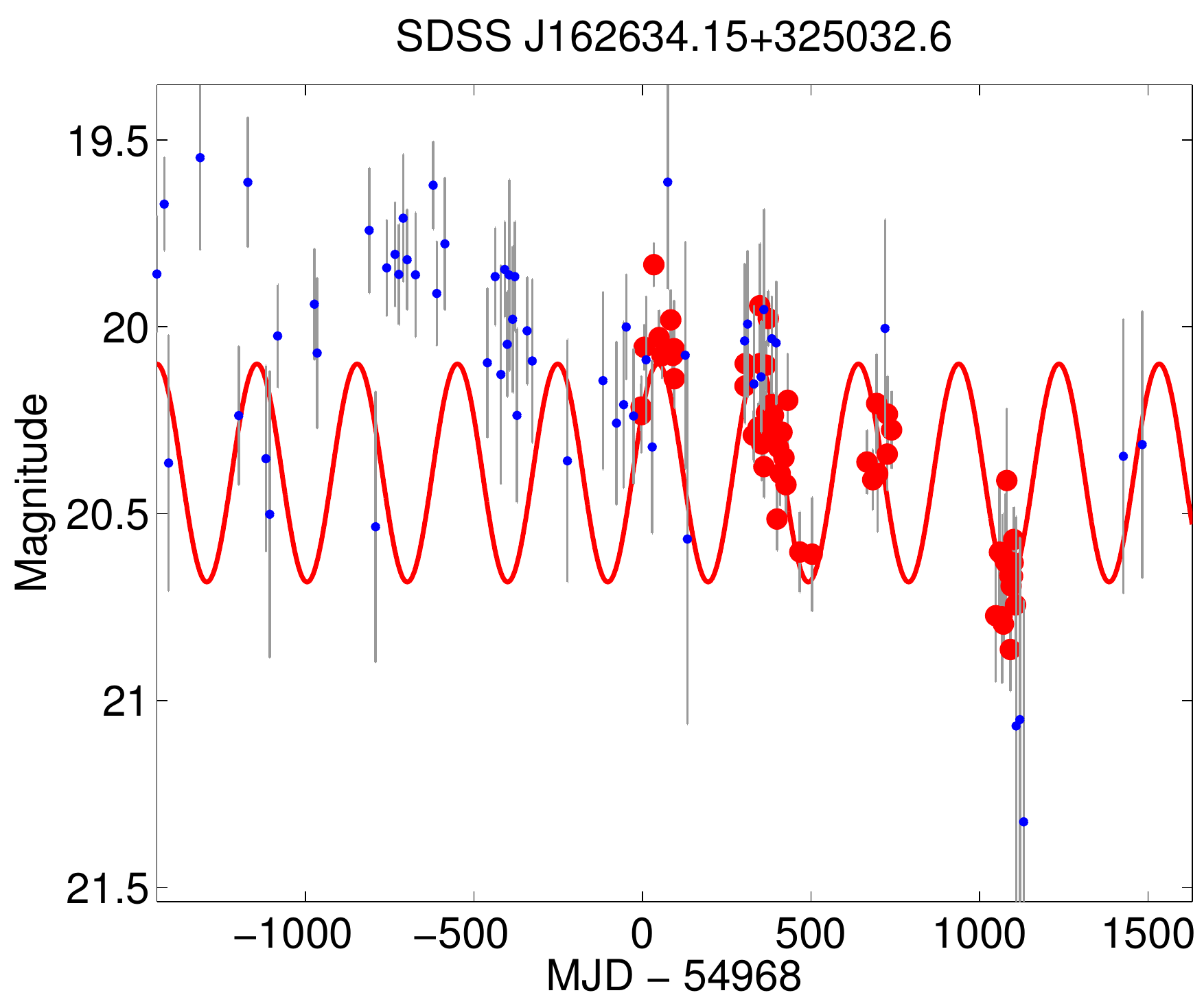}
\end{subfigure} \hspace{0.2cm}
\begin{subfigure}{.45\textwidth}
\centering
\includegraphics[width=8cm,height=4.5cm]{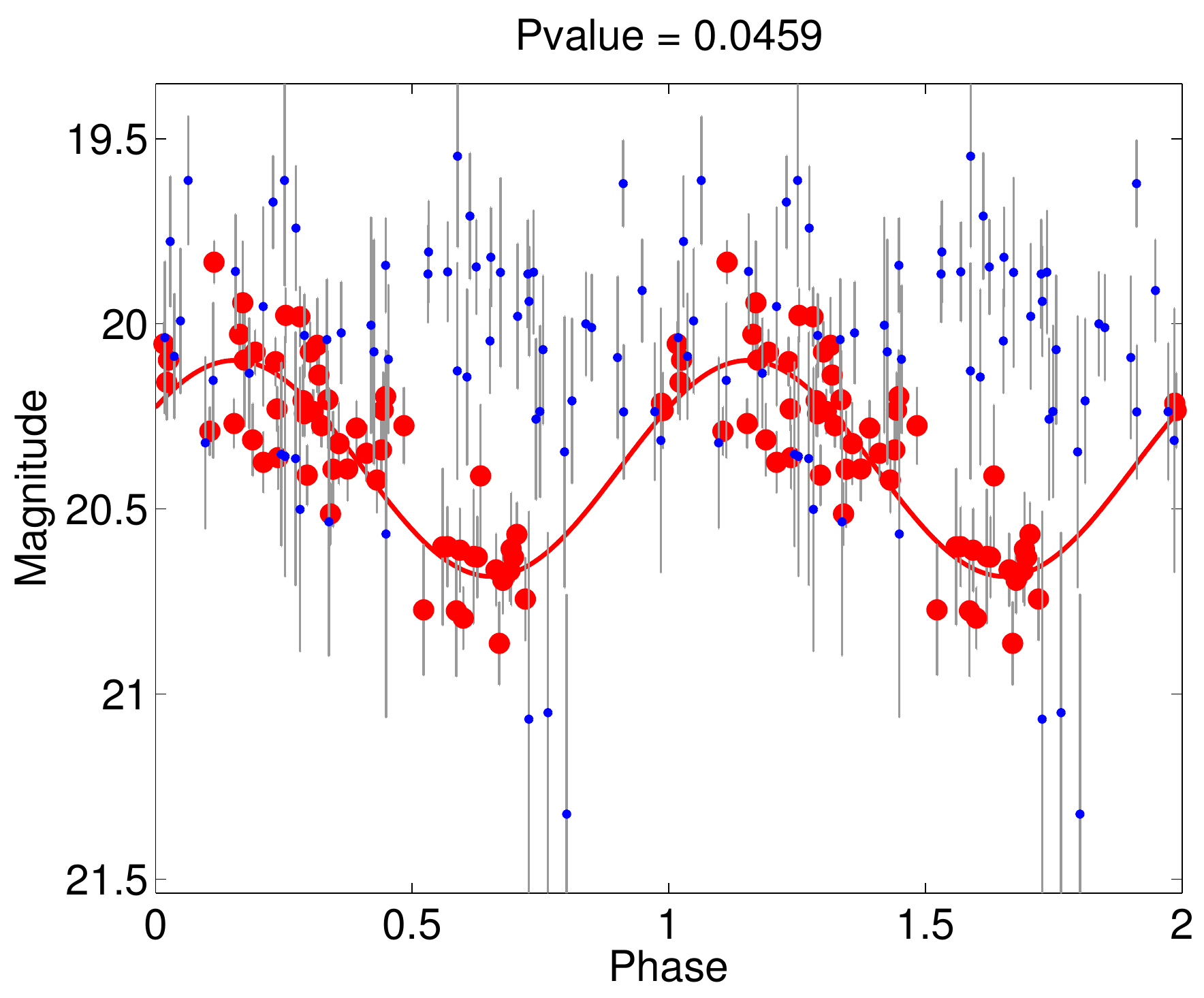}
\end{subfigure} \hspace{0.2cm}
\phantomcaption
\end{figure*}
\begin{figure*}
\ContinuedFloat
\begin{subfigure}{.45\textwidth}
\centering
\includegraphics[width=8cm,height=4.5cm]{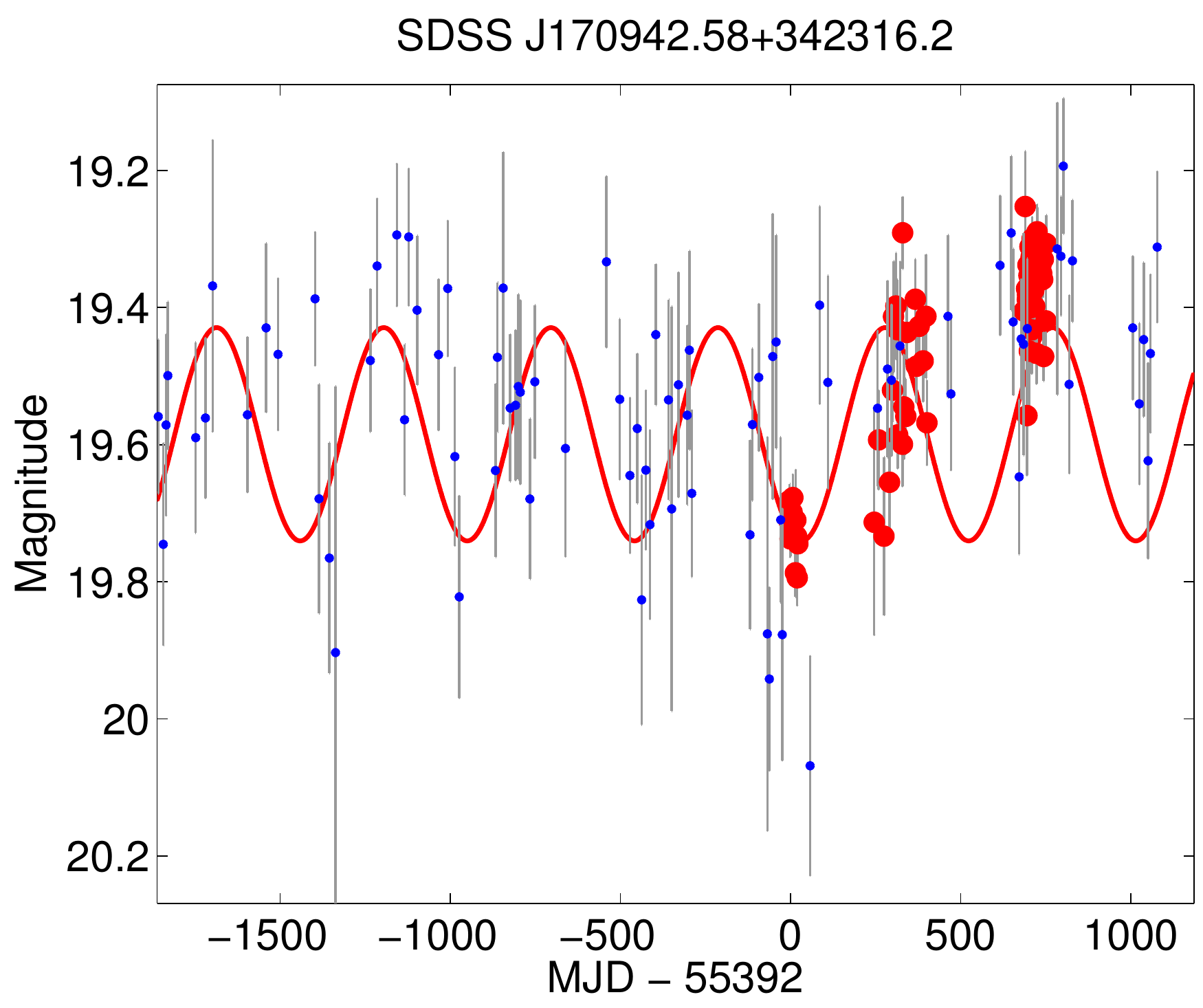}
\end{subfigure} \hspace{0.2cm}
\begin{subfigure}{.45\textwidth}
\centering
\includegraphics[width=8cm,height=4.5cm]{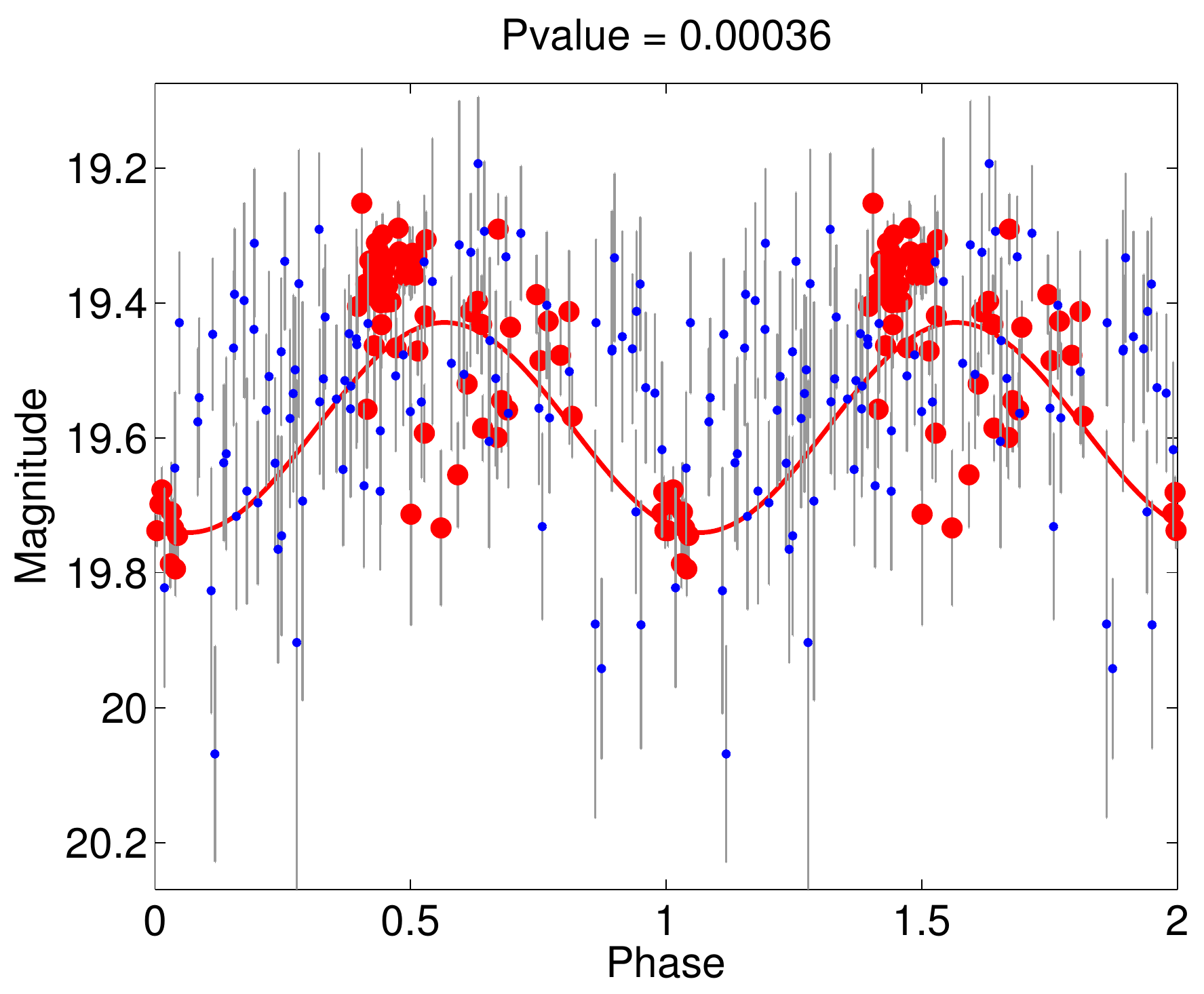}
\end{subfigure} \hspace{0.2cm}
\begin{subfigure}{.45\textwidth}
\centering
\includegraphics[width=8cm,height=4.5cm]{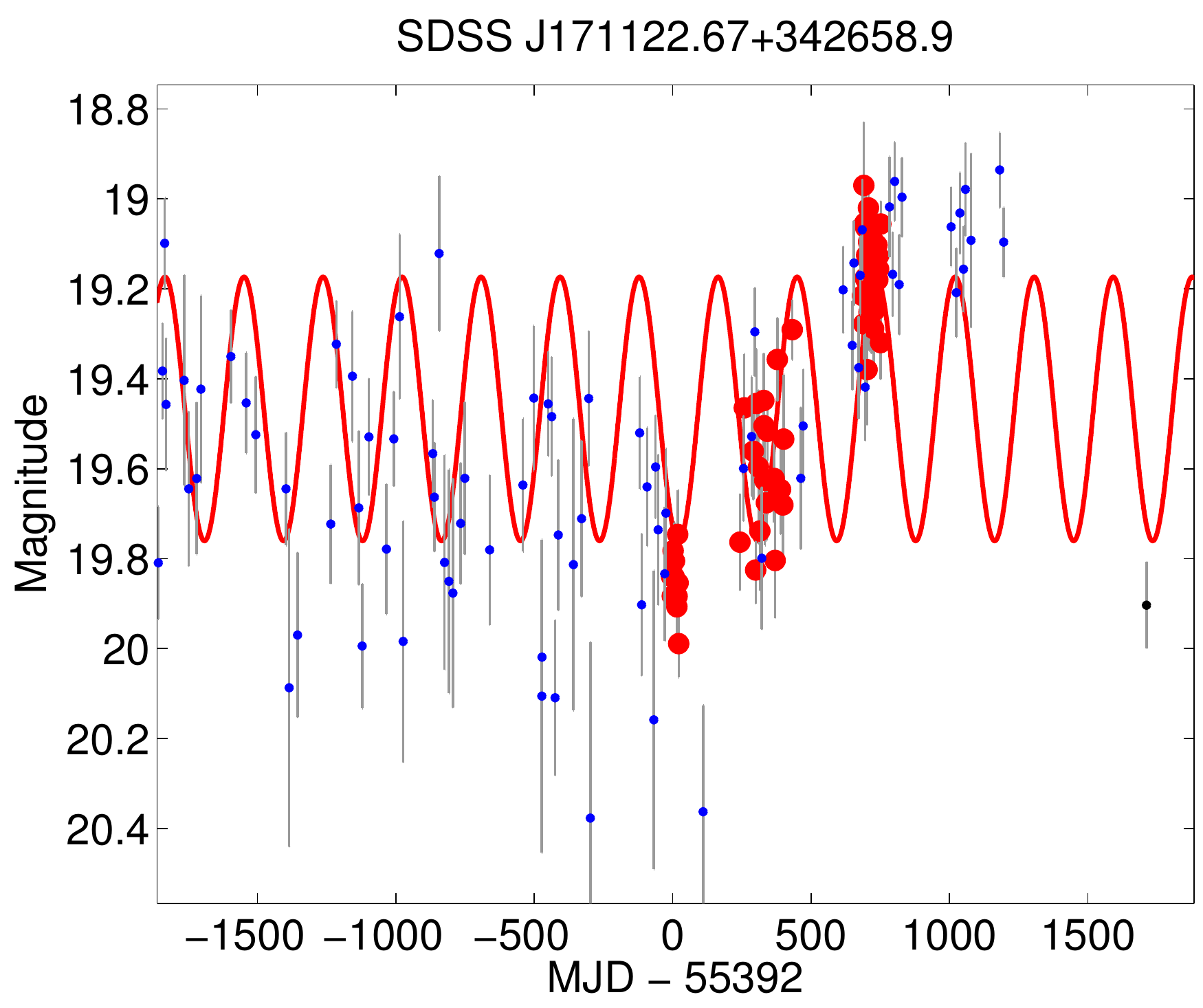}
\end{subfigure} \hspace{0.2cm}
\begin{subfigure}{.45\textwidth}
\centering
\includegraphics[width=8cm,height=4.5cm]{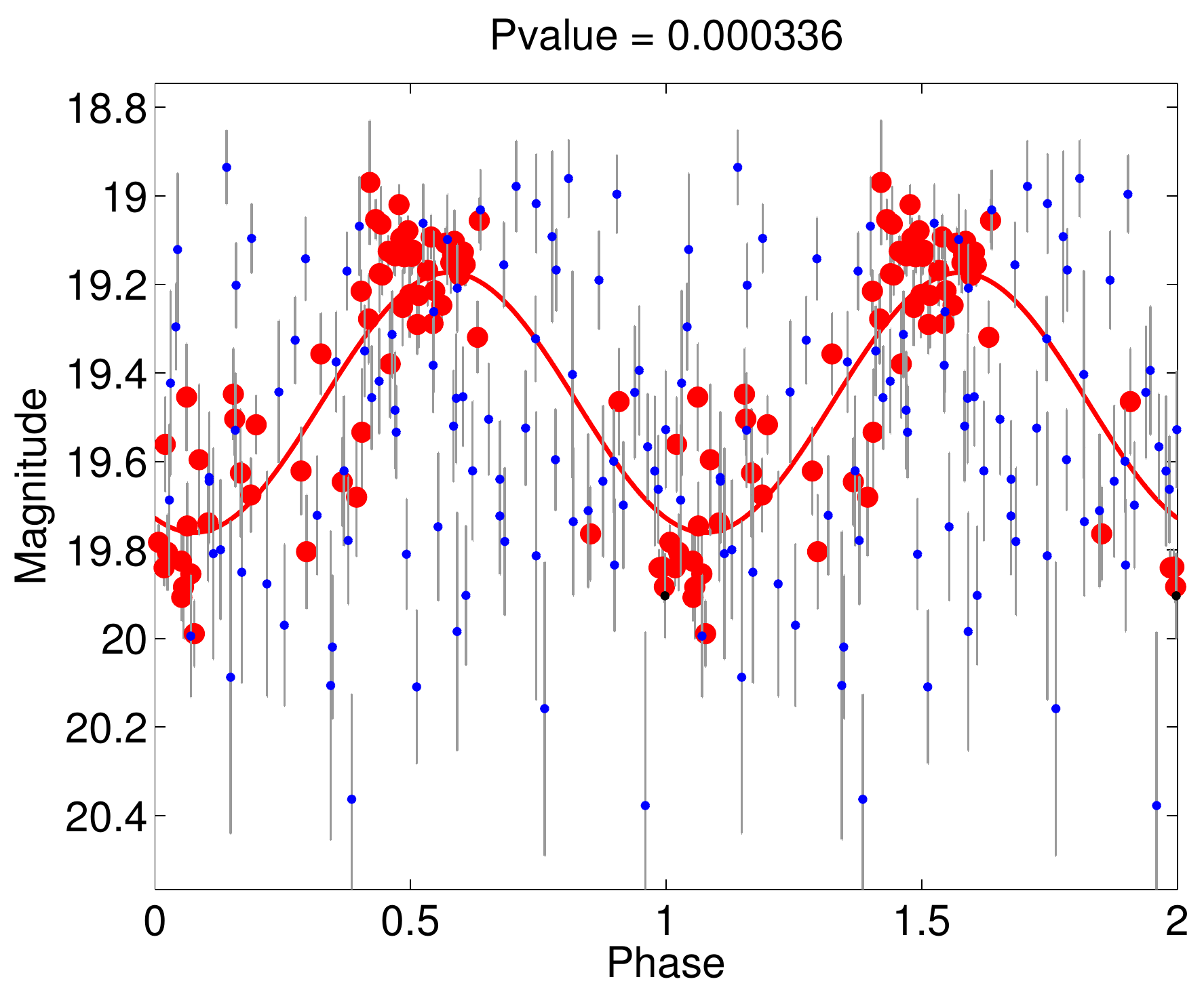}
\end{subfigure} \hspace{0.2cm}
\begin{subfigure}{.45\textwidth}
\centering
\includegraphics[width=8cm,height=4.5cm]{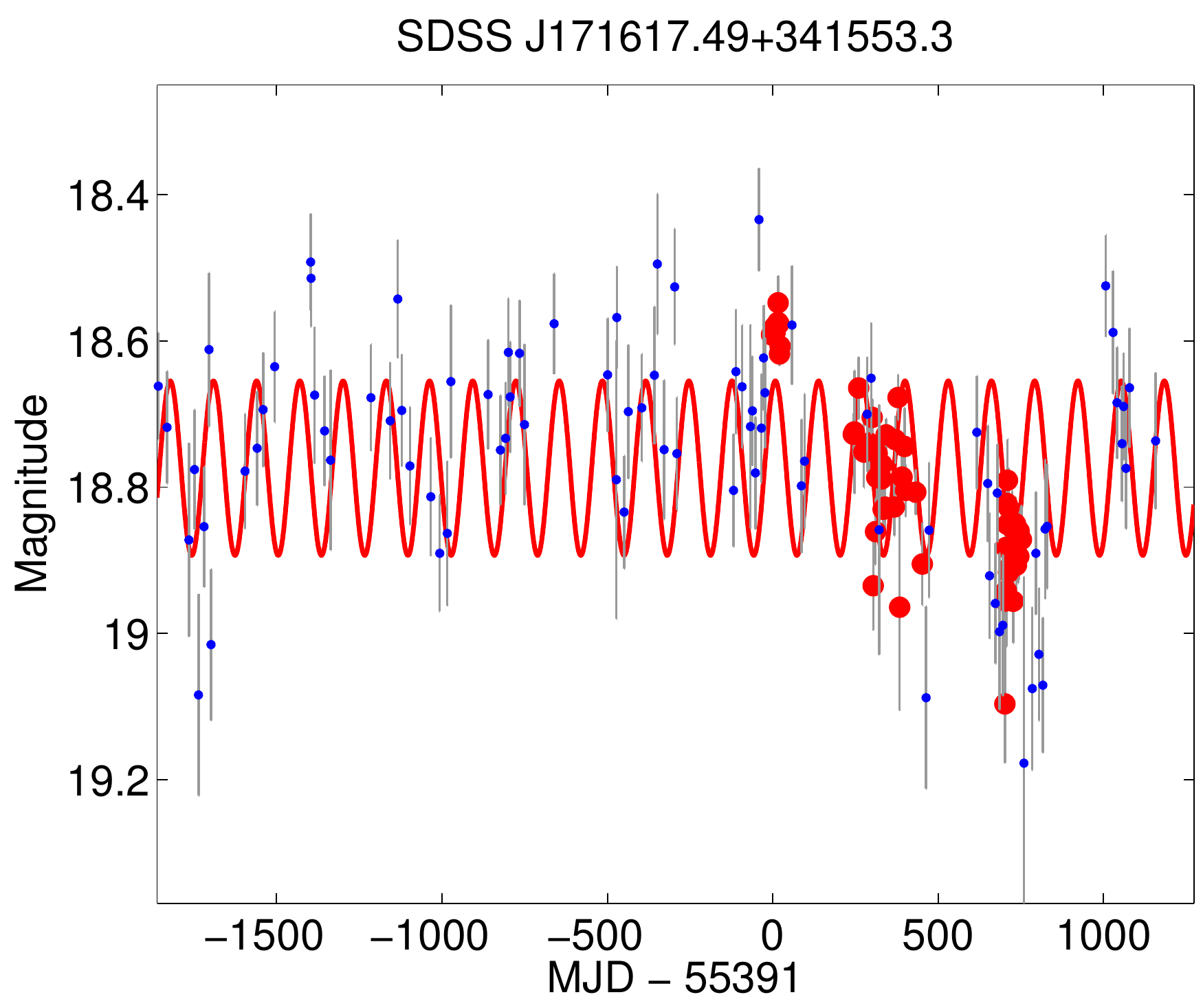}
\end{subfigure} \hspace{0.2cm}
\begin{subfigure}{.45\textwidth}
\centering
\includegraphics[width=8cm,height=4.5cm]{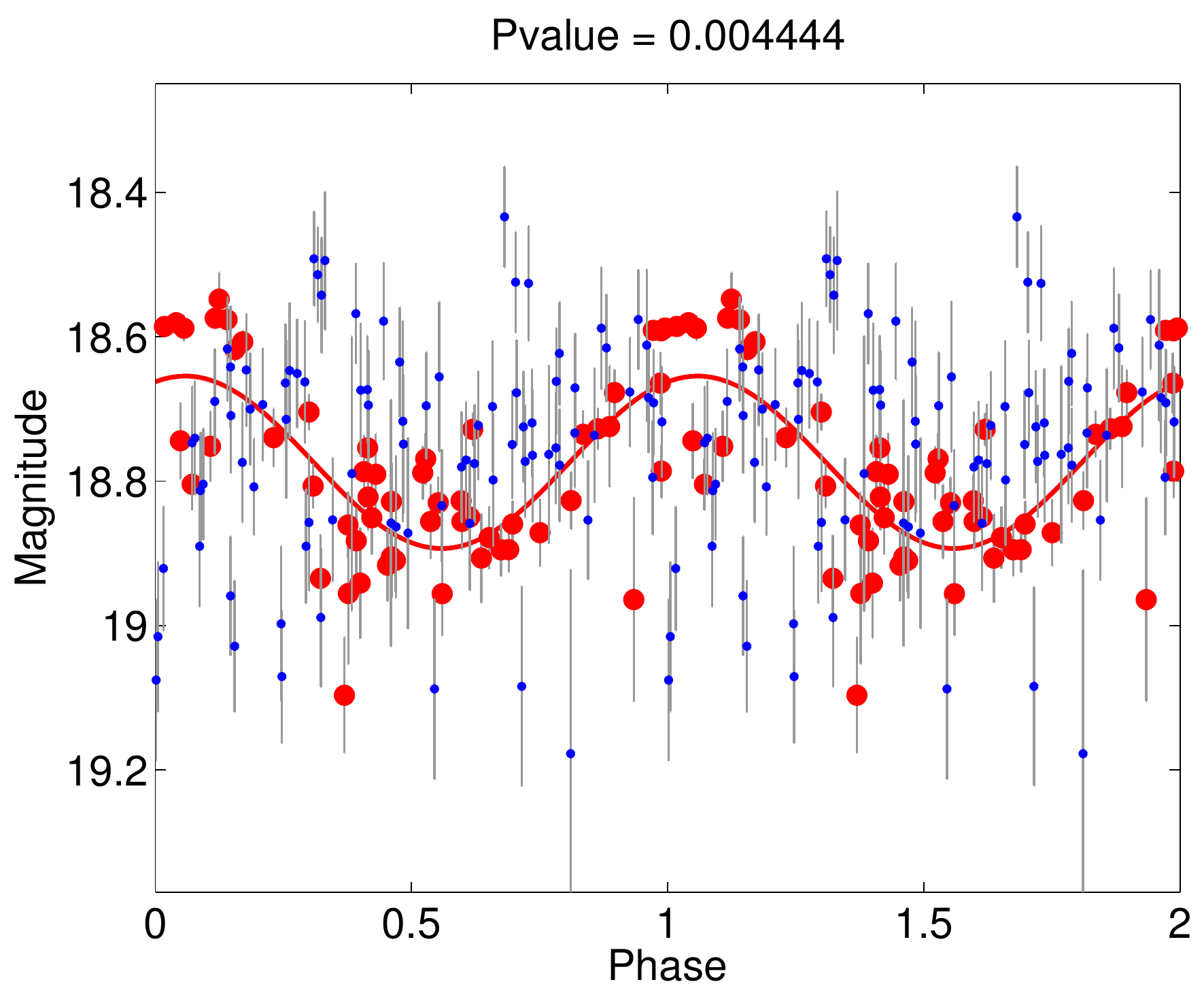}
\end{subfigure} \hspace{0.2cm}
\begin{subfigure}{.45\textwidth}
\centering
\includegraphics[width=8cm,height=4.5cm]{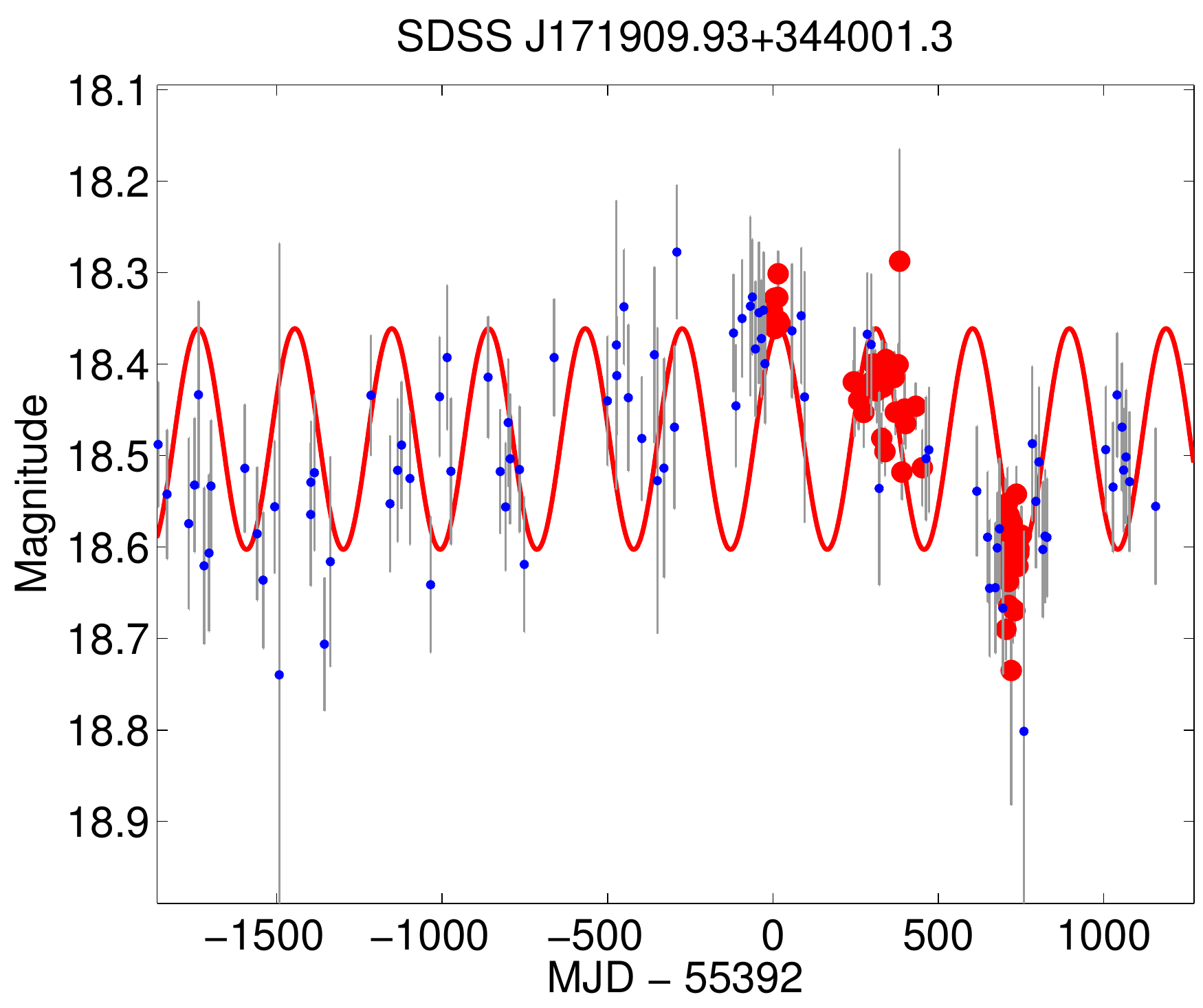}
\end{subfigure} \hspace{0.2cm}
\begin{subfigure}{.45\textwidth}
\centering
\includegraphics[width=8cm,height=4.5cm]{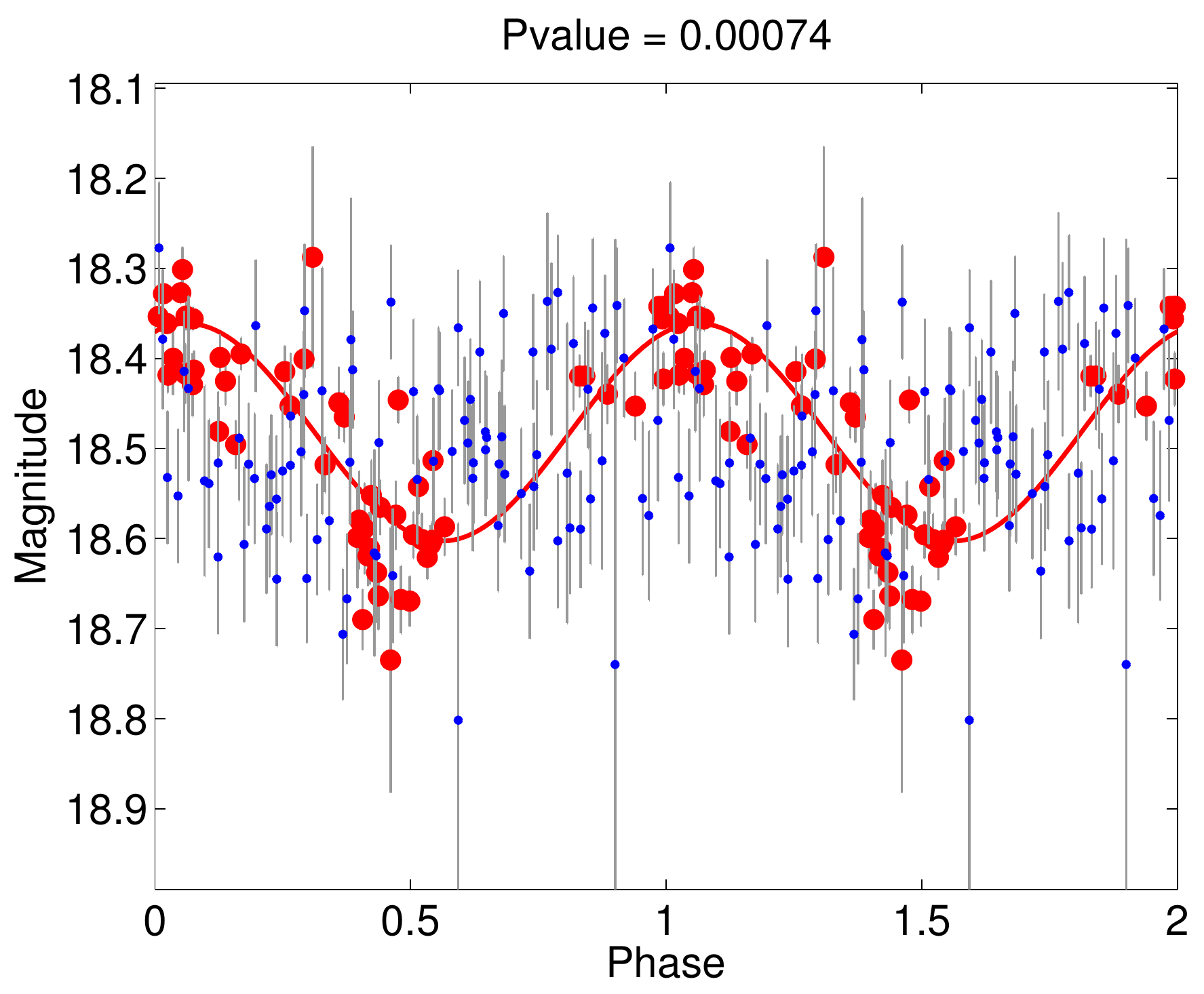}
\end{subfigure} \hspace{0.2cm}
\begin{subfigure}{.45\textwidth}
\centering
\includegraphics[width=8cm,height=4.5cm]{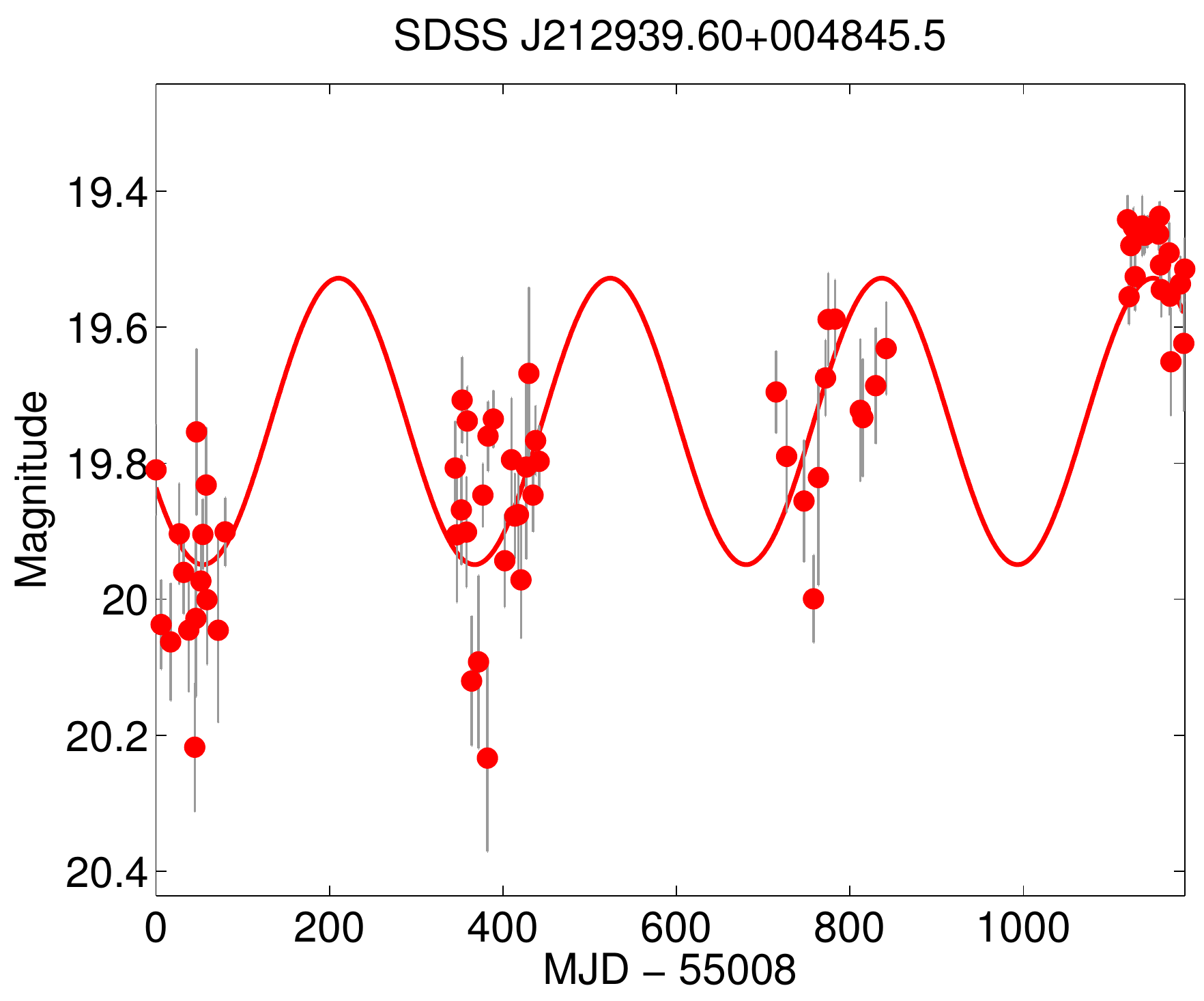}
\end{subfigure} \hspace{0.2cm}
\begin{subfigure}{.45\textwidth}
\centering
\includegraphics[width=8cm,height=4.5cm]{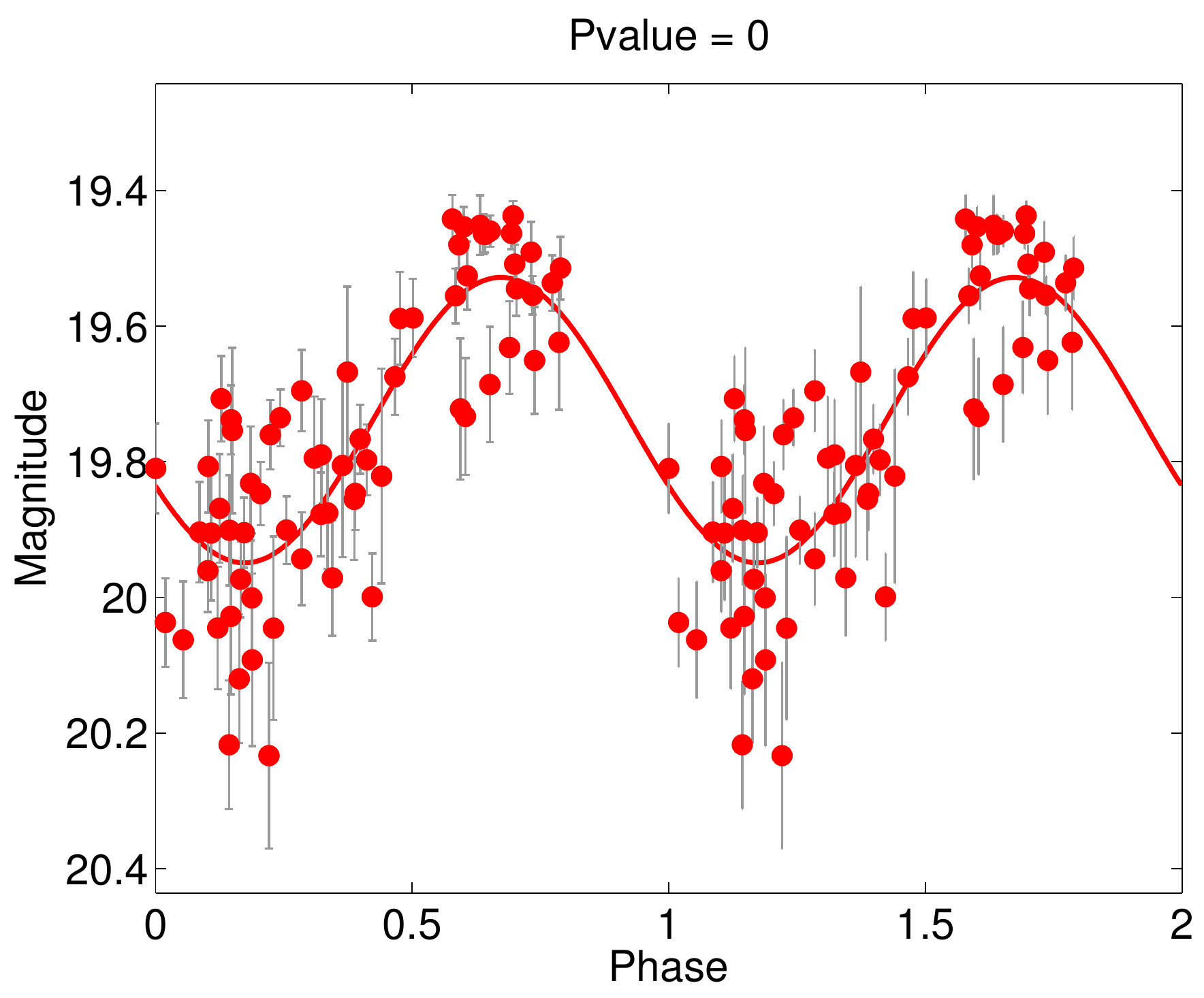}
\end{subfigure} \hspace{0.2cm}
\phantomcaption
\end{figure*}

\begin{figure*}
\ContinuedFloat
\begin{subfigure}{.45\textwidth}
\centering
\includegraphics[width=8cm,height=4.5cm]{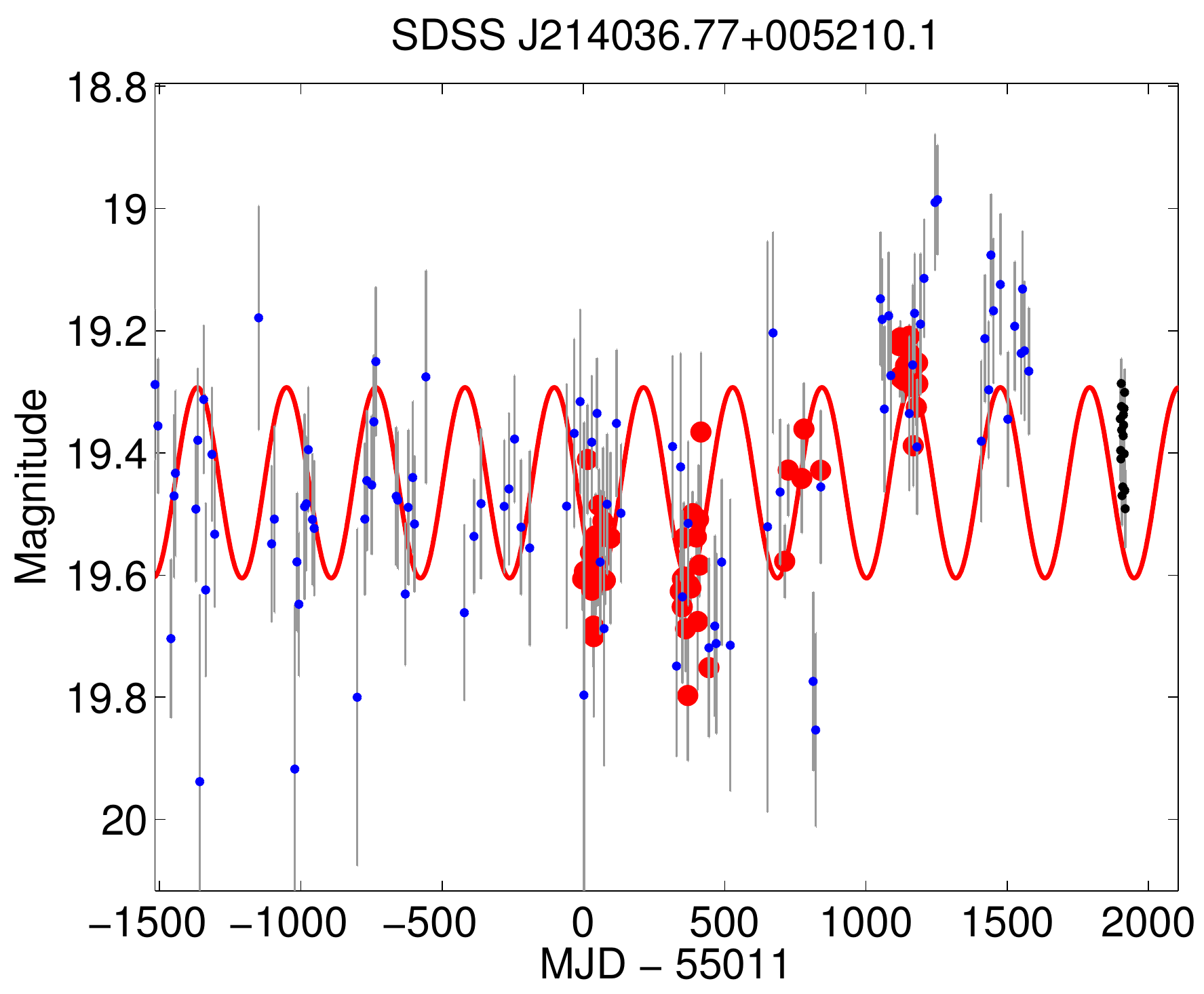}
\end{subfigure} \hspace{0.2cm}
\begin{subfigure}{.45\textwidth}
\centering
\includegraphics[width=8cm,height=4.5cm]{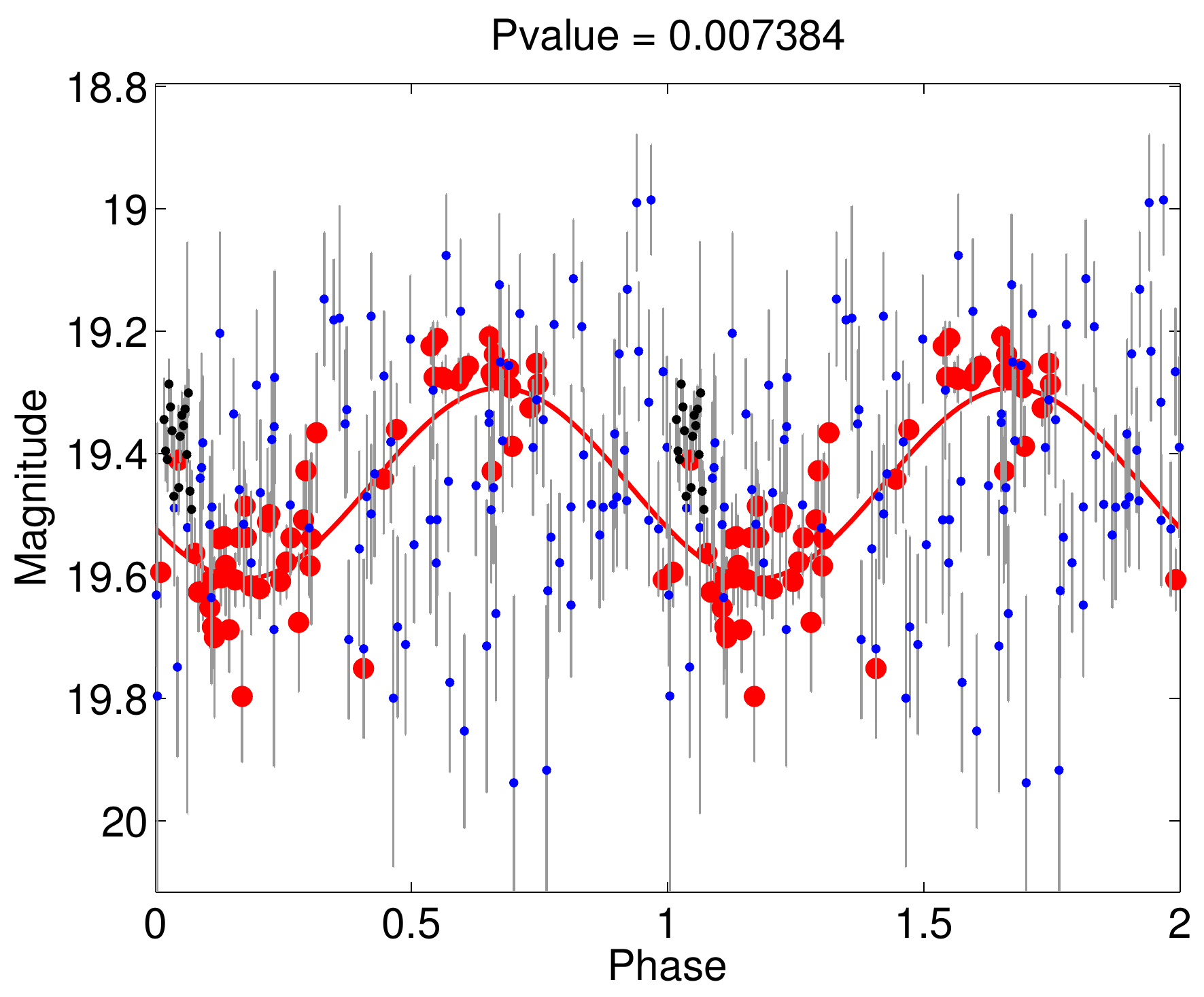}
\end{subfigure} \hspace{0.2cm}
\begin{subfigure}{.45\textwidth}
\centering
\includegraphics[width=8cm,height=4.5cm]{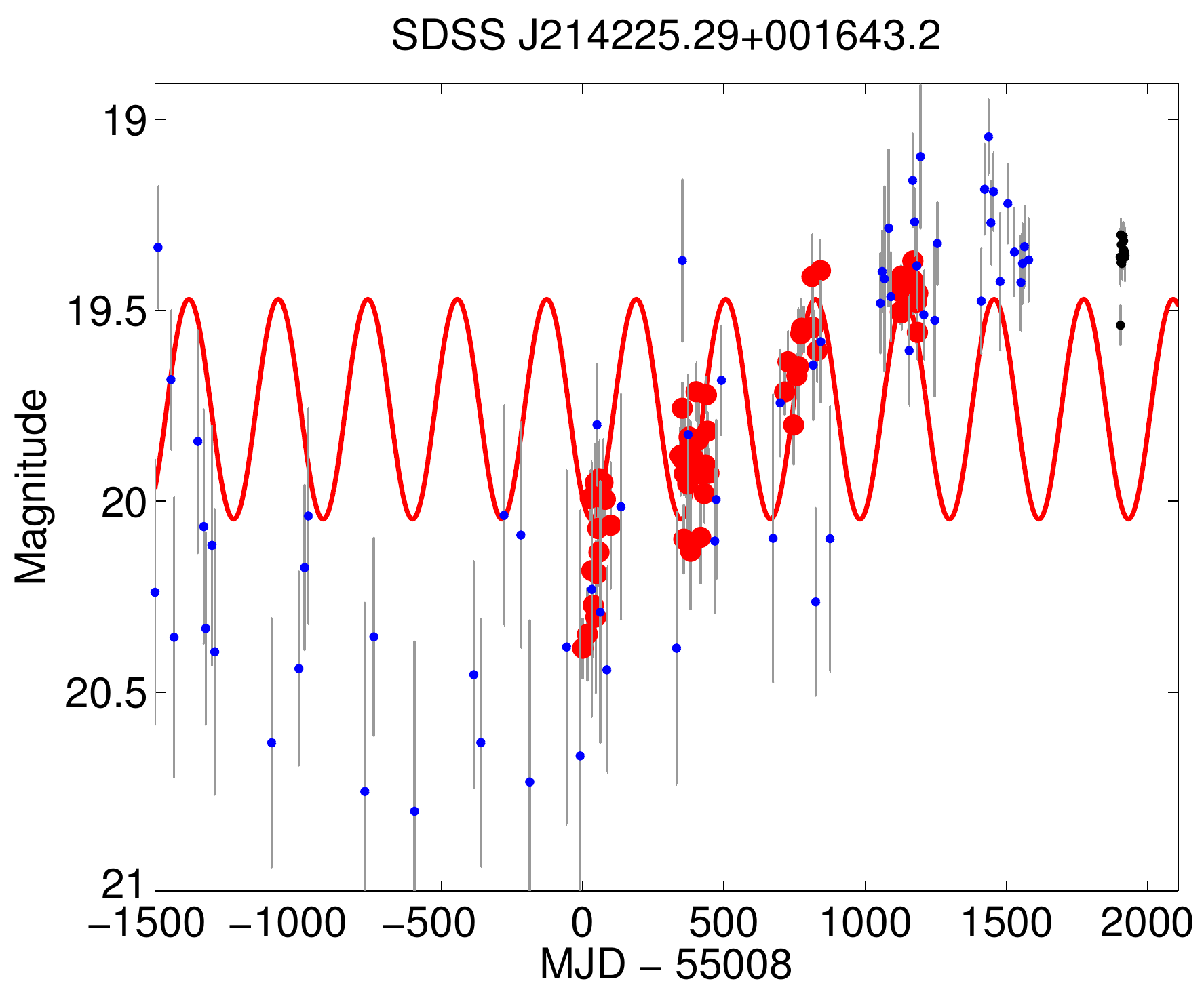}
\end{subfigure} \hspace{0.2cm}
\begin{subfigure}{.45\textwidth}
\centering
\includegraphics[width=8cm,height=4.5cm]{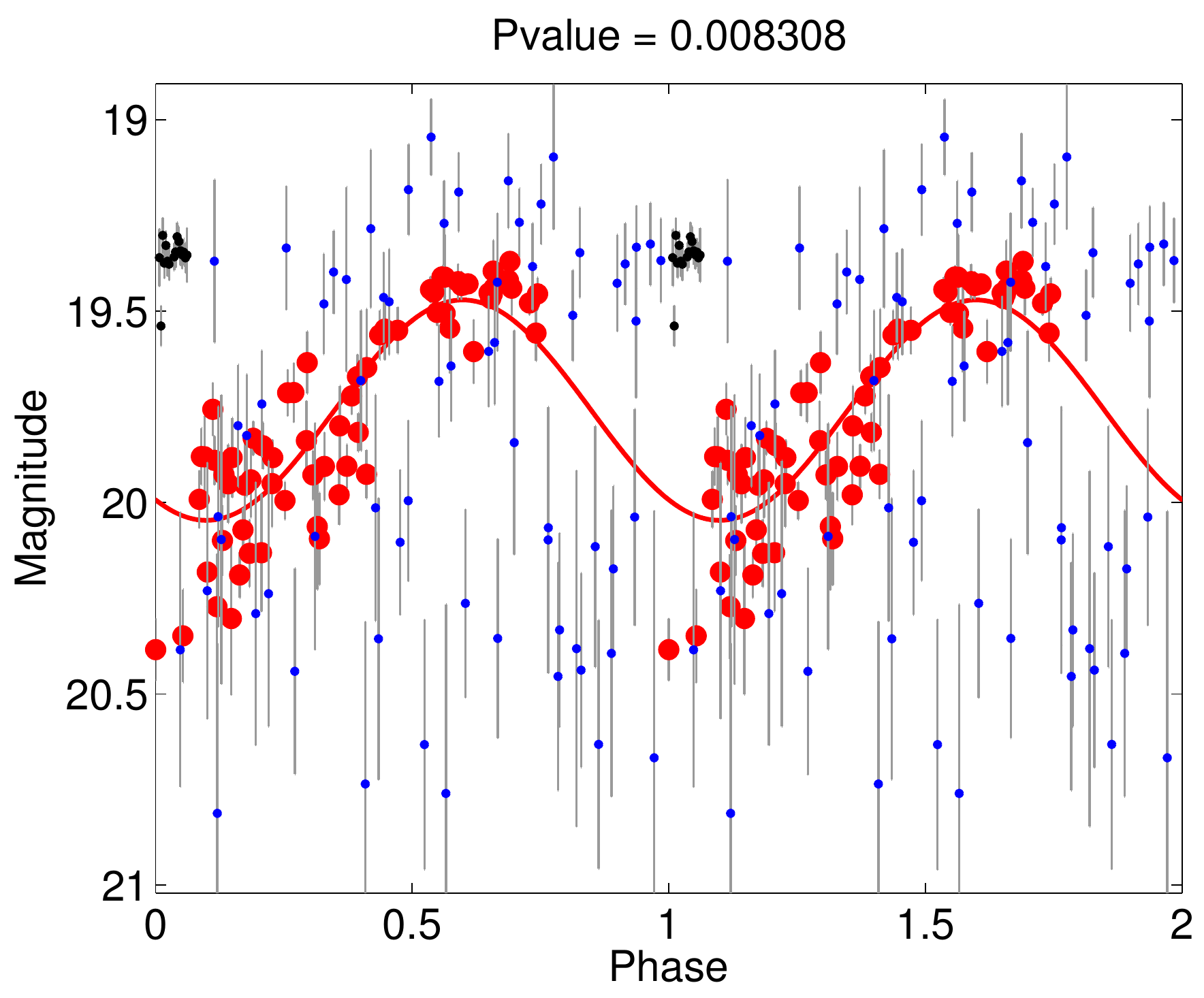}
\end{subfigure} \hspace{0.2cm}
\begin{subfigure}{.45\textwidth}
\centering
\includegraphics[width=8cm,height=4.5cm]{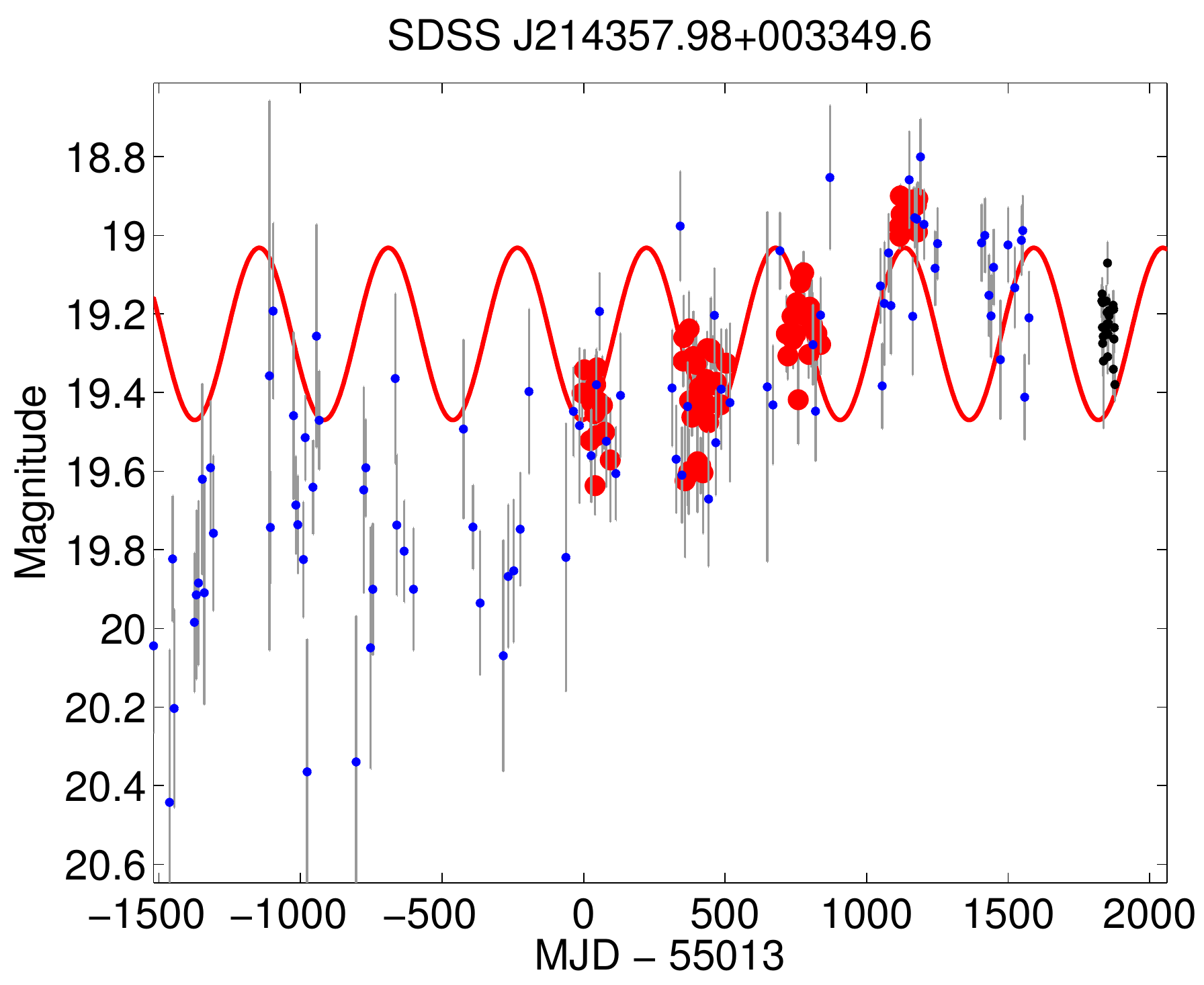}
\end{subfigure} \hspace{0.2cm}
\begin{subfigure}{.45\textwidth}
\centering
\includegraphics[width=8cm,height=4.5cm]{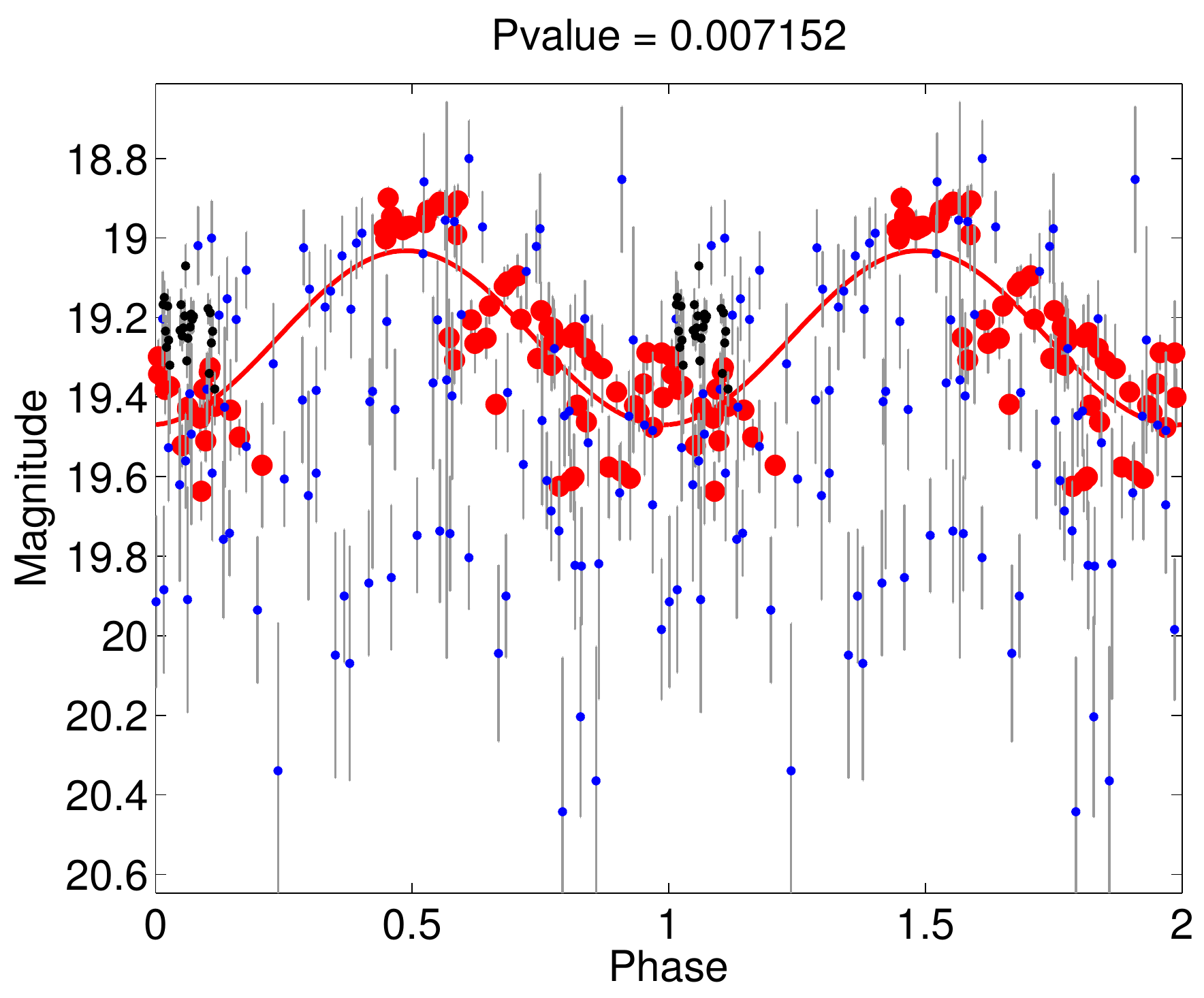}
\end{subfigure} \hspace{0.2cm}
\begin{subfigure}{.45\textwidth}
\centering
\includegraphics[width=8cm,height=4.5cm]{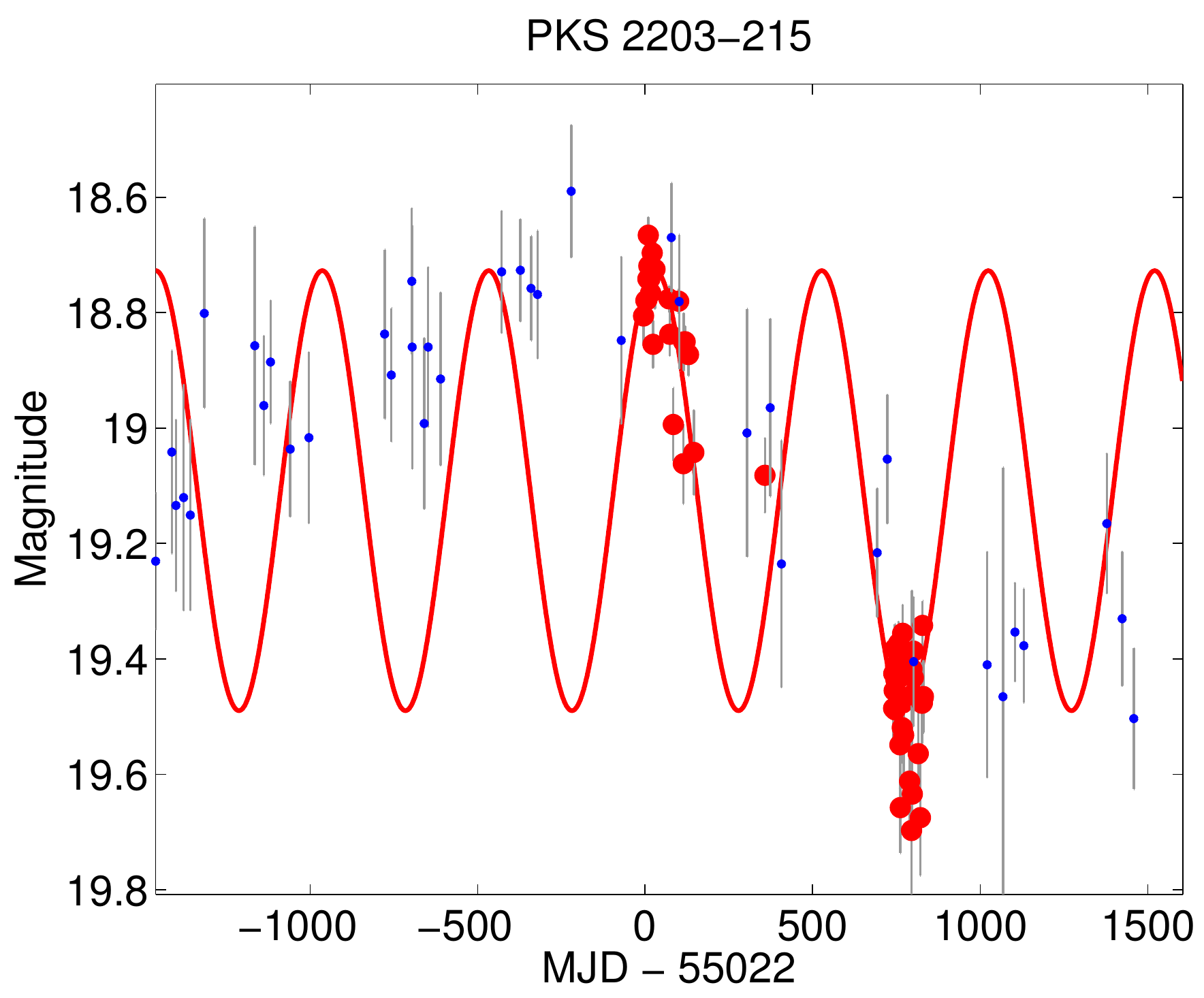}
\end{subfigure} \hspace{0.2cm}
\begin{subfigure}{.45\textwidth}
\centering
\includegraphics[width=8cm,height=4.5cm]{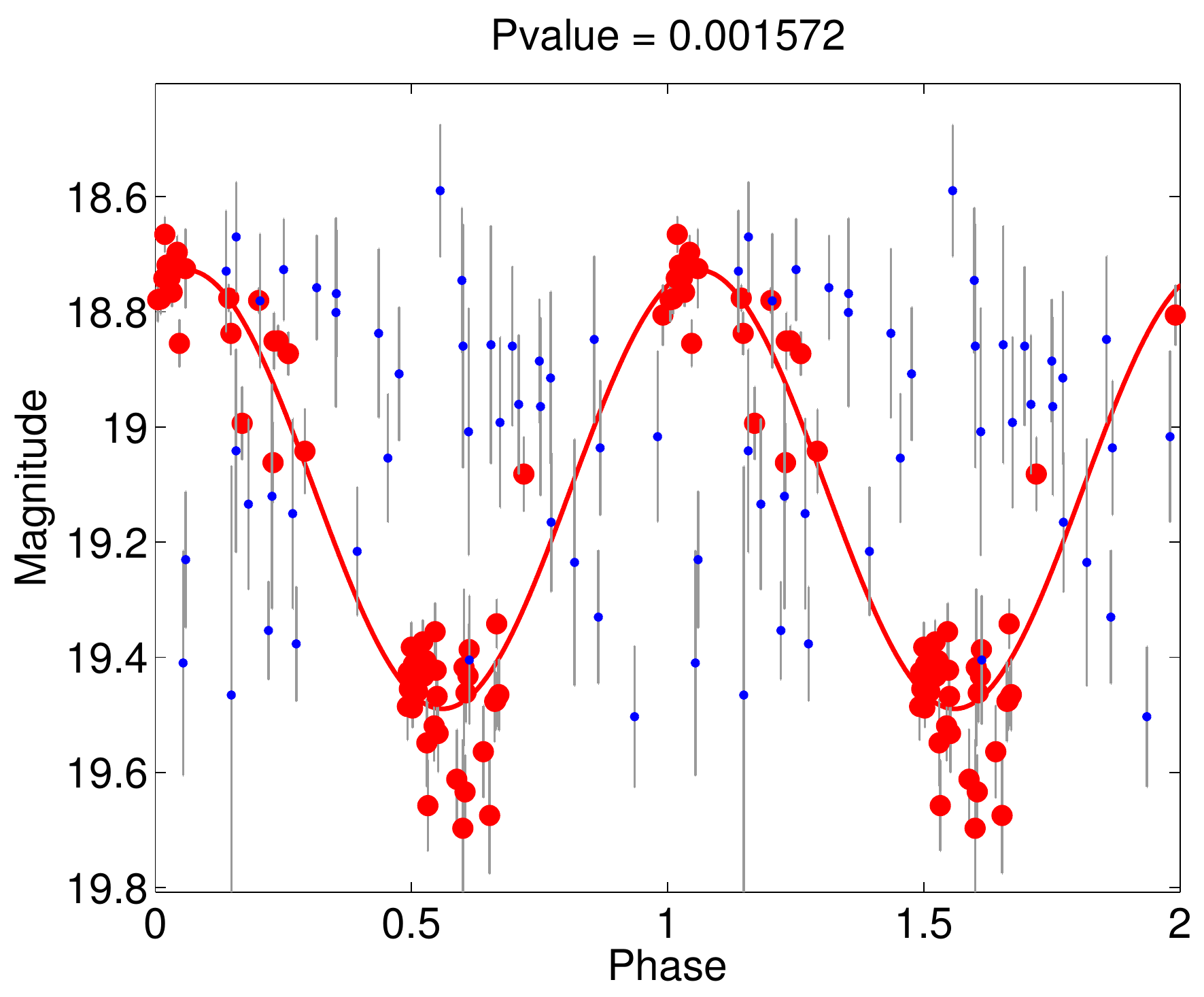}
\end{subfigure} \hspace{0.2cm}
\begin{subfigure}{.45\textwidth}
\centering
\includegraphics[width=8cm,height=4.5cm]{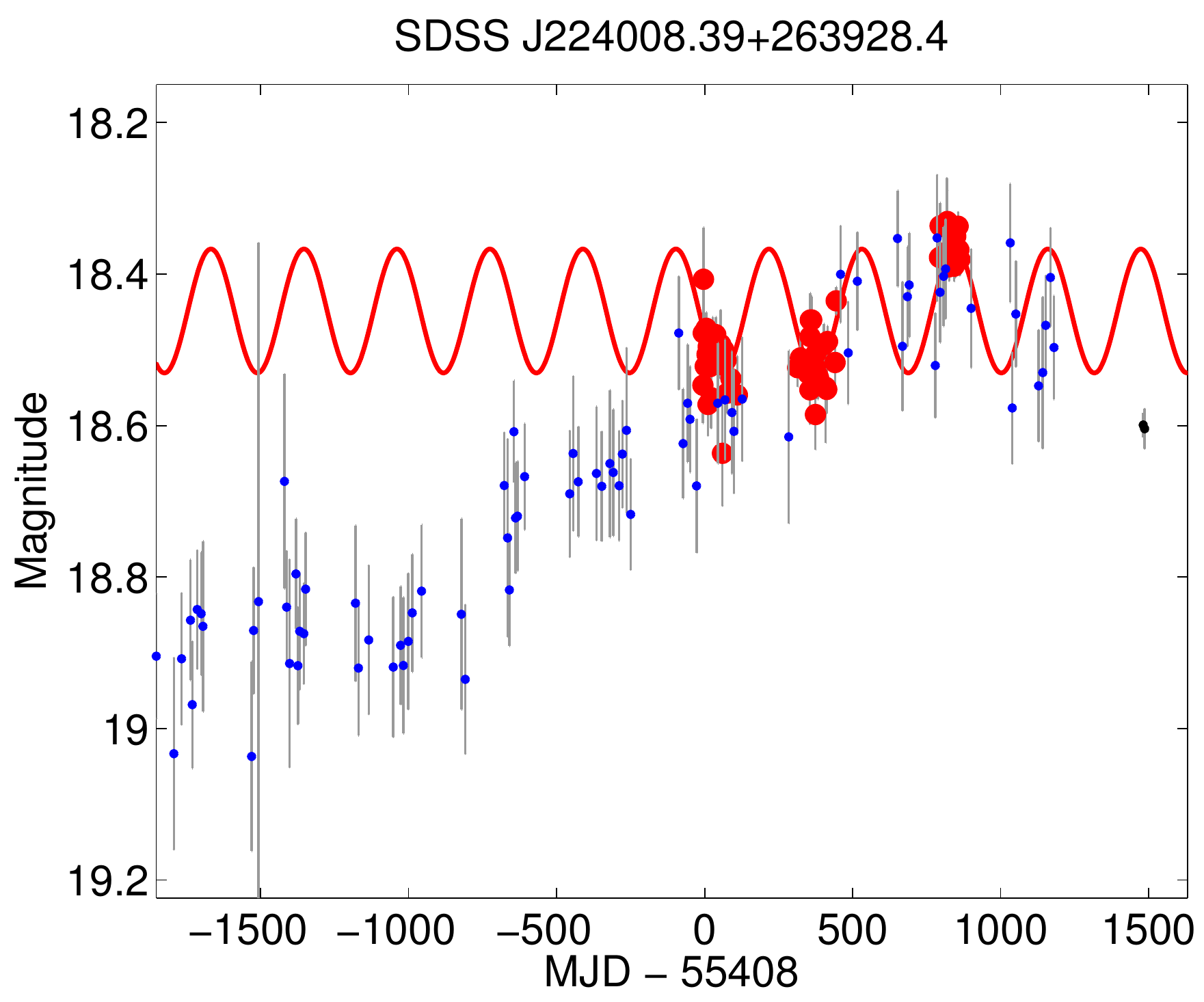}
\end{subfigure} \hspace{0.2cm}
\begin{subfigure}{.45\textwidth}
\centering
\includegraphics[width=8cm,height=4.5cm]{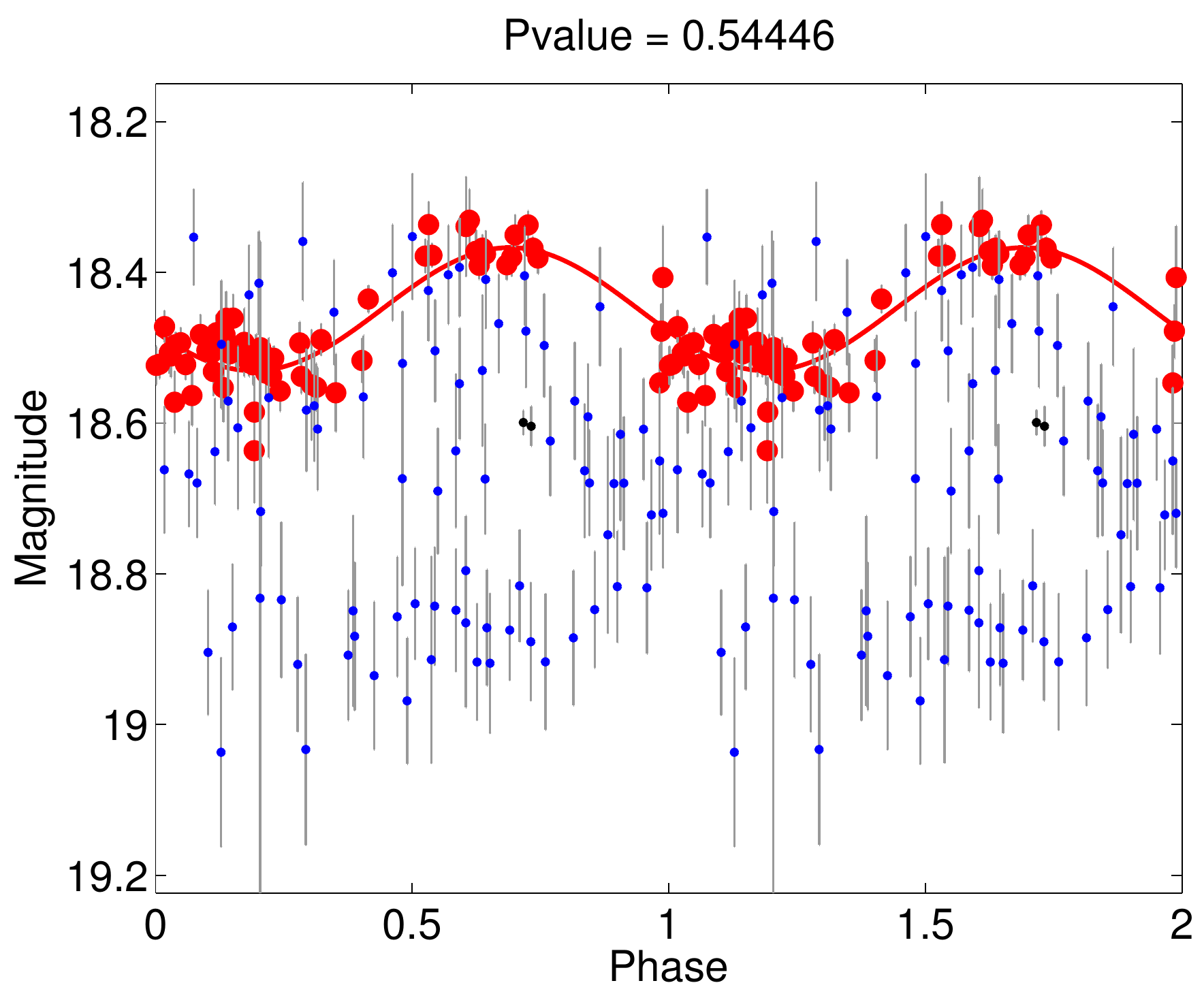}
\end{subfigure} \hspace{0.2cm}
\phantomcaption
\end{figure*}
\begin{figure*}
\ContinuedFloat
\begin{subfigure}{.45\textwidth}
\centering
\includegraphics[width=8cm,height=4.5cm]{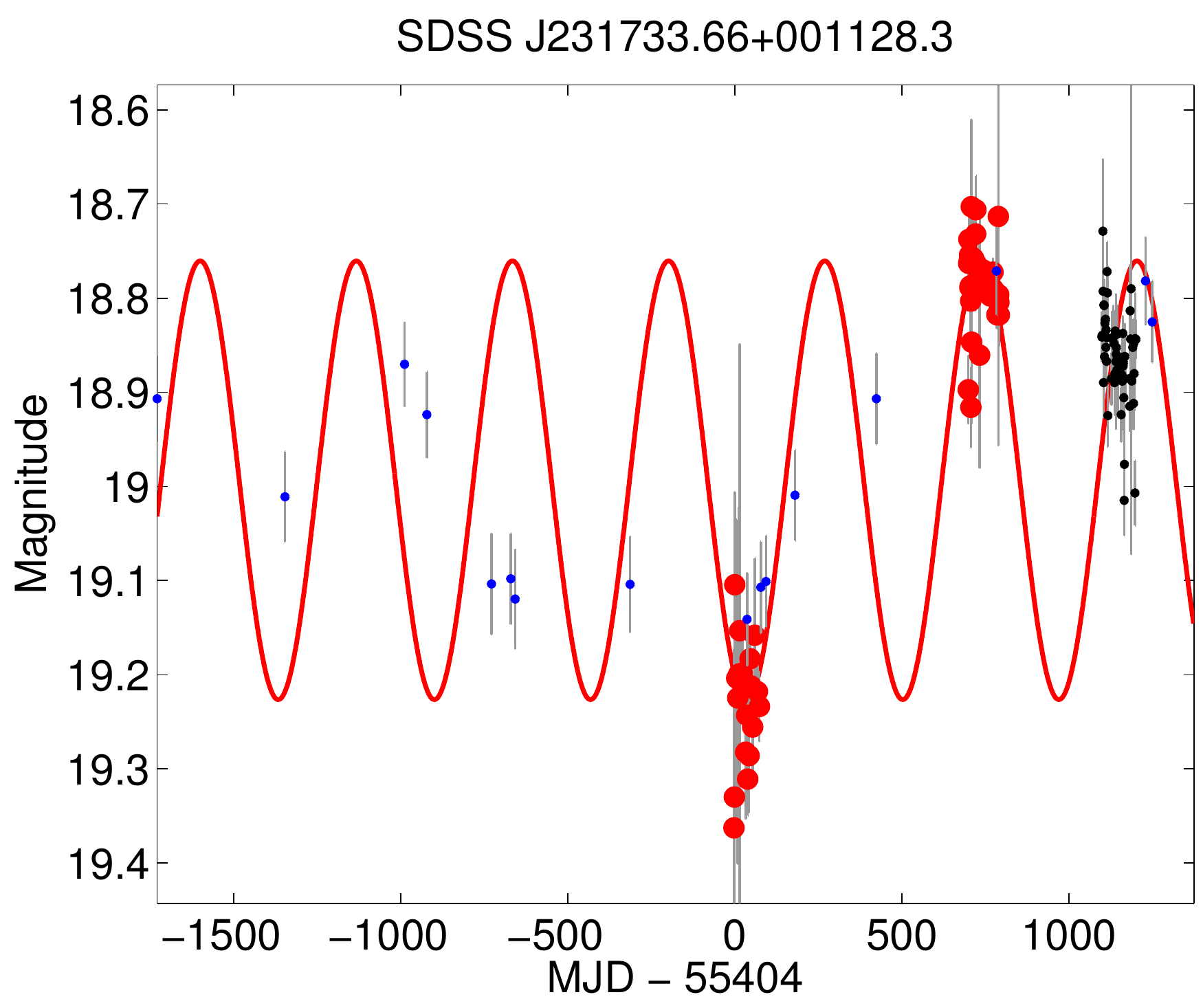}
\end{subfigure} \hspace{0.2cm}
\begin{subfigure}{.45\textwidth}
\centering
\includegraphics[width=8cm,height=4.5cm]{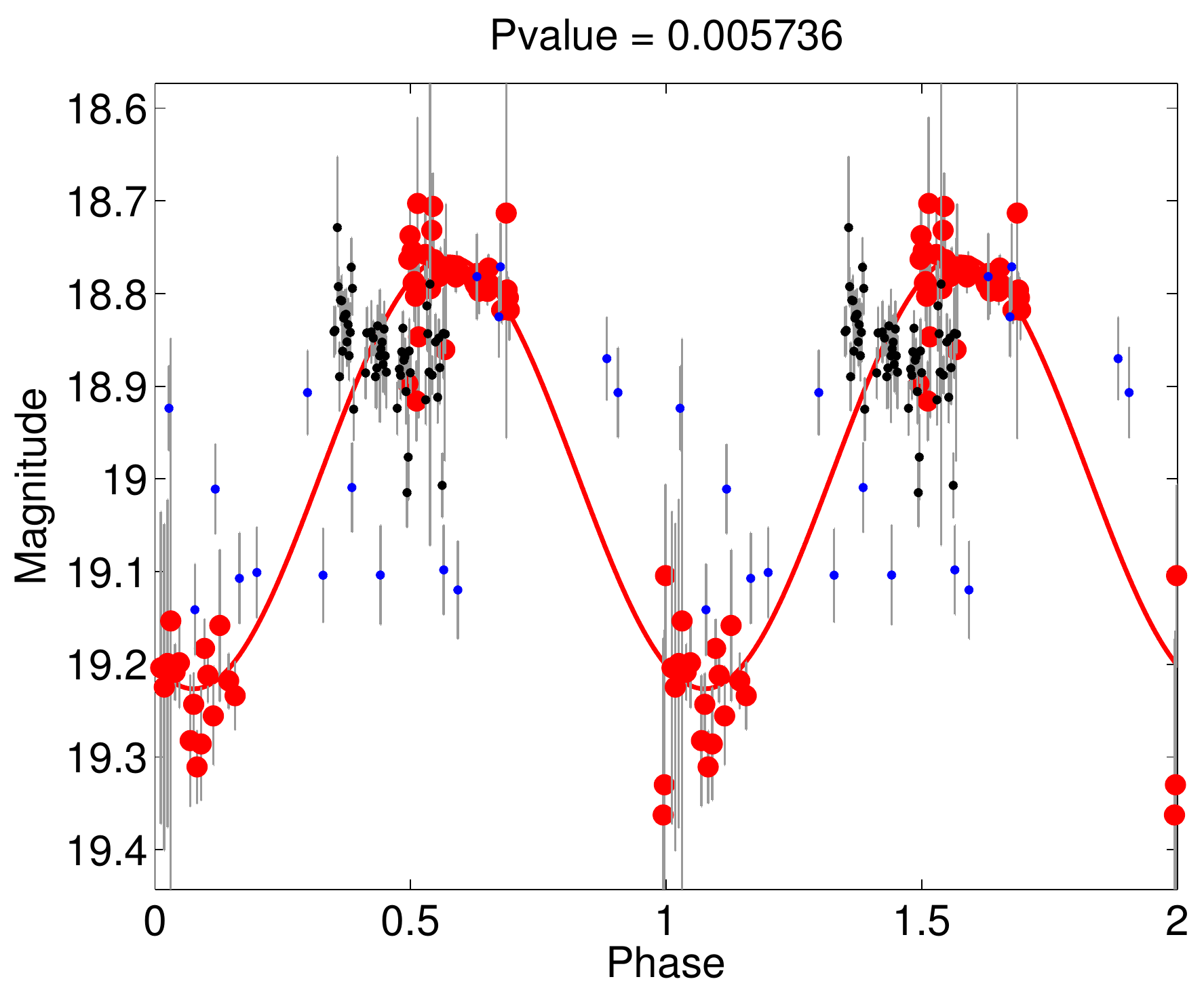}
\end{subfigure} \hspace{0.2cm}
\begin{subfigure}{.45\textwidth}
\centering
\includegraphics[width=8cm,height=4.5cm]{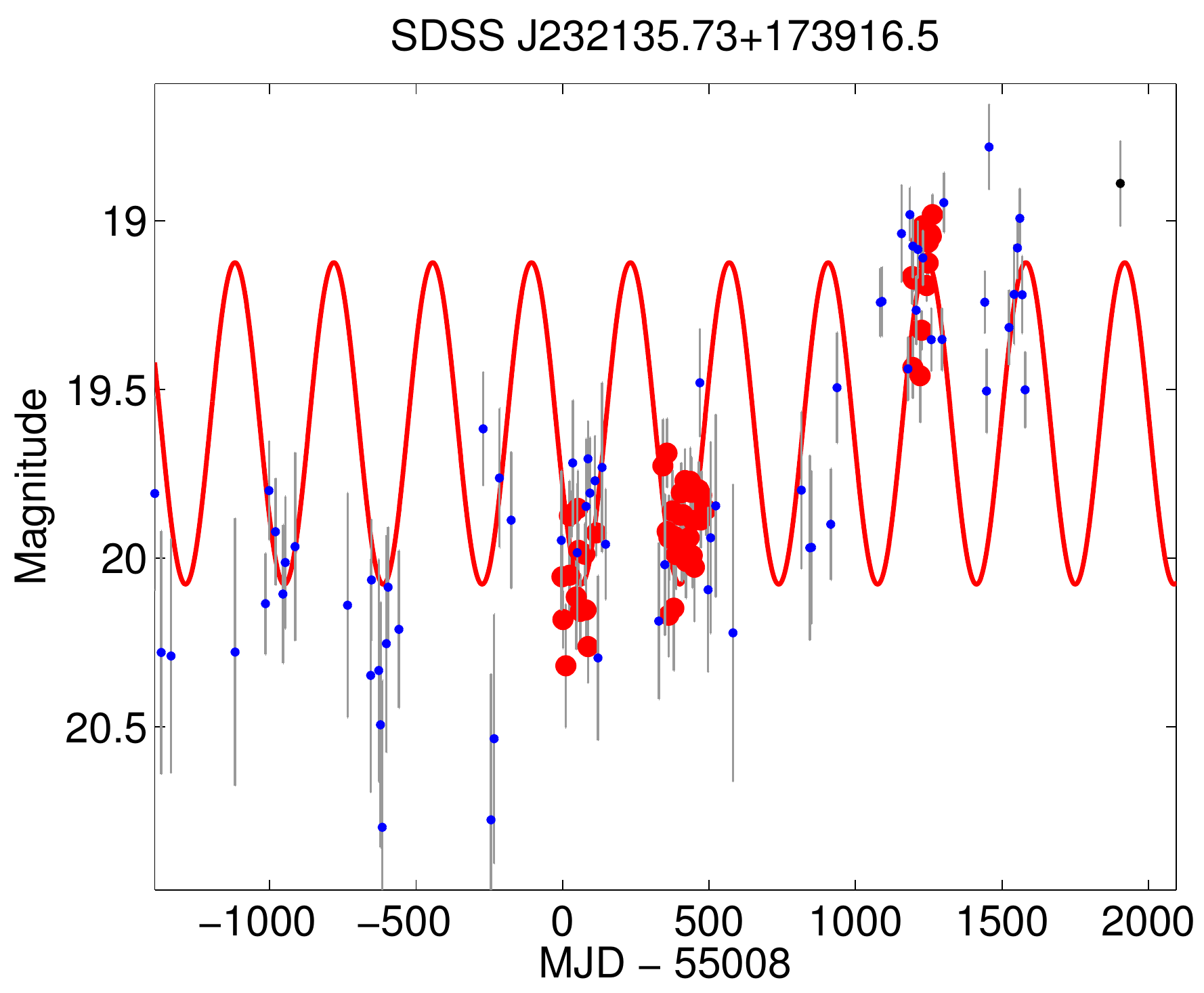}
\end{subfigure} \hspace{0.2cm}
\begin{subfigure}{.45\textwidth}
\centering
\includegraphics[width=8cm,height=4.5cm]{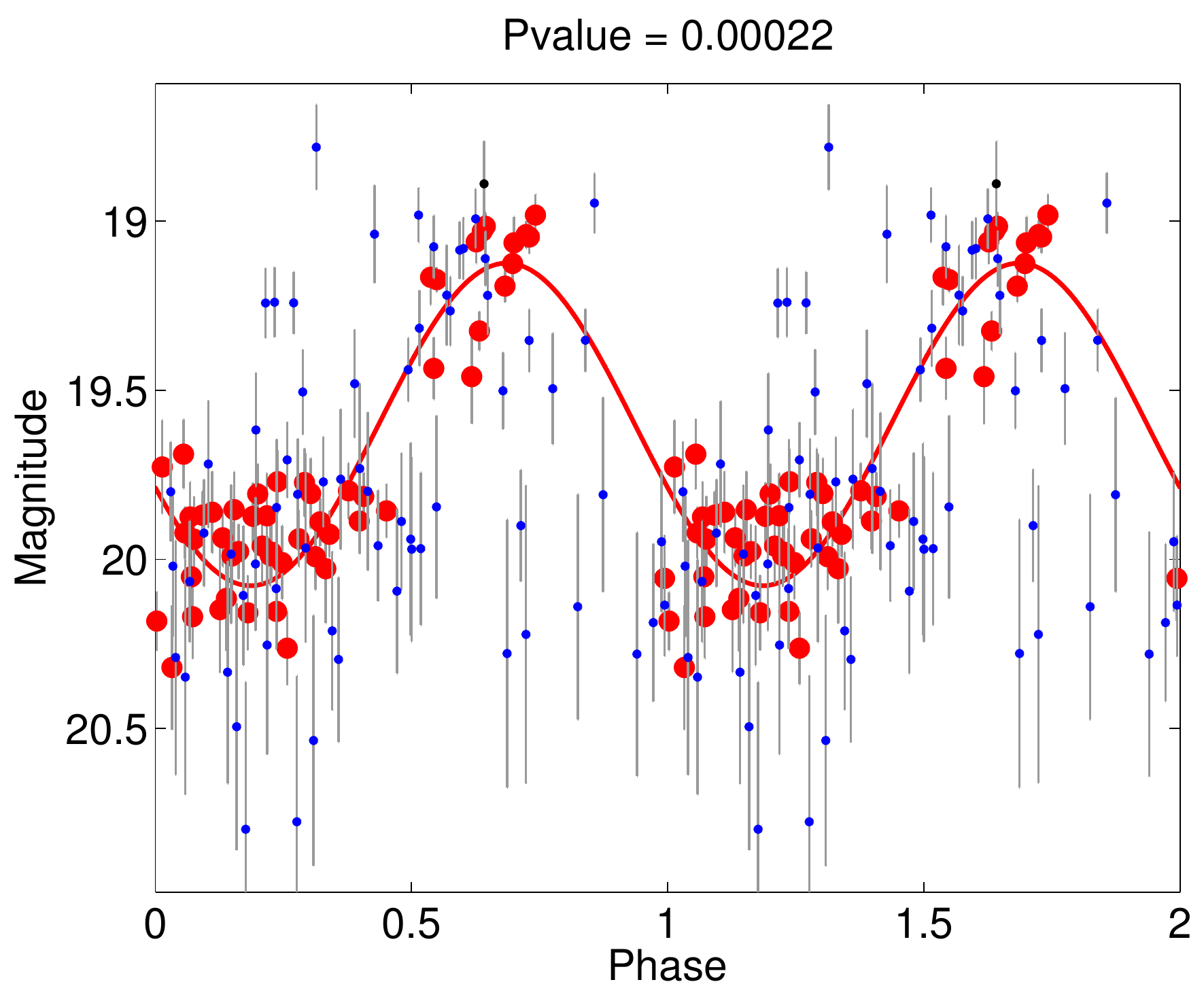}
\end{subfigure} \hspace{0.2cm}
\begin{subfigure}{.45\textwidth}
\centering
\includegraphics[width=8cm,height=4.5cm]{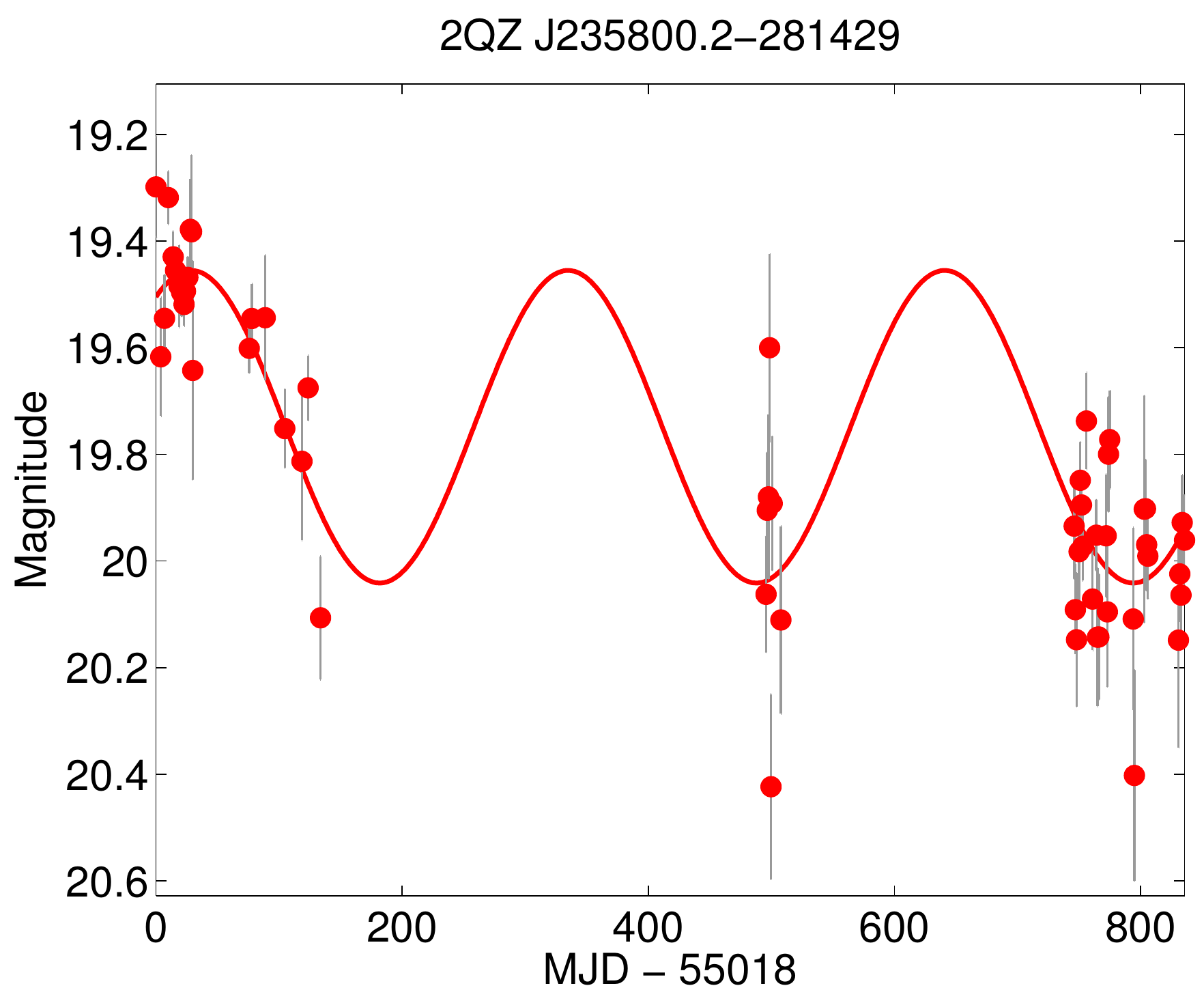}
\end{subfigure} \hspace{0.2cm}
\begin{subfigure}{.45\textwidth}
\centering
\includegraphics[width=8cm,height=4.5cm]{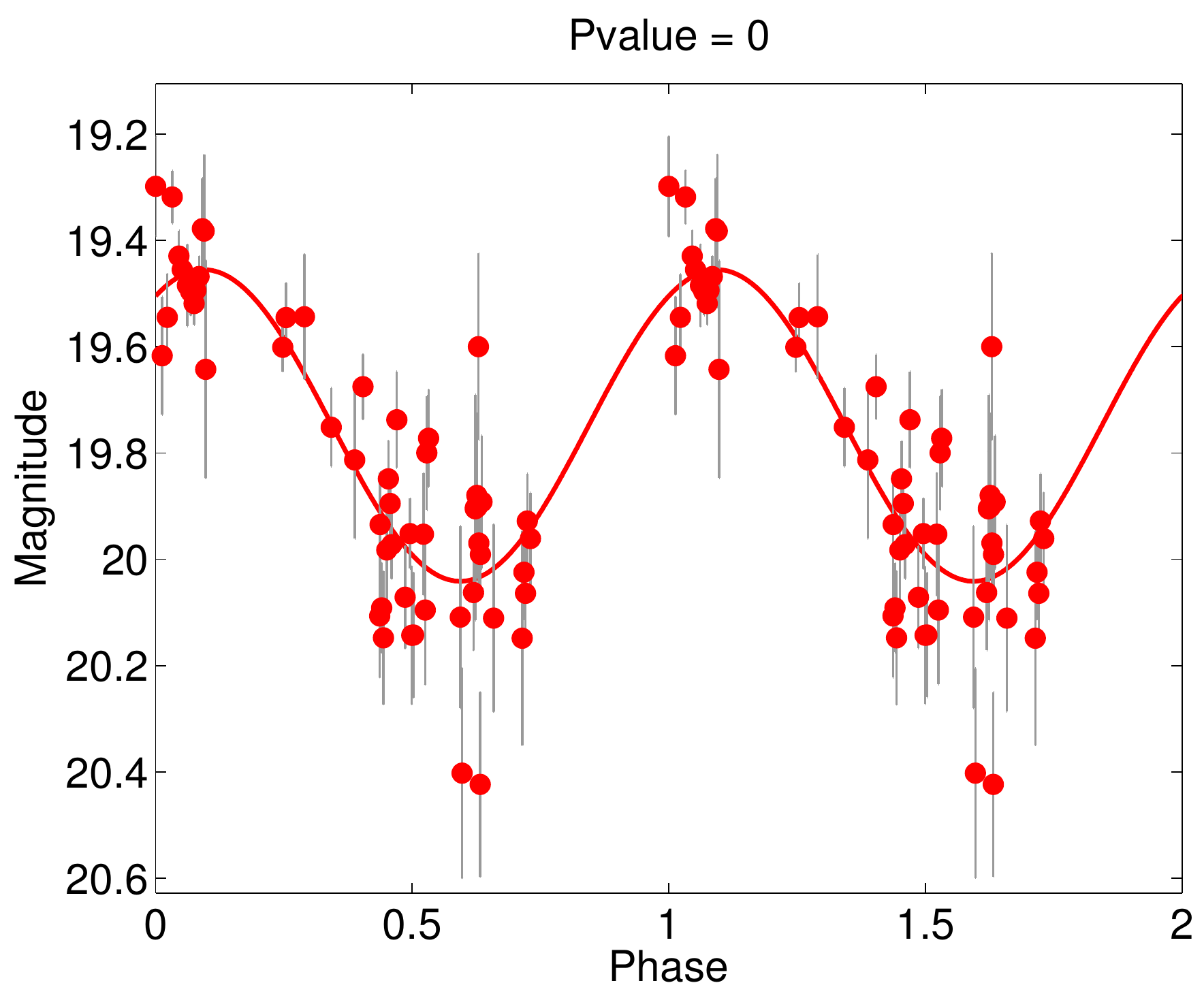}
\end{subfigure} \hspace{0.2cm}
\begin{subfigure}{.45\textwidth}
\centering
\includegraphics[width=8cm,height=4.5cm]{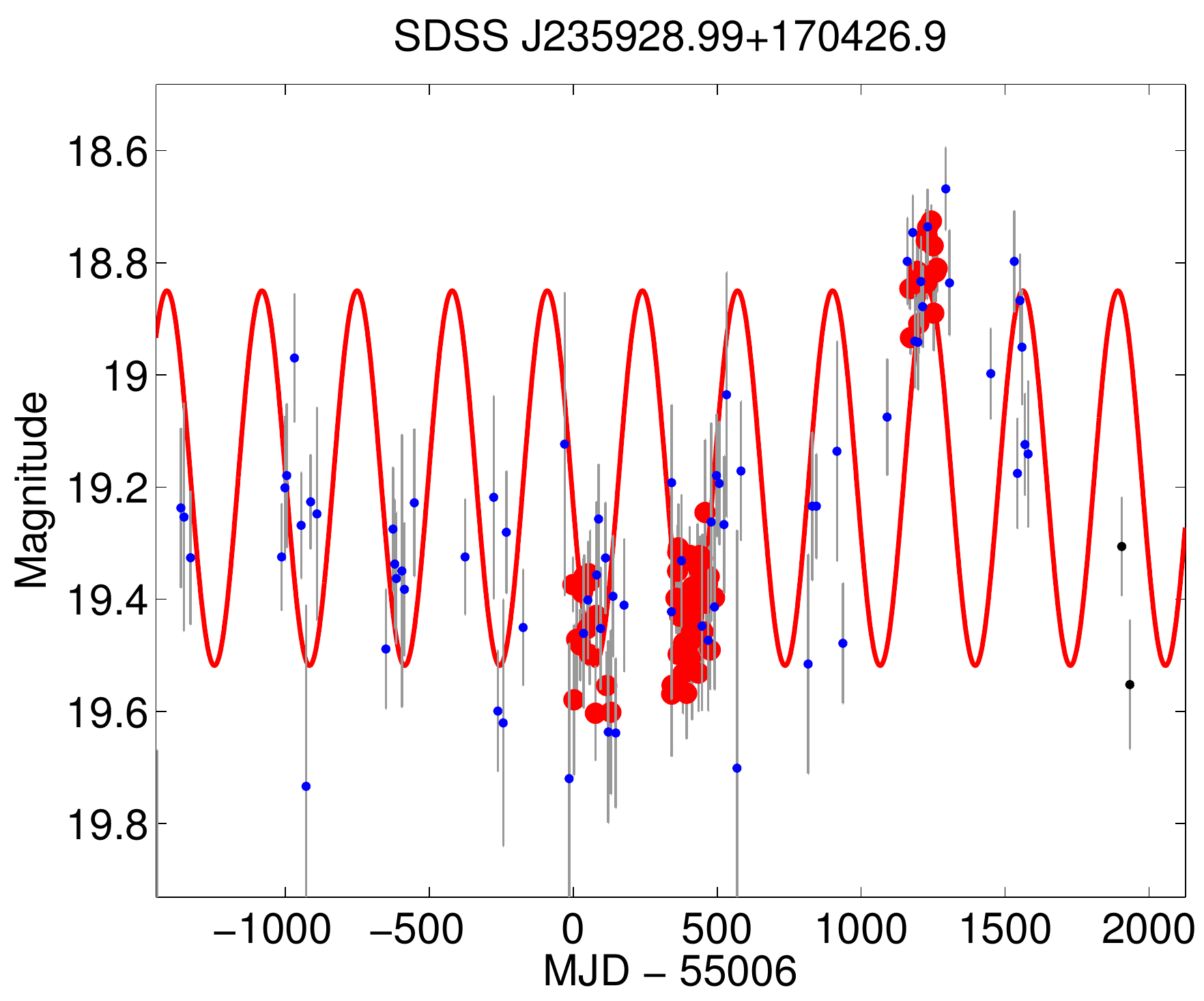}
\end{subfigure} \hspace{0.2cm}
\begin{subfigure}{.45\textwidth}
\centering
\includegraphics[width=8cm,height=4.5cm]{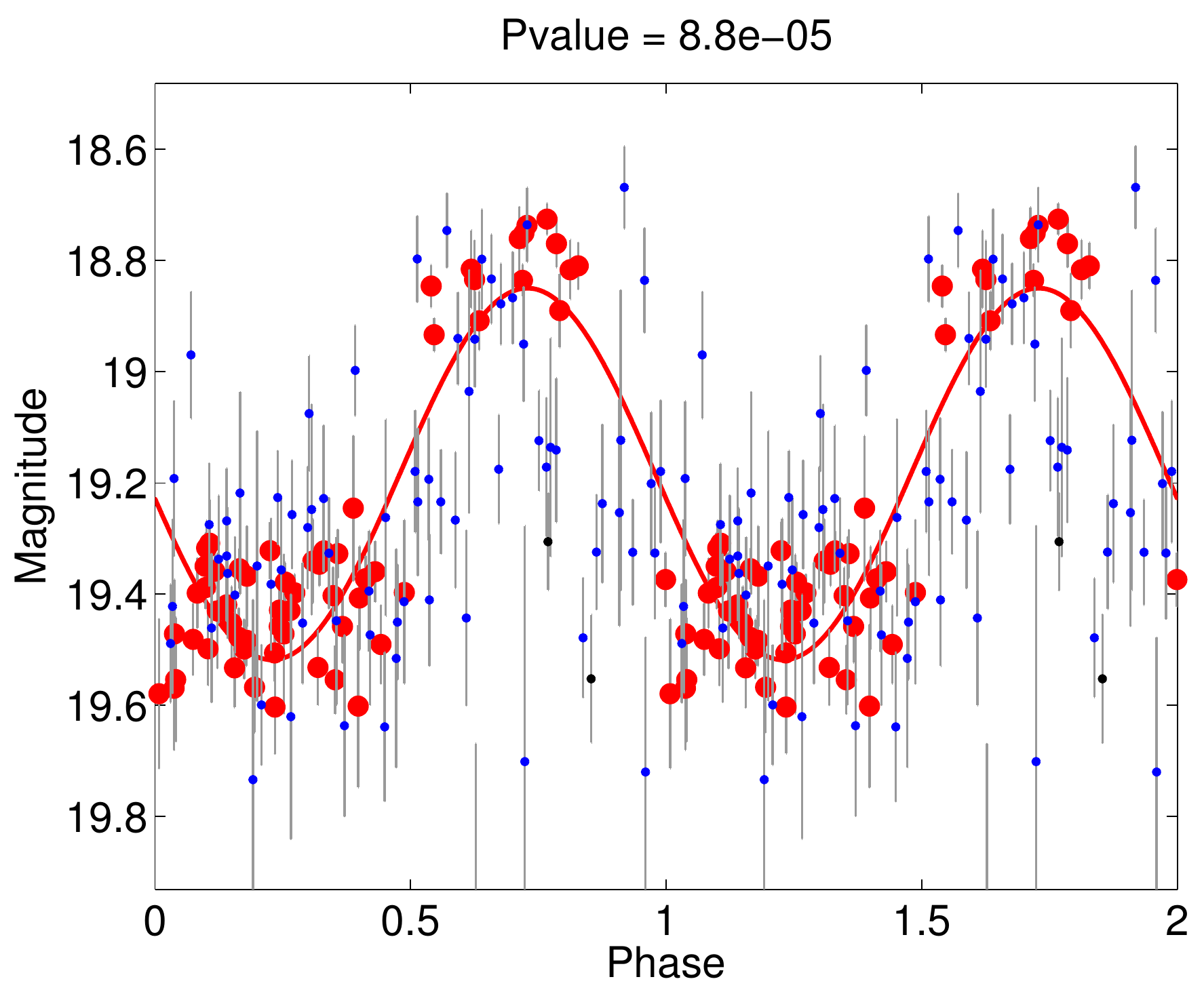}
\end{subfigure} \hspace{0.2cm}
\begin{subfigure}{.45\textwidth}
\centering
\includegraphics[width=8cm,height=4.5cm]{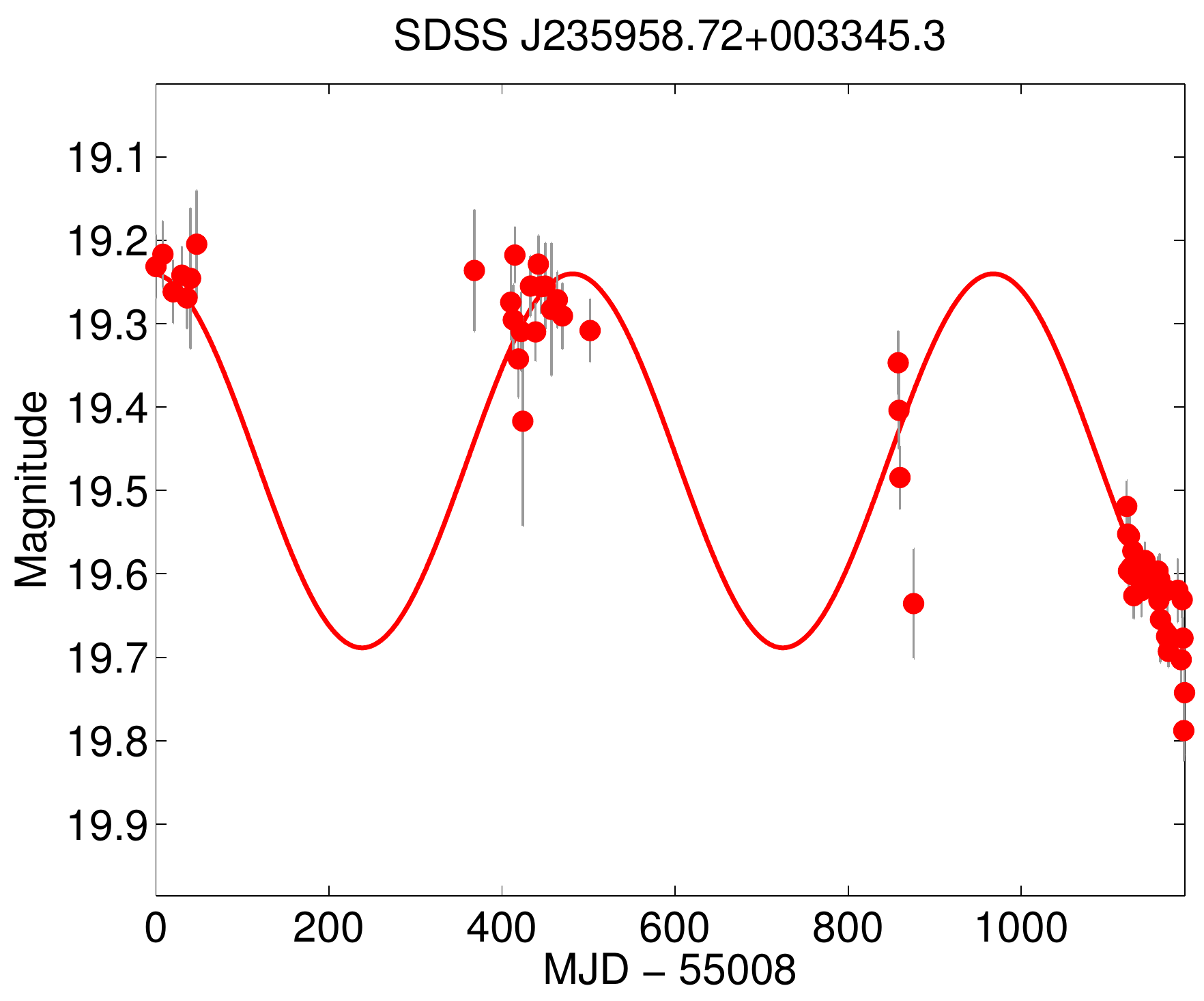}
\end{subfigure} \hspace{0.2cm}
\begin{subfigure}{.45\textwidth}
\centering
\includegraphics[width=8cm,height=4.5cm]{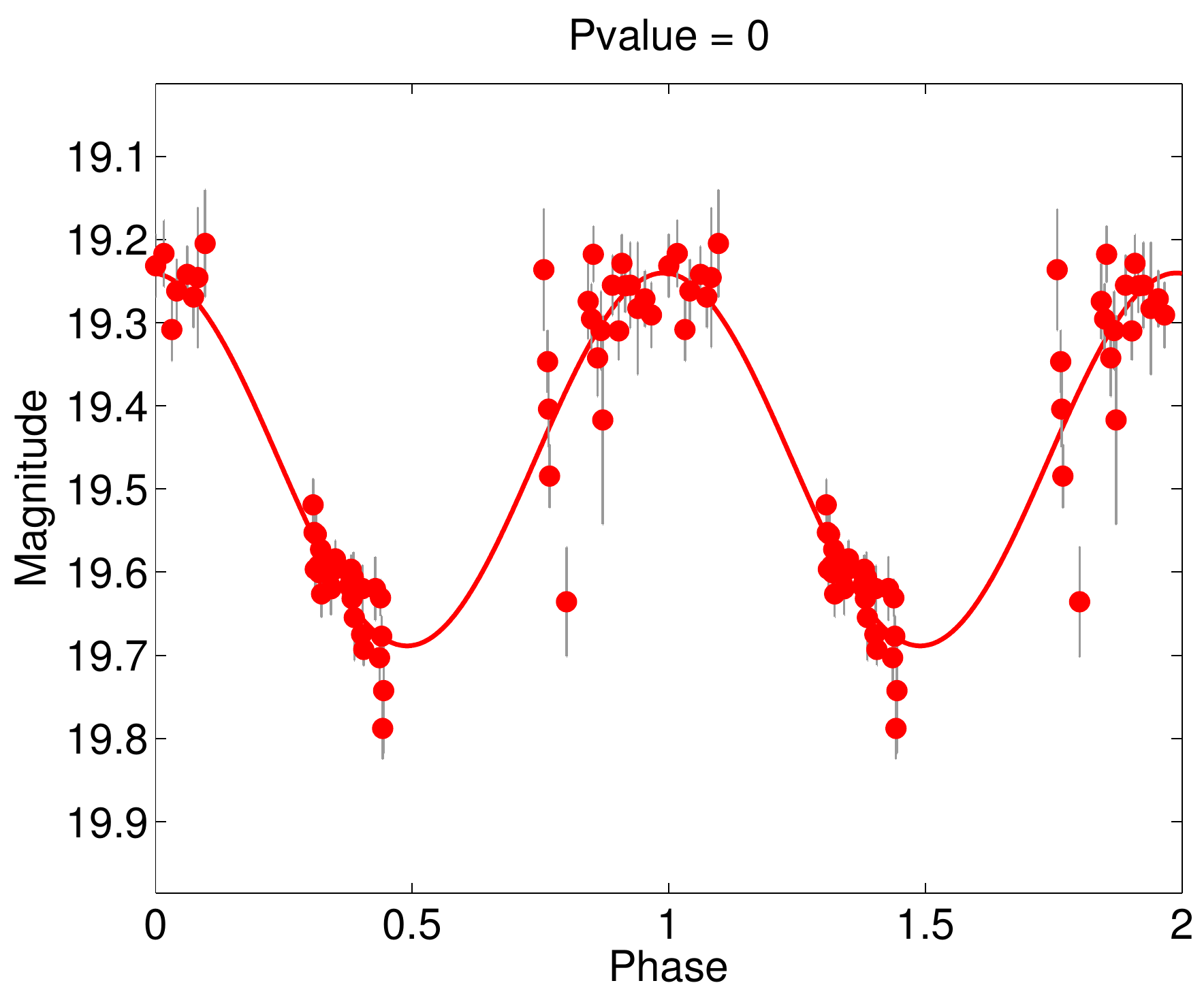}
\end{subfigure} \hspace{0.2cm}

\caption{Light curves of the 50 quasars, in which significant periodicity was identified (red PTF observations, black iPTF observations and blue CRTS observations). The red lines show the best fit sinusoid. The figures on the right show the phase folded light curves. The P-values calculated from the analysis of the composite light curves is also shown.}
\label{Fig:LightCurves}
\end{figure*}

\end{document}